\documentclass[12pt]{article}
\usepackage{amssymb}
\usepackage{amsmath}
\usepackage{amsfonts}
\usepackage{graphicx}
\usepackage{natbib}
\usepackage[nodisplayskipstretch]{setspace}
\usepackage{lscape}
\usepackage{geometry}
\usepackage{rotating}
\usepackage{subfigure}
\usepackage{url}
\usepackage{caption}
\usepackage{version}
\usepackage{titletoc}
\usepackage{afterpage}
\usepackage{placeins}
\usepackage[bottom]{footmisc}
\usepackage{titlesec}%
\providecommand{\U}[1]{\protect\rule{.1in}{.1in}}
\pdfoutput=1

\makeatletter
\g@addto@macro\normalsize{
\setlength\abovedisplayskip{7pt}
\setlength\belowdisplayskip{7pt}
\setlength\abovedisplayshortskip{7pt}
\setlength\belowdisplayshortskip{7pt}
}
\makeatother
\begin{document}

\title{Matching Theory and Evidence on Covid-19 using a Stochastic Network SIR
Model\thanks{We acknowledge helpful comments and suggestions from Thierry
Magnac (the co-editor), three anonymous referees, Alexander Chudik, Cheng
Hsiao, Alessandro Rebucci, Ron Smith, Anastasia Semykina, and participants at
the IAAE 2020 webinar series, the IAAE 2021 Annual Conference, and seminars at
the University of Southern California, Johns Hopkins University, Michigan
State University, University of California, Riverside, and Purdue University.
We would also like to thank Mahrad Sharifvaghefi for helping us compile the
data and Minsoo Cho for helping with the literature review. Correspondence to:
C.F. Yang, Department of Economics, Florida State University, 281 Bellamy
Building, Tallahassee, FL 32306, USA. E-mail: cynthia.yang@fsu.edu.}%
\textbf{{\Large {}}}}
\author{M. Hashem Pesaran\\{\small {}{}University of Southern California, USA, and Trinity College,
Cambridge, UK}\\Cynthia Fan Yang\\{\small {}{}{}{}Florida State University}%
\vspace*{-0.2cm}%
}
\date{December 18, 2021}
\maketitle

\begin{abstract}
This paper develops an individual-based stochastic network SIR model for the
empirical analysis of the Covid-19 pandemic. It derives moment conditions for
the number of infected and active cases for single as well as multigroup
epidemic models. These moment conditions are used to investigate the
identification and estimation of the transmission rates. The paper then
proposes a method that jointly estimates the transmission rate and the
magnitude of under-reporting of infected cases. Empirical evidence on six
European countries matches the simulated outcomes once the under-reporting of
infected cases is addressed. It is estimated that the number of actual cases
could be between 4 to 10 times higher than the reported numbers in October
2020 and declined to 2 to 3 times in April 2021. The calibrated models are
used in the counterfactual analyses of the impact of social distancing and
vaccination on the epidemic evolution, and the timing of early interventions
in the UK and Germany.

%

\vspace{0.4cm}%



\end{abstract}

\pagenumbering{gobble}

\baselineskip0.245in

\newpage\pagenumbering{arabic} \setcounter{page}{1} \doublespacing

\section{Introduction}

Since the outbreak of Covid-19, many researchers in epidemiology, behavioral
sciences, and economics have applied various forms of compartmental models to
study the disease transmission and potential outcomes under different
intervention policies. The compartmental models are a major group of
epidemiological models that categorize a population into several types or
groups, such as susceptible (S), infected (I), and removed (recovered or
deceased, R). Compartmental models owe their origin to the well-known SIR
model pioneered by Kermack and McKendrick (1927) and have been developed in a
number of important directions, allowing for multi-category (multi-location),
under a variety of contact networks and transmission channels.\footnote{See
Section \ref{Sup: literature} of the online supplement for a review of the
related literature. Comprehensive reviews can be found in Hethcote (2000) and
Thieme (2013).}

In this paper, we develop a new stochastic network SIR model in which
individual-specific infection and recovery processes are modelled, allowing
for group heterogeneity and latent individual characteristics that distinguish
individuals in terms of their degrees of resilience to becoming infected. The
model is then used to derive individual-specific conditional probabilities of
infection and recovery. In this respect, our modelling approach is to be
distinguished from the individual-based models in epidemiology that specify
the transition probabilities of individuals from one state to
another.\footnote{See, for example, Rocha and Masuda (2016), Willem et al.
(2017), and Nepomuceno, Resende, and Lacerda (2018).} In modelling the
infection process, we consider an individual's contact pattern with others in
the network, plus an individual-specific latent factor assumed to be
exponentially distributed. The time from infection to recovery (or death) is
assumed to be geometrically distributed. The individual processes are shown to
aggregate up to the familiar multigroup SIR model. We allow for group
heterogeneity and, in line with the literature, assume contact probabilities
are homogeneous within groups but could differ across groups.

We then derive the probabilities of individuals within a given group being in
a particular state at a given time, conditional on contact patterns, exposure
intensities, and unobserved characteristics. These conditional probabilities
are aggregated up to form a set of moment conditions that can be taken to the
data on the number of infected and active cases both at the aggregate and
group (or regional) levels. We make use of the moment conditions to
investigate the identification of the underlying structural parameters. Most
importantly, we show that whilst one cannot distinguish between average
contact numbers and the degree of exposure to the virus upon contact, it is
nevertheless possible to identify the basic and effective reproduction numbers
from relatively short time series observations on infections and recoveries.
Using Monte Carlo simulations, the small sample properties of the proposed
estimator are shown to be satisfactory, with a high degree of precision even
when using two and three weeks of rolling observations.

However, in practice, estimation of the transmission rate must take account of
the well-known measurement problem where the number of infected cases is often
grossly under-reported. This problem is further complicated since the degree
of under-reporting varies over time and tends to fall as society becomes more
familiar with the disease and testing becomes more widespread. To deal with
this mismeasurement problem, we propose a new method that jointly estimates
the transmission rates and the multiplication factor that measures the degree
of under-reporting. Equipped with daily estimates of the transmission rates,
we are then able to calibrate our epidemic model and investigate its
properties under different network topology, group numbers, and different
social distancing and vaccination strategies.

We apply the proposed joint estimation approach to examine how well the
outcome of the proposed epidemic model matches the Covid-19 evidence in the
case of six European countries (Austria, France, Germany, Italy, Spain, and
the UK) from March 2020 to April 2021. We provide rolling estimates of the
transmission rates and related effective reproduction numbers, as well as
estimates of multiple factors. We then use the estimated transmission rates to
calibrate the model parameters across the six countries. The stochastically
simulated outcomes are shown to be reasonably close to the reported cases once
the under-reporting issue has been addressed. We estimate that the degree of
under-reporting declined from a multiple of $4$--$10$ to $2$--$3$ times during
the study period across the countries considered.

Finally, we illustrate the use of our model for two different counterfactual
exercises. First, we consider the effects of vaccination on the evolution of
the epidemic using a multigroup setup, where we also evaluate the implications
of age-based vaccine prioritization on the outcomes. Our model allows each
individual to have their own degree of immunity, with vaccination increasing
this individual-specific immunity by a factor of $20$ in the case of Moderna
or Pfizer-BioNTech that are shown to be $94$--$95\%$ effective (Oliver et al.,
2020, 2021). The multigroup model is particularly helpful in examining the
implications of different vaccine prioritization strategies. Second, we
investigate the potential outcomes if the first lockdown in Germany had been
delayed for one or two weeks; and if the lockdown in the UK had started one or
two weeks earlier. Such counterfactual analyses can be achieved by shifting
the estimated transmission rates forward or backward. We show that early
intervention is critical in managing the infection and controlling the total
number of infected and active cases.

The problem of how to balance the public health risks from the spread of the
epidemic with the economic costs associated with lockdowns and other mandatory
social-distancing regulations will not be addressed in this paper. However,
the proposed network SIR model with its individual-based architecture is
eminently suited to this purpose. The proposed model can be combined with
behavioral assumptions about how individuals trade off infection risk and
economic well-being, thus generalizing the aggregate framework proposed in
Chudik, Pesaran, and Rebucci (2021) to individual-based SIR models.

The rest of the paper is organized as follows. Section \ref{Sec: concepts}
introduces the basic concepts and the classical multigroup SIR\ model. Section
\ref{Sec: model} lays out our individual based stochastic model on a network.
Section \ref{Sec: calibration R} explains the calibration of our model to
basic and effective reproduction numbers. Section \ref{Sec: properties}
documents the properties of the model. Sections \ref{Sec: estimation R}
discusses the estimation of the transmission rate. Section
\ref{Sec: empirical} presents the calibration of the model to Covid-19
evidence. Section \ref{Sec: counterfactual} concentrates on the counterfactual
analyses, and Section \ref{Sec: conclusion} concludes.

To save space, a detailed review of the related literature is given in an
online supplement, where we also provide supplementary theoretical
derivations, additional estimation results, counterfactual outcomes, and data sources.

\section{Basic concepts and the multigroup SIR model\label{Sec: concepts}}

We consider a population of $n$ individuals susceptible to the spread of a
disease with some initially infected individuals. We suppose that the
susceptible population can be categorized into $L$ groups of size $n_{\ell}$,
$\ell=1,2,\ldots,L$, with $L$ fixed such that $n=\sum_{\ell=1}^{L}n_{\ell}$.
It is further assumed that the group shares, $w_{\ell}=n_{\ell}/n>0,$ for all
$n$ and as $n\rightarrow\infty$. The grouping could be based on demographic
factors (age and/or gender), or other observed characteristics such as contact
locations and/or schedules, mode of transmission, genetic susceptibility,
group-specific vaccination coverage, as well as socioeconomic factors (see,
e.g., Hethcote, 2000). Individual $i$ in group $\ell$ will be referred to as
individual $\left(  i,\ell\right)  $, with $i=1,2,\ldots,n_{\ell}$ and
$\ell=1,2,\ldots,L$. It is assumed that $n_{\ell}$ is relatively large but
remains fixed over the course of the epidemic measured in days.

Suppose that individual $\left(  i,\ell\right)  $ becomes infected on day
$t=t_{i\ell}^{\ast}$, and let $x_{i\ell,t}$ take the value of unity for all
$t\geq t_{i\ell}^{\ast}$, and zero otherwise. In this way, we follow the
convention that once an individual becomes infected, he/she is considered as
infected thereafter, irrespective of whether that individual recovers or dies.
Specifically, we set%
\begin{equation}
x_{i\ell,t}=0,\text{ for all }t<t_{i\ell}^{\ast};\text{ \ and \ }x_{i\ell
,t}=1,\text{ for all }t\geq t_{i\ell}^{\ast}. \label{infected}%
\end{equation}
The event of recovery or death of individual $\left(  i,\ell\right)  $ at time
$t$ will be represented by $y_{i\ell,t}$, which will be equal to zero unless
the individual is "removed" (recovered or dead). An individual $(i,\ell)$ is
considered to be "active" if he/she is infected and not yet recovered. We
denote the active indicator by $z_{i\ell,t},$ which is formally defined by%
\begin{equation}
z_{i\ell,t}=\left(  1-y_{i\ell,t}\right)  x_{i\ell,t}. \label{active}%
\end{equation}
$z_{i\ell,t}$ takes the value of $0$ if individual $(i,\ell)$ has not been
infected, or has been infected but recovered/dead. It takes the value of $1$
if he/she is infected and not yet recovered. Any individual $(i,\ell)$ who has
not been infected is viewed "susceptible" and indicated by $s_{i\ell,t}=1$,
where%
\begin{equation}
s_{i\ell,t}=1-z_{i\ell,t}-y_{i\ell,t}. \label{susceptible}%
\end{equation}

It then readily follows that the total (cumulative) number of those "infected"
in group $\ell$ at the end of day $t$ is given by%
\begin{equation}
C_{\ell t}=\sum_{i=1}^{n_{\ell}}x_{i\ell,t},\text{ }\ell=1,2,\ldots,L,
\label{Cl}%
\end{equation}
where the summation is over all individuals in group $\ell$. The total number
of "recovered" in group $\ell$ in day $t$ is given by%
\begin{equation}
R_{\ell t}=\sum_{i=1}^{n_{\ell}}y_{i\ell,t},\text{ }\ell=1,2,\ldots,L.
\label{Rl}%
\end{equation}
The total number of "active" cases (individuals who are infected and not yet
removed) in group $\ell$ in day $t$ is
\begin{equation}
I_{\ell t}=\sum_{i=1}^{n_{\ell}}z_{i\ell,t}=\sum_{i=1}^{n_{\ell}}x_{i\ell
,t}-\sum_{i=1}^{n_{\ell}}x_{i\ell,t}y_{i\ell,t}=C_{\ell t}-R_{\ell t}.
\label{Il}%
\end{equation}
The number of "susceptible" individuals in group $\ell$ in day $t$ is%
\begin{equation}
S_{\ell t}=\sum_{i=1}^{n_{\ell}}s_{i\ell,t}=n_{\ell}-I_{\ell t}-R_{\ell
t}\text{.} \label{Sl}%
\end{equation}
Our model does not distinguish between recovery and death. Once an individual
is removed (recovered or dead), following the SIR literature, we assume that
he/she cannot be infected again. Under this assumption, recovery and death
have the same effects on the evolution of the epidemic, and accordingly in
what follows we shall not differentiate between recovery and death and refer
to their total as "removed".

The classic multigroup SIR\ model in discretized form can be written
as\footnote{See, e.g., Guo, Li, and Shuai (2006), Zhang et al. (2020), and the
references therein.}%
\begin{align}
S_{\ell,t+1}-S_{\ell t}  &  =-S_{\ell t}\sum_{\ell^{^{\prime}}=1}^{L}%
\beta_{\ell\ell^{\prime}}I_{\ell^{^{\prime}}t},\label{SIR-S}\\
I_{\ell,t+1}-I_{\ell t}  &  =S_{\ell t}\sum_{\ell^{^{\prime}}=1}^{L}%
\beta_{\ell\ell^{\prime}}I_{\ell^{^{\prime}}t}-\gamma_{\ell}I_{\ell
t}\label{SIR-I}\\
R_{\ell,t+1}-R_{\ell t}  &  =\gamma_{\ell}I_{\ell t}, \label{SIR-R}%
\end{align}
for $\ell=1,2,\ldots,L$ and $t=1,2,\ldots,T$, where $S_{\ell t}$, $I_{\ell t}$
and $R_{\ell t}$ are defined as above, $\gamma_{\ell}$ is the recovery rate
which is assumed to be time-invariant and the same for all people in group
$\ell$, and $\beta_{\ell\ell^{\prime}}$ is the transmission coefficient
between $S_{\ell t}$ and $I_{\ell^{^{\prime}}t}$. Note that individuals in
group $\ell^{^{\prime}}$ may transmit the disease to individuals in group
$\ell$, with the new infections in group $\ell$ given by $S_{\ell t}\sum
_{\ell^{^{\prime}}=1}^{L}\beta_{\ell\ell^{\prime}}I_{\ell^{^{\prime}}t}$.

\section{An individual based stochastic epidemic model on a
network\label{Sec: model}}

We now depart from the literature by explicitly modelling the individual
indicators, $x_{i\ell,t}$ and $y_{i\ell,t}$, (and hence $z_{i\ell,t}$) and
then simulate and aggregate up to match the theoretical predictions with
realized aggregated outcomes. In this section, we first describe the infection
and recovery processes at the individual level, we then show how they lead to
the moment conditions at group levels, and finally derive the relation between
aggregated outcomes from our model and the multigroup SIR\ deterministic model.

\subsection{Modelling the infection and recovery processes}

As an attempt to better integrate individual decisions to mitigate their
health risk within the epidemic model, we propose to directly model
$x_{i\ell,t}$ for each individual $(i,\ell)$, as compared to modelling the
group aggregates $S_{\ell t}$, $R_{\ell t}$, and $I_{\ell t}$. We follow the
micro-econometric literature and model the infection process using the latent
variable, $x_{i\ell,t+1}^{\ast},$ which determines whether individual
$(i,\ell)$ becomes infected. Specifically, we begin with the following Markov
switching process for individual $(i,\ell)$:%
\begin{equation}
x_{i\ell,t+1}=x_{i\ell,t}+\left(  1-x_{i\ell,t}\right)  I\left(  x_{i\ell
,t+1}^{\ast}>0\right)  , \label{xil+}%
\end{equation}
where $I\left(  \mathcal{A}\right)  $ is the indicator function that takes the
value of unity if $\mathcal{A}$ holds and zero otherwise. We suppose that
$x_{i\ell,t+1}^{\ast}$ is composed of two different components. The first
component relates to the contact pattern of individual $(i,\ell)$ with all
other individuals in the active set, denoted by $z_{j\ell^{\prime},t},$ both
within (when $\ell^{\prime}=\ell$) and outside of his/her group (when
$\ell^{\prime}\neq\ell$). The second component is an unobserved
individual-specific infection threshold variable, $\xi_{i\ell,t+1}>0$.
Formally, we set%
\begin{equation}
x_{i\ell,t+1}^{\ast}=\tau_{i\ell}\sum_{\ell^{^{\prime}}=1}^{L}\sum
_{j=1}^{n_{\ell^{^{\prime}}}}d_{i\ell,j\ell^{^{\prime}}}\left(  t\right)
z_{j\ell^{\prime},t}-\xi_{i\ell,t+1}, \label{xstar}%
\end{equation}
where the first component depends on the pattern of contacts, $d_{i\ell
,j\ell^{^{\prime}}}\left(  t\right)  $, whether the contacted individuals are
infectious, $z_{j\ell^{^{\prime}},t}$, and the exposure intensity parameter,
denoted by $\tau_{i\ell}$. $\mathbf{D}(t)=\left[  d_{i\ell,j\ell^{^{\prime}}%
}\left(  t\right)  \right]  $ is the contact network matrix, such that
$d_{i\ell,j\ell^{^{\prime}}}\left(  t\right)  =1$ if individual $\left(
i,\ell\right)  $ is in contact with individual $\left(  j,\ell^{^{\prime}%
}\right)  $ at time $t.$ $z_{j\ell^{\prime},t}=\left(  1-y_{j\ell^{\prime}%
,t}\right)  x_{j\ell^{\prime},t}$ is an infectious indicator, already defined
by (\ref{active}), and takes the value of unity if individual $\left(
j,\ell^{\prime}\right)  $ is infected and not yet recovered, zero otherwise.
The exposure intensity parameter, $\tau_{i\ell}$, is group-specific and
depends on the average duration of contacts, whether the contacting
individuals are wearing facemasks, and if they follow other recommended precautions.

The multigroup structure of the first component of (\ref{xstar}) covers a wide
range of observable characteristics, and can be extended to allow for
differences in age, location, and medical pre-conditions. There are also many
unobservable characteristics that lead to different probabilities of
infection, even for individuals with the same contact patterns and exposure
intensities. To allow for such latent factors, we have introduced the
individual-specific positive random variable ($\xi_{i\ell,t+1}>0$) which
represents the individual's degree of resilience to becoming infected and
varies across $(i,\ell)$ and $t.$ \textit{Ceteris paribus}, an individual with
a low value of $\xi_{i\ell,t+1}$ is more likely to become infected.
$\xi_{i\ell,t+1}$ is assumed to be independently distributed over $i$, $\ell$
and $t$, and follows an exponential distribution with the cumulative
distribution function given by
\begin{equation}
\Pr\left(  \xi_{i\ell,t+1}<a\right)  =1-\exp\left(  -\mu_{i\ell}^{-1}a\right)
,\text{ \ for }a>0,\label{xi}%
\end{equation}
where $\mu_{i\ell}=E\left(  \xi_{i\ell,t+1}\right)  $.

To complete the specification of the infection process we also need to model
$y_{i\ell,t}$, namely the recovery indicator. We assume that recovery depends
on the number of days since infection. Specifically, the recovery process for
individual $i$ is defined by%
\begin{equation}
y_{i\ell,t+1}=y_{i\ell,t}+z_{i\ell,t}\text{ }\zeta_{i\ell,t+1}\left(
t_{i\ell}^{\ast}\right)  , \label{Reci}%
\end{equation}
where $z_{i\ell,t}=\left(  1-y_{i\ell,t}\right)  x_{i\ell,t}$, $\zeta
_{i\ell,t+1}\left(  t_{i\ell}^{\ast}\right)  =1$ if individual $\left(
i,\ell\right)  $ recovers at time $t+1$, having been infected exactly at time
$t_{i\ell}^{\ast}$ and not before, and $\zeta_{i\ell,t+1}\left(  t_{i\ell
}^{\ast}\right)  =0,$ otherwise. The analysis of recovery simplifies
considerably if we assume time to removal, denoted by $T_{i\ell,t}^{\ast
}=t-t_{i\ell}^{\ast},$ follows the geometric distribution (for $t-t_{i\ell
}^{\ast}=1,2,\ldots$)%
\begin{equation}
\Pr\left[  \zeta_{i\ell,t+1}\left(  t_{i\ell}^{\ast}\right)  =1\right]
=\Pr\left(  T_{i\ell,t}^{\ast}=t-t_{i\ell}^{\ast}\right)  =\gamma_{\ell
}\left(  1-\gamma_{\ell}\right)  ^{t-t_{i\ell}^{\ast}-1}. \label{Pr_Tstar}%
\end{equation}
Then the probability of recovery at time $t+1$\ having remained infected for
$t-t_{i\ell}^{\ast}-1$ days (also known as the "hazard function") is given by
\begin{equation}
E\left[  \zeta_{i\ell,t+1}\left(  t_{i\ell}^{\ast}\right)  \left\vert
x_{i\ell,t},y_{i\ell,t},y_{i\ell,t-1},\ldots,y_{i\ell,t_{i\ell}^{\ast}%
}\right.  \right]  =\frac{\Pr\left(  T_{i\ell,t}^{\ast}=t-t_{i\ell}^{\ast
}\right)  }{\Pr\left(  T_{i\ell,t}^{\ast}>t-t_{i\ell}^{\ast}-1\right)
}=\gamma_{\ell}, \label{haz}%
\end{equation}
which is the same across all individuals within a given group and, most
importantly, does not depend on the number of days since infection.\footnote{A
more general specification that allows the recovery probability to depend on
the number of days being infected is considered in Section
\ref{Sup: GeomTruncated} of the online supplement.} Therefore, using
(\ref{haz}) in (\ref{Reci}), the recovery micro-moment condition simplifies to%
\begin{equation}
E\left(  y_{i\ell,t+1}\left\vert y_{i\ell,t},z_{i\ell,t}\right.  \right)
=y_{i\ell,t}+\gamma_{\ell}z_{i\ell,t}\text{,} \label{MomReci}%
\end{equation}
which implies that
\begin{equation}
E\left(  y_{i\ell,t+1}\left\vert \mathbf{y}_{\ell,t},\mathbf{z}_{\ell
,t}\right.  \right)  =y_{i\ell,t}+\gamma_{\ell}z_{i\ell,t}\text{,}
\label{MomReciB}%
\end{equation}
where $\mathbf{y}_{\ell,t}=(y_{1\ell,t},y_{2\ell,t},....,y_{n_{\ell}\ell
,t})^{\prime}$ and $\mathbf{z}_{\ell,t}=(z_{1\ell,t},z_{2\ell,t}%
,....,z_{n_{\ell}\ell,t})^{\prime}.$

We assume that $d_{i\ell,j\ell^{^{\prime}}}\left(  t\right)  $, the elements
of the $n\times n$ network matrix $\mathbf{D}\left(  t\right)  =\left[
d_{i\ell,j\ell^{^{\prime}}}\left(  t\right)  \right]  ,$ are independent draws
with $E\left[  d_{i\ell,j\ell^{^{\prime}}}\left(  t\right)  \right]
=p_{\ell\ell^{^{\prime}}}$, namely, the probability of contacts is homogeneous
within groups but differs across groups. Let $\boldsymbol{d}_{i}^{^{\prime}%
}\left(  t\right)  $ be the $i^{th}$ row of $\mathbf{D}\left(  t\right)  $.
Also let $\mathbf{z}_{t}$ be a column vector consisting of $z_{j\ell,t}$, for
$j=1,2,\ldots,n_{\ell}$ and $\ell=1,2,\ldots,L$. Then using (\ref{xil+}) we
have
\begin{align*}
E\left[  x_{i\ell,t+1}|x_{i\ell,t},\mathbf{z}_{t},\boldsymbol{d}_{i}\left(
t\right)  ,\tau_{i\ell},\mu_{i\ell}\right]   &  =x_{i\ell,t}+\left(
1-x_{i\ell,t}\right)  \Pr\left[  x_{i\ell,t+1}^{\ast}>0|\text{ }\mathbf{z}%
_{t}\text{, }\boldsymbol{d}_{i}\left(  t\right)  \right] \\
&  =x_{i\ell,t}+\left(  1-x_{i\ell,t}\right)  \Pr\left[  \xi_{i\ell,t+1}%
<\tau_{i\ell}\sum_{\ell^{^{\prime}}=1}^{L}\sum_{j=1}^{n_{\ell^{^{\prime}}}%
}d_{i\ell,j\ell^{^{\prime}}}\left(  t\right)  z_{j\ell^{^{\prime}},t}\right]
\\
&  =x_{i\ell,t}+\left(  1-x_{i\ell,t}\right)  \left\{  1-\exp\left[
-\theta_{i\ell}\sum_{\ell^{^{\prime}}=1}^{L}\sum_{j=1}^{n_{\ell^{^{\prime}}}%
}d_{i\ell,j\ell^{^{\prime}}}\left(  t\right)  z_{j\ell^{^{\prime}},t}\right]
\right\}  ,
\end{align*}
where $\theta_{i\ell}=\tau_{i\ell}/\mu_{i\ell}$, which represents the net
exposure effect. Since in general individual contact patterns are not
observed, we also need to derive $E\left(  x_{i\ell,t+1}|x_{i\ell
,t},\mathbf{z}_{t},\theta_{i\ell}\right)  $. To this end, we first note that%
\begin{align*}
E\left\{  \exp\left[  -\theta_{i\ell}d_{i\ell,j\ell^{^{\prime}}}\left(
t\right)  z_{j\ell^{^{\prime}},t}\right]  |\text{ }\mathbf{z}_{t}%
,\theta_{i\ell}\right\}   &  =\Pr\left[  d_{i\ell,j\ell^{^{\prime}}}\left(
t\right)  =0\right]  +\exp\left(  -\theta_{i\ell}z_{j\ell^{^{\prime}}%
,t}\right)  \Pr\left[  d_{i\ell,j\ell^{^{\prime}}}\left(  t\right)  =1\right]
\\
&  =\left(  1-p_{\ell\ell^{^{\prime}}}\right)  +\exp\left(  -\theta_{i\ell
}z_{j\ell^{^{\prime}},t}\right)  p_{\ell\ell^{^{\prime}}},
\end{align*}
and since by assumption $d_{i\ell,j\ell^{^{\prime}}}\left(  t\right)  $\ are
independently distributed, we then have%
\begin{align*}
E\left(  x_{i\ell,t+1}|x_{i\ell,t},\mathbf{z}_{t},\theta_{i\ell}\right)   &
=x_{i\ell,t}+\left(  1-x_{i\ell,t}\right)  \left\{  1-E\left[  \exp\left(
-\theta_{i\ell}\sum_{\ell^{^{\prime}}=1}^{L}\sum_{j=1}^{n_{\ell^{\prime}}%
}d_{i\ell,j\ell^{^{\prime}}}\left(  t\right)  z_{j\ell^{^{\prime}},t}\right)
\right]  \right\} \\
&  =1-\left(  1-x_{i\ell,t}\right)  \prod_{\ell^{^{\prime}}=1}^{L}\prod
_{j=1}^{n_{\ell^{\prime}}}\left[  1-p_{\ell\ell^{^{\prime}}}+\exp\left(
-\theta_{i\ell}z_{j\ell^{^{\prime}},t}\right)  p_{\ell\ell^{^{\prime}}%
}\right]  .
\end{align*}
However, recall that $z_{j\ell^{\prime},t}=1$ if individual $\left(
j,\ell^{\prime}\right)  $ is currently infected (namely if at time $t$ he/she
is a member of the active set, $\mathcal{I}_{t}$), otherwise $z_{j\ell
^{\prime},t}=0$. In the latter case $1-p_{\ell\ell^{^{\prime}}}+\exp\left(
-\theta_{i\ell}z_{j\ell^{^{\prime}},t}\right)  p_{\ell\ell^{^{\prime}}}=1$,
and hence
\begin{align}
E\left(  x_{i\ell,t+1}|x_{i\ell,t},\mathbf{z}_{t},\theta_{i\ell}\right)   &
=1-\left(  1-x_{i\ell,t}\right)  \prod_{\ell^{^{\prime}}=1}^{L}\left[
\prod_{j=1,z_{j\ell^{\prime},t}=1}^{n_{\ell^{\prime}}}\left(  1-p_{\ell
\ell^{^{\prime}}}+p_{\ell\ell^{^{\prime}}}e^{-\theta_{i\ell}}\right)  \right]
\nonumber\\
&  =1-\left(  1-x_{i\ell,t}\right)  \prod_{\ell^{^{\prime}}=1}^{L}\left(
1-p_{\ell\ell^{^{\prime}}}+p_{\ell\ell^{^{\prime}}}e^{-\theta_{i\ell}}\right)
^{I_{\ell^{\prime}t}}, \label{MomInfi}%
\end{align}
where $I_{\ell^{\prime}t}=\sum_{j=1}^{n_{\ell^{\prime}}}z_{j\ell^{\prime}%
,t}=C_{\ell^{\prime}t}-R_{\ell^{\prime}t}$. See also (\ref{Il}).

\subsection{Moment conditions at group and aggregate levels}

We will first derive the moment conditions at the group level. Denote the per
capita infected and recovered in group $\ell$ by $c_{\ell t}=C_{\ell
t}/n_{\ell}$ and $r_{\ell t}=R_{\ell t}/n_{\ell}$, respectively, and note that
$i_{\ell t}=c_{\ell t}-r_{\ell t}=I_{\ell t}/n_{\ell}$. Let $p_{\ell
\ell^{^{\prime}}}=k_{\ell\ell^{\prime}}/n_{\ell^{\prime}},$ where $k_{\ell
\ell^{\prime}}$ is the mean daily contacts from group $\ell^{\prime}$\ for an
individual in group $\ell$.\footnote{The parameters of the model, including
$\tau_{\ell}$, $k_{\ell\ell^{\prime}}$, $p_{\ell\ell^{^{\prime}}}$,
$\beta_{\ell\ell^{\prime}},$ and $\mu_{i\ell}$ can be time-varying due to
behavioral changes, vaccination, or other reasons. However, we suppress the
time subscript $t$ to simplify the exposition.}\ To preserve the symmetry of
contact probabilities, the mean contact numbers must satisfy the so-called
reciprocity condition, $n_{\ell}k_{\ell\ell^{\prime}}=n_{\ell^{\prime}}%
k_{\ell^{\prime}\ell}$ (see, e.g., Willem et al., 2020). That is, the total
number of contacts that people in group $\ell$ have with people in group
$\ell^{\prime}$ must be the same as the number of contacts that people in
group $\ell^{\prime}$ have with those in group $\ell$. In practice, $n_{\ell}$
is often quite large, with $k_{\ell\ell^{\prime}}$ relatively small (often
less than $30$). Therefore, it is reasonable to assume that $k_{\ell
\ell^{\prime}}$ is fixed in $n_{\ell^{\prime}}$ and hence $p_{\ell
\ell^{^{\prime}}}=O\left(  n_{\ell^{\prime}}^{-1}\right)  $. Then we have%
\begin{align}
&  \quad\ln\left[  \left(  1-p_{\ell\ell^{^{\prime}}}+p_{\ell\ell^{^{\prime}}%
}e^{-\theta_{i\ell}}\right)  ^{n_{\ell^{\prime}}i_{\ell^{\prime}t}}\right]
=n_{\ell^{\prime}}i_{\ell^{\prime}t}\ln\left[  1-p_{\ell\ell^{^{\prime}}%
}\left(  1-e^{-\theta_{i\ell}}\right)  \right]  \nonumber\\
&  =-n_{\ell^{\prime}}i_{\ell^{\prime}t}p_{\ell\ell^{^{\prime}}}\left(
1-e^{-\theta_{i\ell}}\right)  +O\left(  n_{\ell^{\prime}}p_{\ell\ell
^{^{\prime}}}^{2}\right)  =-\left(  1-e^{-\theta_{i\ell}}\right)
i_{\ell^{\prime}t}k_{\ell\ell^{\prime}}+O\left(  n_{\ell^{\prime}}%
^{-1}\right)  .\label{ln_approx}%
\end{align}
Suppose that $\theta_{i\ell}$ is small enough such that $1-e^{-\theta_{i\ell}%
}\approx\theta_{i\ell}$ is sufficiently accurate. Also recall that $w_{\ell
}=n_{\ell}/n>0,$ for all $\ell$, then $n_{\ell}$ rises at the same rate as
$n$. It follows that%
\begin{equation}
\left(  1-p_{\ell\ell^{^{\prime}}}+p_{\ell\ell^{^{\prime}}}e^{-\theta_{i\ell}%
}\right)  ^{n_{\ell^{^{\prime}}}i_{\ell^{\prime}t}}=\exp\left(  \theta_{i\ell
}i_{\ell^{\prime}t}k_{\ell\ell^{\prime}}\right)  +O\left(  n^{-1}\right)
,\label{pll'_approx}%
\end{equation}
and hence%
\begin{equation}
\prod_{\ell^{^{\prime}}=1}^{L}\left(  1-p_{\ell\ell^{^{\prime}}}+p_{\ell
\ell^{^{\prime}}}e^{-\theta_{i\ell}}\right)  ^{I_{\ell^{\prime}t}}=\exp\left[
\theta_{i\ell}\left(  \sum_{\ell^{\prime}=1}^{L}i_{\ell^{\prime}t}k_{\ell
\ell^{\prime}}\right)  \right]  +O\left(  n^{-1}\right)  .\label{product_pll'}%
\end{equation}
Let $\chi_{\ell t}=\sum_{\ell^{\prime}=1}^{L}i_{\ell^{\prime}t}k_{\ell
\ell^{\prime}}=\mathbf{i}_{t}^{\prime}\mathbf{k}_{\ell\circ}$, where
$\mathbf{i}_{t}=(i_{1t},i_{2t},\ldots,i_{Lt})^{\prime}$ and $\mathbf{k}%
_{\ell\circ}=\left(  k_{\ell1},k_{\ell2},\ldots,k_{\ell L}\right)  ^{\prime}.$
Let $H\left(  \chi_{\ell t},\boldsymbol{\varphi}_{\ell}\right)  =E\left(
e^{\chi_{\ell t}\theta_{i\ell}}|\chi_{\ell t}\right)  $, where the
expectations are taken with respect to the distribution of $\theta_{i\ell}$
for a given $\ell$, and $\boldsymbol{\varphi}_{\ell}$ refers to the parameters
of the distribution of $\theta_{i\ell}$ over $i$ in group $\ell$. Note that
$H\left(  \chi_{\ell t},\boldsymbol{\varphi}_{\ell}\right)  $ is the moment
generating function of $\theta_{i\ell}$, assumed to be the same across all
individuals. Then using the above result in the micro infection moment
conditions, (\ref{MomInfi}), gives%
\begin{equation}
E\left(  x_{i\ell,t+1}|x_{i\ell,t},\mathbf{i}_{t}\right)  =1-\left(
1-x_{i\ell,t}\right)  H\left(  \chi_{\ell t},\boldsymbol{\varphi}_{\ell
}\right)  +O\left(  n^{-1}\right)  .\label{Exilt}%
\end{equation}
Let $\mathbf{x}_{\ell t}=(x_{1\ell,t},x_{2\ell,t},...,x_{n_{\ell}\ell
,t})^{\prime}$ and note that since $x_{i\ell,t}$ is a subset of $\mathbf{x}%
_{\ell t}$, then
\[
E\left[  E\left(  x_{i\ell,t+1}|\mathbf{x}_{\ell t},\mathbf{i}_{t}\right)
\left\vert x_{i\ell,t}\right.  \right]  =E\left(  x_{i\ell,t+1}|x_{i\ell
,t},\mathbf{i}_{t}\right)  ,
\]
and the moment condition (\ref{Exilt}) also implies that
\begin{equation}
E\left(  x_{i\ell,t+1}|\mathbf{x}_{\ell t},\chi_{\ell t}\right)  =1-\left(
1-x_{i\ell,t}\right)  H\left(  \chi_{\ell t},\boldsymbol{\varphi}_{\ell
}\right)  +O\left(  n^{-1}\right)  ,\label{ExiltB}%
\end{equation}
for $i=1,2,...,n$. Averaging the above conditions over $i$ for a given group
$\ell$, and recalling that $c_{\ell,t+1}=\sum_{i=1}^{n}x_{i\ell,t+1}/n_{\ell}%
$, we obtain\
\begin{equation}
E\left(  c_{_{\ell,t+1}}|c_{\ell t},\mathbf{i}_{t}\right)  =1-\left(
1-c_{\ell t}\right)  H\left(  \chi_{\ell t},\boldsymbol{\varphi}_{\ell
}\right)  +O\left(  n^{-1}\right)  .\label{clmoments_general}%
\end{equation}

We will return to the heterogeneous $\theta_{i\ell}$ in the counterfactual
analysis of vaccination to be discussed in Section \ref{Sec: counterfactual},
where $\theta_{i\ell}$ is associated with the vaccine effectiveness for
individual $(i,\ell)$. In order to derive analytical results and achieve
identification in estimation, in what follows, we assume $\theta_{i\ell
}=\theta_{\ell}=\tau_{\ell}/\mu_{\ell}$ for all $i$ in group $\ell$. Also note
that $\tau_{\ell}$ and $\mu_{\ell}$ are not separately identified. Without
loss of generality, we normalize $\mu_{\ell}=1$. Under these conditions,
$\theta_{i\ell}=\tau_{\ell}$, the group-level infection moment condition can
be written as
\[
E\left(  \left.  \sum_{i=1}^{n_{\ell}}x_{i\ell,t+1}\right\vert C_{\ell
t},\mathbf{I}_{t}\right)  =E\left(  C_{\ell,t+1}|\text{ }\mathbf{z}%
_{t}\right)  =n_{\ell}-\left(  n_{\ell}-C_{\ell t}\right)  \prod
_{\ell^{^{\prime}}=1}^{L}\left(  1-p_{\ell\ell^{^{\prime}}}+p_{\ell
\ell^{^{\prime}}}e^{-\tau_{\ell}}\right)  ^{I_{\ell^{\prime}t}},
\]
which can be written equivalently as (recall that $I_{\ell t}=C_{\ell
t}-R_{\ell t}$)
\begin{equation}
E\left(  C_{\ell,t+1}-C_{\ell t}|\text{ }C_{\ell t},\mathbf{I}_{t}\right)
=\left(  n_{\ell}-C_{\ell t}\right)  \left[  1-\prod_{\ell^{^{\prime}}=1}%
^{L}\left(  1-p_{\ell\ell^{^{\prime}}}+p_{\ell\ell^{^{\prime}}}e^{-\tau_{\ell
}}\right)  ^{C_{\ell^{\prime}t}-R_{\ell^{\prime}t}}\right]  ,\text{ for }%
\ell=1,2,\ldots,L, \label{Clagg}%
\end{equation}
Also aggregating the micro recovery moment conditions, (\ref{MomReciB}), we
have%
\begin{equation}
E\left(  R_{\ell,t+1}|R_{\ell t},C_{\ell t}\right)  =\left(  1-\gamma_{\ell
}\right)  R_{\ell,t}+\gamma_{\ell}C_{\ell t},\text{ for }\ell=1,2,\ldots,L.
\label{MomRec2}%
\end{equation}

To sum up, in per capita terms, we obtain the following $2L\,$dimensional
system of moment conditions (for $\ell=1,2,\ldots,L$)%
\begin{align}
E\left(  \left.  \frac{1-c_{\ell,t+1}}{1-c_{\ell t}}\right\vert \text{
}\mathbf{i}_{t}\right)   &  =\prod_{\ell^{^{\prime}}=1}^{L}\left(
1-p_{\ell\ell^{^{\prime}}}+p_{\ell\ell^{^{\prime}}}e^{-\tau_{\ell}}\right)
^{n_{\ell^{^{\prime}}}\text{ }i_{\ell^{\prime}t}},\label{clmoments}\\
E\left(  r_{\ell,t+1}|r_{\ell t},c_{\ell t}\right)   &  =\left(
1-\gamma_{\ell}\right)  r_{\ell,t}+\gamma_{\ell}c_{\ell t}, \label{rlmoments}%
\end{align}
Given time series data on $\mathbf{c}_{t}=(c_{1t},c_{2t},\ldots,c_{Lt}%
)^{\prime}$ and $\mathbf{r}_{t}=(r_{1t},r_{2t},\ldots,r_{Lt})^{\prime}$, the
above moment conditions can be used to estimate the structural parameters,
$\gamma_{\ell}$, $\tau_{\ell}$ and $p_{\ell\ell^{\prime}}=p_{\ell^{\prime}%
\ell}$.

In relating the theory to the data, one may need to further aggregate across
groups to the population level if group-level data are unavailable or
unreliable. It is interesting to note that the multigroup model does not lead
to a model for the aggregates, $C_{t}=\sum_{\ell=1}^{L}C_{\ell t}$ and
$I_{t}=\sum_{\ell=1}^{L}I_{\ell t}$, without additional restrictions. To see
this, using (\ref{product_pll'}) in (\ref{clmoments}) and under the assumption
that $\theta_{i\ell}=\tau_{\ell}$, we obtain%
\begin{equation}
E\left(  1-c_{\ell,t+1}|c_{\ell t},\mathbf{i}_{t}\right)  =\left(  1-c_{\ell
t}\right)  \exp\left(  -\sum_{\ell^{\prime}=1}^{L}\beta_{\ell\ell^{\prime}%
}i_{\ell^{\prime}t}\right)  +O\left(  n^{-1}\right)  . \label{clmoments3}%
\end{equation}
where $\beta_{\ell\ell^{\prime}}=\left(  1-e^{-\tau_{\ell}}\right)
k_{\ell\ell^{\prime}}\approx\tau_{\ell}k_{\ell\ell^{\prime}}$. The
approximation holds since $\tau_{\ell}$ is small. Notice that $\sum_{\ell
=1}^{L}w_{\ell}c_{\ell}=C_{t}/n=c_{t}$ and $\sum_{\ell=1}^{L}w_{\ell}i_{\ell
}=I_{t}/n=i_{t}$. If we multiply both sides of (\ref{clmoments3}) by $w_{\ell
}$ and sum across $\ell=1,2,\ldots,L$, we obtain%
\begin{equation}
E\left(  1-c_{t+1}|c_{\ell t},\mathbf{i}_{t}\right)  =\sum_{\ell=1}^{L}%
w_{\ell}\left(  1-c_{\ell t}\right)  \exp\left(  -\sum_{\ell^{\prime}=1}%
^{L}\beta_{\ell\ell^{\prime}}i_{\ell^{\prime}t}\right)  +O\left(
n^{-1}\right)  . \label{cagg}%
\end{equation}
It is now clear that the group moment condition for infected cases,
(\ref{clmoments}), does not aggregate up to the moment conditions in terms of
$c_{t}$ and $i_{t}$, unless $\beta_{\ell\ell^{\prime}}/n_{\ell^{\prime}}$ is
the same across all $\ell$ and $\ell^{^{\prime}}$. It is also straightforward
to see that the group moment condition for recovery, (\ref{rlmoments}), does
not aggregate up either unless $\gamma_{\ell}=\gamma$ for all $\ell$.

In the case of a single group, we have $\tau_{\ell}=\tau$, $k_{\ell
\ell^{\prime}}=k$, and $\beta_{\ell\ell^{\prime}}=\beta\approx\tau k,$ for all
$\ell$ and $\ell^{^{\prime}}$. Then (\ref{cagg}) simplifies to%
\begin{equation}
E\left(  \left.  \frac{1-c_{t+1}}{1-c_{t}}\right\vert \text{ }i_{t}\right)
=e^{-\beta i_{t}}+O\left(  n^{-1}\right)  . \label{cagg_L1}%
\end{equation}
Also, if $\gamma_{\ell}=\gamma$ for all $\ell$, the recovery moment condition,
(\ref{rlmoments}), becomes
\begin{equation}
E\left(  r_{t+1}|r_{t},c_{t}\right)  =\left(  1-\gamma\right)  r_{t}+\gamma
c_{t}. \label{ragg_L1}%
\end{equation}
Given aggregate data on $c_{t}$, $i_{t},$ and $r_{t}$, one can estimate
$\beta$ and $\gamma$ using the moment conditions (\ref{cagg_L1}) and
(\ref{ragg_L1}), respectively. Interestingly, it can be shown that the
multigroup SIR model given by (\ref{SIR-S})--(\ref{SIR-R}) is a
linearized-deterministic version of the above moment conditions. The
relationship between our model and the classical SIR model is set out in
Section \ref{Sec: Relation to SIR} of the online supplement.

\section{Basic and effective reproduction numbers \label{Sec: calibration R}}

In this section, we consider the calibration of our model to a given basic
reproduction number assuming no intervention, and derive the effective
reproduction numbers in terms of mean contact patterns, exposure intensities,
and the recovery rate. We also consider the problem of identifying contact
patterns from the exposure rates in single and multigroup contexts.

\subsection{Basic reproduction number}

The basic reproduction number, denoted by $\mathcal{R}_{0}$, is defined as
"\textit{the average number of secondary cases produced by one infected
individual during the infected individual's entire infectious period assuming
a fully susceptible population}" (Del Valle, Hyman, and Chitnis, 2013). By
construction, $\mathcal{R}_{0}$ measures the degree to which an infectious
disease spreads when left unchecked. The infection spreads if $\mathcal{R}%
_{0}>1$ and abates if $\mathcal{R}_{0}<1$.

In order to derive $\mathcal{R}_{0}$ for our multigroup model, we suppose that
on day $1$ a fraction $w_{\ell}=n_{\ell}/n$ of each group $\ell$ becomes
infected, which represents the equivalent of one individual becoming infected
as required by the definition of $\mathcal{R}_{0}$. That is, on day $1$,
$R_{\ell1}=0$, $I_{\ell1}=C_{\ell1}=w_{\ell}$, for $\ell=1,2,\ldots,L$. Also,
in view of our model of the recovery, the probability that an individual
infected on day $1$ remains infected on day $s\geq1$ is given by $\Pr\left(
T_{i\ell,t}^{\ast}\geq s\right)  =\left(  1-\gamma_{\ell}\right)  ^{s-1}$, for
$s=1,2,\ldots$.\footnote{Using (\ref{Pr_Tstar}), note that $\Pr\left(
T_{i\ell,t}^{\ast}\geq s\right)  =1-\Pr\left(  T_{i\ell,t}^{\ast}<s\right)
=1-\sum_{j=1}^{s-1}\Pr\left(  T_{i\ell,t}^{\ast}=j\right)  =1-\sum_{j=1}%
^{s-1}\gamma_{\ell}\left(  1-\gamma_{\ell}\right)  ^{j-1}=\left(
1-\gamma_{\ell}\right)  ^{s-1}.$} Hence, we have
\begin{equation}
\mathcal{R}_{0}=\sum_{\ell=1}^{L}\sum_{s=1}^{\infty}\left(  1-\gamma_{\ell
}\right)  ^{s-1}E\left(  C_{_{\ell,s+1}}|\mathbf{w}\right)  , \label{R0}%
\end{equation}
where $\mathbf{w=(}w_{1},w_{2},\ldots,w_{L})^{\prime}$. Now using
(\ref{Clagg}) for $s=2$ we have%
\begin{equation}
E\left(  C_{\ell2}|\mathbf{w}\right)  =n_{\ell}\left[  1-\prod_{\ell
^{^{\prime}}=1}^{L}\left(  1-p_{\ell\ell^{^{\prime}}}+p_{\ell\ell^{^{\prime}}%
}e^{-\tau_{\ell}}\right)  ^{w_{\ell^{^{\prime}}}}\right]  . \label{C_l2}%
\end{equation}
Due to the large number of possibilities that follow after the second day of
the epidemic, it is not possible to derive similar analytical expressions for
$E\left(  C_{_{\ell,s}}|\mathbf{w}\right)  $, $s=3,4,\ldots$. But since the
weights of these future expected values decay geometrically, and at the start
of the epidemic the number of infected is likely to be very small relative to
the susceptible population, we think it is reasonable to follow the literature
(Farrington and Whitaker, 2003; Elliott and Gourieroux, 2020) and assume that
$E\left(  C_{_{\ell,s+1}}|\mathbf{w}\right)  \approx E\left(  C_{\ell
2}|\mathbf{w}\right)  $, for $s\geq1$.\footnote{Elliott and Gourieroux (2020)
make a similar assumption that the expected number of susceptibles is constant
over time in deducing the reproduction numbers (p. 7). The constant recovery
intensity is a standard assumption in the SIR\ literature. In contrast, our
calibration to the reproduction numbers differs significantly from Elliott and
Gourieroux (2020) in that we directly consider individual $i$ in group $\ell$
and his/her contact network, rather than modelling population groups
classified by their S, I, or R\ status.} Under this assumption, the following
approximate expression for $\mathcal{R}_{0}$ obtains:%
\[
\mathcal{R}_{0}\approx\sum_{\ell=1}^{L}\sum_{s=1}^{\infty}\left(
1-\gamma_{\ell}\right)  ^{s-1}E\left(  C_{\ell2}|\mathbf{w}\right)
=\sum_{\ell=1}^{L}\gamma_{\ell}^{-1}E\left(  C_{\ell2}|\mathbf{w}\right)  .
\]
In the case where the recovery rates are the same across the groups
($\gamma_{\ell}=\gamma$), the above expression simplifies further and we have
$\mathcal{R}_{0}\approx\gamma^{-1}\sum_{\ell=1}^{L}E\left(  C_{\ell
2}|\mathbf{w}\right)  $.\footnote{In the case of Covid-19, it is universally
assumed that $\gamma_{\ell}=\gamma=1/14$, which we also adopt in our empirical
analysis.} Now using (\ref{C_l2}) gives
\begin{equation}
\mathcal{R}_{0}\approx\gamma^{-1}\left[  n-\sum_{\ell=1}^{L}n_{\ell}%
\prod_{\ell^{^{\prime}}=1}^{L}\left(  1-p_{\ell\ell^{^{\prime}}}+p_{\ell
\ell^{^{\prime}}}e^{-\tau_{\ell}}\right)  ^{w_{\ell^{^{\prime}}}}\right]  .
\label{R0_exact}%
\end{equation}

To see how the above result relates to the well-known expression
$\mathcal{R}_{0}=\beta/\gamma$, consider the case of a single group with
$p=k/(n-1)\thickapprox k/n$ as $n$ is large. Then the expression in
(\ref{R0_exact}) reduces to%
\begin{equation}
\gamma\mathcal{R}_{0}\approx n\left[  1-\left(  1-p+pe^{-\tau}\right)
\right]  =np(1-e^{-\tau})\thickapprox\tau k,\label{R0_one group}%
\end{equation}
where the last result follows by $1-e^{-\tau}\approx\tau$. Hence the model can
be calibrated to any choice of $\mathcal{R}_{0}$ and $\gamma$ by setting the
average number of contacts, $k$, and/or the exposure intensity parameter,
$\tau$. It is clear that $\tau$ and $k\,$are not separately identified---only
their product is identified. In addition, we would obtain the standard result
$\gamma\mathcal{R}_{0}=\beta$ for SIR\ models if we set $\beta=np(1-e^{-\tau
})\thickapprox\tau k$.

Returning to the multigroup case, expression (\ref{R0_one group}) continues to
apply if the population is homogeneous in the sense that $p_{\ell
\ell^{^{\prime}}}=p$, $\tau_{\ell}=\tau$, for all $\ell$ and $\ell^{^{\prime}%
}$. But in the more realistic case of group heterogeneity, we can use
(\ref{R0_exact}) to calibrate $\tau_{\ell}$ and/or $p_{\ell\ell^{^{\prime}}}$
for given choices of $\mathcal{R}_{0}$ and $\gamma$. Since we have assumed
that $n_{\ell}$ is large and $L$ is fixed, (\ref{R0_exact}) can be further
simplified with a linear approximation derived as follows. Let $A_{n,\ell
\ell^{^{\prime}}}=\left(  1-p_{\ell\ell^{^{\prime}}}+p_{\ell\ell^{^{\prime}}%
}e^{-\tau_{\ell}}\right)  ^{w_{\ell^{^{\prime}}}}$, and use a similar argument
as in (\ref{ln_approx}) to obtain (recall that $p_{\ell\ell^{\prime}}%
=O(n^{-1})$)%
\begin{align*}
\ln A_{n,\ell\ell^{^{\prime}}}  &  =w_{\ell^{^{\prime}}}\ln\left[
1-p_{\ell\ell^{^{\prime}}}\left(  1-e^{-\tau_{\ell}}\right)  \right]
=-w_{\ell^{^{\prime}}}p_{\ell\ell^{^{\prime}}}\left(  1-e^{-\tau_{\ell}%
}\right)  +O\left(  p_{\ell\ell^{^{\prime}}}^{2}\right) \\
&  \thickapprox-\tau_{\ell}w_{\ell^{^{\prime}}}p_{\ell\ell^{^{\prime}}%
}+O\left(  n^{-2}\right)  .
\end{align*}
Then $A_{n,\ell\ell^{^{\prime}}}=\exp\left(  -\tau_{\ell}w_{\ell^{^{\prime}}%
}p_{\ell\ell^{^{\prime}}}\right)  +O\left(  n^{-2}\right)  $. Using this in
(\ref{R0_exact}) gives
\begin{align}
\gamma\mathcal{R}_{0}  &  =n-\sum_{\ell=1}^{L}n_{\ell}\left\{  \exp\left(
-\tau_{\ell}\sum_{\ell^{^{\prime}}=1}^{L}w_{\ell^{^{\prime}}}p_{\ell
\ell^{^{\prime}}}\right)  +O\left(  n^{-2}\right)  \right\}  =\sum_{\ell
=1}^{L}n_{\ell}\tau_{\ell}\sum_{\ell^{^{\prime}}=1}^{L}w_{\ell^{^{\prime}}%
}p_{\ell\ell^{^{\prime}}}+O\left(  n^{-1}\right) \nonumber\\
&  =n\sum_{\ell=1}^{L}\sum_{\ell^{^{\prime}}=1}^{L}w_{\ell}w_{\ell^{^{\prime}%
}}\left(  \tau_{\ell}p_{\ell\ell^{^{\prime}}}\right)  +O\left(  n^{-1}\right)
. \label{R0_pll'}%
\end{align}
As before setting $p_{\ell\ell^{^{\prime}}}=k_{\ell\ell^{\prime}}%
/n_{\ell^{\prime}}$, for $n$ sufficiently large, the above expression can be
written equivalently as
\begin{equation}
\gamma\mathcal{R}_{0}=\beta=\sum_{\ell=1}^{L}w_{\ell}\beta_{\ell}\text{,
\ with }\beta_{\ell}=\sum_{\ell^{^{\prime}}=1}^{L}\tau_{\ell}k_{\ell
\ell^{\prime}}\text{,} \label{R0_multigroup}%
\end{equation}
where $\beta$ and $\beta_{\ell}$ are the aggregate and group-specific
transmission rates, respectively.

Similar to the case of a single group, equation (\ref{R0_pll'}) implies that
$\tau_{\ell}$ and $p_{\ell\ell^{^{\prime}}}$ are not separately identified;
only their products are identified (or equivalently, $\beta_{\ell\ell^{\prime
}}\approx\tau_{\ell}k_{\ell\ell^{\prime}}$ are identified). To see this more
formally, consider the simple case of two groups $(L=2)$. Then for
sufficiently large $n$, using (\ref{R0_pll'}) with $L=2$ we have%
\begin{align}
\gamma\mathcal{R}_{0}  &  \thickapprox nw_{1}\left(  w_{1}\tau_{1}p_{11}%
+w_{2}\tau_{1}p_{12}\right)  +nw_{2}\left(  w_{1}\tau_{2}p_{21}+w_{2}\tau
_{2}p_{22}\right) \nonumber\\
&  =nw_{1}^{2}\left(  \tau_{1}p_{11}\right)  +nw_{1}w_{2}\left(  \tau_{1}%
+\tau_{2}\right)  p_{12}+nw_{2}^{2}\left(  \tau_{2}p_{22}\right)  ,
\label{ID_L2}%
\end{align}
where the last line follows by the symmetry of contact probabilities:
$p_{\ell\ell^{^{\prime}}}=p_{\ell^{^{\prime}}\ell}$. It is clear from
(\ref{ID_L2}) that only $\tau_{1}p_{11}$, $\left(  \tau_{1}+\tau_{2}\right)
p_{12}$, and $\tau_{2}p_{22}$ can be identified given $w_{1}$, $w_{2}$, $n$
and $\gamma$. More generally, for finite $L\geq2$, $\tau_{\ell}p_{\ell
\ell^{^{\prime}}}$ are identified for any $\ell$ and $\ell^{^{\prime}%
}=1,2,\ldots,L$.

\subsection{Effective reproduction numbers and mitigation policies}

In reality, the average number of secondary cases will vary over time as a
result of the decline in the number of susceptible individuals (due to
immunity or death) and/or changes in behavior (due to mitigation strategies
such as social distancing, quarantine measures, travel restrictions and
wearing of facemasks). The effective reproduction number, which we denote by
$\mathcal{R}_{et}$,\footnote{We use this notation in order to clearly
distinguish the effective reproduction number from the number of removed
cases, $R_{t}$.} is the expected number of secondary cases produced by one
infected individual in a population that includes both susceptible and
non-susceptible individuals at time $t.$ In a multigroup setting, we represent
"one infected individual" by the vector of population proportions,
$\mathbf{w=(}w_{1},w_{2},\ldots,w_{L})^{\prime}$. The evolution of
$\mathcal{R}_{et}$ is determined by the remaining number of susceptibles by
groups, $S_{\ell t}=n_{\ell}-C_{\ell t},$ for $\ell=1,2,\ldots,L$. Formally,
$\mathcal{R}_{et}$ is defined by
\begin{equation}
\gamma\mathcal{R}_{et}=\sum_{\ell=1}^{L}E\left(  C_{\ell,t+1}-C_{\ell
t}|\text{ }\mathbf{I}_{t}=\mathbf{w}\right)  .\label{Ret_def}%
\end{equation}
In the absence of any interventions, using (\ref{Clagg}) we have
\begin{equation}
\gamma\mathcal{R}_{et}=\sum_{\ell=1}^{L}\left(  n_{\ell}-C_{\ell t}\right)
\left[  1-\prod_{\ell^{^{\prime}}=1}^{L}\left(  1-p_{\ell\ell^{^{\prime}}%
}+p_{\ell\ell^{^{\prime}}}e^{-\tau_{\ell}}\right)  ^{w_{\ell^{\prime}}%
}\right]  .\label{RetExat}%
\end{equation}
Recalling that $\left(  1-p_{\ell\ell^{^{\prime}}}+p_{\ell\ell^{^{\prime}}%
}e^{-\tau_{\ell}}\right)  ^{w_{\ell^{^{\prime}}}}=\exp\left(  -\tau_{\ell
}w_{\ell^{^{\prime}}}p_{\ell\ell^{^{\prime}}}\right)  +O\left(  n^{-2}\right)
$, then for $n$ sufficiently large we have the following approximate
expression for $\mathcal{R}_{et}$:
\[
\gamma\mathcal{R}_{et}=\sum_{\ell=1}^{L}S_{\ell t}\left(  \sum_{\ell^{\prime
}=1}^{L}\tau_{\ell}w_{\ell^{^{\prime}}}p_{\ell\ell^{^{\prime}}}\right)
+O\left(  n^{-1}\right)  ,
\]
Setting $p_{\ell\ell^{^{\prime}}}=k_{\ell\ell^{\prime}}/n_{\ell^{\prime}}$ we
can alternatively write $\gamma\mathcal{R}_{et}$ as (recall that
$w_{\ell^{\prime}}=n_{\ell^{\prime}}/n$ and $s_{\ell t}=S_{\ell t}/n_{\ell}$)
\begin{equation}
\gamma\mathcal{R}_{et}=\sum_{\ell=1}^{L}w_{\ell}\beta_{\ell}s_{\ell
t}+O\left(  n^{-1}\right)  ,\label{Ret}%
\end{equation}
where $\beta_{\ell}$ is already defined by (\ref{R0_multigroup}).

In the case of a single group or when $\beta_{\ell}=\beta$ is homogeneous
across groups, the above expression simplifies to $\gamma\mathcal{R}%
_{et}=\beta\left(  \sum_{\ell=1}^{L}w_{\ell}s_{\ell t}\right)  =\beta s_{t}$,
which can be written equivalently as $\mathcal{R}_{et}=(1-c_{t})\mathcal{R}%
_{0}$. In the absence of any interventions $\mathcal{R}_{et}$ declines as
$c_{t}$ rises, and $\mathcal{R}_{et}$ falls below $1$ when $c_{t}%
>(\mathcal{R}_{0}-1)/\mathcal{R}_{0}$. The value $(\mathcal{R}_{0}%
-1)/\mathcal{R}_{0}$ is often referred to as the herd immunity threshold. For
the multigroup case, using (\ref{R0_multigroup}) and (\ref{Ret}), the
condition for herd immunity is more complicated and is given by (for $n$
sufficiently large)
\[
\frac{\sum_{\ell=1}^{L}w_{\ell}\beta_{\ell}s_{\ell t}}{\gamma}=\left[
\frac{\sum_{\ell=1}^{L}w_{\ell}\beta_{\ell}\left(  1-c_{\ell t}\right)  }%
{\sum_{\ell=1}^{L}w_{\ell}\beta_{\ell}}\right]  \mathcal{R}_{0}<1,
\]
and the herd immunity threshold becomes%
\[
\frac{\sum_{\ell=1}^{L}w_{\ell}\beta_{\ell}c_{\ell t}}{\sum_{\ell=1}%
^{L}w_{\ell}\beta_{\ell}}>\frac{\mathcal{R}_{0}-1}{\mathcal{R}_{0}}.
\]
This formula clearly shows that for herd immunity to apply, the group-specific
infection rate, $c_{\ell t}$, must be sufficiently large -- shielding one
group requires higher infection rates in other groups with larger population
weights. To see this, let us consider a simple example of two groups ($L=2$)
with a homogeneous transmission rate across the two groups ($\beta_{1}%
=\beta_{2}=\beta$), and note that $0<w_{1},w_{2}<1$. Suppose that policymakers
want to shield Group 1, which may comprise elderly people, from infection. In
the extreme case where all individuals in Group 1 are protected, namely,
$c_{1t}=0$, then herd immunity requires $c_{2t}>(\mathcal{R}_{0}-1)/\left(
\mathcal{R}_{0}w_{2}\right)  $, which is higher than the threshold value of
$(\mathcal{R}_{0}-1)/\mathcal{R}_{0}$ where the population groups are treated symmetrically.

Social intervention might be necessary if the herd immunity threshold is too
high and could lead to significant hospitalization and deaths. In such cases,
intervention becomes necessary to reduce the transmission rates $\beta_{\ell}%
$, thus introducing independent policy-induced reductions in the transmission
rates. In the presence of social policy interventions, the effective
reproduction number for the multigroup can be written as%
\begin{equation}
\gamma\mathcal{R}_{et}=\sum_{\ell=1}^{L}w_{\ell}\beta_{\ell t}\left(
1-c_{\ell t}\right)  +O\left(  n^{-1}\right)  .\label{Ret_pll'}%
\end{equation}
where (using (\ref{R0_multigroup})) $\beta_{\ell t}=\sum_{\ell^{^{\prime}}%
=1}^{L}\tau_{\ell t}k_{\ell\ell^{\prime},t}.$ Reductions in $\beta_{\ell t}$
can come about either by reducing the average number of contacts within and
across groups, $k_{\ell\ell^{\prime},t}$, or by reducing the group-specific
exposure intensity parameter, $\tau_{\ell t}$, or both. Since only the product
of $\tau_{\ell t}$ and $k_{\ell\ell^{\prime},t}$ is identified, in our
simulations we fix the contact patterns and calibrate the desired value of
$\beta_{\ell t}$ by setting the value of $\tau_{\ell t}$ for each $\ell$ to
achieve a desired $\mathcal{R}$ number. Of course, one would obtain equivalent
results if the average number of contacts is assumed to be time-varying and
the exposure intensity parameter is assumed constant. In the case of a single
group or when $\beta_{\ell t}=\beta_{t}$ for all $\ell$, we have
\begin{equation}
\mathcal{R}_{et}=(1-c_{t})\frac{\beta_{t}}{\gamma},\label{Ret_one_group}%
\end{equation}
where $(1-c_{t})$ is the herding component. It is also worth bearing in mind
that at the outset of epidemic outbreaks the value of $c_{t}$ is close to zero
which ensures that $\mathcal{R}_{e0}=\beta_{0}/\gamma=\mathcal{R}_{0}$.

\section{Calibration and simulation of the model\label{Sec: properties}}

Although it is difficult to obtain an analytical solution to the
individual-based stochastic epidemic model, we can study its properties by
simulations. This section focuses on the baseline scenario of no containment
measures or mutation of the virus so that the transmission rate is constant.
We will discuss simulation results with time-varying transmission rate under
social distancing and vaccination in Section \ref{Sec: counterfactual}. In
light of the recent studies on the value of $\mathcal{R}_{0}$ for Covid-19, we
set $\mathcal{R}_{0}=3$.\footnote{A summary of published estimates of
$\mathcal{R}_{0}$ is provided in Table 1 of D'Arienzo and Coniglio (2020).}
For the recovery rate, in view of the World Health Organization guidelines of
two weeks self-isolation, we set $\gamma=1/14$.\footnote{\label{gamma14}%
Similar guidelines issued by the US and the UK can be found at
\url{https://www.cdc.gov/coronavirus/2019-ncov/if-you-are-sick/quarantine.html}
and
\url{https://www.nhs.uk/conditions/coronavirus-covid-19/self-isolation-and-treatment/how-long-to-self-isolate/},
respectively (last accessed October 2020).} It follows that $\beta
=\gamma\mathcal{R}_{0}=3/14.$

We consider dividing the population into $L=5$ age groups: $[0,15)$,
$[15,30)$, $[30,50)$, $[50,65)$, and $65$+ years old, and, of course, one can
readily consider a different number of groups based on other characteristics
if such data are available. We use the data on Germany as an illustration. The
social contact surveys by Mossong et al. (2008) provide rich data on the
contact patterns in Germany. We update the contact matrix by age with the most
recent population data such that the reciprocity condition, $n_{\ell}%
k_{\ell\ell^{\prime}}=n_{\ell^{\prime}}k_{\ell^{\prime}\ell}$, is satisfied.
The population shares for the five age groups are $\mathbf{w}=(0.13,$ $0.17,$
$0.28,$ $0.20,$ $0.21)^{\prime}$, and the resulting (pre-pandemic) contact
matrix is
\begin{footnotesize}
\begin{equation}
\mathbf{K}=\left(  k_{\ell\ell^{^{\prime}}}\right)  =\left(
\begin{tabular}
[c]{rrrrr}%
3.43 & 1.10 & 2.34 & 0.67 & 0.47\\
0.87 & 4.55 & 2.72 & 1.14 & 0.41\\
1.11 & 1.64 & 3.74 & 1.42 & 0.78\\
0.45 & 0.96 & 1.99 & 2.30 & 0.92\\
0.31 & 0.34 & 1.08 & 0.91 & 1.70
\end{tabular}
\right)  ,\label{K0}%
\end{equation}%
\end{footnotesize}%
where the element, $k_{\ell\ell^{^{\prime}}}$, represents the average number
of daily contacts reported by participants in group $\ell$ with someone of
group $\ell^{\prime}$. The larger diagonal values in (\ref{K0}) indicate that
people tend to mix more with others of the same age group---a phenomenon well
documented by contact surveys across different countries. In order to
calibrate $\tau_{\ell}$ across groups, we match the ratio of infection
probabilities of groups with the ratio of reported cases. Specifically,
denoting the reference group by $\ell_{0}$ and the ratio of reported
infections of group $\ell$ to group $\ell_{0}$ by $\lambda_{\ell\ell_{0}}$,
then $\lambda_{\ell\ell_{0}}$ should match the ratio of the related
probabilities, namely,%
\[
\lambda_{\ell\ell_{0}}=\frac{E\left(  x_{i\ell,t+1}|\mathbf{z}_{t}\right)
}{E\left(  x_{i\ell_{0},t+1}|\mathbf{z}_{t}\right)  }=\frac{1-\prod
_{\ell^{^{\prime}}=1}^{L}\left(  1-p_{\ell\ell^{^{\prime}}}+p_{\ell
\ell^{^{\prime}}}e^{-\tau_{\ell}}\right)  ^{n_{\ell^{\prime}}i_{\ell^{\prime
}t}}}{1-\prod_{\ell^{^{\prime}}=1}^{L}\left(  1-p_{\ell_{0}\ell^{^{\prime}}%
}+p_{\ell_{0}\ell^{^{\prime}}}e^{-\tau_{\ell_{0}}}\right)  ^{n_{\ell^{\prime}%
}i_{\ell^{\prime}t}}}.
\]
Using (\ref{product_pll'}), we now have%
\[
\lambda_{\ell\ell_{0}}\approx\frac{1-\exp\left(  -\tau_{\ell}\sum
_{\ell^{\prime}=1}^{L}i_{\ell^{\prime}t}k_{\ell\ell^{\prime}}\right)  }%
{1-\exp\left(  -\tau_{\ell_{0}}\sum_{\ell^{\prime}=1}^{L}i_{\ell^{\prime}%
t}k_{\ell_{0}\ell^{\prime}}\right)  }\approx\frac{\tau_{\ell}}{\tau_{\ell_{0}%
}}\frac{\sum_{\ell^{\prime}=1}^{L}i_{\ell^{\prime}t}k_{\ell\ell^{\prime}}%
}{\sum_{\ell^{\prime}=1}^{L}i_{\ell^{\prime}t}k_{\ell_{0}\ell^{\prime}}}.
\]
For the purpose of calibration, we further assume that $i_{\ell t}=w_{\ell
}f_{t}$, where $f_{t}$ can be viewed as the latent common driver of the
epidemic at time $t$. It then follows that
\begin{equation}
\lambda_{\ell\ell_{0}}=\frac{\tau_{\ell}}{\tau_{\ell_{0}}}\frac{\sum
_{\ell^{\prime}=1}^{L}w_{\ell^{\prime}}k_{\ell\ell^{\prime}}}{\sum
_{\ell^{\prime}=1}^{L}w_{\ell^{\prime}}k_{\ell_{0}\ell^{\prime}}%
}.\label{ratio_infection}%
\end{equation}
We now use data on infected cases in Germany by the five age groups at the end
of 2020 (before the rollout of Covid-19 vaccines) to calibrate the relative
transmission rates by groups. Setting the first age group as the reference
group ($\ell_{0}=1$), we obtain $\boldsymbol{\lambda}=\left(  \lambda
_{\ell\ell_{0}}\right)  =\left(  1,\text{ }2.83,\text{ }3.81,\text{
}2.94,\text{ }2.39\right)  ^{^{\prime}}$, which in conjunction with
(\ref{ratio_infection}) yields $\boldsymbol{\tau}=\left(  \tau_{\ell}\right)
=\tau_{1}\left(  1,\text{ }2.21,\text{ }3.04,\text{ }3.15\text{ },3.93\right)
^{\prime}$. To calibrate $\tau_{1}$, we use (\ref{R0_pll'}) and obtain
$\tau_{1}=0.011$ setting $\mathcal{R}_{0}=3$ and $\gamma=1/14$.

For each replication, the simulation begins with $1/1000$ of total population
randomly infected on day $t=1$, that is, $c_{1}^{\left(  b\right)  }%
=i_{1}^{\left(  b\right)  }=0.001$ and $r_{1}^{\left(  b\right)  }=0$, where
$b$ denotes the $b^{th}$ replication, for $b=1,2,\ldots,B$.\footnote{We find
that there will not be outbreaks in many replications if the simulation begins
with less than $1/1000$ of the population initially infected.} Then from $t=2$
onwards, the infection and recovery processes follow (\ref{xil+}) and
(\ref{Reci}), respectively, for $\ell=1,2,\ldots,L.$ The proportion of
infections for each age group is computed as $c_{\ell t}^{\left(  b\right)
}=C_{\ell t}^{\left(  b\right)  }/n_{\ell}=\sum_{i=1}^{n_{\ell}}x_{i\ell
,t}^{\left(  b\right)  }/n_{\ell}$, and the daily new cases are computed by
$\Delta c_{\ell t}^{\left(  b\right)  }=c_{\ell t}^{\left(  b\right)
}-c_{\ell,t-1}^{\left(  b\right)  }$. The aggregate infections and new cases
are computed as $c_{t}^{\left(  b\right)  }=\sum_{\ell=1}^{L}w_{\ell}c_{\ell
t}^{\left(  b\right)  }$ and $\Delta c_{t}^{\left(  b\right)  }=c_{t}^{\left(
b\right)  }-c_{t-1}^{\left(  b\right)  }$, respectively. Details on the
generation of random networks are given in Section \ref{Sup: Power Law} of the
online supplement. Note that the contact network randomly changes every day
(and also across replications). This feature captures the random nature of
many encounters an individual has on a daily basis. We consider $B=1,000$
replications and set the population size to $n=10,000$. We also tried larger
population sizes, but, as will be seen below, the interquartile range of the
simulated new cases is already very tight when $n=10,000$. Some simulation
results for $n=50,000$ and $n=100,000$ are provided in Figure
\ref{fig: diff_n} of the online supplement.

\begin{figure}[!hp]%
\caption
{Simulated group-specific and aggregate new cases when there are no containment measures with ${\mathcal
{R}_0}=3$}%
\vspace{-0.2cm}%
\footnotesize
\label{fig: multigroup}

\begin{center}%
\begin{tabular}
[c]{ccc}%
\textbf{Group 1: [0, 15)} &  & \textbf{Group 2: [15, 30)}\\%
{\includegraphics[
height=2.1465in,
width=2.6783in
]%
{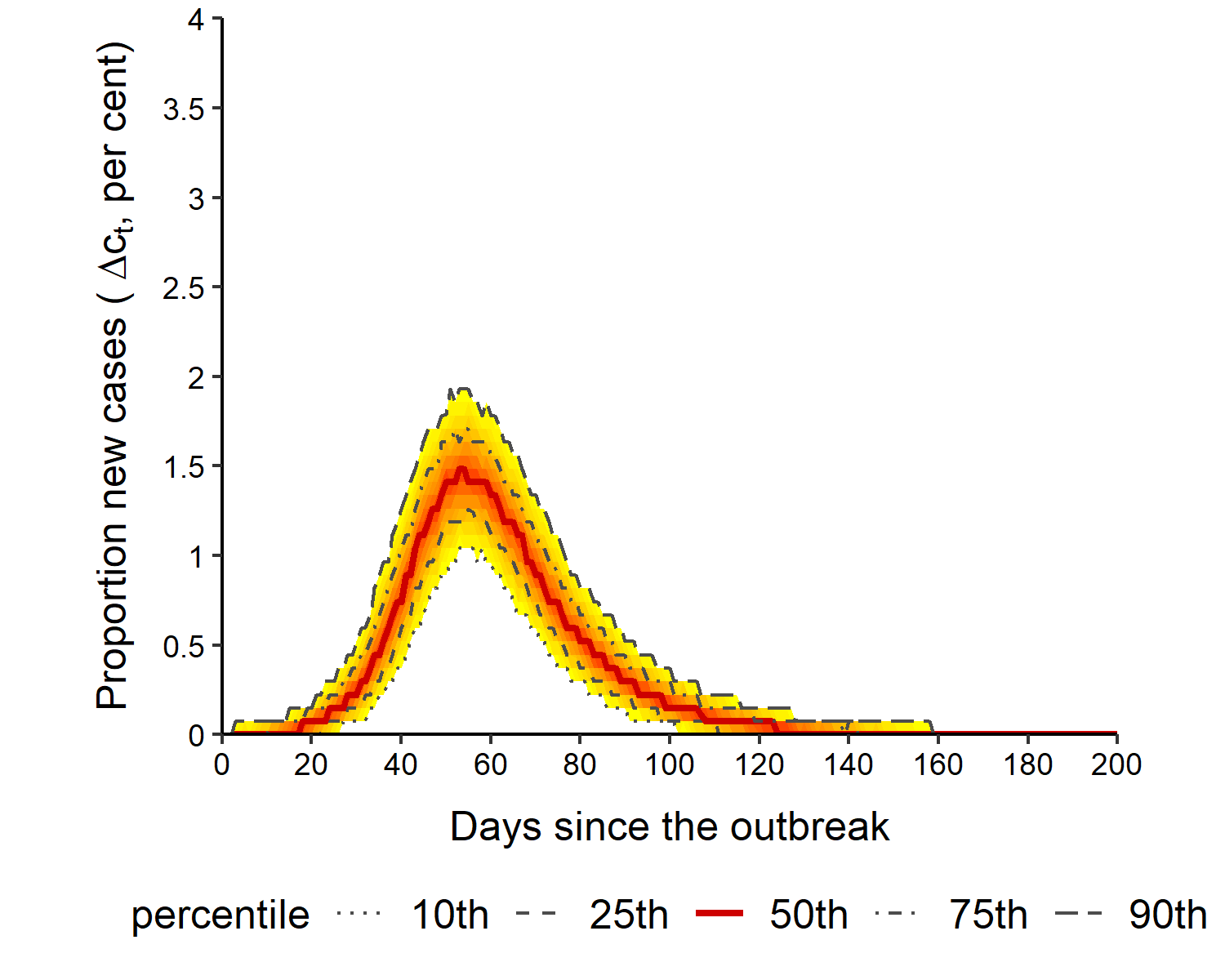}%
}
&  &
{\includegraphics[
height=2.1465in,
width=2.6783in
]%
{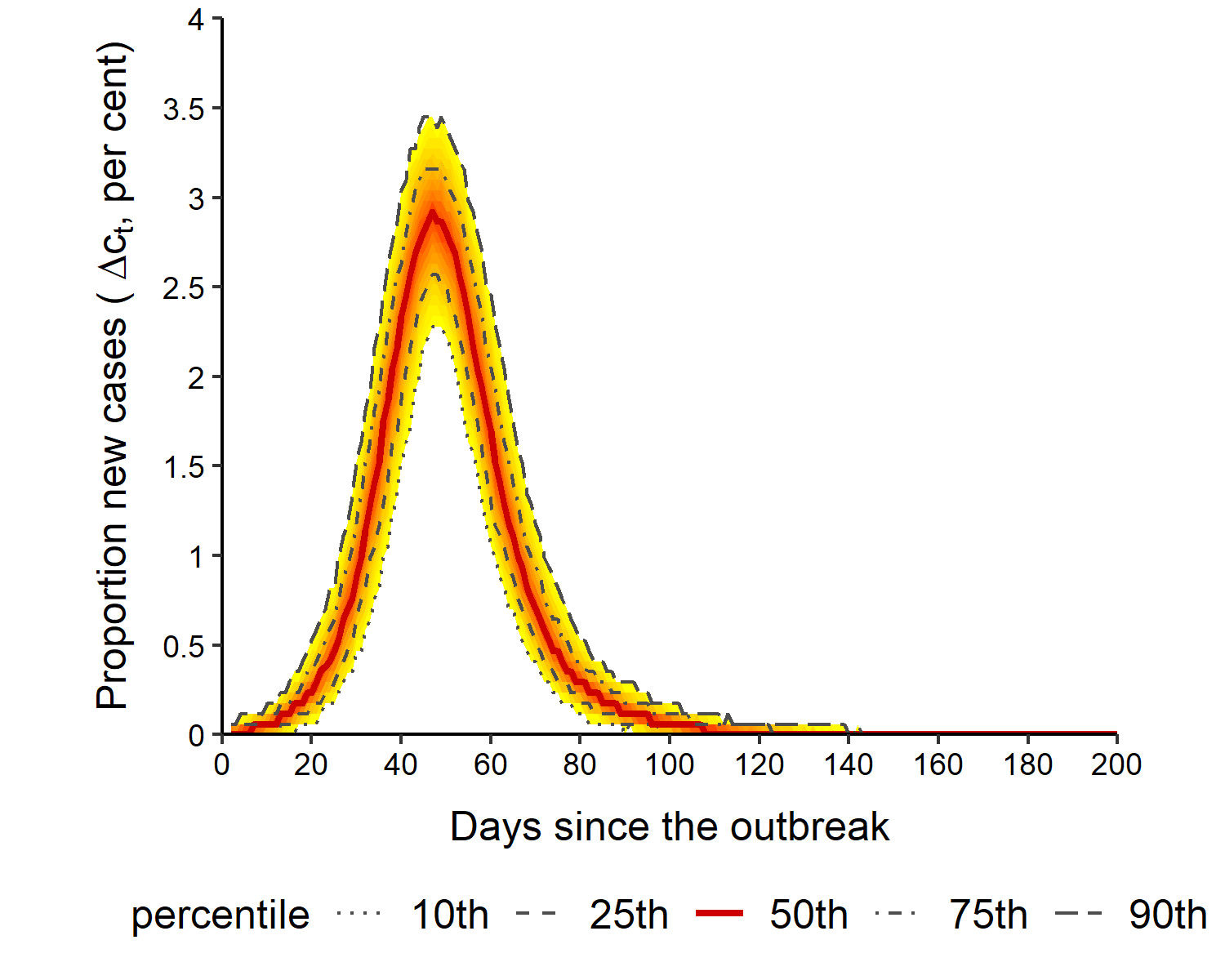}%
}
\\
$c_{1}^{\ast}=0.62$ &  & $c_{2}^{\ast}=0.95$\\
&  & \\
&  & \\
\textbf{Group 3: [30, 50)} &  & \textbf{Group 4: [50, 65)}\\%
{\includegraphics[
height=2.1465in,
width=2.6783in
]%
{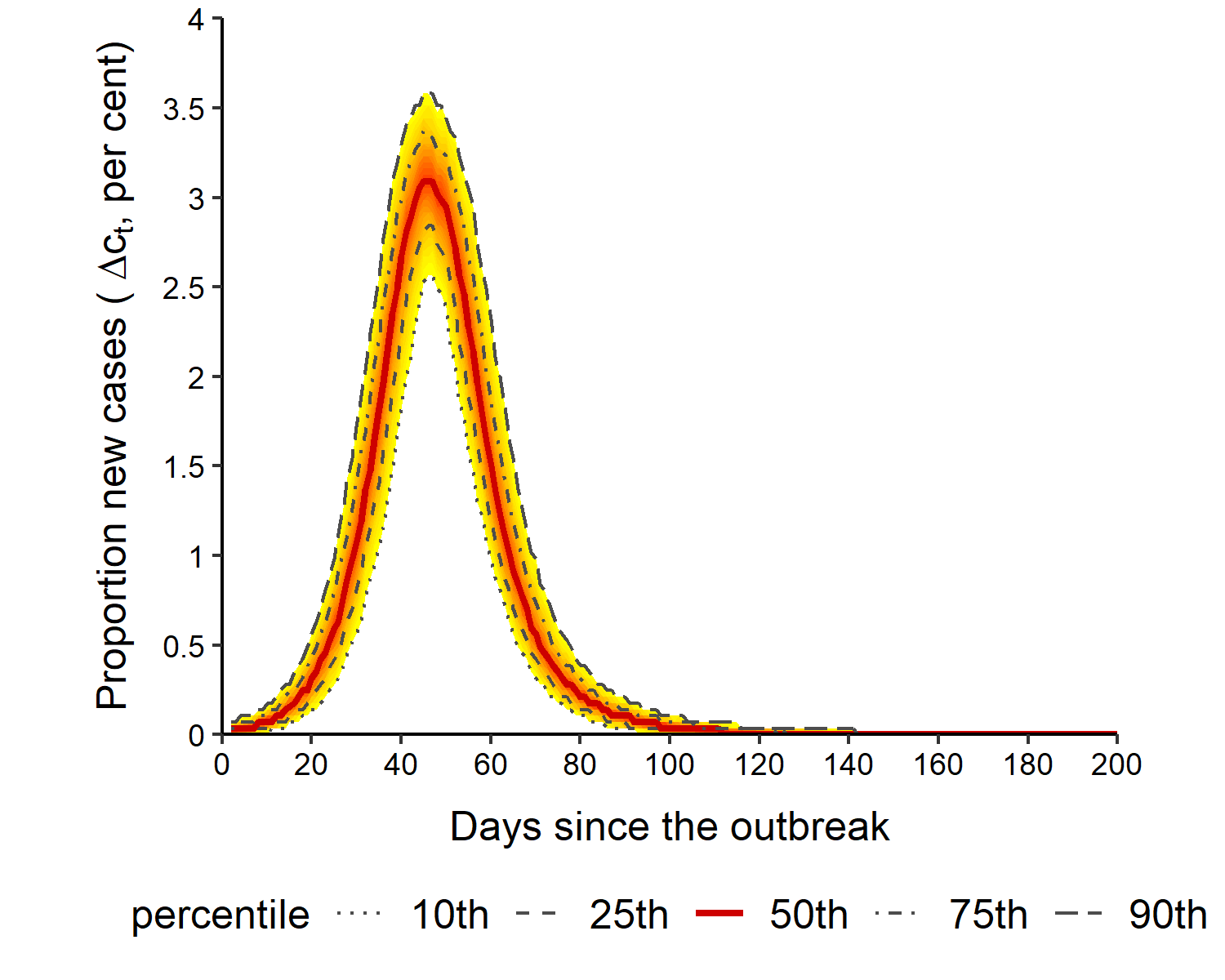}%
}
&  &
{\includegraphics[
height=2.1465in,
width=2.6783in
]%
{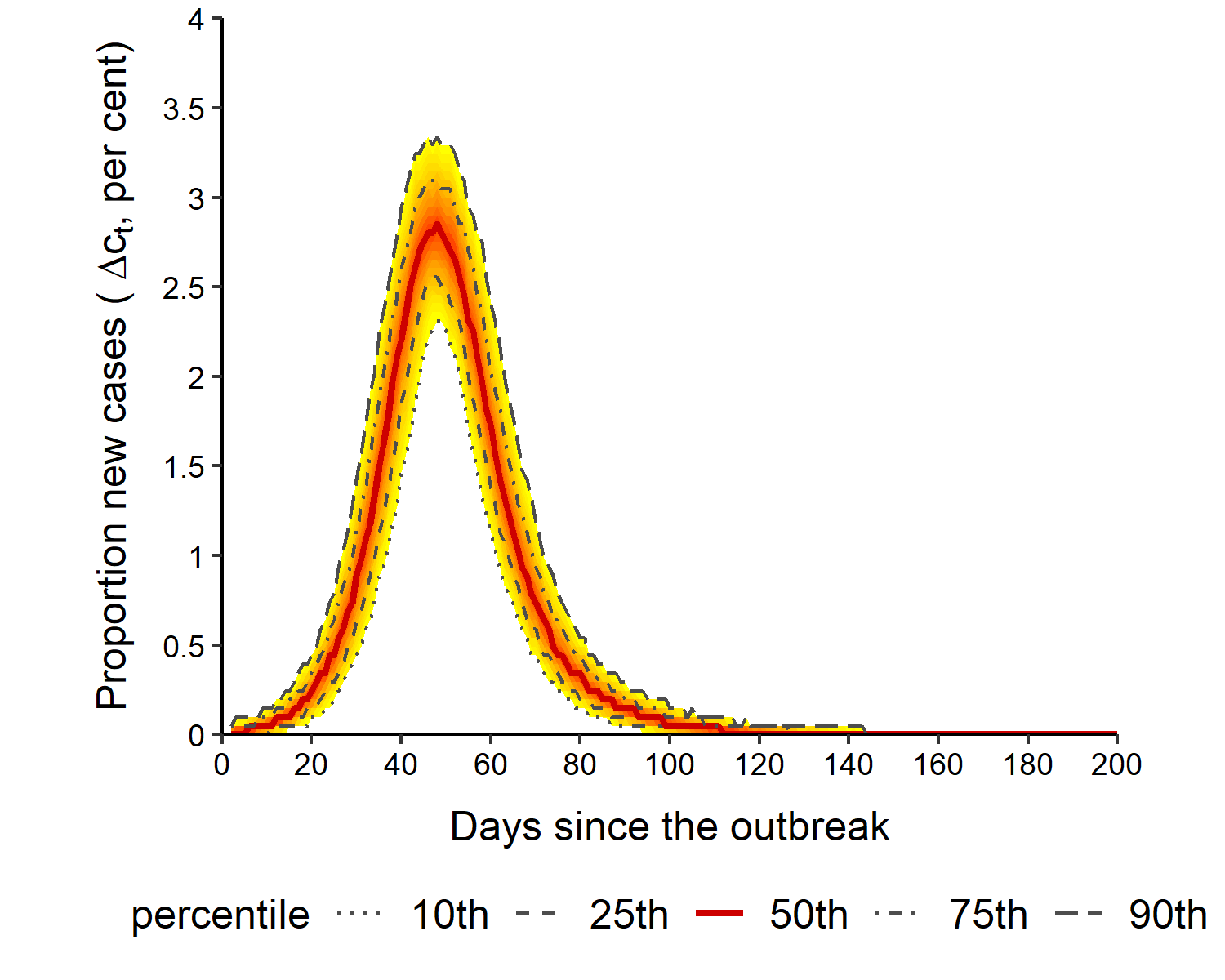}%
}
\\
$c_{3}^{\ast}=0.97$ &  & $c_{4}^{\ast}=0.94$\\
&  & \\
&  & \\
\textbf{Group 5: 65+} &  & \textbf{Aggregate}\\%
{\includegraphics[
height=2.1465in,
width=2.6783in
]%
{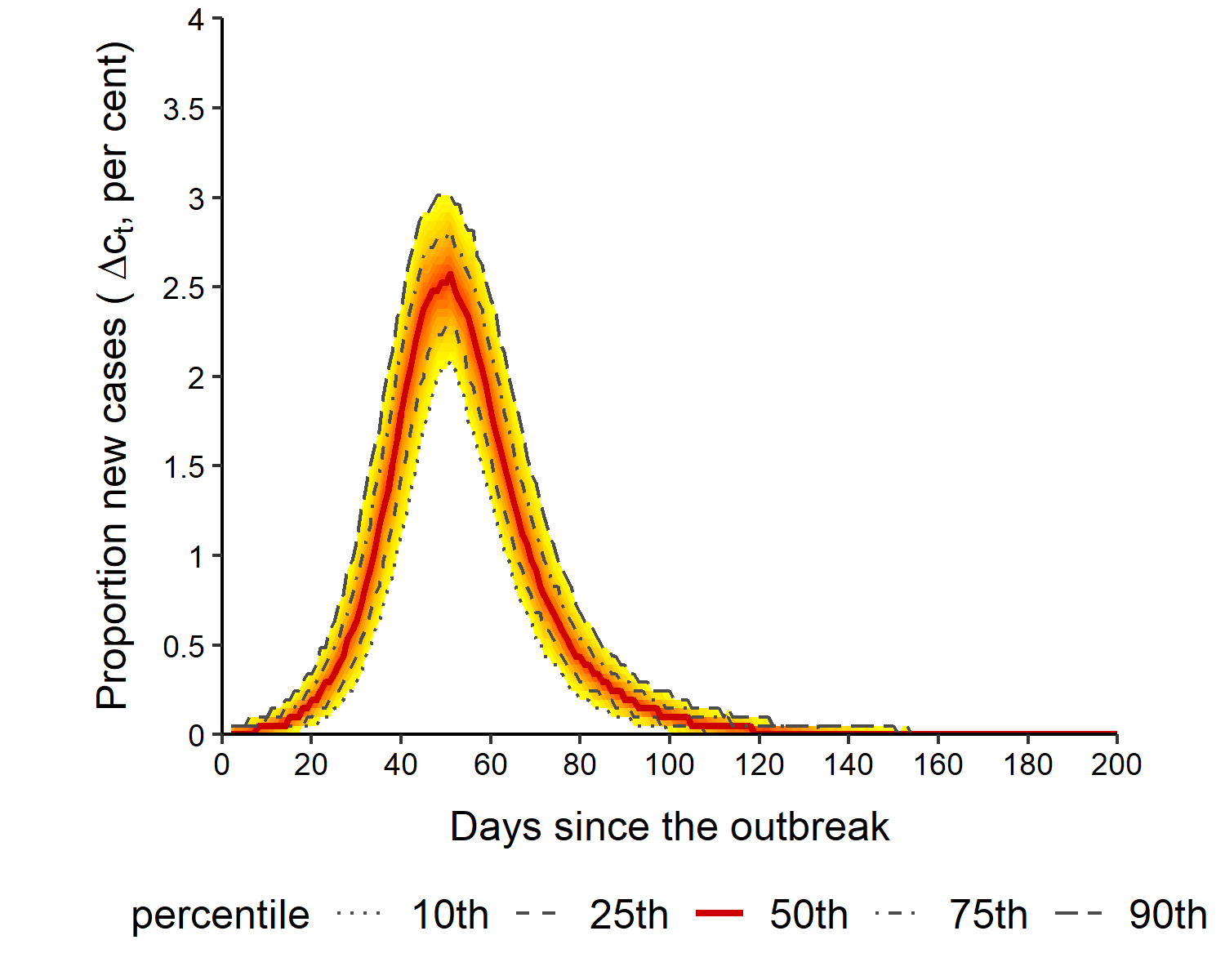}%
}
&  &
{\includegraphics[
height=2.1465in,
width=2.6783in
]%
{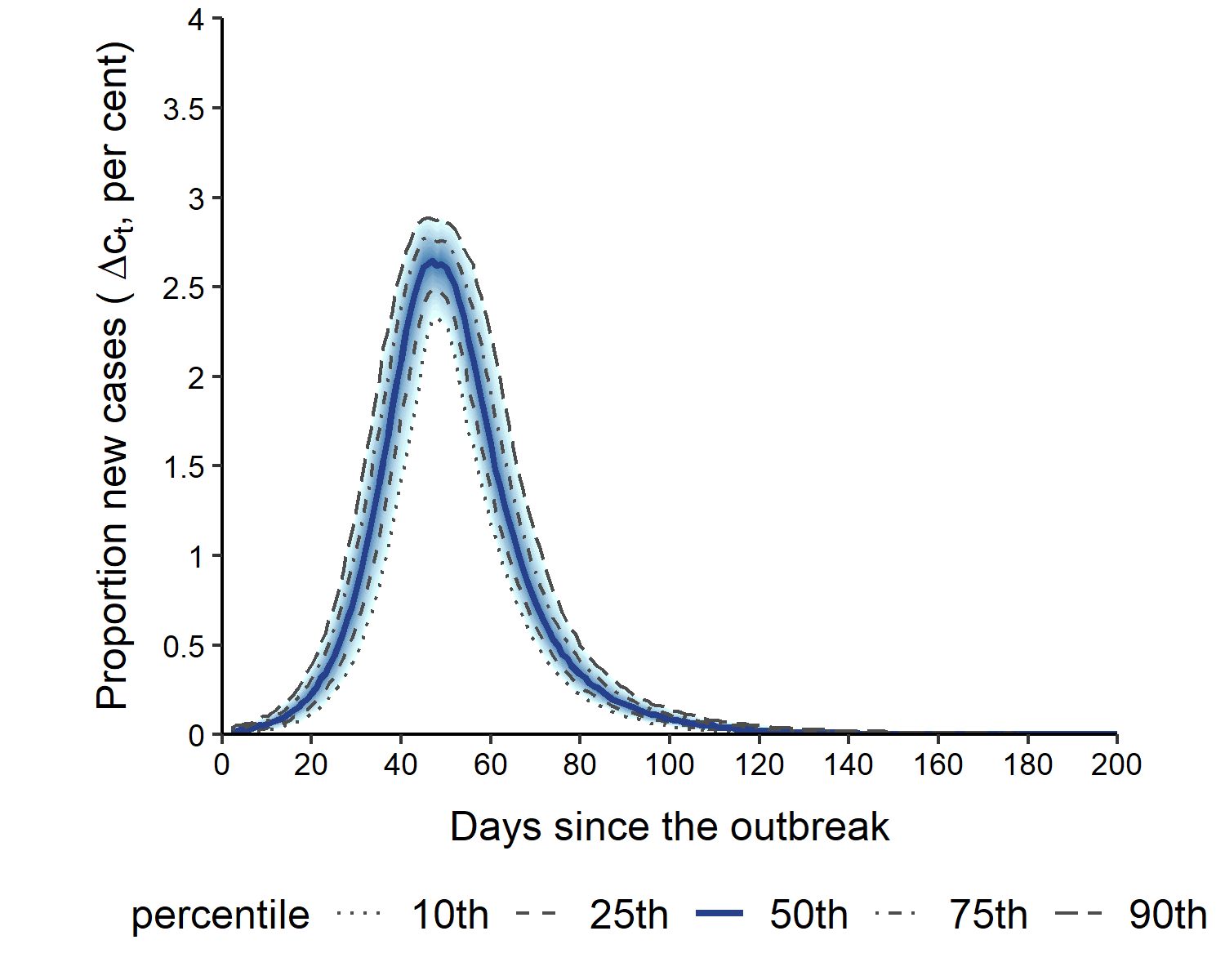}%
}
\\
$c_{5}^{\ast}=0.90$ &  & $c^{\ast}=0.90$%
\end{tabular}

\end{center}

%

\vspace{-0.2cm}%
Notes: The duration of the epidemic is $T^{\ast}=215$ days$.$ $c_{\ell}^{\ast
}=B^{-1}\sum_{b=1}^{B}\max_{t}c_{\ell t}^{(b)}$, for $\ell=1,2,\ldots,L$, and
$c^{\ast}=B^{-1}\sum_{b=1}^{B}\max_{t}c_{t}^{(b)}.$ The number of replications
is $B=1,000$. Population size is $n=10,000$.%

\end{figure}

Figure \ref{fig: multigroup} displays the simulated proportion of
group-specific and aggregate new cases in fan chart style with the $10^{th}$,
$25^{th}$, $50^{th}$, $75^{th},$ and $90^{th}$ percentiles over $1,000$
replications. The mean values are very close to the median and not shown. We
also report the maximum proportion of infected for each group averaged across
replications, i.e., $c_{\ell}^{\ast}=B^{-1}\sum_{b=1}^{B}\max_{t}c_{\ell
t}^{(b)}$, and the maximum proportion of aggregate infected, $c^{\ast}%
=B^{-1}\sum_{b=1}^{B}\max_{t}c_{t}^{(b)}$. The duration of the epidemic,
denoted by $T^{\ast}$, is computed as the number of days to reach zero active
cases averaged across replications.\footnote{Note that the model implies that
the disease will not spread again once $i_{t}$ becomes zero.}

Figure \ref{fig: multigroup} shows that if the disease transmits at a fixed
$\mathcal{R}_{0}=3$, the youngest age group will have the lowest maximum
proportion of infections, ending up with $62$ percent infected in comparison
to over $90$ percent infected in the other groups. The uncontrolled epidemic
is expected to end about $215$ days after the outbreak. The daily new cases
for the five groups peak around the same time (about $50$ days on average),
with the highest daily infection ranging from $2.0$ percent in the youngest
group to $3.7$ percent in the middle-aged group (Group 3). As a whole, the
maximum aggregate infection rate will reach $90$ percent, with daily new cases
peaking at $2.9$ percent of the population.%

In order to examine whether the number of groups affects the aggregate
outcomes, we carry out simulations using a single group model and compare the
results with the aggregate outcomes using the multigroup model. When there is
only one group, the contact network reduces to the Erd\H{o}s-R\'{e}nyi random
network (simply referred to as the random network below), where each pair of
the nodes (or individuals) are connected at random with a uniform probability
$p=k/\left(  n-1\right)  \approx k/n,$ where $k$ is the mean degree of the
network (or the mean number of contacts per individual).\footnote{The degree
of a node in a network is the number of connections it has (or the number of
edges attached to it).} We set the average number of contacts to $k=10$ based
on the literature on social contacts in the pre-Covid period, and then set the
exposure intensity parameter to $\tau=\beta/k=\gamma\mathcal{R}_{0}/k$, where,
as before, $\gamma=1/14$ and $\mathcal{R}_{0}=3$. Figure
\ref{fig: compare_multi_agg} of the online supplement shows that the simulated
aggregate outcomes are very close under the single- and multi-group models.
This finding is reasonable because the simulations were performed with the
same fixed $\beta=\gamma\mathcal{R}_{0}$. The heterogeneity in $\tau_{\ell}$
and $p_{\ell\ell^{^{\prime}}}$ affects the epidemic curves for each group but
does not seem to impact the aggregate outcomes. This result suggests that an
aggregate analysis may be justified if the primary focus is on the spread of
the infection across the population as a whole rather than on particular
age/type groups.

Lastly, to investigate whether the results are robust to different network
topologies, we considered another widely used contact network---the power law
random network, in which a small number of nodes (individuals) may have a
relatively high number of links (contacts). Figure \ref{fig: network_compare}
of the online supplement shows that the simulation outcomes obtained by the
Erd\H{o}s-R\'{e}nyi and the power law networks with the same average number of
contacts are very similar.

\section{Estimation of transmission rates \label{Sec: estimation R}}

The previous section investigates the properties of the model, assuming the
transmission rates are given. This section turns to detailing how to estimate
the transmission rate using data on infected cases. We first derive the method
of moments estimation of the transmission rate when there are no measurement
errors and present the finite sample properties of the estimators using Monte
Carlo techniques. We then allow for under-reporting of infected cases and
propose a recursive joint estimation of the transmission rate and the degree
of under-reporting by a simulated method of moments.

\subsection{Estimation without measurement
errors\label{Sec: estimation R no errors}}

Let us first consider the case of a single group, and recall that the moment
condition for this case is given by (\ref{cagg_L1}), which is replicated here
for convenience%
\begin{equation}
E\left(  \left.  \frac{1-c_{t+1}}{1-c_{t}}\right\vert \text{ }i_{t}\right)
=e^{-\beta i_{t}}+O\left(  n^{-1}\right)  . \label{cagg_L1_copy}%
\end{equation}
We can estimate the transmission rate, $\beta$, using (\ref{cagg_L1_copy}) by
nonlinear least squares (NLS) given time series data on $\left\{  c_{t}%
,i_{t}\right\}  $. The recovery rate, $\gamma$, can be estimated using the
recovery equation, (\ref{rlmoments}), $E\left(  r_{t+1}|r_{t},c_{t}\right)
=\left(  1-\gamma\right)  r_{t}+\gamma c_{t}$. Nevertheless, in reality,
$r_{t}$ is often not recorded in a timely manner and $\gamma$ is estimated
from the hospitalization data. We therefore set $\gamma=1/14$ in our
estimation and calibration exercises,\footnote{For the rationale behind
setting $\gamma=1/14$, see Footnote \ref{gamma14}.} and discuss the properties
of the moment estimator of $\gamma$ in the online supplement.

In the absence of any interventions (voluntary or mandatory), we have
$\beta=\gamma\mathcal{R}_{0}$, where, as before, $\mathcal{R}_{0}$ is the
basic reproduction number. It follows that $\mathcal{R}_{0}$ can be estimated
by $\mathcal{\hat{R}}_{0}=\hat{\beta}/\gamma$, where $\hat{\beta}$ is the NLS
estimate of $\beta$ using (\ref{cagg_L1_copy}). Under social interventions,
the recovery equation holds (since $\gamma$ is unaffected), but the moment
condition for $c_{t+1}$ now depends on the time-varying transmission rate,
$\beta_{t}$. For $\gamma$ in the range of $1/14$ to $1/21$, it is reasonable
to use two or three weeks rolling windows when estimating $\beta_{t}$. For a
window of size $W$, we have
\begin{equation}
\hat{\beta}_{t}\left(  W\right)  =\text{Argmin}_{\beta}\sum_{\tau=t-W+1}%
^{t}\left(  \frac{1-c_{\tau}}{1-c_{\tau-1}}-e^{-\beta\text{ }i_{\tau-1}%
}\right)  ^{2}. \label{betathat}%
\end{equation}
Note that even though\ the time series $\left\{  c_{t},i_{t}\right\}  $ over
the course of the epidemic are non-stationary, the rolling estimation is based
on short-$T$ series ($T=14$ or $21$).

To examine the finite sample performance of $\hat{\beta}_{t}\left(  W\right)
$, we estimate $\beta$ using the simulated data generated from the single
group stochastic SIR\ model on a random network with mean contact $k=10$ and
assuming $1/1000$ of the population is randomly infected on day $1$. The true
value of the transmission rate is set to $\beta=3/14$ such that $\mathcal{R}%
_{0}=\beta/\gamma=3$. We consider population sizes $n=10,000$, $50,000,$ and
$100,000$, and set the the number of replications to $B=1,000$. Recall that
$T$ is fixed and $n\rightarrow\infty$. To alleviate noise induced by zero and
near-zero observations at the start and final stages of the epidemic, the
rolling estimation of $\beta$ is carried out over the $4^{th}-15^{th}$ weeks
after the outbreak.

Since the value of $\beta$ is quite small, we present the estimation results
in terms of $\mathcal{\hat{R}}_{0}\left(  W\right)  =\hat{\beta}_{t}\left(
W\right)  /\gamma$. Table \ref{tab: R0_2W} summarizes the bias and root mean
square error (RMSE) of the 2-weekly rolling estimates of $\mathcal{R}_{0},$
averaged over the four non-overlapping 3-weekly sub-samples, for different
population sizes. The bias is computed as $B^{-1}\sum_{b=1}^{B}\left[
\mathcal{\hat{R}}_{0}^{\left(  b\right)  }\left(  W\right)  -\mathcal{R}%
_{0}\right]  $, and the RMSE is computed by $\sqrt{B^{-1}\sum_{b=1}^{B}\left[
\mathcal{\hat{R}}_{0}^{\left(  b\right)  }\left(  W\right)  -\mathcal{R}%
_{0}\right]  ^{2}}$, where $\mathcal{\hat{R}}_{0}^{\left(  b\right)  }\left(
W\right)  =\hat{\beta}_{t}^{(b)}\left(  W\right)  /\gamma$ and $\hat{\beta
}^{(b)}\left(  W\right)  $ is the estimate of $\beta$ in the $b^{th}$
simulated sample. As can be seen from Table \ref{tab: R0_2W}, although
$\mathcal{\hat{R}}_{0}$ tends to slightly underestimate $\mathcal{R}_{0}$, its
bias and RMSE\ are quite small in all experiments and sub-samples. The RMSE
declines as the population size $n$ increases, but $\mathcal{R}_{0}$ can be
estimated reasonably well even with $n=10,000$. Comparing the results over
different epidemic stages, the RMSE is relatively larger at the early and late
stages of the epidemic. This finding is not surprising since it is difficult
to obtain precise estimates when $c_{t}$ and $i_{t}$ are near zero. We also
considered the 3-weekly rolling estimates, reported in Table \ref{tab: R0_3W}
of the online supplement, and as can be seen are very close to the 2-weekly
estimates, with slightly better performance in the early and late stages of
the epidemic. We will hereafter mainly focus on the 2-weekly rolling estimation.%

\begin{table}[t]%
\caption
{Finite sample properties of the 2-weekly rolling estimates of $\mathcal
{R}_0$, in the case where it is fixed at ${\mathcal{R}_0}=3$}%
\label{tab: R0_2W}%
\renewcommand{\arraystretch}{1.1}%
\vspace{-0.35cm}%

\begin{center}%
\begin{tabular}
[c]{llcccc}\hline
&  & \multicolumn{4}{c}{3-weekly sub-samples}\\\cline{3-6}%
\multicolumn{2}{r}{Weeks since the outbreak} & $4^{th}-6^{th}$ &
$7^{th}-9^{th}$ & $10^{th}-12^{th}$ & $13^{th}-15^{th}$\\\hline
Population &  &  &  &  & \\
$n=10,000$ & Bias & \multicolumn{1}{r}{-0.0108} & \multicolumn{1}{r}{-0.0038}
& \multicolumn{1}{r}{-0.0006} & \multicolumn{1}{r}{0.0014}\\
& RMSE & \multicolumn{1}{r}{0.0988} & \multicolumn{1}{r}{0.0563} &
\multicolumn{1}{r}{0.1048} & \multicolumn{1}{r}{0.2436}\\
$n=50,000$ & Bias & \multicolumn{1}{r}{-0.0017} & \multicolumn{1}{r}{-0.0001}
& \multicolumn{1}{r}{-0.0010} & \multicolumn{1}{r}{-0.0012}\\
& RMSE & \multicolumn{1}{r}{0.0405} & \multicolumn{1}{r}{0.0251} &
\multicolumn{1}{r}{0.0481} & \multicolumn{1}{r}{0.1070}\\
$n=100,000$ & Bias & \multicolumn{1}{r}{-0.0002} & \multicolumn{1}{r}{0.0005}
& \multicolumn{1}{r}{-0.0003} & \multicolumn{1}{r}{-0.0009}\\
& RMSE & \multicolumn{1}{r}{0.0282} & \multicolumn{1}{r}{0.0172} &
\multicolumn{1}{r}{0.0335} & \multicolumn{1}{r}{0.0771}\\\hline
\end{tabular}

\end{center}

%

\footnotesize\flushleft
{}Notes: The true value of $\mathcal{R}_{0}$ is set to $\beta/\gamma$, where
$\beta=3/14$ and $\gamma=1/14$ so that $\mathcal{R}_{0}=3$. We fix $\gamma$
and estimate $\beta$ using (\ref{betathat}). The number of replications is
$B=1,000$.%

\end{table}%

Similar moment conditions can also be used to estimate the parameters of the
multigroup model. If time series data on $\left\{  c_{\ell t},i_{\ell
t}\right\}  $, for $\ell=1,2,\ldots,L$ ($L$ is finite) are available, we can
estimate $\beta_{\ell\ell^{\prime}}$ using the moment conditions
(\ref{clmoments3}), namely,%
\[
E\left(  \left.  \frac{1-c_{\ell,t+1}}{1-c_{\ell t}}\right\vert \text{
}\mathbf{i}_{t}\right)  =\exp\left(  -\sum_{\ell^{\prime}=1}^{L}\beta
_{\ell\ell^{\prime}}i_{\ell^{\prime}t}\right)  +O\left(  n^{-1}\right)  .
\]
Then, as we have discussed in Section \ref{Sec: calibration R}, $\beta
_{\ell\ell^{\prime}}$ is identifiable from (\ref{R0_pll'}) for given values of
$\gamma$ and $\mathbf{w=(}w_{1},w_{2},\ldots,w_{L})^{\prime}.$

\subsection{Estimation allowing for measurement
errors\label{Sec: estimation R with MF}}

It is widely recognized that in practice $c_{t}$ and $r_{t}$ are
under-reported. The magnitude of under-reporting is measured by the
multiplication factor (MF) in the literature (see, e.g., Gibbons et al.,
2014). It is expected that the MF will decline over time since data quality
will improve as more testing is conducted, but, in any case, MF is certainly
greater than one. Denoting the multiplication factor by $m_{t},$ and denoting
the observed values of $c_{t}$ and $i_{t}$ by $\tilde{c}_{t}$ and
$\tilde{\imath}_{t}$, respectively, we have $c_{t}=m_{t}\tilde{c}_{t}$ and
$i_{t}=c_{t}-r_{t}=m_{t}\tilde{\imath}_{t}$ (assuming that $r_{t}%
=R_{t}/n=m_{t}\tilde{r}_{t}$, where $\tilde{R}_{t}$ is the observed value of
$R_{t}$). Then the moment condition in terms of the observed values
($\tilde{c}_{t}$ and $\tilde{\imath}_{t}$) can be written as%
\begin{equation}
E\left(  \left.  \frac{1-m_{t+1}\tilde{c}_{t+1}}{1-m_{t}\tilde{c}_{t}%
}\right\vert \text{ }\tilde{\imath}_{t},\tilde{c}_{t}\right)  =e^{-\beta
_{t}m_{t}\tilde{\imath}_{t}}+O\left(  n^{-1}\right)  . \label{cm1}%
\end{equation}
It can be seen from (\ref{cm1}) that $m_{t}$ is not identified when $\tilde
{c}_{t}$ and $\tilde{\imath}_{t}$ are very small in the early stage of the
epidemic. When $\tilde{c}_{t}$ becomes large enough, we can estimate $m_{t}$
by the simulated method of moments based on (\ref{cm1}). In practice, $m_{t}$
varies slowly, and it is reasonable to assume $m_{t}=m_{t-1}$ within a short
time interval (two or three weeks). Then we have%
\begin{equation}
\frac{1-m_{t}\tilde{c}_{t}}{1-m_{t}\tilde{c}_{t-1}}=B^{-1}\sum_{b=1}%
^{B}e^{-\beta_{t-1}i_{t-1}^{\left(  b\right)  }}\text{,} \label{SMM}%
\end{equation}
where $i_{t}^{\left(  b\right)  }$ denotes the simulated value of $i_{t}$ in
the $b^{th}$ replication and $B$ is the total number of replications. Solving
(\ref{SMM}) for $m_{t}$ yields%
\begin{equation}
m_{t}=\frac{1-B^{-1}\sum_{b=1}^{B}e^{-\beta_{t}i_{t-1}^{\left(  b\right)  }}%
}{\tilde{c}_{t}-\left(  B^{-1}\sum_{b=1}^{B}e^{-\beta_{t-1}\imath
_{t-1}^{\left(  b\right)  }}\right)  \tilde{c}_{t-1}}. \label{mt}%
\end{equation}
It is now clear that one can estimate $m_{t}$ by (\ref{mt}) for given values
of $\beta_{t}$, and estimate $\beta_{t}$ by (\ref{betathat}) if $m_{t}$ is
known. Accordingly, we propose a method that estimates $\beta_{t}$ and $m_{t}$
jointly. The algorithm is described in detail in Section \ref{Sup: estimation}
of the online supplement. We apply the procedure recursively using 2- and
3-weekly rolling windows in the next section to examine how the transmission
rates and under-reporting of cases changed over time in a number of countries
and evaluate how our model matches the Covid-19 evidence.

\section{Matching the model with evidence from a number of European countries
\label{Sec: empirical}}

We now assess how our model matches with the recorded cases in six European
countries: Austria, France, Germany, Italy, Spain, and the United Kingdom,
while taking account of under-reporting of infections.\footnote{We also
examined how our model matches the Covid-19 evidence in the US. The estimates
of $\mathcal{R}_{et}$ for the country as a whole and for each of the $48$
mainland states and the District of Columbia are displayed in Section
\ref{Sup: empirical_Re_US} of the online supplement. The estimates of the
multiplication factor and the comparison between the realized and calibrated
new cases are presented in Figure \ref{fig: US_MF} of the online supplement.}
The Covid-19 outbreak in Europe began with Italy in early February 2020, with
the recorded number of infections accelerating rapidly from February 21
onward. A rapid rise in infections took place about one week later in Spain,
France, and Germany, followed by UK and Austria at the end of February.

To estimate $\beta_{t}$, we need observations on per capita infected and
active cases, $c_{t}$ and $i_{t}$. Using the recorded number of infected
cases, $\tilde{C}_{t}$, and population data, the per capita cases, $\tilde
{c}_{t}$, are readily available, where as before we use the tilde symbol to
indicate observed values. Since $\tilde{I}_{t}=\tilde{C}_{t}-\tilde{R}_{t}$,
we can obtain $\tilde{I}_{t}$ if the number of removed (the sum of those who
recovered and the deceased) cases, $\tilde{R}_{t}$, is available.
Unfortunately, the recovery data is either not reported or is subject to
severe measurement error/reporting issues in many countries. For all six
countries, we therefore estimate the number of removed using the recursion
$\tilde{R}_{t}=\left(  1-\gamma\right)  \tilde{R}_{t-1}+\gamma\tilde{C}_{t-1}%
$, for $t=2,3,\ldots,T$, where the recovery rate $\gamma$ is set to $1/14,$
and values of $\tilde{R}_{t}$ are generated starting with $\tilde{R}%
_{1}=\tilde{C}_{1}=0$. We then compute $\tilde{I}_{t}$ by subtracting the
estimated $\tilde{R}_{t}$ from the recorded $\tilde{C}_{t}$%
.\footnote{Specifically, among the six countries, the recorded data on
recovery are unavailable for Spain and UK; they are of poor quality for France
and Italy; they are relatively close to our estimated recovery for Austria and
Germany. We have also calibrated our model using the recorded recovery data
for Austria and Germany as a robustness check and obtained similar results.}
That is, in this empirical exercise, we only need data on Covid-19 cases per
capita, $\tilde{c}_{t}$. To alleviate the wide fluctuations in the data due to
irregular update schedules and reporting/recording delays, we smooth the
series by taking the 7-day moving average before they are used in the
estimation and calibration.

We adopt the joint estimation approach proposed in the last section to
calibrate and evaluate our stochastic network model. The procedure is applied
recursively using 2- and 3-weekly rolling windows. Recall that in the early
stage of an epidemic, $\tilde{c}_{t}$ is small, and MF\ is not identified. We
choose $\tilde{c}^{0}=0.01$ as the threshold value.\footnote{For the countries
we considered, there is virtually no difference in the estimates of $\beta
_{t}$ when $c_{t}\leq0.01$ if MF takes the value from $2$ to $7$.} When
$\tilde{c}_{t}\leq\tilde{c}^{0}$, we use an initial guess of the
multiplication factor, MF $=5$, in the estimation of $\beta_{t}$. Since the
early Covid data are quite noisy, we start the rolling estimation when the
daily new cases exceed one per $100,000$ people and use the estimates below
$\mathcal{R}_{0}=3$ in the calibration. Specifically, the simulations begin
with $1/1000$ of the population randomly infected on day $1$. To render the
calibrations comparable across the countries in our sample, during the first
week after the outbreak we set the value of $\beta$ such that $\mathcal{R}%
_{0}$ equals its first estimate, $\hat{\beta}_{t}/\gamma$, that is less than
$3$. Then, from the second week onwards, we set $\beta_{t}$ to the rolling
estimates computed from the realized data (with MF $=5$) until $\tilde{c}_{t}$
reaches $\tilde{c}^{0}$ on day $t^{0}$. As shown in Section
\ref{Sec: properties}, it makes little difference to the aggregate outcomes
whether we carry out the simulations using single- or multi-group models.
Since we are interested in comparing the calibrated outcomes with realized
cases, we conduct simulations using the single group model with the
Erd\H{o}s-R\'{e}nyi random network in this exercise. The first estimate of MF
is computed as the ratio of the average calibrated cases to realized cases on
day $t^{0}$. When $\tilde{c}_{t}>\tilde{c}^{0}$, we perform the joint rolling
estimation of $\beta_{t}$ and $m_{t}$ using (\ref{joint_est_beta}) and
(\ref{joint_est_MF}). We present the 2-weekly estimation and calibration
results in the main paper. The results using the 3-weekly rolling windows are
very close and are given in the online supplement.\footnote{Figures
\ref{fig: Euro_Re_cmp_2W_3W} and \ref{fig: Euro_MF_cmp_2W_3W} of the online
supplement compares the 2- and 3-weekly estimates of $\mathcal{R}_{et}$ and
MF, respectively. We find that the 2- and 3-weekly estimates of $\mathcal{R}%
_{et}$ are quite similar. The 3-weekly estimates of MF tend to be slightly
higher than the 2-weekly estimates, but overall they are very close.} Since
the moment condition, (\ref{cm1}), used in the joint estimation was derived
assuming no vaccination, we end the joint estimation when the recorded share
of the population fully vaccinated reaches $10$ percent. The population size
in simulations is set to $n=50,000$. To ease the computational burden, the
number of replications is set to $B=500$.%

\begin{figure}[htp]%
\caption{Rolling estimates of the effective reproduction numbers ($\mathcal
{R}_{et}%
$) for selected European countries, with start and end dates of the respective lockdowns}%
\vspace{-0.2cm}%
\label{fig: Euro_Re}%

\begin{footnotesize}%

\begin{center}%
\begin{tabular}
[c]{ccc}%
Austria &  & France\\%
{\includegraphics[
height=1.9951in,
width=2.6524in
]%
{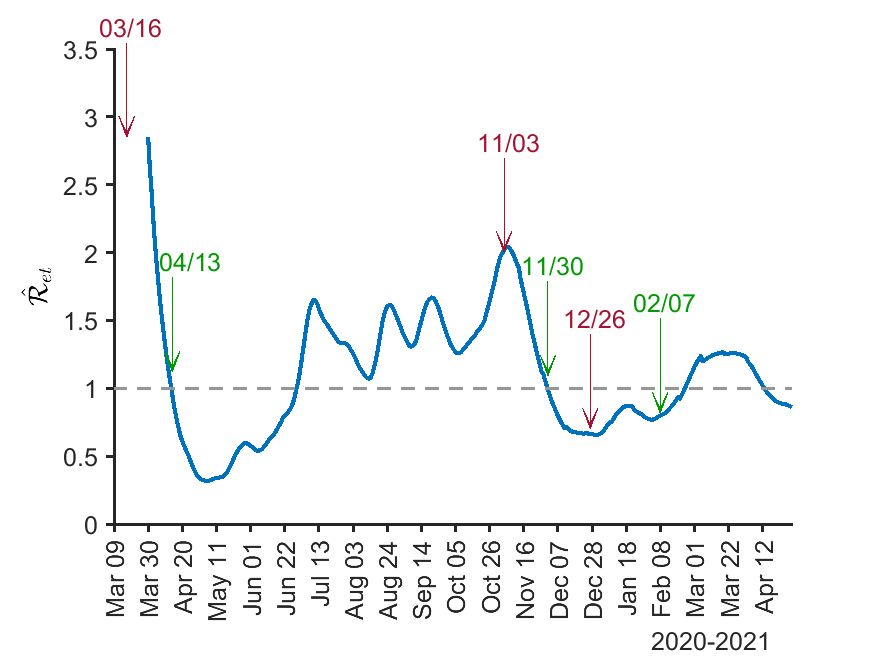}%
}
&  &
{\includegraphics[
height=1.9951in,
width=2.6524in
]%
{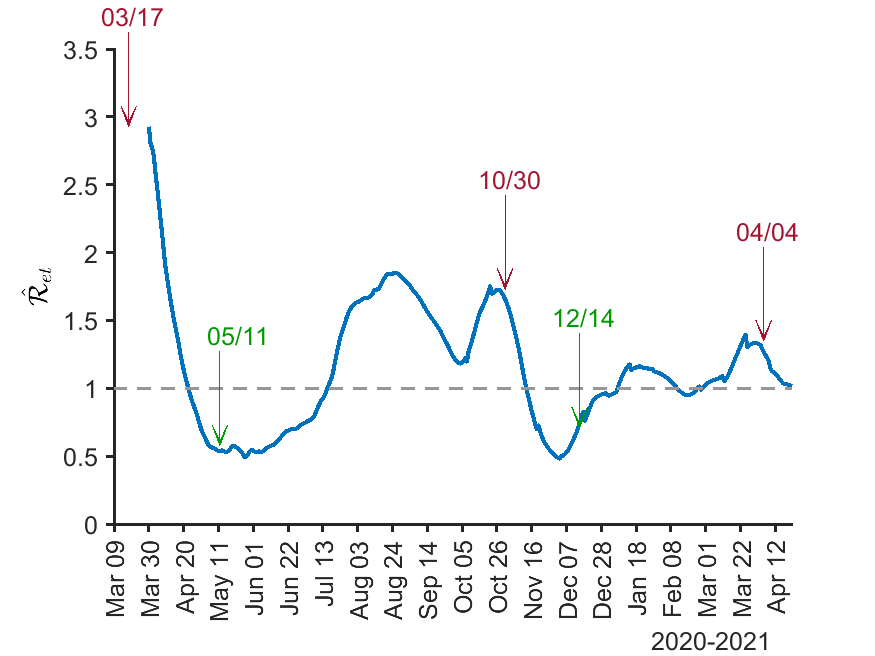}%
}
\\
&  & \\
Germany &  & Italy\\%
{\includegraphics[
height=1.9951in,
width=2.6524in
]%
{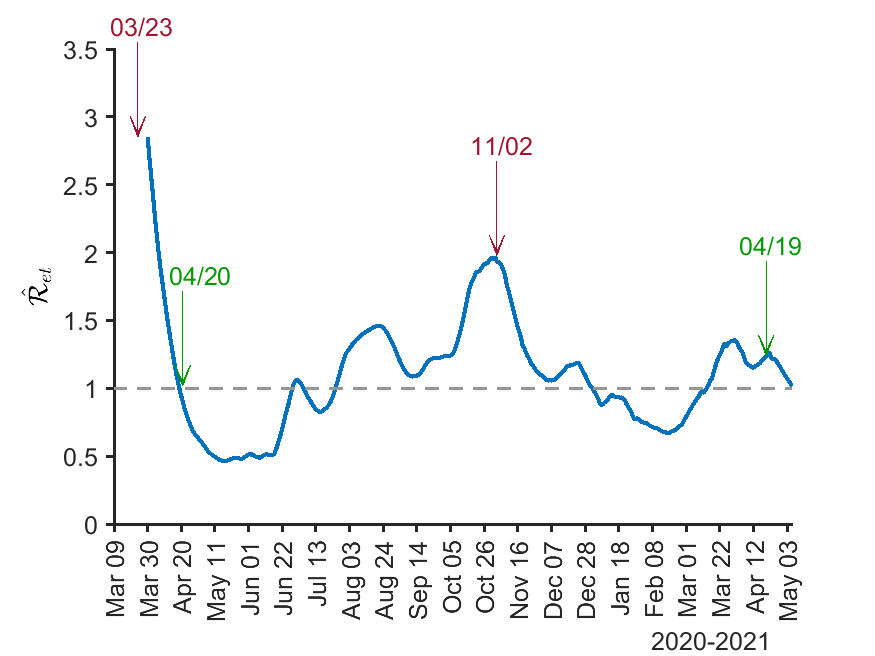}%
}
&  &
{\includegraphics[
height=1.9951in,
width=2.6524in
]%
{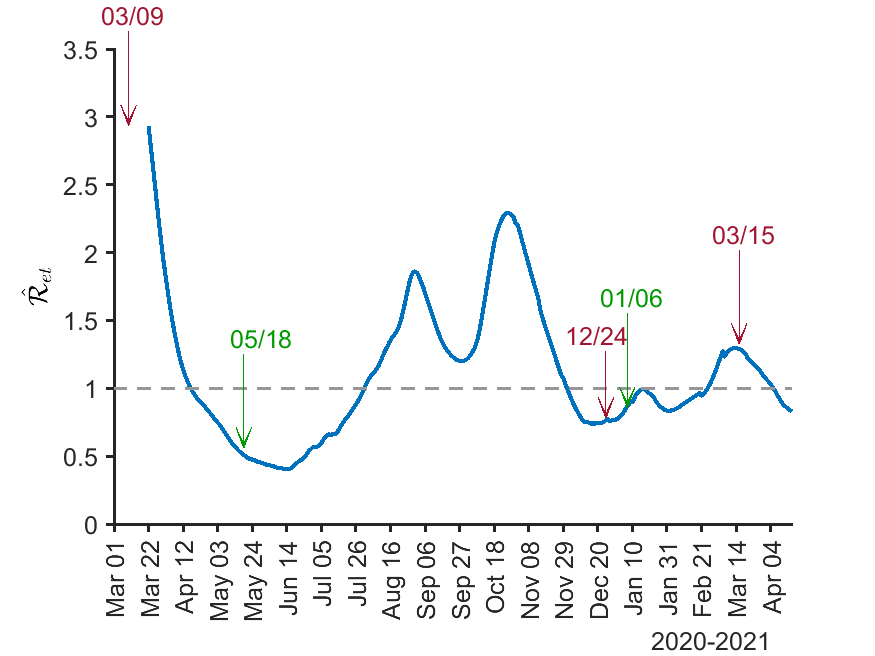}%
}
\\
&  & \\
Spain &  & UK\\%
{\includegraphics[
height=1.9951in,
width=2.6524in
]%
{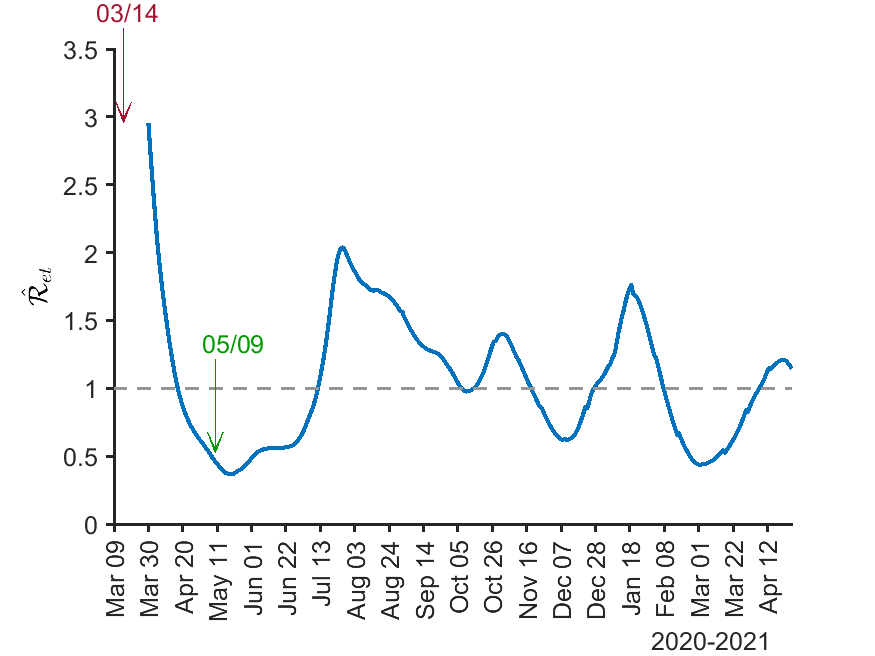}%
}
&  &
{\includegraphics[
height=1.9951in,
width=2.6524in
]%
{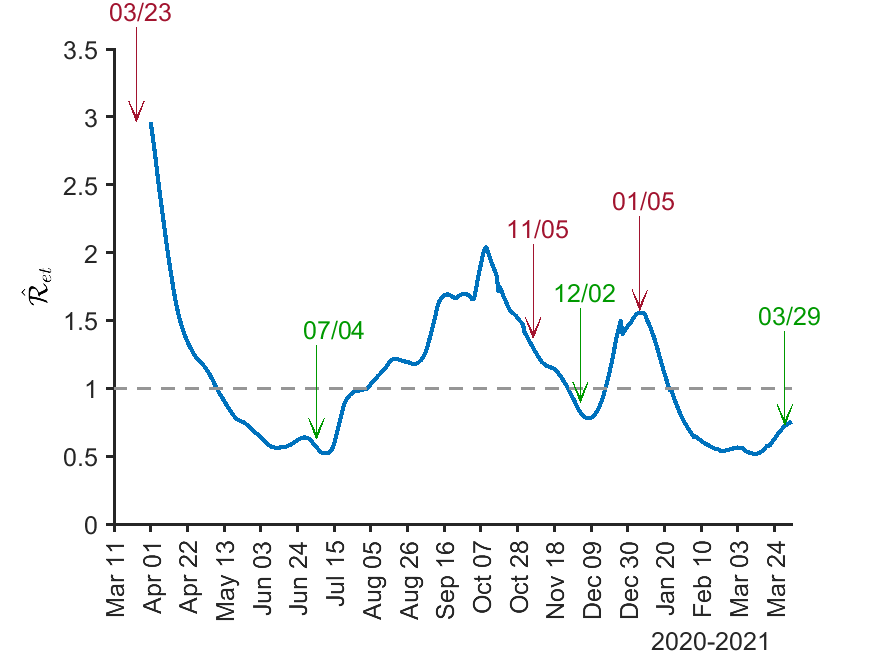}%
}
\end{tabular}

\end{center}

%

\vspace{0cm}%
\footnotesize
Notes: $\mathcal{\hat{R}}_{et}=\left(  1-\hat{m}_{t}\tilde{c}_{t}\right)
\hat{\beta}_{t}/\gamma$, where $\tilde{c}_{t}$ is the reported number of
infections per capita and $\gamma=1/14$. $W_{\beta}=W_{m}=2$ weeks. The joint
estimation starts when $\tilde{c}_{t}>0.01$. The initial guess estimate of the
multiplication factor is $5$. The simulation uses the single group model with
the random network and population size $n=50,000$. The number of replications
is $500$. The number of removed (recoveries + deaths) is estimated recursively
using $\tilde{R}_{t}=\left(  1-\gamma\right)  \tilde{R}_{t-1}+\gamma\tilde
{C}_{t-1}$ for all countries, with $\tilde{C}_{1}=\tilde{R}_{1}=0$. Red
(green) arrows indicate the start (end) dates of the respective lockdowns.%

\end{footnotesize}%
%

\end{figure}%

\begin{figure}[htp]%
\caption
{Rolling estimates of the multiplication factor for selected European countries}%
\vspace{-0.2cm}%
\label{fig: Euro_MF}%

\begin{footnotesize}%

\begin{center}%
\begin{tabular}
[c]{ccc}%
Austria &  & France\\%
{\includegraphics[
height=1.9951in,
width=2.6524in
]%
{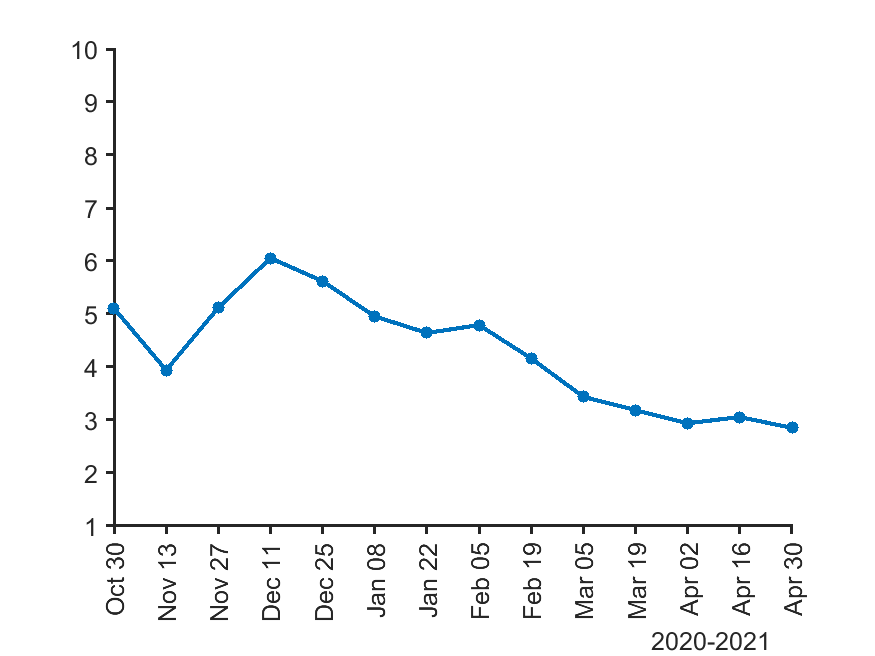}%
}
&  &
{\includegraphics[
height=1.9951in,
width=2.6524in
]%
{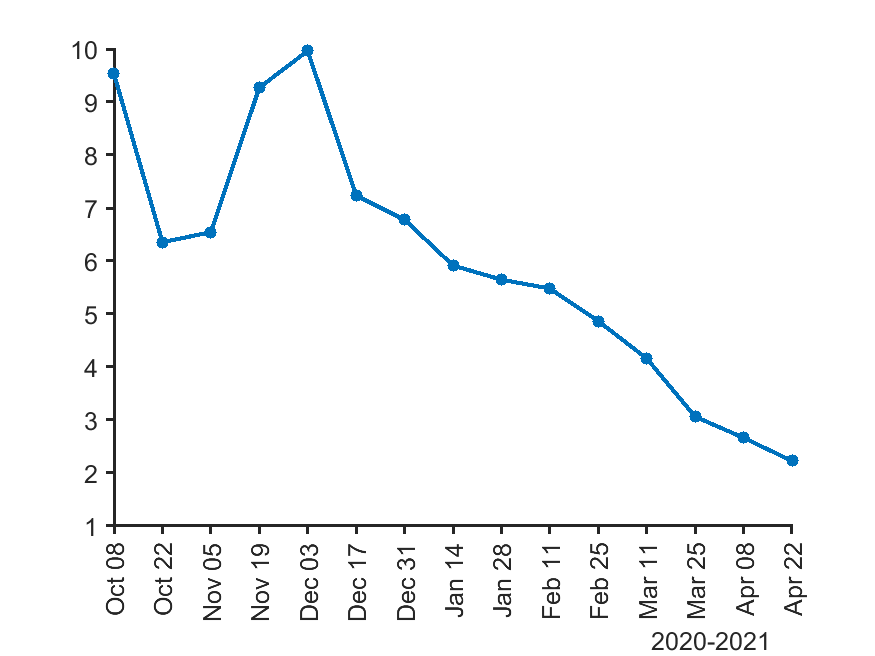}%
}
\\
&  & \\
Germany &  & Italy\\%
{\includegraphics[
height=1.9951in,
width=2.6524in
]%
{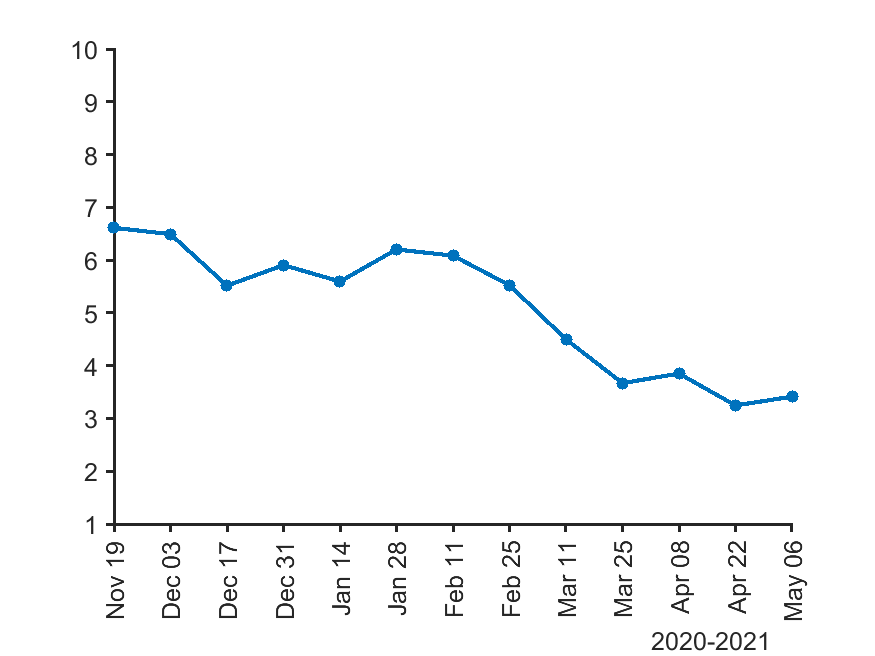}%
}
&  &
{\includegraphics[
height=1.9951in,
width=2.6524in
]%
{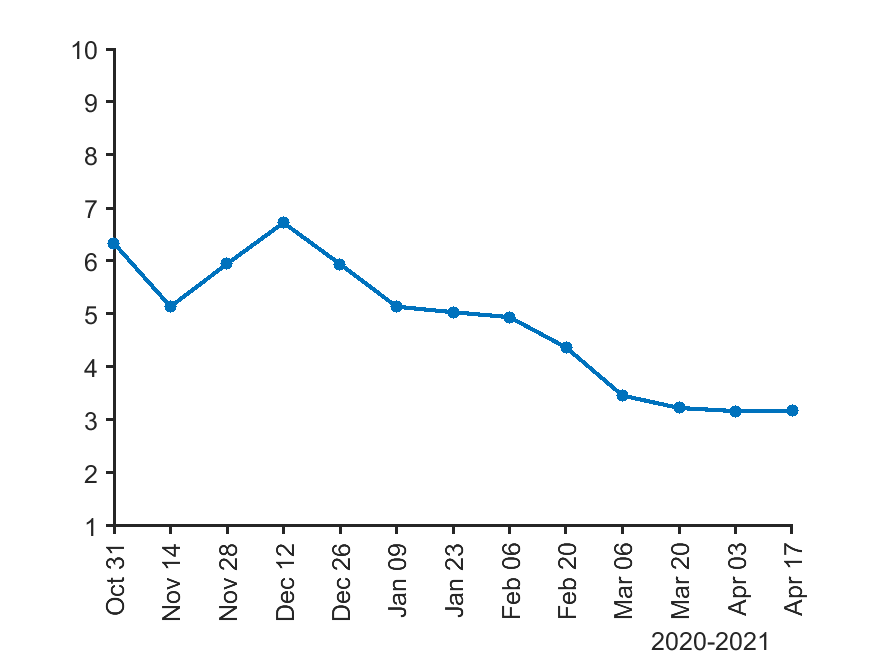}%
}
\\
&  & \\
Spain &  & UK\\%
{\includegraphics[
height=1.9951in,
width=2.6524in
]%
{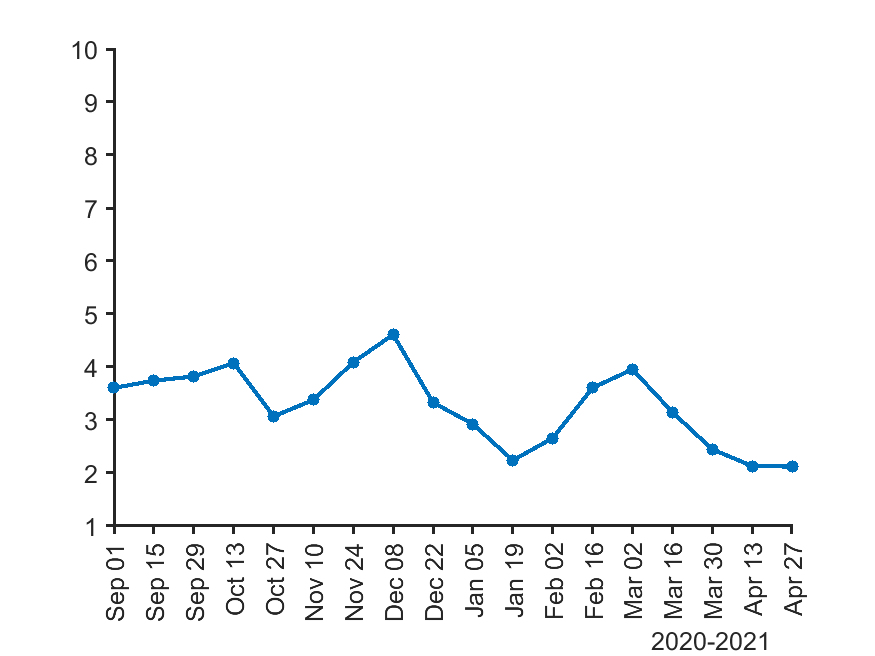}%
}
&  &
{\includegraphics[
height=1.9951in,
width=2.6524in
]%
{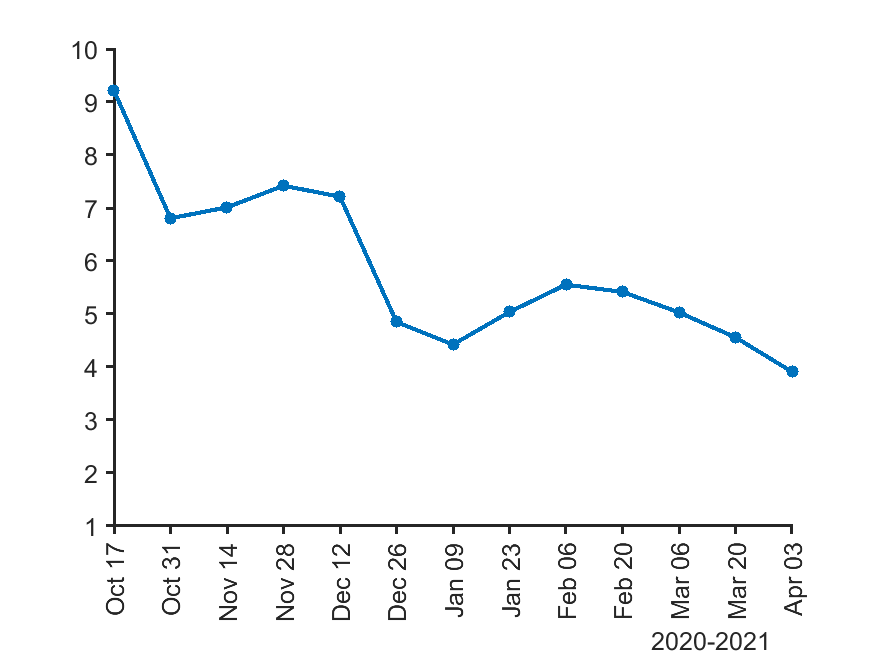}%
}
\end{tabular}

\end{center}

%

\vspace{0cm}%
\footnotesize
Notes: See the notes to Figure \ref{fig: Euro_Re}.%

\end{footnotesize}%
%

\end{figure}

Figure \ref{fig: Euro_Re} shows the evolution of $\mathcal{\hat{R}}_{et}$ over
the period March 2020--April 2021 for the six countries, where $\mathcal{\hat
{R}}_{et}=\left(  1-\hat{m}_{t}\tilde{c}_{t}\right)  \hat{\beta}_{t}/\gamma$
and $\hat{m}_{t}$ is displayed in Figure \ref{fig: Euro_MF}.\footnote{In the
early stages of the epidemic when $c_{t}$ is small, estimates of
$\mathcal{R}_{et}$ and $\beta_{t}/\gamma$ are very close, even if we set MF to
$10$. See Figure \ref{fig: Euro_TR_Re} in the online supplement where
$\mathcal{\hat{R}}_{et}$ are compared with $\hat{\beta}_{t}/\gamma$ for the
six countries.} It also marks the start and end dates of the respective
lockdowns.\footnote{The lockdown dates across countries can be found at
\url{https://en.wikipedia.org/wiki/COVID-19_pandemic_lockdowns}. One could
consider more accurate measures of the strictness of the lockdowns. For
example, Chudik, Pesaran, and Rebucci (2021) studied the impact of the
OxCGRT's stringency index on the estimated $\mathcal{R}_{et}$.} It should be
noted that the epidemic tends to expand (contract) if $\mathcal{\hat{R}}_{et}$
is above (below) unity.\footnote{See also Figure \ref{fig: Euro_dc_TR_fixMF}
of the online supplement for graphs of recorded daily new cases for these
countries.} Among the six countries, Italy started the first national lockdown
on March 9, 2020, followed by Spain, Austria, and France about a week later,
and Germany and the UK two weeks later (on March 23, 2020). As can be seen
from Figure \ref{fig: Euro_Re}, $\mathcal{\hat{R}}_{et}$ fell below one in mid
to late April 2020 in all these countries except for the UK, which took a bit
longer before falling below unity in early May. On average, it took $36$ days
to bring $\mathcal{\hat{R}}_{et}$ down below one from the start of the
lockdown, with Germany being the fastest ($27$ days) and the UK being the
slowest ($47$ days). By the end of May 2020, $\mathcal{\hat{R}}_{et}$ were
brought down below $0.5$ in all six countries except for the UK, where the
lowest value of $\mathcal{\hat{R}}_{et}$ occurred in early July. As lockdowns
eased, not surprisingly, the transmission rates started to rise. This new
surge in estimates of $\mathcal{R}_{et}$ led some of the countries to announce
their second lockdowns in early November 2020. The estimates of $\mathcal{R}%
_{et}$ fell below one again in December 2020, but the second trough in
$\mathcal{\hat{R}}_{et}$ is higher than the first in all countries except
France. As the pandemic progressed, $\mathcal{\hat{R}}_{et}$ displayed
different patterns (timing and magnitudes of peaks and troughs) across the six
countries due to different containment measures. By late April 2021 (the end
of our sample), $\mathcal{\hat{R}}_{et}$ is estimated to be close to one in
all these countries, but they appear to be rising in Spain and the UK and
falling in the other countries.%

Figure \ref{fig: Euro_MF} plots the estimated MF for the six countries. The
results offer evidence of substantial under-reporting in the pandemic's early
stages, with the magnitude of under-reporting falling over time in all these
countries. Closer inspection of the figure shows that the number of cases in
Austria, Germany, and Italy in late October-mid November 2020 was
underestimated by $5$--$6$ times, which declined to $2$--$3$ times in late
April 2021. About a quarter of actual infections were recorded in Spain in
September 2020, compared to about a half being detected in late April 2021.
France and the UK have a greater level of under-reporting in the early
stages---the number of cases was underestimated by as much as a factor of
$9$--$10$, which fell to $2$--$4$ during the study period. Overall, the
magnitude of these estimated MF seems reasonable and comparable to the
estimates obtained by other approaches in the literature.\footnote{See, for
example, Jagodnik et al. (2020), Li et al. (2020), Havers et al. (2020),
Kalish et al. (2021), and Rahmandad, Lim, and Sterman (2021). See also Section
\ref{Sup: literature} of the online supplement.}%

\begin{figure}[htp]%
\caption
{Realized and calibrated number of new cases of Covid-19 for selected European countries}%
\vspace{-0.2cm}%
\label{fig: Euro_dc}

\begin{center}%
\begin{tabular}
[c]{ccc}%
Austria &  & France\\%
{\includegraphics[
height=2.1465in,
width=2.6783in
]%
{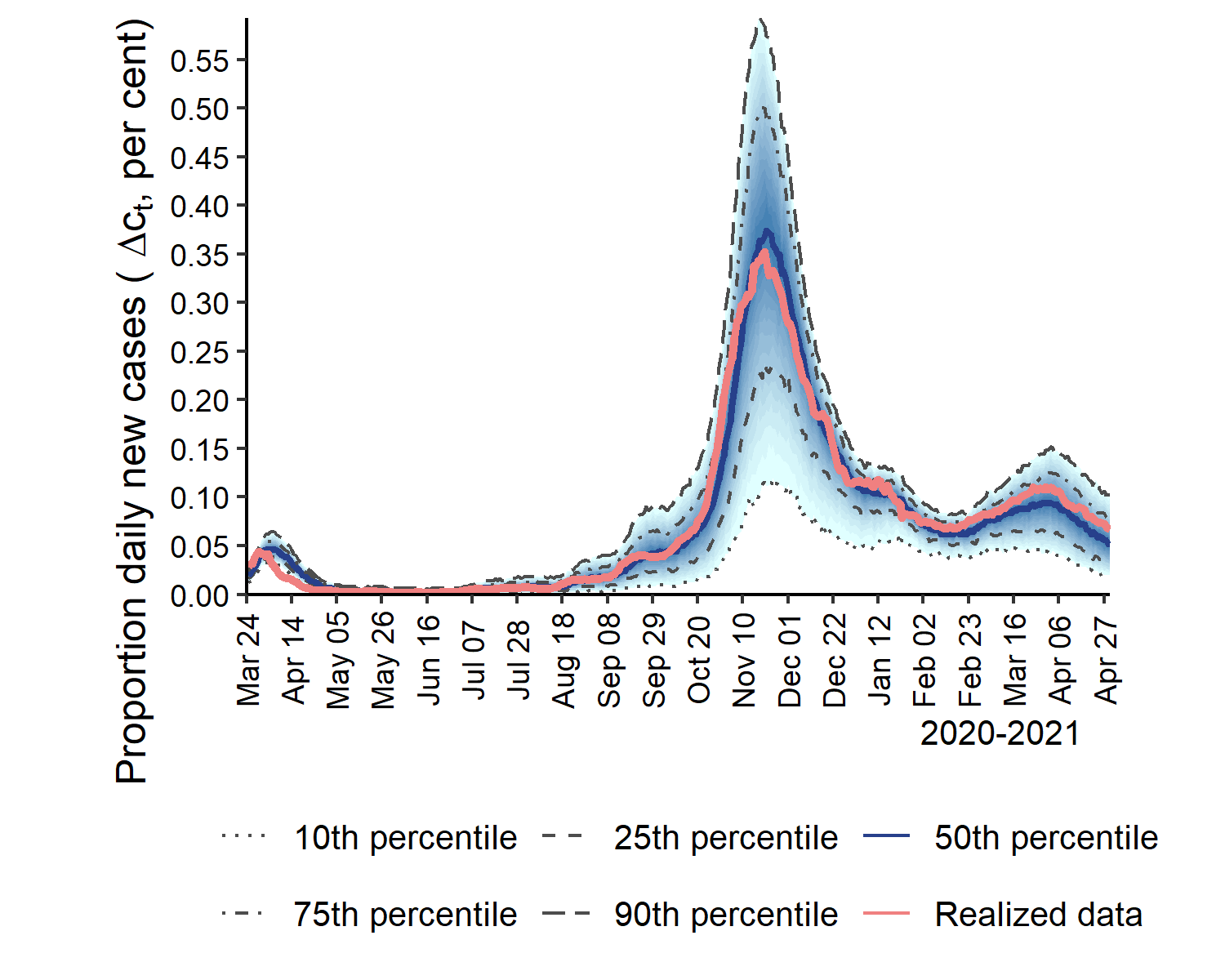}%
}
&  &
{\includegraphics[
height=2.1465in,
width=2.6783in
]%
{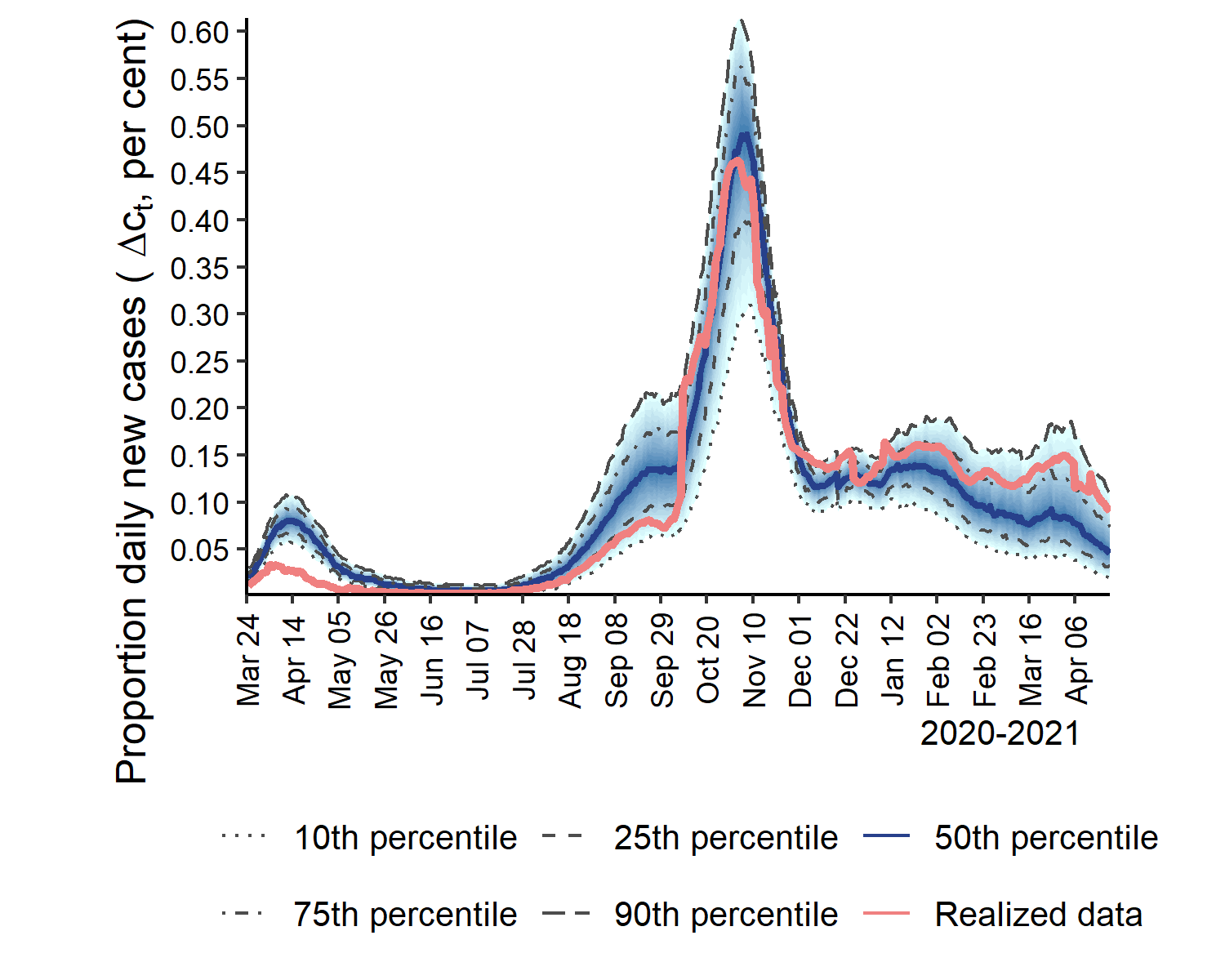}%
}
\\
&  & \\
Germany &  & Italy\\%
{\includegraphics[
height=2.1465in,
width=2.6783in
]%
{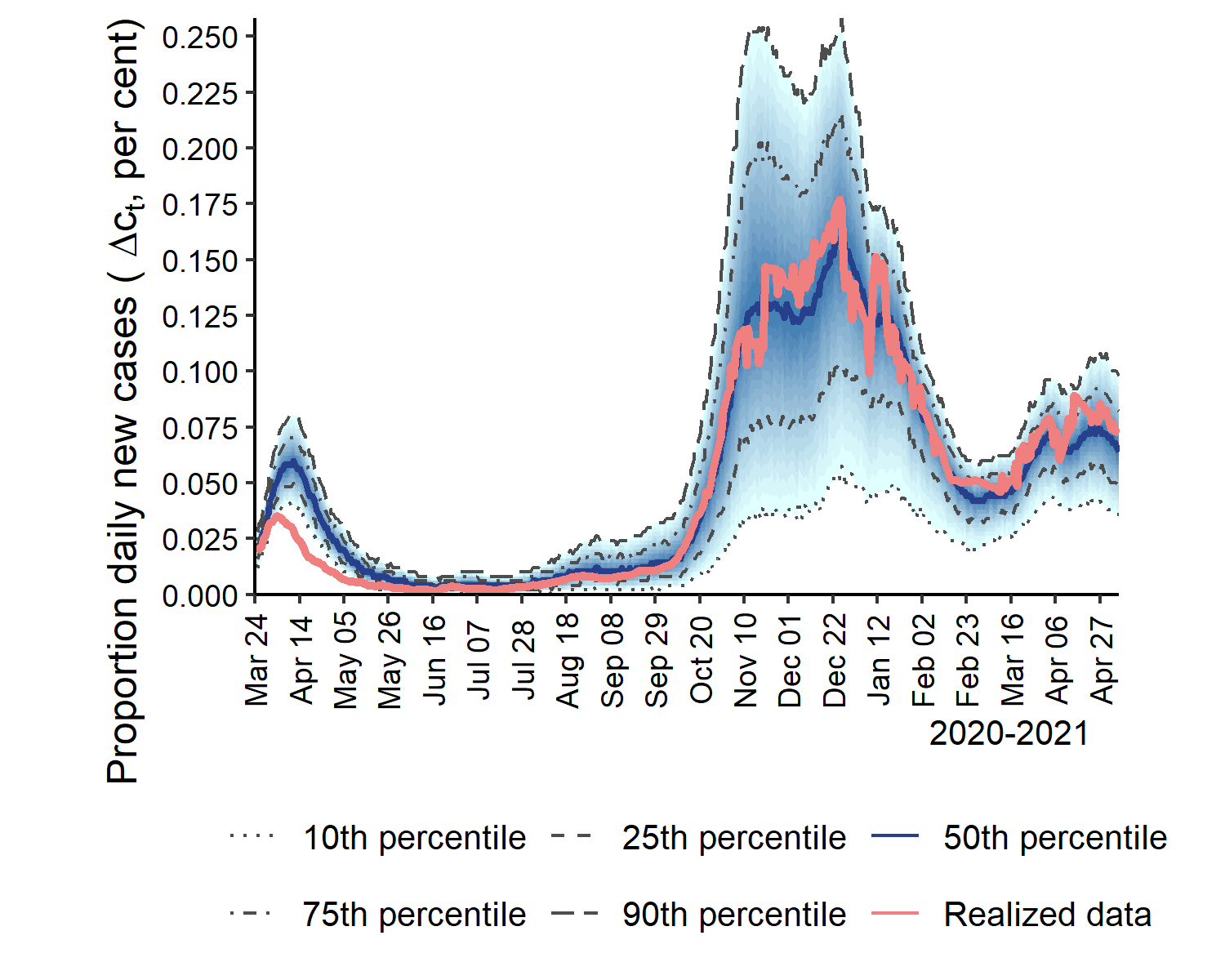}%
}
&  &
{\includegraphics[
height=2.1465in,
width=2.6783in
]%
{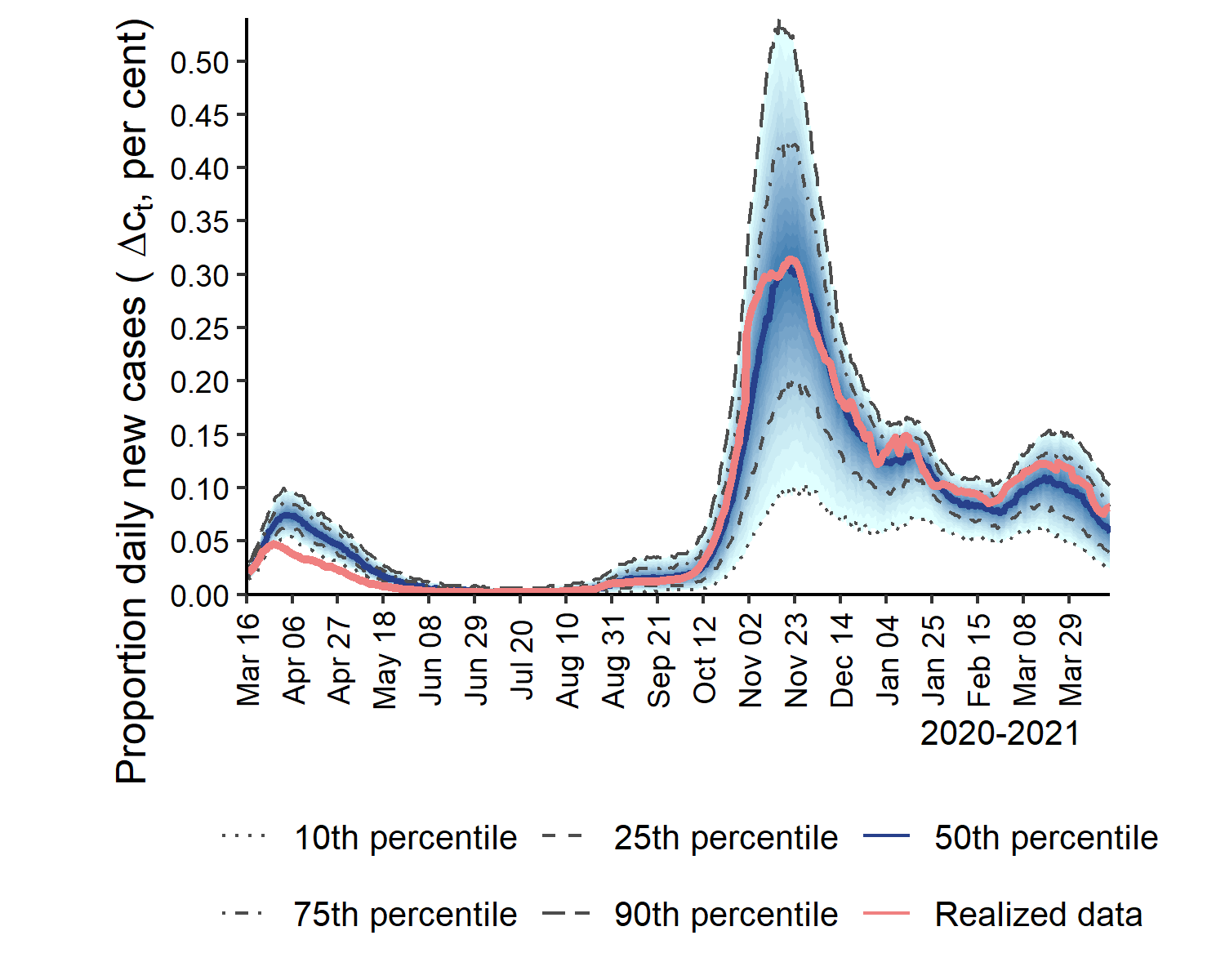}%
}
\\
&  & \\
Spain &  & UK\\%
{\includegraphics[
height=2.1465in,
width=2.6783in
]%
{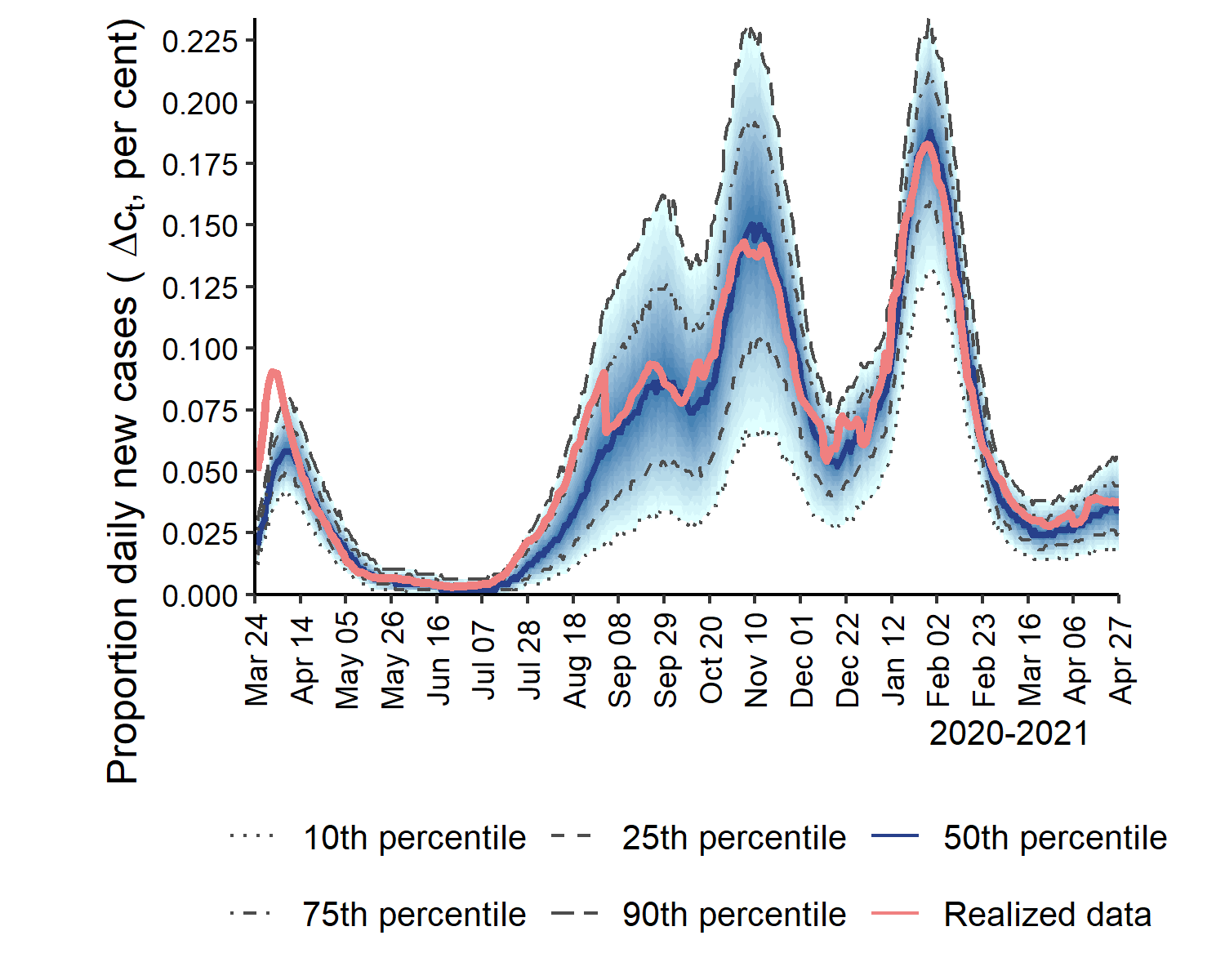}%
}
&  &
{\includegraphics[
height=2.1465in,
width=2.6783in
]%
{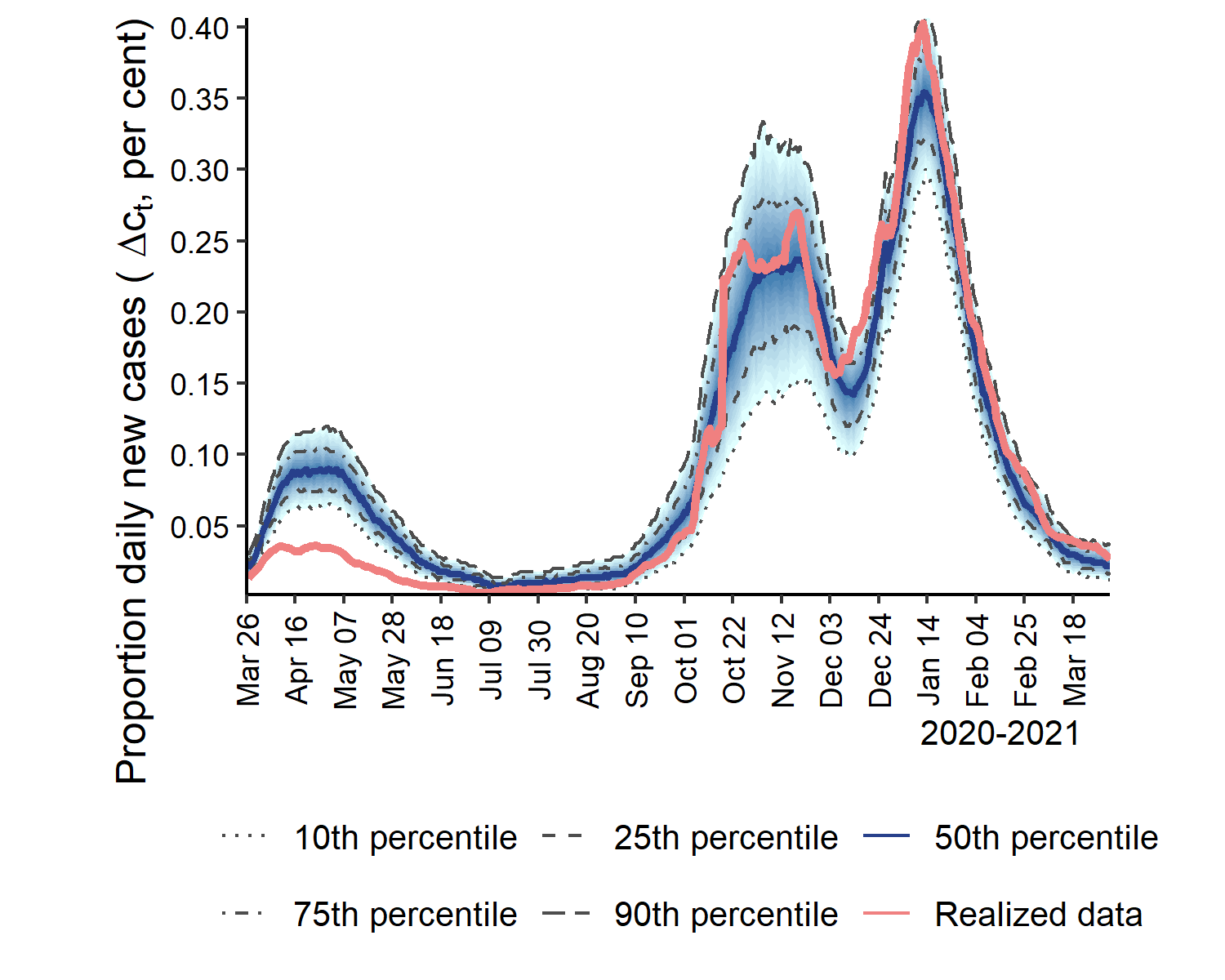}%
}
\end{tabular}

\end{center}

%

\vspace{-0.4cm}%
\footnotesize
{}Notes: Realized series (7-day moving average) multiplied by the estimated
multiplication factor is displayed in red. See also the notes to Figure
\ref{fig: Euro_Re}.%

\end{figure}%

Figure \ref{fig: Euro_dc} presents the calibrated new cases and the 7-day
moving average of the reported new cases multiplied by the estimated MF. The
fan charts depict the $10^{th}$ through the $90^{th}$ percentiles of the
calibrated data. It can be seen that once we have taken account of
under-reporting, the calibrated cases match with the recorded cases fairly
well. It is noteworthy that our model is able to catch the multiple waves\ of
Covid-19 cases over the course of the epidemic. Lastly, it is interesting to
see how the total cases per capita compare across the six countries, with and
without adjustments for under-reporting. Figure \ref{fig: Euro_ct} of the
online supplement displays the reported total cases and the case numbers after
adjusting for under-reporting using the MF estimates. The results show that
the number of total cases could have been underestimated three to five times
in these countries as of early August 2021. These comparisons clearly show the
importance of adjusting the number of infected cases due to under-reporting,
which can be reasonably estimated using our joint estimation procedure.

\section{Counterfactual exercises \label{Sec: counterfactual}}

Having shown that the outcomes of the calibrated model closely match the
evidence, we now demonstrate how the model can be used for two counterfactual
analyses of interest. First, we investigate the impact of social distancing
and vaccination on the evolution of the epidemic. To simplify the exposition,
we consider an epidemic with two waves and investigate if the second wave can
be avoided by vaccination. Second, we consider counterfactual outcomes that
could have resulted from different timing of the early interventions in
Germany and the UK.

\subsection{Social distancing and vaccination \label{Sec: vacc}}

In order to understand the impact of non-pharmaceutical interventions and
vaccination on controlling the epidemic, we perform counterfactual analyses
using the multigroup model with the five age groups introduced in Section
\ref{Sec: properties}. The different age groups in the model also allow us to
consider the implications of prioritizing the Covid-19 vaccine by age.

In reality, the transmission rate varies over time due to both voluntary and
mandatory social distancing as well as other mitigation measures such as
vaccination. Here we use social distancing to refer broadly to all types of
non-pharmaceutical interventions (including lockdown measures). We assume
that, in the absence of a vaccination program, the (scaled) transmission rate,
$\beta_{t}/\gamma$, equals $3$ in the first two weeks after the epidemic
outbreak, falls to $0.9$ linearly over the next three weeks, and remains at
$0.9$ for eight weeks. When social distancing is relaxed, the transmission
rate increases to $1.5$ linearly over the next three weeks and remains at
$1.5$ thereafter. Note that the effective reproduction number, $\mathcal{R}%
_{t}$, could still fall below unity due to herding.\footnote{Figure
\ref{fig: TR_dist} of the online supplement displays the time profile of the
transmission rate under this social distancing policy.} Also note that
$\beta_{t}=\sum_{\ell=1}^{L}w_{\ell}\beta_{\ell t}$, where $\beta_{\ell
t}=\sum_{\ell^{\prime}=1}^{L}\tau_{\ell t}k_{\ell\ell^{\prime},t}$. We assume
that $\beta_{\ell t}$ has the same rate of change as $\beta_{t}$, for all
$\ell$.

To model vaccination, we assume, for simplicity, that a single-dose shot
vaccine with efficacy of $\epsilon_{v}$ becomes available when $i_{t}=i^{0}$.
The vaccine takes full effect immediately, and its immune protection does not
wane over time.\footnote{On August 5, 2021, Moderna reported that its vaccine
efficacy remained almost the same through six months after the second shot.
\url{https://time.com/6087722/moderna-vaccine-long-term-efficacy/}} The
effectiveness of vaccination can be measured by the parameter $\mu_{i\ell}$,
which is the mean of the infection threshold variable, $\xi_{it}$, defined in
(\ref{xi}). We assume that $\mu_{i\ell}$ takes the value $\mu^{0}$ if
individual $(i,\ell)$ is not vaccinated and takes the value $\mu^{1}$ after
vaccination. In the case of a single group, the probability of an individual
getting infected when the proportion of active cases is $i^{0}$, for any given
value of $\mu_{i}$, is
\begin{equation}
E\left(  x_{i,t+1}\left\vert x_{it}=0,\mu_{i},i_{t}=i^{0}\right.  \right)
=1-\left(  1-p+pe^{-\frac{\tau}{\mu_{i}}}\right)  ^{ni^{0}},
\label{prob_infection}%
\end{equation}
which declines with $\mu_{i}$. By definition, the vaccine efficacy should
equal the percentage reduction in the probability of infection. Then the value
of $\mu_{i}$ associated with efficacy $\epsilon_{v}$ is given by
\begin{align}
&  E\left(  x_{i,t+1}\left\vert x_{it}=0,\mu_{i}=\mu^{1},i_{t}=i^{0}\right.
\right)  -E\left(  x_{i,t+1}\left\vert x_{it}=0,\mu_{i}=\mu^{0},i_{t}%
=i^{0}\right.  \right) \label{vacc_efficacy}\\
&  =-\epsilon_{v}E\left(  x_{i,t+1}\left\vert x_{it}=0,\mu_{i}=\mu^{0}%
,i_{t}=i^{0}\right.  \right) \nonumber
\end{align}
Using the result in (\ref{pll'_approx}), we have%
\begin{equation}
1-\left(  1-p+pe^{-\frac{\tau}{\mu_{i}}}\right)  ^{ni^{0}}\approx1-\exp\left(
-\frac{\tau ki^{0}}{\mu_{i}}\right)  \approx\frac{\tau ki^{0}}{\mu_{i}}.
\label{prob_approx}%
\end{equation}
Combining (\ref{prob_infection}), (\ref{vacc_efficacy}), and
(\ref{prob_approx}), we obtain $\tau ki^{0}/\mu^{1}\approx\left(
1-\epsilon_{v}\right)  \tau ki^{0}/\mu^{0}$, which simplifies to\footnote{The
exact solution is very close to its approximation given by (\ref{mu_sol}).}%
\begin{equation}
\mu^{1}/\mu^{0}\approx1/(1-\epsilon_{v}). \label{mu_sol}%
\end{equation}
This result also holds in the multigroup model.\footnote{A\ proof is given in
Section \ref{Sup: VE} of the online supplement.} Intuitively, (\ref{mu_sol})
states that an individual becomes $1/\left(  1-\epsilon_{v}\right)  $ times
more immune relative to his/her level of immunity after vaccination.

In simulations, an individual's degree of resilience, $\xi_{it}$, are i.i.d.
draws from an exponential distribution with mean $\mu^{0}$ $(\mu^{1})$ before
(after) vaccination. Without loss of generality, we normalize $\mu^{0}=1.$ The
Pfizer-BioNTech and Moderna vaccines have been shown to have $95\%$ and
$94.1\%$ efficacy rates in preventing symptomatic laboratory-confirmed
Covid-19 infection, respectively (Oliver et al., 2020, 2021). Accordingly,
using (\ref{mu_sol}) we have $\mu^{1}=20$ for $\epsilon_{v}=0.95$, namely
Pfizer and Moderna vaccines increase the level of immunity by a factor of
$20$.\footnote{We also considered $\epsilon_{v}=0.66$, which is in line with a
$66.3\%$ efficacy rate of the Johnson \& Johnson vaccine (Oliver et al.,
2021). The results are presented in Section \ref{Sup: vacc} of the online
supplement.}

We suppose that $75$ percent of the population is vaccinated over $12$ weeks,
with an equal number of people vaccinated each day.\footnote{We chose to
consider constant daily vaccination rate and a relatively short period as an
example. Of course, one could consider increasing daily rate over a more
extended period if desired.} We consider two vaccination schemes---random
vaccination and vaccination in decreasing age order. Under the former, people
are randomly selected without replacement for vaccination irrespective of
their age. In the latter, older people are vaccinated first. Individuals
within an age group are randomly selected for vaccination on each day when
their group is eligible. After all people in the oldest group have been
vaccinated, the second-oldest group becomes eligible. This process continues
until $75$ percent of the population is vaccinated. In both schemes, we assume
that vaccination eligibility does not depend on whether an individual is
susceptible, infected, or recovered.

\begin{figure}[!htb]%
\caption
{Simulated number of new cases when there are (a) no containment measures, (b) social distancing only, (c) vaccination only, and (d) combined social distancing and vaccination}%
\label{fig: dist_vacc}%
\vspace{-0.2cm}%
\footnotesize

\begin{center}%
\begin{tabular}
[c]{ccc}%
\textbf{(a) No containment} &  & \textbf{(b) Social distancing only}\\%
{\includegraphics[
height=1.7781in,
width=2.3609in
]%
{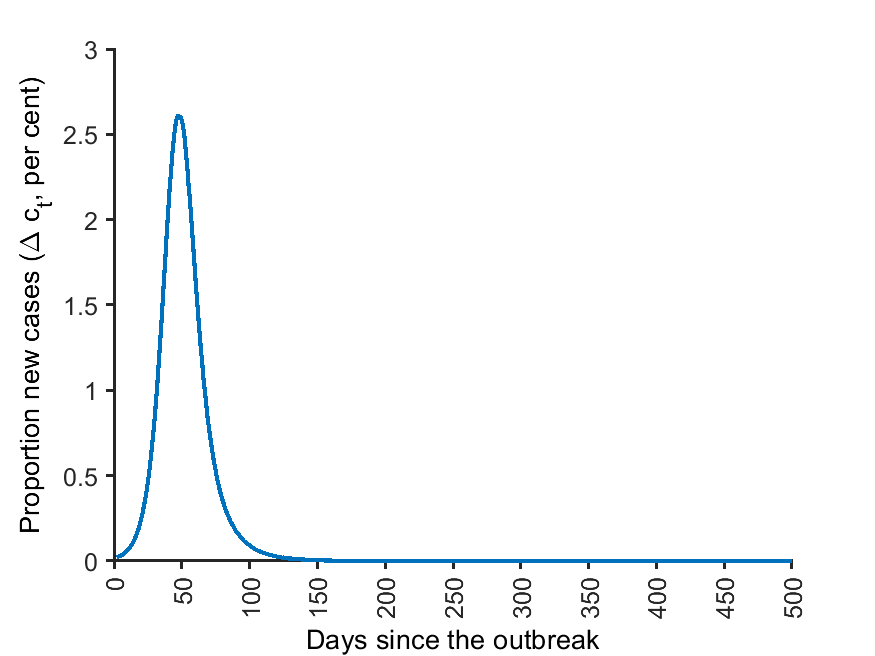}%
}
&  &
{\includegraphics[
height=1.7781in,
width=2.3609in
]%
{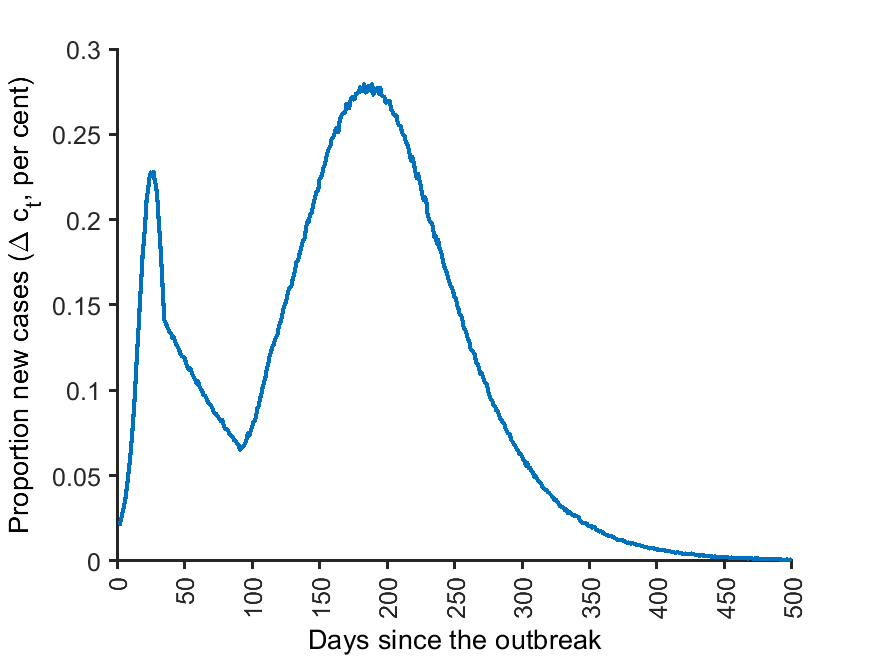}%
}
\\
$c^{\ast}=0.90$, $T^{\ast}=215$ days &  & $c^{\ast}=0.50$, $T^{\ast}=486$
days\\
&  & \\
\textbf{(c) Vaccination only} &  & \textbf{(d) Social distancing and
vaccination}\\%
{\includegraphics[
height=1.7781in,
width=2.3609in
]%
{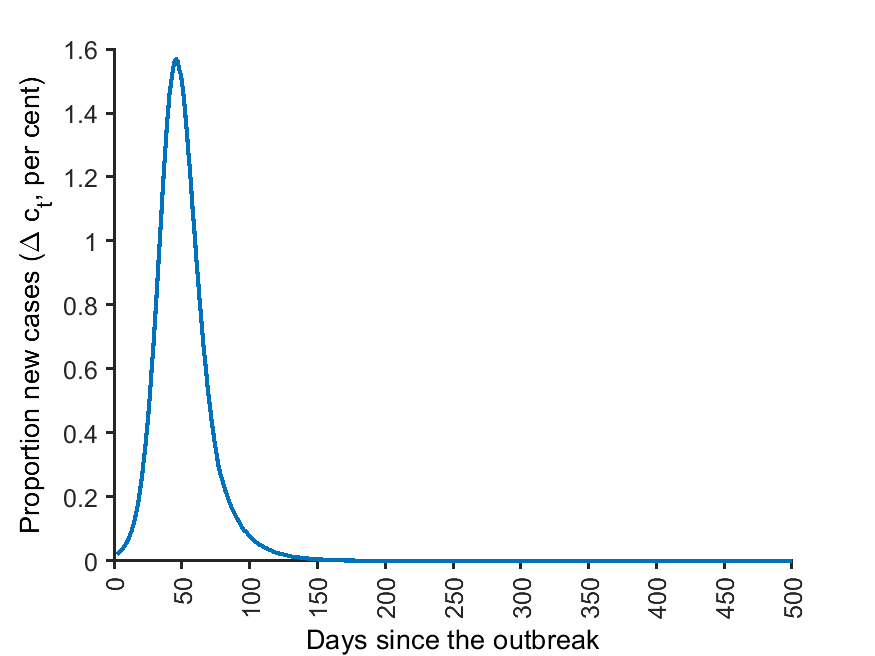}%
}
&  &
{\includegraphics[
height=1.7781in,
width=2.3609in
]%
{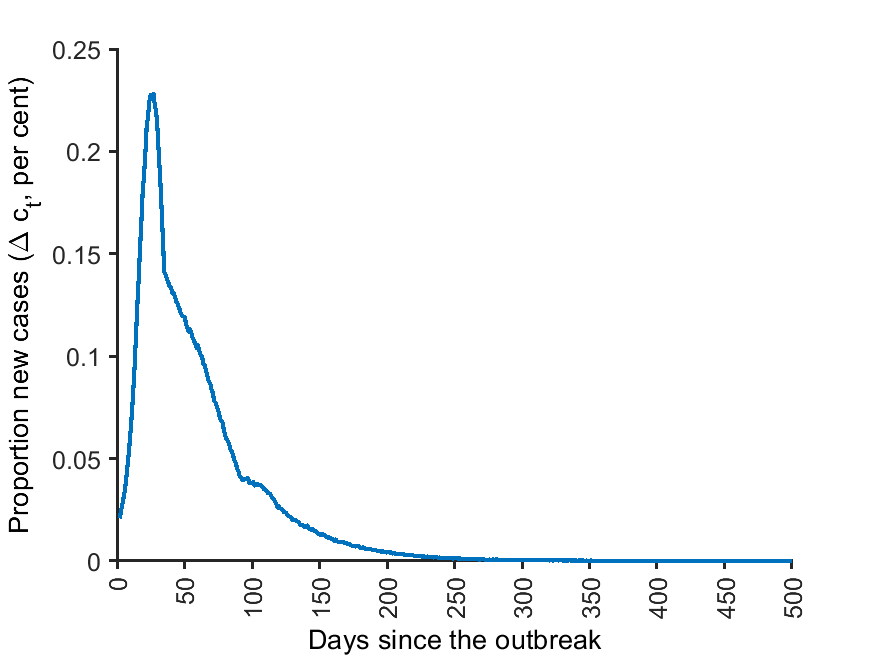}%
}
\\
$c^{\ast}=0.59$, $T^{\ast}=201$ days &  & $c^{\ast}=0.12$, $T^{\ast}=215$ days
\end{tabular}

\end{center}

%

\vspace{-0.3cm}
Notes: The average number of new cases over $1,000$ replications is displayed.
The simulations use the multigroup model with the five age groups as described
in Section \ref{Sec: properties}. Population size is $n=10,000$. Under social
distancing, the transmission rate, $\beta_{t}/\gamma$, equals $3$ in the first
two weeks after the outbreak, falls to $0.9$ linearly over the next three
weeks, and remains at $0.9$ for eight weeks. When social distancing is
relaxed, the transmission rate increases to $1.5$ linearly over the next three
weeks and remains at $1.5$ afterward. In the absence of social distancing
(i.e., in cases (a) and (c)), the transmission rate, $\beta_{t}/\gamma$, is
fixed at $3$. In case (c), the vaccination starts from the $4^{th}$ week after
the outbreak$.$ In case (d), the vaccination starts during the last month of
social distancing (i.e., the $10^{th}$ week after the outbreak). In cases (c)
and (d), $75$ percent of the population is randomly vaccinated over $12$
weeks, and the vaccine efficacy is $\epsilon_{v}=0.95$. $c^{\ast}$ denotes the
maximum proportion of infected and is computed as $B^{-1}\sum_{b=1}^{B}%
\max_{t}c_{t}^{(b)},$ with $B=1,000$ replications. $T^{\ast}$ is the duration
of the epidemic.%

\end{figure}

Let us first consider the random vaccination scheme. Figure
\ref{fig: dist_vacc} compares the aggregate outcomes when there are (a) no
containment measures, (b) social distancing only, (c) vaccination only, and
(d) combined social distancing and vaccination. Specifically, the transmission
rate, $\beta_{t}/\gamma$, is fixed at $3$ in the absence of social distancing
(i.e., in cases (a) and (c)). In case (c), the vaccination starts from the
$4^{th}$ week after the outbreak$.$ In case (d), the vaccination starts during
the last month of social distancing (i.e., the $10^{th}$ week after the
outbreak). In cases (c) and (d), $75$ percent of the population is randomly
vaccinated over $12$ weeks, and the vaccine efficacy is set at $\epsilon
_{v}=0.95$.%

The results show that social distancing can quickly bring down the daily case
rate and thus reduce the demands on the healthcare system. However, as social
distancing restrictions are relaxed, a second wave is expected to emerge. The
second wave may have a higher peak than the first wave if the transmission
rates rise too fast due to increasing contact rates or the exposure intensity.
Comparing graphs (a) and (b) reveal that social distancing alone can reduce
the maximum proportion of cases from $90$ percent in an uncontrolled epidemic
to $50$ percent. Nonetheless, the duration of the epidemic could more than
double, and an enduring epidemic may entail high social and economic costs. If
vaccination is the only containment tool, it must be implemented soon enough
to curb the spread of the disease, especially for a highly contagious disease
with $\mathcal{R}_{0}$ about $3$ (or even higher as evidenced by the new
variants). Graph (c) shows that even if (in a very unlikely scenario) a highly
effective vaccine becomes available during the $4^{th}$ week after the
outbreak, $59$ percent of the population could end up getting infected. In
reality, developing a new vaccine takes considerable time. Therefore,
vaccination is not a substitute for non-pharmaceutical interventions, which
are necessary to slow the spread of the disease, allowing more time for
vaccine development. Vaccination can effectively prevent the second wave if it
is rolled out during the last month of social distancing, as shown in Graph
(d). Under the assumption that $75$ percent of the population end up getting
vaccinated with efficacy of $\epsilon_{v}=0.95$, the maximum proportion of
infected could reduce to $12$ percent, and the number of highest daily new
infections could be more than $10$ and $7$ times lower than that in cases (a)
and (c), respectively. We also examined the implications of different
vaccination coverages, start times, speeds of delivery, and vaccine
efficacies. The results are summarized in Section \ref{Sup: vacc} of the
online supplement, where we provide counterfactual outcomes assuming (i) 50
percent \textit{versus} 75 percent vaccination coverage, (ii) vaccination
starts at the end of social distancing \textit{versus} during the last month
of social distancing, (iii) 75 percent of the population getting vaccinated
over 8 \textit{versus} 12 weeks, and (iv) $\epsilon_{v}=0.95$ \textit{versus}
$0.66$.%

\begin{figure}[htp]%
\caption
{Simulated average number of group-specific and aggregate new cases using the multigroup model with social distancing, assuming random vaccination or vaccination in decreasing age order}%
\label{fig: vacc_priority}%
\vspace{-0.2cm}%
\footnotesize

\begin{center}%
\begin{tabular}
[c]{ccc}%
\textbf{Group 1: [0, 15)} &  & \textbf{Group 2: [15, 30)}\\%
{\includegraphics[
height=1.7772in,
width=2.3618in
]%
{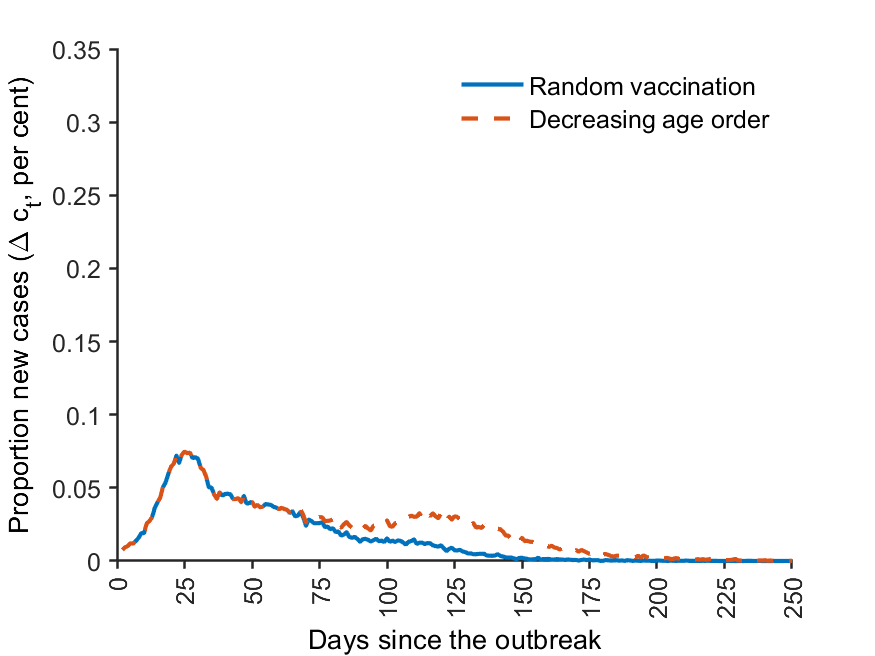}%
}
&  &
{\includegraphics[
height=1.7772in,
width=2.3618in
]%
{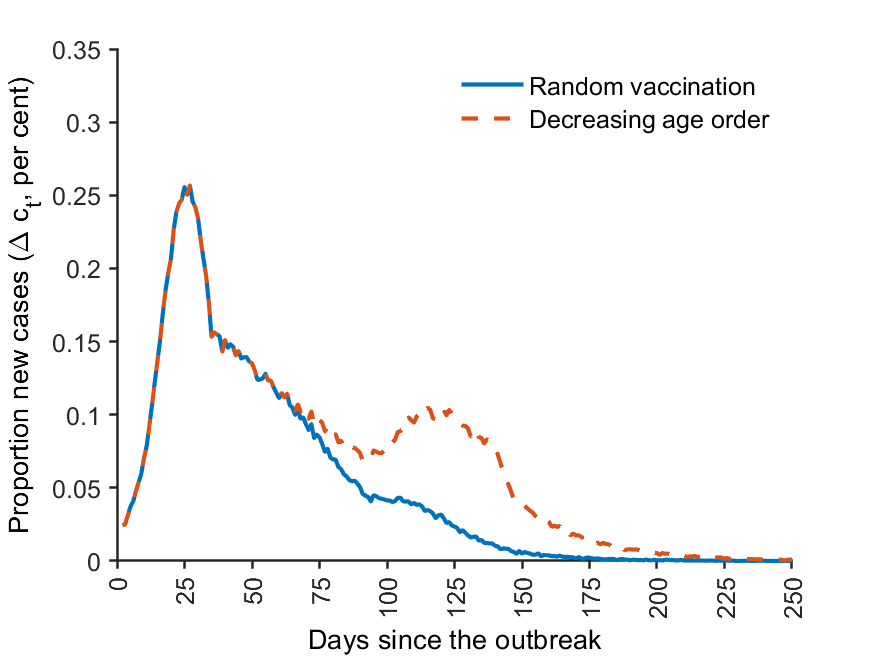}%
}
\\
Random: $c_{1}^{\ast}=0.04$ &  & Random: $c_{2}^{\ast}=0.13$\\
By age: \ $\ c_{1}^{\ast}=0.05$ &  & By age: \ $\ c_{2}^{\ast}=0.17$\\
&  & \\
&  & \\
\textbf{Group 3: [30, 50)} &  & \textbf{Group 4: [50, 65)}\\%
{\includegraphics[
height=1.7772in,
width=2.3618in
]%
{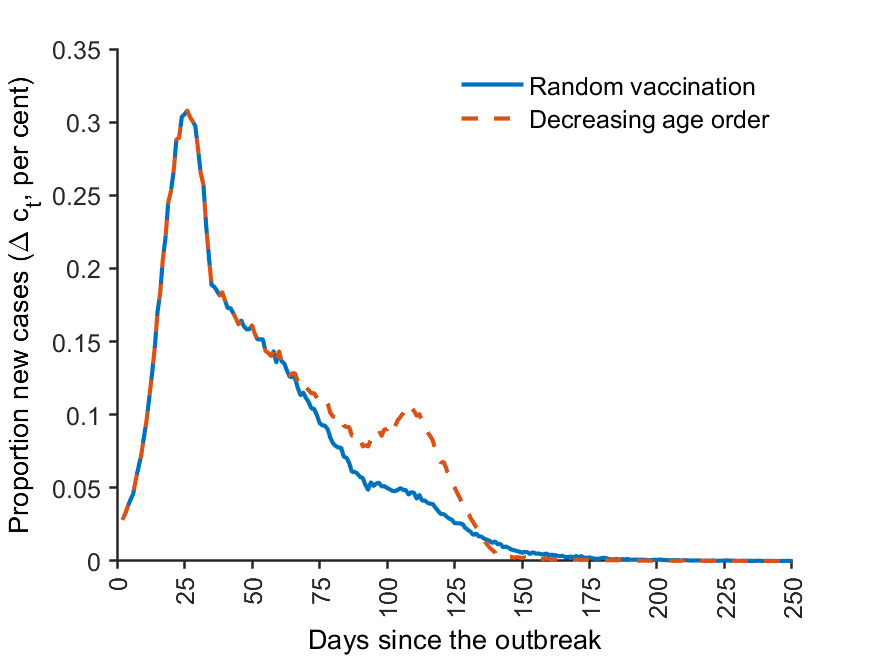}%
}
&  &
{\includegraphics[
height=1.7772in,
width=2.3618in
]%
{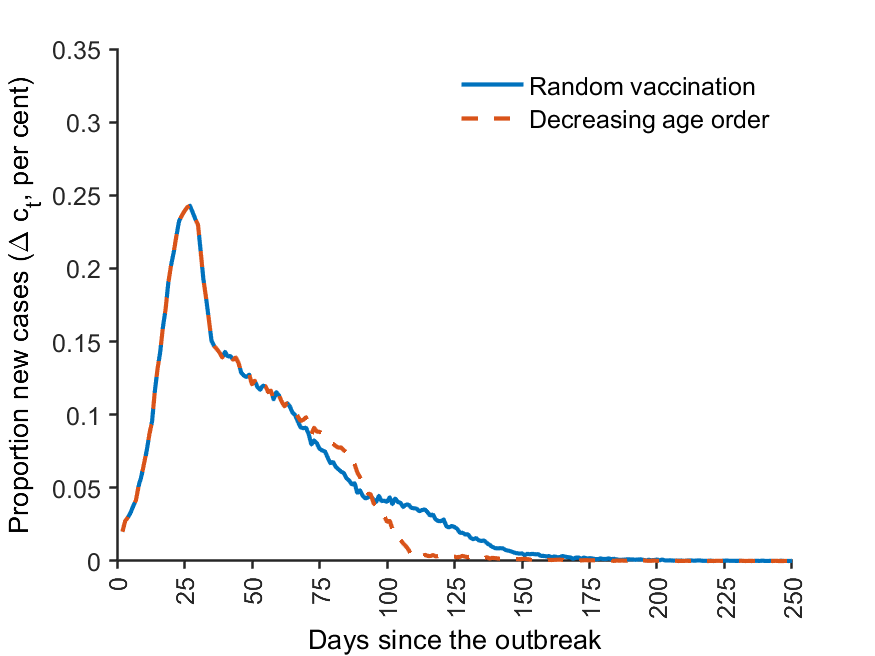}%
}
\\
Random: $c_{3}^{\ast}=0.15$ &  & Random: $c_{4}^{\ast}=0.12$\\
By age: \ $\ c_{3}^{\ast}=0.17$ &  & By age: \ $\ c_{4}^{\ast}=0.12$\\
&  & \\
&  & \\
\textbf{Group 5: 65+} &  & \textbf{Aggregate}\\%
{\includegraphics[
height=1.7772in,
width=2.3618in
]%
{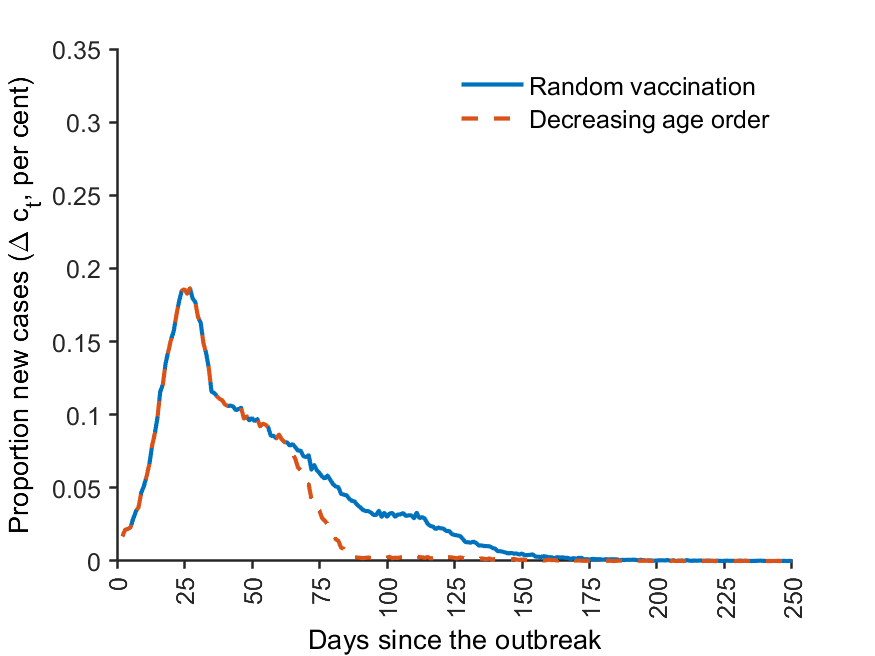}%
}
&  &
{\includegraphics[
height=1.7772in,
width=2.3618in
]%
{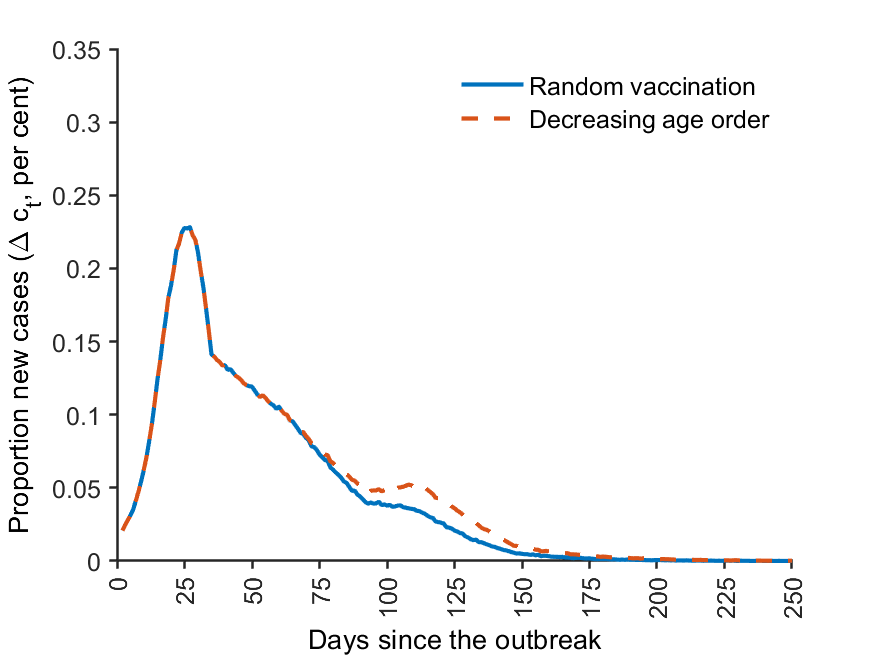}%
}
\\
Random: $c_{5}^{\ast}=0.10$ &  & Random: $c^{\ast}=0.12$\\
By age: \ $\ c_{5}^{\ast}=0.08$ &  & By age: \ $\ c^{\ast}=0.13$\\
&  &
\end{tabular}

\end{center}

%

\vspace{-0.2cm}%
Notes: The average number of new cases over $B=1,000$ replications is
displayed. Population size is $n=10,000$. The social distancing policy is the
same as that considered in Figure \ref{fig: dist_vacc}. The vaccination starts
during the last month of social distancing (i.e., the $10^{th}$ week after the
outbreak). $75$ percent of the population is vaccinated over $12$ weeks. The
vaccine efficacy is $\epsilon_{v}=0.95.$ The duration of the epidemic is
$T^{\ast}=215$ days under random vaccination and $T^{\ast}=233$ days under
vaccination by decreasing age order. $c_{\ell}^{\ast}=B^{-1}\sum_{b=1}^{B}%
\max_{t}c_{\ell t}^{(b)},$for $\ell=1,2,\ldots,5,$ and $c^{\ast}=B^{-1}%
\sum_{b=1}^{B}\max_{t}c_{t}^{(b)}$.%

\end{figure}%

Figure \ref{fig: vacc_priority} compares the simulation outcomes under random
vaccination and vaccination in decreasing age order for each age group and the
entire population, assuming the same social distancing policy as described
above. In this experiment, the vaccination starts during the last month of
social distancing. $75$ percent of the population is vaccinated over $12$
weeks, and the vaccine efficacy is $\epsilon_{v}=0.95$.\footnote{We also
examined the case if $50$ percent of the population is vaccinated over eight
weeks. See Figure \ref{fig: vacc_priority_pct50} of the online supplement.}
The results show that the maximum proportion of infected in the oldest group
is reduced by $2$ percentage points if the old gets vaccinated first, compared
to random vaccination. Not surprisingly, the cost of protecting the elderly is
reflected in the higher infection rates in the younger age groups, increasing
the maximum cases by $1$, $4$, and $2$ percentage points for age groups 1 to
3, respectively. The vaccine effectively curbs the spread of the disease and
prevents the second surge of cases in the two senior groups. A comparison of
the aggregate outcomes reveals that prioritizing the old results in a higher
level of overall infections and a longer duration of the epidemic, owning to
higher contact rates of the younger population. Of course, how to prioritize
vaccines is a complex decision requiring further information on the rates of
hospitalization and death in each age group. It also requires evaluating the
social and economic costs of high infection rates and lockdown measures among
young people.

\subsection{Counterfactual outcomes of early interventions in UK and Germany}

\begin{figure}[p]%
\caption
{Counterfactual number of infected and active cases for Germany and UK under different lockdown scenarios}%
\vspace{-0.2cm}%
\label{fig: Germany_UK_1w}

\begin{center}%
\hspace*{-0.6cm}%
\begin{tabular}
[c]{ccc}%
\multicolumn{3}{c}{%
\begin{footnotesize}%
\textbf{What if the German lockdown was delayed one week?}%
\end{footnotesize}%
\vspace{0.2cm}%
}\\%
\hspace*{-0.2cm}%
{\footnotesize Infected cases} & {\footnotesize Active cases} &
\hspace*{-0.8cm}%
\begin{footnotesize}%
$\mathcal{\hat{R}}_{et}$%
\end{footnotesize}%
\\%
{\includegraphics[
height=1.8273in,
width=2.8193in
]%
{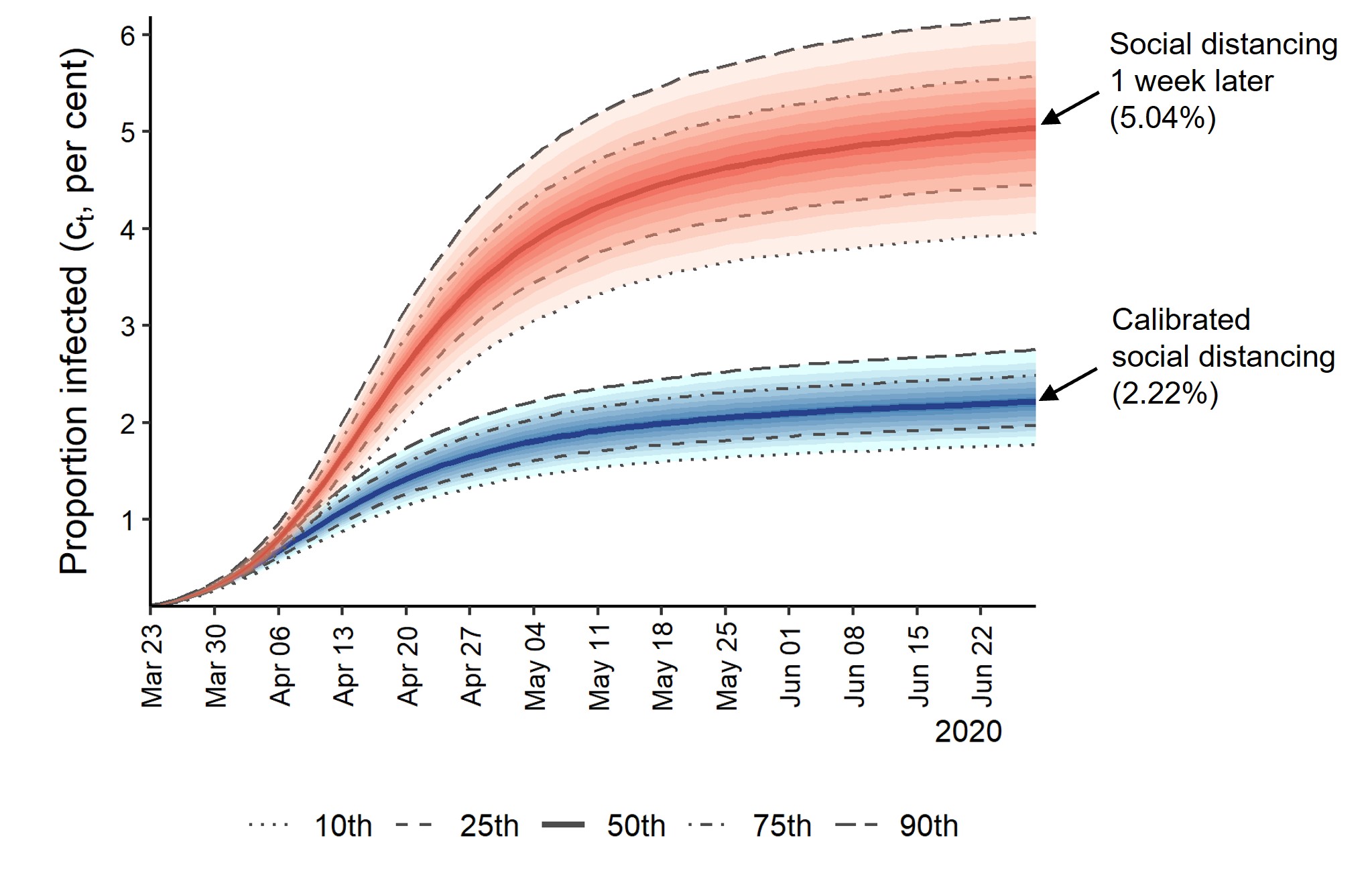}%
}
&
{\includegraphics[
trim=0.223018in 0.000000in 0.000000in 0.000000in,
height=1.8412in,
width=2.1776in
]%
{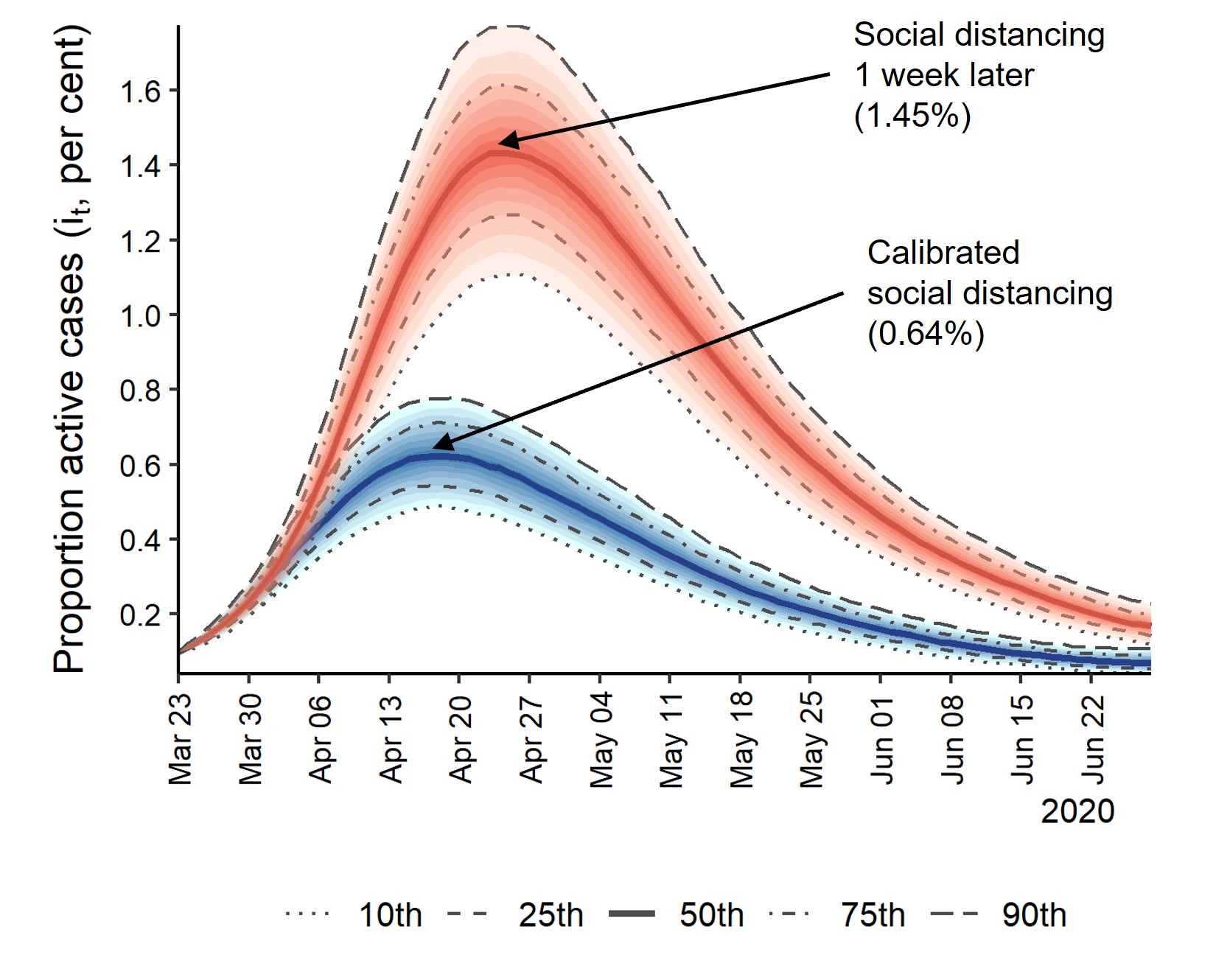}%
}
&
\hspace*{-0.5cm}%
{\includegraphics[
trim=0.000000in -0.660383in 0.000000in 0.000000in,
height=1.9406in,
width=2.2451in
]%
{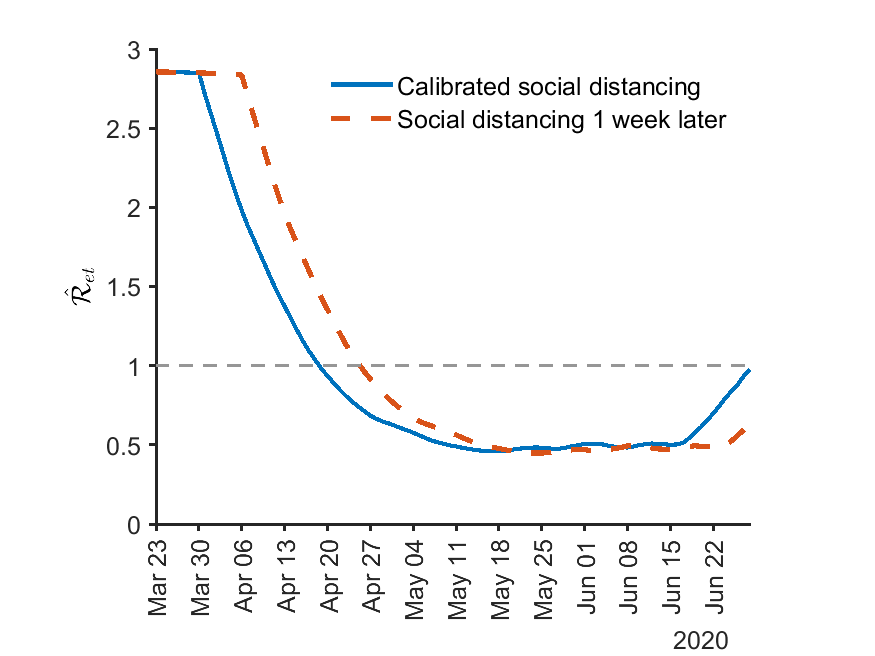}%
}
\\
&  & \\
\multicolumn{3}{c}{%
\begin{footnotesize}%
\textbf{What if the UK lockdown was brought forward one week?}%
\end{footnotesize}%
\vspace{0.2cm}%
}\\%
\hspace*{-0.2cm}%
{\footnotesize Infected cases} & {\footnotesize Active cases} &
\hspace*{-0.8cm}%
\begin{footnotesize}%
$\mathcal{\hat{R}}_{et}$%
\end{footnotesize}%
\\%
{\includegraphics[
height=1.8273in,
width=2.8288in
]%
{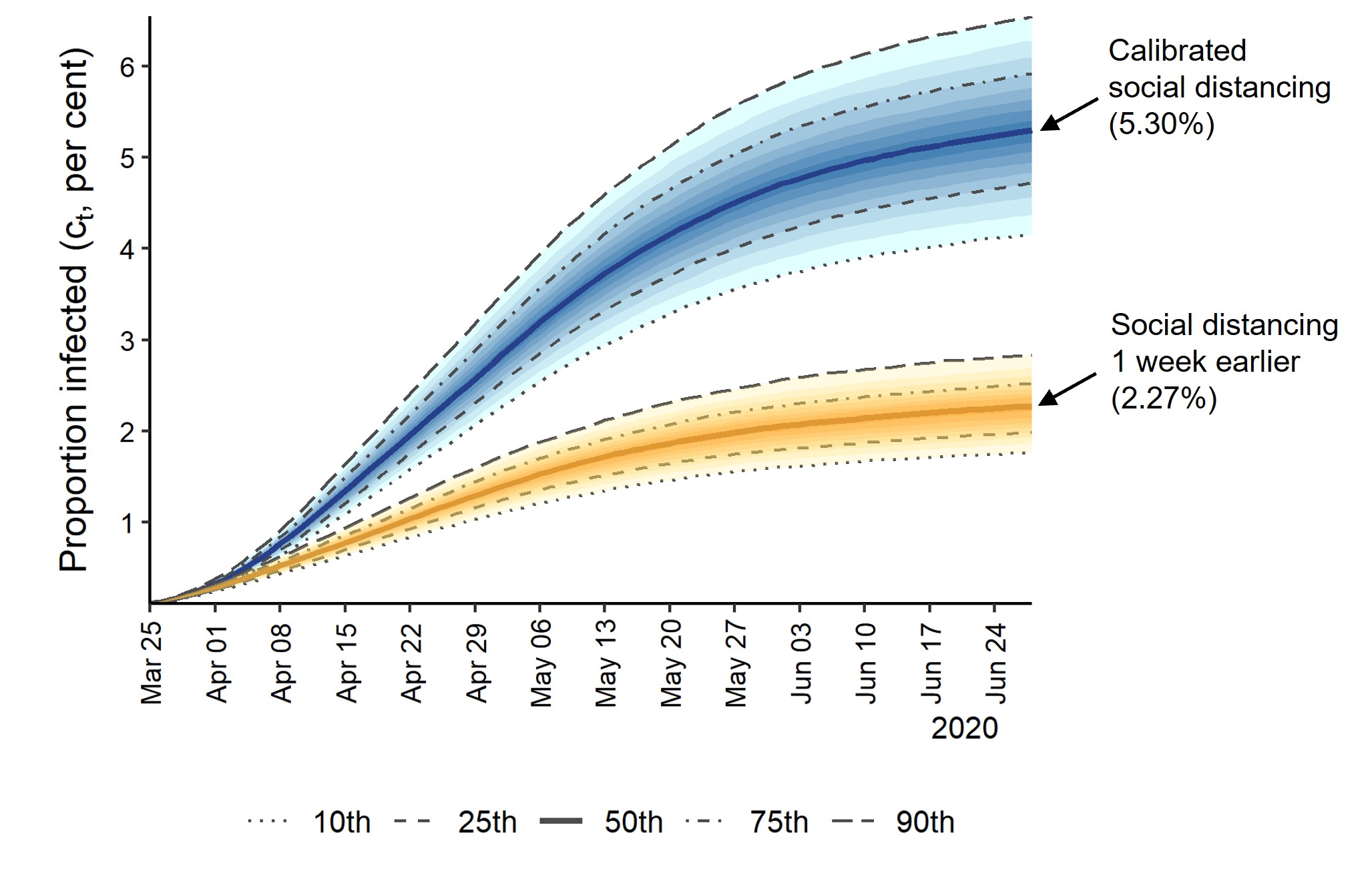}%
}
&
{\includegraphics[
trim=0.222955in 0.000000in 0.000000in 0.000000in,
height=1.8273in,
width=2.2883in
]%
{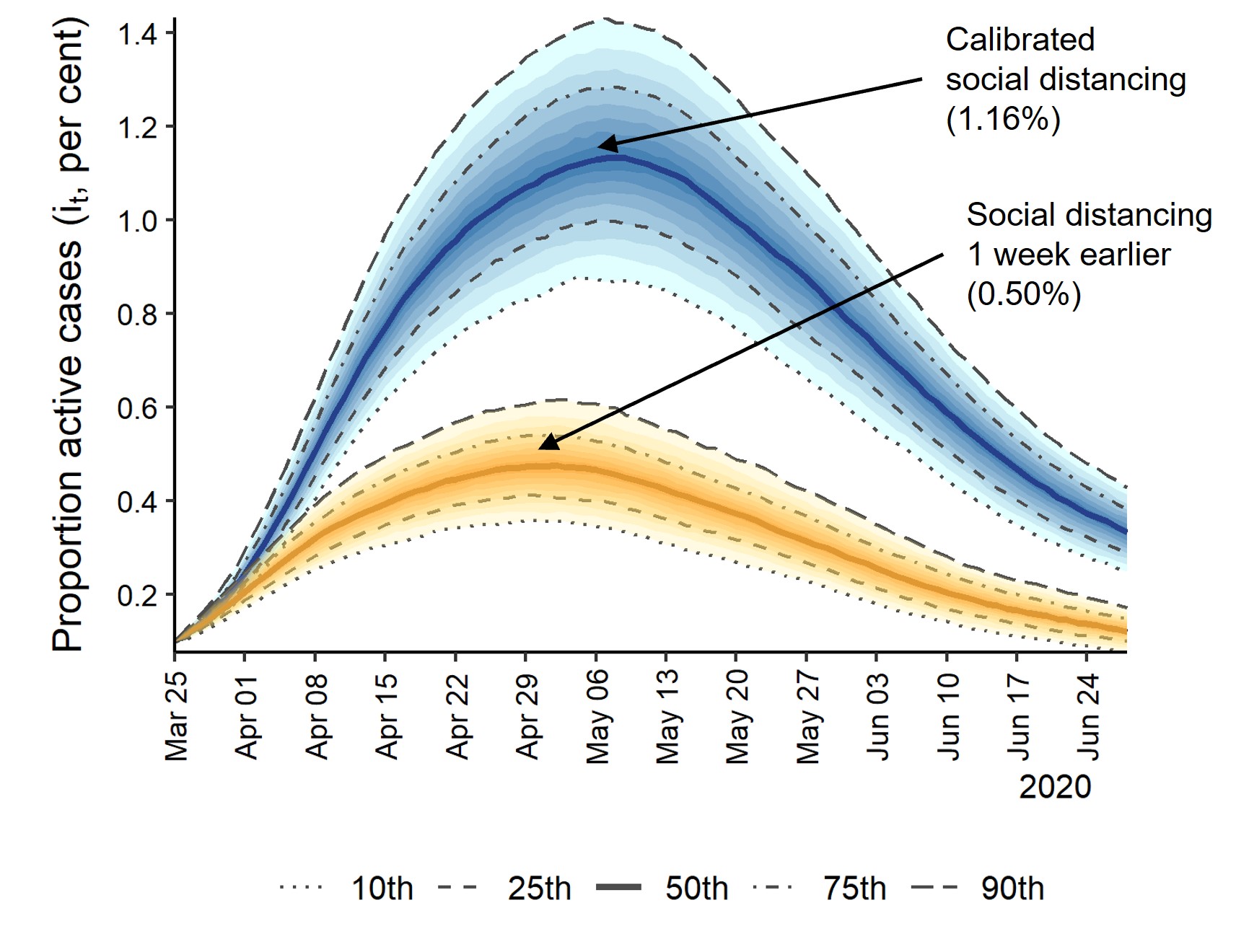}%
}
&
\hspace*{-0.5cm}%
{\includegraphics[
trim=0.000000in -0.660383in 0.000000in 0.000000in,
height=1.9406in,
width=2.2451in
]%
{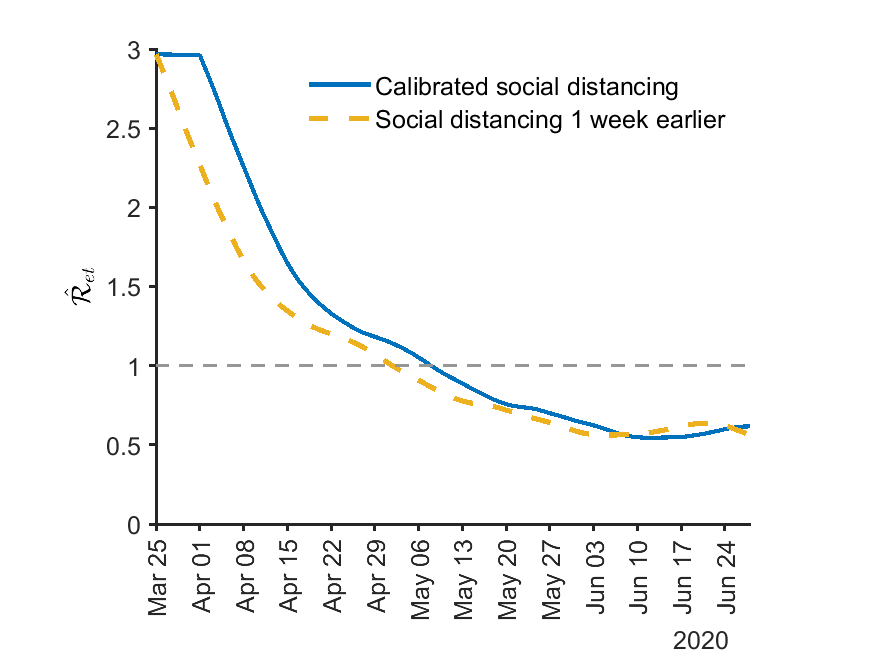}%
}
\end{tabular}

\end{center}

%

\vspace{-0.3cm}%
\footnotesize
{}Notes: The simulation uses the single group model with the
Erd\H{o}s-R\'{e}nyi random network and begins with $1/1000$ of the population
randomly infected on day $1$. The population size used in the simulation is
$n=50,000$. The recover rate is $\gamma=1/14$. The number of removed
(recoveries + deaths) is estimated recursively using $\tilde{R}_{t}=\left(
1-\gamma\right)  \tilde{R}_{t-1}+\gamma\tilde{C}_{t-1}$ for both countries,
with $\tilde{C}_{1}=\tilde{R}_{1}=0$, where $\tilde{C}_{t}$ is the reported
number of infections. $\hat{\beta}_{t}$ is the $2$-weekly rolling estimate
computed by (\ref{betathat}) assuming MF $=5$. The mean of $\mathcal{\hat{R}%
}_{et}^{(b)}=\left(  1-c_{t}^{(b)}\right)  \hat{\beta}_{t}/\gamma$, for
$b=1,2,\ldots,1000$ replications, is displayed in the last column.%

\end{figure}

We now turn to different counterfactual outcomes that could have resulted from
different timing of the first lockdowns in Germany and the UK, focusing on the
first wave\ of Covid-19 that leveled off at the end of June 2020 in both
countries.\footnote{For example, Neil Ferguson, once an advisor to the UK
government, stated on June 10, 2021, that "Had we introduced lockdown measures
a week earlier, we would have reduced the final death toll by at least a
half". See
\url{https://www.politico.com/news/2020/06/10/boris-johnson-britain-coronavirus-response-312668}.}
In particular, we investigate the quantitative effect of bringing forward the
lockdown in the UK on the number of infected cases, as compared to the effect
of delaying the lockdown in Germany. To this end, we shift the estimated
$\beta_{t}$ values backward or forward for one or two weeks. As shown in
Figure \ref{fig: Germany_UK_1w}, if the German lockdown had been delayed by
one week, the maximum proportion of infected cases would have increased from
$2.2$ to $5.0$ percent, and the maximum proportion of active cases would have
risen from $0.6$ to $1.5$ percent. In contrast, if the UK lockdown had been
brought forward by one week, the model predicts that the maximum proportion of
infected cases would have reduced from $5.3$ to $2.3$ percent, and the maximum
number of active cases would have reduced from $1.2$ to $0.5$ percent. These
results suggest that the UK could have achieved a similarly low level of
infected cases per capita as Germany if it had implemented social distancing
sooner. The maximum proportion of infected (active cases) is estimated to rise
further to $10.8$ ($3.2$) percent if the German lockdown was delayed by two
weeks, and the maximum proportion of infected (active cases) is estimated to
decrease further to $1.1$ ($0.3$) percent if the UK lockdown was brought
forward by two weeks.\footnote{See Figure \ref{fig: Germany_UK_2w} of the
online supplement.} In summary, this counterfactual exercise shows that it is
critical to take measures to lower the effective reproduction number as early
as possible if a policymaker aims to control the number of infected and active cases.%

\section{Concluding remarks \label{Sec: conclusion}}

This paper has developed a stochastic network SIR model for empirical analyses
of the Covid-19 pandemic across countries or regions. Moment conditions are
derived for the number of infected and active cases for the single group as
well as multigroup models. It is shown how these moment conditions can be used
to identify the structural parameters and provide rolling estimates of the
transmission rate in different phases of the epidemic. To allow for
time-varying under-reporting of cases, it proposes a method that jointly
estimates the transmission rate and the multiplication factor using a
simulated method of moments approach. In empirical applications to six
European countries, the estimates of the transmission rate are used to
calibrate the proposed epidemic model. It is shown that the simulated outcomes
are reasonably close to the reported cases once the under-reporting of cases
is addressed. The multiplication factors are found to be declining over the
course of the pandemic. It is estimated that the actual number of infections
could be between $4$--$10$ times higher than the number of reported cases
around October 2020, whereas only $2$--$3$ times higher in April 2021. The
multigroup model is used for counterfactual analyses of the impact of social
distancing and vaccination on the evolution of the epidemic. It is shown that
lockdown measures are needed to slow down the spread of a highly contagious
disease such as Covid-19, buying time for the development of vaccines and
treatments. Vaccination can prevent additional waves of epidemics as social
distancing is eased after lockdowns if it is introduced early enough. The
calibrated model is also used for empirically-based counterfactual analyses of
the first lockdowns in Germany and the UK. It is shown that the UK could have
achieved an outcome similar to that experienced by Germany during the first
wave if she had started the lookdown just one week earlier. Almost
symmetrically, Germany would have experienced much higher infection rates
(similar to the UK's experience) if she had started the lockdown one week later.%

\vspace{0.3cm}%
%

\noindent\large
\textbf{References}%

\begin{singlespace}
\begin{footnotesize}%
\vspace{-0.3cm}%
%

\begin{list}{}{\setlength{\topsep}{8pt}\setlength{\leftmargin}{0.1in}%
\setlength{\listparindent}{-0.1in}\setlength{\itemindent}{-0.1in}%
\setlength{\parsep}{0.6em}}%
%

\item
Chudik, A., M. H. Pesaran, and A. Rebucci (2021). COVID-19 time-varying
reproduction numbers worldwide: An empirical analysis of mandatory and
voluntary social distancing. NBER working paper No. 28629.%

\item
D'Arienzo, M. and A. Coniglio (2020). Assessment of the SARS-CoV-2 basic
reproduction number, R0, based on the early phase of COVID-19 outbreak in
Italy. \textit{Biosafety and Health 2}(2), 57--59.%

\item
Del Valle, S. Y., J. M. Hyman, and N. Chitnis (2013). Mathematical models of
contact patterns between age groups for predicting the spread of infectious
diseases. \textit{Mathematical Biosciences and Engineering 10}, 1475.%

\item
Elliott, S. and C. Gourieroux (2020). Uncertainty on the reproduction ratio in
the SIR model. arXiv preprint: 2012.11542.%

\item
Farrington, P., and H., Whitaker (2003). Estimation of Effective Reproduction
Numbers for Infectious Diseases Using Serological Survey Data.
\textit{Biostatistics}, 4, 621--632.%

\item
Gibbons, C. L., M.-J. J. Mangen, D. Plass, A. H. Havelaar, R. J. Brooke, P.
Kramarz, ... M. E. Kretzschmar (2014). Measuring underreporting and
underascertainment in infectious disease datasets: A comparison of methods.
\textit{BMC Public Health 14}(1), 147.%

\item
Guo, H., M. Y. Li, and Z. Shuai (2006). Global stability of the endemic
equilibrium of multigroup SIR epidemic models. \textit{Canadian Applied
Mathematics Quarterly 14}(3), 259--284.%

\item
Havers, F. P., C. Reed, T. Lim, J. M. Montgomery, J. D. Klena, A. J. Hall, ...
N. J. Thornburg (2020). Seroprevalence of antibodies to SARS-CoV-2 in 10 sites
in the United States, March 23-May 12, 2020. \textit{JAMA Internal Medicine},
\textit{180}(12), 1576--1586.%

\item
Hethcote, H. W. (2000). The mathematics of infectious diseases. \textit{SIAM
Review 42}(4), 599--653.%

\item
Jagodnik, K., F. Ray, F. M. Giorgi, and A. Lachmann (2020). Correcting
under-reported COVID-19 case numbers: estimating the true scale of the
pandemic. medRxiv preprint doi: 10.1101/2020.03.14.20036178.%

\item
Kalish, H., C. Klumpp-Thomas, S. Hunsberger, H. A. Baus, M. P. Fay, N.
Siripong, ... K. Sadtler (2021). Undiagnosed SARS-CoV-2 seropositivity during
the first six months of the COVID-19 pandemic in the United States.
\textit{Science Translational Medicine} \textit{13}(601), 1--11.%

\item
Kermack, W. and A. McKendrick (1927). A contribution to the mathematical
theory of epidemics. \textit{Proceedings of the Royal Society of London.
Series A. 115}(772), 700--721.%

\item
Li, R., S. Pei, B. Chen, Y. Song, T. Zhang, W. Yang, and J. Shaman (2020).
Substantial undocumented infection facilitates the rapid dissemination of
novel coronavirus (SARS-CoV-2). \textit{Science 368}(6490), 489--493.%

\item
Mossong, J., N. Hens, M. Jit, P. Beutels, K. Auranen, R. Mikolajczyk, ... W.
J. Edmunds (2008). Social contacts and mixing patterns relevant to the spread
of infectious diseases. \textit{PLoS Med 5}(3), e74.%

\item
Nepomuceno, E., D. F. Resende, and M. J. Lacerda (2018). A survey of the
individual-based model applied in biomedical and epidemiology. \textit{Journal
of Biomedical Research and Reviews,} \textit{1}(1): 11--24.%

\item
Oliver, S., J. Gargano, M. Marin, M. Wallace, K. G. Curran, M. Chamberland,
... K. Dooling (2020). The advisory committee on immunization practices'
interim recommendation for use of Pfizer-BioNTech COVID-19 vaccine --- United
States, December 2020. \textit{MMWR. Morbidity and Mortality Weekly Report
69}(50), 1922--1924.%

\item
Oliver, S., J. Gargano, M. Marin, M. Wallace, K. G. Curran, M. Chamberland,
... K. Dooling (2021). The advisory committee on immunization practices'
interim recommendation for use of Moderna COVID-19 vaccine --- United States,
December 2020. \textit{MMWR. Morbidity and Mortality Weekly Report 69}(5152), 1653--1656.%

\item
Oliver, S. E., J. W. Gargano, H. Scobie, M. Wallace, S. C. Hadler, J. Leung,
... K. Dooling (2021). The advisory committee on immunization practices'
interim recommendation for use of Janssen COVID-19 vaccine --- United States,
February 2021. \textit{MMWR. Morbidity and Mortality Weekly Report 70}(9), 329--332.%

\item
Rahmandad, H., T. Y. Lim, and J. Sterman (2021). Behavioral dynamics of
COVID-19: estimating underreporting, multiple waves, and adherence fatigue
across 92 nations. \textit{System Dynamics Review 37}(1), 5--31.%

\item
Rocha, L. E. and N. Masuda (2016). Individual-based approach to epidemic
processes on arbitrary dynamic contact networks. \textit{Scientific Reports
6}, 31456.%

\item
Thieme, H. R. (2013). \textit{Mathematics in population biology}, Volume 12 of
\textit{Princeton Series in Theoretical and Computational Biology}. Princeton
University Press.%

\item
Willem, L., F. Verelst, J. Bilcke, N. Hens, and P. Beutels (2017). Lessons
from a decade of individual-based models for infectious disease transmission:
A systematic review (2006--2015). \textit{BMC infectious diseases}
\textit{17}(1), 612.%

\item
Willem, L., T. Van Hoang, S. Funk, P. Coletti, P. Beutels, and N. Hens (2020).
SOCRATES: An online tool leveraging a social contact data sharing initiative
to assess mitigation strategies for COVID-19. \textit{BMC Research Notes
13}(1), 1--8.%

\item
Zhang, J., M. Litvinova, Y. Liang, Y. Wang, W. Wang, S. Zhao, ... H. Yu
(2020). Supplementary materials for "Changes in contact patterns shape the
dynamics of the COVID-19 outbreak in China". Available at: science.sciencemag.org/content/368/6498/1481/suppl/DC1.%

\end{list}%
%

\end{footnotesize}
\end{singlespace}%
\pagebreak%

\clearpage
%

\appendix
%

\onehalfspacing
%

\vspace*{0.3cm}%

\begin{center}
{\Large Online Supplement to "Matching Theory and Evidence on }%

\vspace*{0.3cm}%
{\Large Covid-19 using a Stochastic Network SIR Model"\bigskip}

{\large M. Hashem Pesaran}

{\small University of Southern California, USA, and Trinity College,
Cambridge, UK\bigskip}

{\large Cynthia Fan Yang}

{\small Florida State University\bigskip}%

\large{December 18, 2021}

{\Large \bigskip}
\end{center}

%

\doublespacing
\normalsize
%

\renewcommand\theequation{S.\arabic{equation}}
\setcounter{equation}{0}
\renewcommand\thesection{S\arabic{section}}
\setcounter{section}{0}
\renewcommand\thepage{S\arabic{page}}
\setcounter{page}{1}
\setcounter{theorem}{1}
\renewcommand{\thefootnote}{S\arabic{footnote}}
\setcounter{footnote}{0}
\renewcommand{\thetable}{S.\arabic{table}}
\setcounter{table}{0}
\renewcommand\thefigure{S.\arabic{figure}}
\setcounter{figure}{0}
\renewcommand\thelemma{S.\arabic{lemma}}
\setcounter{lemma}{0}
\renewcommand\theremark{S.\arabic{remark}}
\setcounter{remark}{0}%

This online supplement is set out in eight sections. Section
\ref{Sup: literature} reviews the literature. Section \ref{Sup: theory}
establishes the classical multigroup SIR model as a linearized version of the
moment conditions we have derived for our proposed model. This section also
generalizes the proposed model to allow for truncated geometric recovery and
provides a derivation of vaccine efficacy in the multigroup version of the
model. Section \ref{Sup: calibration} discusses the edge probability and how
the random networks were generated in our simulation exercises. This section
also compares the simulated models across different population sizes, the
number of groups, and network types. Section \ref{Sup: estimation both}
reports additional Monte Carlo results on estimation of the transmission rate
and details the algorithm used to jointly estimate the transmission rate and
multiplication factor. It also discusses the estimation of the recovery rate.
Section \ref{Sup: empirical_Re} presents additional estimates of the
reproduction numbers for selected European countries and the US. Section
\ref{Sup: empirical_MF} reports further estimates of the multiplication factor
for the European countries and the US. It also compares the reported total
cases without and with adjustment for under-reporting. Section
\ref{Sup: counterfactual} provides results of additional counterfactual
exercises. Finally, Section \ref{Sup: data} gives the details of data sources.

\section{Related literature \label{Sup: literature}}

Our modelling approach relates to two important strands of the literature on
mathematical modelling of infectious diseases, namely the classical SIR\ model
due to \cite{Kermack1927SIR} and its various extensions to multigroup
SIR\ models, and the individual-based network models. The multigroup
SIR\ model allows for a heterogeneous population where each compartment (S, I,
or R) is further partitioned into multiple groups based on one or more
factors, including age, gender, location, contact patterns, and a number of
economic and social factors. One of the earliest multigroup models was
pioneered by \cite{Lajmanovich1976multigroupSIS}, who developed a class of
SIS\ (susceptible-infected-susceptible) models for the transmission of
gonorrhea. Subsequent extensions to the multigroup SIR model and its variants
include \cite{Hethcote1978multi},
\citeauthor{Thieme1983global}
(%
\citeyear{Thieme1983global}%
,
\citeyear{Thieme1985local}%
), \cite{Beretta1986multiSIR}, and many others. Reviews of multigroup models
can be found in \cite{Hethcote2000SIAM} and \cite{Thieme2013book}. For some of
the recent contributions on the multigroup SIR\ models and their stability
conditions, see, for example,
\citet*{Hyman1999differential}%
,
\citet*{Guo2006global}%
,
\citet*{Li2010global}%
,
\citet*{Ji2011multiSIR}%
,
\citet*{Ding2015lyapunov}
and
\citet*{Zhou2017stability}%
. In contrast, we do not model the progression of epidemics at the compartment
level; instead, we develop an individual-based stochastic model from which we
derive a set of aggregate moment conditions. Interestingly, we are able to
show that the multigroup SIR model can be derived as a
linearized-deterministic version of our individual-based stochastic model.

Our analysis also relates to the more recent literature on mathematical models
of epidemics on networks, whereby the spread of the epidemic is modelled via
networks (or graphs), with nodes representing single individuals or groups of
individuals and links (or edges) representing contacts. The adoption of
networks in epidemiology has opened up a myriad of possibilities, using more
realistic contact patterns to investigate the impact of network structure on
epidemic outcomes and to design network-based interventions.
\citet*{Kiss2017book}
provide a systematic treatment of this literature, with related reviews in
\cite{Miller2014review} and \cite{Pastor2015review}.

Being based on individual outcomes, our approach is more closely related to
the individual-based models surveyed by \cite{Willem2017review} and
\citet*{Nepomuceno2018survey}%
. These models consider the transition probability of individuals from one
state (S, I, R) to another
\citep{Rocha2016IBA, Gourieroux2020AES}%
. In contrast, as noted in the introduction, we do not model the transition
probabilities, but rather we model the contact probabilities and unobserved
individual-specific probability of becoming infected, and then derive
individual-specific transition probabilities. Like the individual-based
models, our approach also allows for considerable group heterogeneity and has
the advantage that aggregates up to the multigroup SIR\ model.

In order to calibrate the average number of contacts in our model, we drew
upon the literature on social contact patterns relevant to the transmission of
respiratory infectious diseases. Before the outbreak of Covid-19, large-scale
social contact surveys have been conducted in many countries aiming to guide
effective policies on infectious disease control and
prevention.\footnote{Summaries of these social contact surveys are provided by
\cite{Hoang2019review} and Supplementary Table S1 of \cite{Leung2017contactHK}%
.} The POLYMOD study of social contacts in eight European countries by
\cite{Mossong2008} is a notable landmark.\footnote{The eight countries are
Belgium, Germany, Finland, Great Britain, Italy, Luxembourg, The Netherlands,
and Poland.} Many similar surveys have been conducted since. Among them, the
contact studies in Hong Kong
\citep{Leung2017contactHK}
and Shanghai
\citep{Zhang2019contactSH}
provide valuable information about the pre-Covid social contacts in China.
Most of these studies summarize contact patterns based on age groups, contact
locations (e.g., households, schools, workplaces), and time schedules (e.g.,
weekdays or weekends) that can be utilized in multigroup epidemiological
models. With the outbreak of Covid-19, a few recent articles reported
significant changes in contact patterns. For example, \cite{Zhang2020science}
find that the median number of daily contacts in Wuhan went down from $7$ in
normal times to $2$ after the Covid-19 outbreak. The median number of daily
contacts in Shanghai fell from $10$ to $2$. \cite{Jarvis2020UK} find that the
average daily number of contacts declined from $10.8$ in normal times to $2.8$
immediately after the lockdown in the UK. In all these three cases, the
contact number by age flattened after the outbreak.

In this study, we propose a new method of estimating the transmission rate,
$\beta_{t}$, using the moment conditions we derive from our stochastic network
SIR model. The transmission rate is closely connected to the reproduction
numbers, which are epidemiologic metrics used to measure the intensity of an
infectious disease. The basic reproduction number, denoted by $\mathcal{R}%
_{0}$, is the number of new infections expected to result from one infected
individual at the start of the epidemic, and within SIR\ models it is defined
by $\mathcal{R}_{0}=\beta_{0}/\gamma$, where $\beta_{0}$ is the initial
transmission rate, and $\gamma$ is the recovery rate. For the current
Covid-19, estimates of $\mathcal{R}_{0}$ range between $2$ to $3$.\footnote{A
summary of published $\mathcal{R}_{0}$ values is provided in Table 1 of
\cite{DArienzo2020R0Italy}.} Since the disease transmissibility will vary over
time due to changes in immunity and/or mitigation policies, the effective
reproduction number, which we denote by $\mathcal{R}_{et}$, measures the
$\mathcal{R}$ number $t$ periods after the initial outbreak. The effective
$\mathcal{R}$ number is governed by the extent to which the susceptible
population is shrinking and the effectiveness of mitigation policies (whether
mandated or voluntary). In the single group SIR\ model, we have $\mathcal{R}%
_{et}=\left(  1-c_{t}\right)  \beta_{t}/\gamma$, where $c_{t}$ is the per
capita number of infected cases at time $t$.

Various methods are available in the epidemiological literature to estimate
the reproduction numbers at the beginning and/or in real time during
epidemics, but there is no uniform framework. Estimation approaches that are
data-driven and involve simplifying assumptions include the use of the number
of susceptibles at endemic equilibrium, the average age at infection, the
final size equation, and calculation from the intrinsic growth rate of the
number of infections
\citep*{Heffernan2005perspectives}%
. Estimation of reproduction numbers based on different mathematical models
are reviewed by \cite{Chowell2008review},
\citet*{Obadia2012R0review}%
, and \cite{Nikbakht2019comparison}. More recent contributions focusing on
estimation of reproduction numbers for the Covid-19 pandemic include
\citet*
{Atkeson2020estimating}%
, \cite{Baqaee2020secondwave}, \cite{ElliottGourieroux2020},
\cite{Fernandez2020SIRD}, \cite{Korolev2020JoE}, and \cite{Toda2020SIR}.

In this study, we estimate the transmission rate using the moment conditions
derived from our stochastic individual-based network SIR model. We do not use
mortality data due to its unreliability,\footnote{The recorded Covid death
toll has undergone major revisions on several occasions. For example, the UK
death toll was revised downwards by 5,377 on August 12, 2020, after a review
concluded the daily death figure should only include deaths that had occurred
within 28 days of a positive test.} but instead, our method of moment
estimation requires only data on per capita infected cases. Our estimation
method is not only simple to apply but also accounts for the time-varying
under-reporting of cases. It has been widely acknowledged that the reported
infected cases may suffer from considerable under-reporting, especially during
the early stages of the epidemic. \cite{Li2020undocumented} estimate that only
$14$ percent of all infections were documented in China prior to the January
23, 2020 travel restrictions. This translates to a multiplication factor (MF)
of $1/0.14\approx7.1$. \cite{Jagodnik2020correcting} estimate that the
recorded cases were under-reported by a factor in the range of $3$ to $16$
times in seven countries as of March 28, 2020.\footnote{See Table 2 of
\cite{Jagodnik2020correcting}. The seven countries considered are China,
France, Italy, Spain, the US, Germany, and the UK.} In the US, according to
the study by \cite{Havers2020CDC} led by the Centers for Disease Control and
Prevention (CDC), the number of infected cases is likely to be ten times more
than reported based on antibody tests from March through May 2020. A more
recent study based on antibodies from the National Institutes of Health
estimates that $20$ million individuals in the US were infected by mid-July,
2020, about $17$ million more than previously thought
\citep{Kalish2021undiagnosed}%
. This implies that MF is about $20/3\approx6.7.$
\citet*{Rahmandad2020underreporting}
consider $92$ countries through December 22, 2020, and estimate that the
cumulative cases are $7.03$ times the number of officially reported cases,
with $10^{th}$--$90^{th}$ percentile range $3.2$--$18$. They also find that
the magnitude of under-reporting has declined over time as testing has
increased. Another source of measurement errors is reporting delays.
\cite{Harris2020delay} estimates that in New York City, the mean delay in
reporting was five days, with $15$ percent of cases reported after ten or more
days, from June 21--August 1, 2020. Many existing estimation methods of
reproduction numbers do not allow for measurement errors and might not be
robust to acknowledged under-reporting errors. For instance, the
SUR\ estimates developed by \cite{Korolev2020JoE} may be biased downward if
one neglects the under-reporting of confirmed cases.

Our study also contributes to a growing literature on quantitative epidemic
policy analyses. We focus on two counterfactual analyses, but our model can be
used in a variety of other contexts. First, we investigate the impact of
different vaccination strategies in conjunction with social distancing policy
on the evolution of the epidemic. Second, we study the timing of the
lockdowns, comparing the spread of Covid-19 in UK and Germany in March 2020. A
number of studies have used the SIR\ or other compartmental models to consider
the effects of different intervention strategies (such as isolating the
elderly, closing schools and/or workplaces, and alternating work/school
schedules)\ by hypothetically lowering the average number of contacts of some
specific age groups, and/or contact locations/schedules from normal
(pre-Covid) levels. (See \cite{Acemoglu2020optimal},
\cite{Akbarpour2020socioeconomic}, \cite{Ferguson2020report},
\cite{Matrajt2020Washington}, \cite{Willem2020socrates}, among others.)
\citet*{CPR2020distancing}
simulate the trade-off between flattening the epidemic curves and lessening
unemployment loss under different degrees of mandatory and voluntary social
distancing policies using a modified SIR model. \cite{Toda2020SIR} simulates
the effects of different mitigation policies on epidemic curves by reducing
the transmission rate in the SIR\ model from its initial level.
\cite{Atkeson2020estimating} investigate the impact of earlier or later
mitigation measures on the death toll. Our model can be used to investigate
different non-pharmaceutical interventions either by lowering the number of
contacts across age groups and/or by reducing the rate of infection upon
contact. To study the effect of vaccination on controlling infection, we
calibrate the model parameters so that the reduction in the probability of
infection matches a given vaccine efficacy. A large body of literature has
extended the standard SIR models to allow for vaccination. The typical method
is to add an additional compartment, V, for vaccinated individuals and model
its relationship with other compartments by another differential equation.
See, for example,
\citet*{Berkane2021vacc}%
, \cite{Dashtbali2021vacc}, and \cite{Schlickeiser2021vacc}.

\section{Theoretical details and extensions\label{Sup: theory}}

\subsection{Relation to the multigroup SIR model \label{Sec: Relation to SIR}}

In this section, we show that the classical multigroup SIR model given by
(\ref{SIR-S})--(\ref{SIR-R}) of the main paper is a linearized-deterministic
version of our moment conditions. To see this, using the identity $S_{\ell
t}=n_{\ell}-C_{\ell t}$ and $s_{\ell t}=S_{\ell t}/n_{\ell}$,
(\ref{clmoments3}) of the main paper can be expressed as%
\begin{equation}
E\left(  s_{\ell,t+1}|s_{\ell t},\mathbf{i}_{t}\right)  =s_{\ell t}\exp\left(
-\sum_{\ell^{\prime}=1}^{L}\beta_{\ell\ell^{\prime}}i_{\ell^{\prime}t}\right)
+O\left(  n^{-1}\right)  \thickapprox s_{\ell t}\left(  1-\sum_{\ell^{\prime
}=1}^{L}\beta_{\ell\ell^{\prime}}i_{\ell^{\prime}t}\right)  +O\left(
n^{-1}\right)  . \label{slmoment}%
\end{equation}
Let $\Delta s_{\ell,t+1}=s_{\ell,t+1}-s_{\ell t}$. Then (\ref{slmoment}) can
be rewritten as%
\begin{equation}
E\left(  \Delta s_{\ell,t+1}|s_{\ell t},\mathbf{i}_{t}\right)  \thickapprox
-s_{\ell t}\sum_{\ell^{\prime}=1}^{L}\beta_{\ell\ell^{\prime}}i_{\ell^{\prime
}t}+O\left(  n^{-1}\right)  . \label{E(change in sl)}%
\end{equation}
In comparison, dividing both sides of (\ref{SIR-S}) in the multigroup
SIR\ model by $n_{\ell}$ gives%
\begin{equation}
\Delta s_{\ell,t+1}=-s_{\ell t}\sum_{\ell^{\prime}=1}^{L}\beta_{\ell
\ell^{\prime}}i_{\ell^{\prime}t}. \label{SIR_change in sl}%
\end{equation}
To exactly match the deterministic expression of $s_{\ell t}$ given by
(\ref{SIR_change in sl}) with the stochastic process given by
(\ref{E(change in sl)}), we can introduce either an additive or a
multiplicative random error to the right-hand side of (\ref{SIR_change in sl}%
). To ensure that $s_{\ell t}$ is non-negative for all $t$, a multiplicative
error with mean unity would be a more reasonable choice.

Turning to the recovery process, the recovery governed by (\ref{MomRec2}) of
the main paper matches with the deterministic recovery equation,
(\ref{SIR-R}), of the SIR\ model under a geometric recovery process. Finally,
since $I_{\ell t}=n_{\ell}-R_{\ell t}+S_{\ell t}$, the active cases of our
model also match with the infected equation, (\ref{SIR-I}), of the SIR\ model.

\subsection{Truncated geometric model of recovery\label{Sup: GeomTruncated}}

Here we consider a generalization of the recovery model used in the main
paper. Suppose that for all individuals in group $\ell$ ($\ell=1,2,\ldots,L$),
the time to recovery (or infection duration), denoted by $T_{i\ell,t}^{\ast
}=t-t_{i\ell}^{\ast}$, follows a truncated geometric distribution with the
probability mass distribution%
\begin{equation}
\Pr\left(  T_{i\ell,t}^{\ast}=t-t_{i\ell}^{\ast}\right)  =A_{\ell}\left(
1-\gamma_{\ell}\right)  ^{t-t_{i\ell}^{\ast}}\text{, for }t-t_{i\ell}^{\ast
}=1,2,\ldots,\mathfrak{D}_{\ell}, \label{TrunGeom}%
\end{equation}
where $\mathfrak{D}_{\ell}$ is the maximum number of days for an individual to
recover and is assumed to be the same for all individuals in group $\ell$.
$\gamma_{\ell}$ is the probability of recovery on each day if the geometric
distribution is non-truncated (i.e., $\mathfrak{D}_{\ell}\rightarrow\infty$).
$A_{\ell}$ is a normalizing constant such that%
\[
A_{\ell}\sum_{\mathfrak{s}=1}^{\mathfrak{D}_{\ell}}\left(  1-\gamma_{\ell
}\right)  ^{\mathfrak{s}}=A_{\ell}\left[  \frac{1-\left(  1-\gamma_{\ell
}\right)  ^{\mathfrak{D}_{\ell}}}{1-\left(  1-\gamma_{\ell}\right)  }\right]
=1,
\]
which yields%
\begin{equation}
A_{\ell}=\frac{\gamma_{\ell}}{1-\left(  1-\gamma_{\ell}\right)  ^{\mathfrak{D}%
_{\ell}}}. \label{Al}%
\end{equation}
Then the "hazard function", denoted by $h_{\ell}\left(  \mathfrak{s}%
,\mathfrak{D}\right)  $ ($\mathfrak{s}=1,2,\ldots,\mathfrak{D}$), defined as
the probability of individuals in group $\ell$ recovering at time
$\mathfrak{s}$ conditional on having remained infected for $\mathfrak{s}-1$
days is given by%
\begin{align*}
h_{\ell}\left(  \mathfrak{s},\mathfrak{D}_{\ell}\right)   &  =\frac{\Pr\left(
T^{\ast}=\mathfrak{s}\right)  }{\Pr\left(  T^{\ast}>\mathfrak{s}-1\right)
}=\frac{\Pr\left(  T^{\ast}=\mathfrak{s}\right)  }{1-\Pr\left(  T^{\ast}%
\leq\mathfrak{s}-1\right)  }\\
&  =\frac{A_{\ell}\left(  1-\gamma_{\ell}\right)  ^{\mathfrak{s}-1}}%
{1-A_{\ell}\sum_{x=1}^{\mathfrak{s}-1}\left(  1-\gamma_{\ell}\right)  ^{x}}.
\end{align*}
Now using (\ref{Al}) we have
\[
h_{\ell}\left(  \mathfrak{s},\mathfrak{D}_{\ell}\right)  =\frac{\gamma_{\ell
}\left(  1-\gamma_{\ell}\right)  ^{\mathfrak{s}-1}}{\left(  1-\gamma_{\ell
}\right)  ^{\mathfrak{s}-1}-\left(  1-\gamma_{\ell}\right)  ^{\mathfrak{D}%
_{\ell}}},\text{ for }\mathfrak{s}=1,2,\ldots,\mathfrak{D}_{\ell}\mathfrak{.}%
\]
Note that given a finite $\mathfrak{D}_{\ell}$ and $0<\gamma_{\ell}<1$,
$h_{\ell}\left(  \mathfrak{s},\mathfrak{D}_{\ell}\right)  $ monotonically
increases with $\mathfrak{s}$. Hence, by assuming a truncated geometric
distribution for recovery time, we are able to allow for the possibility that
the longer an individual is infected, the more likely s/he will recover. It is
also clear that $h_{\ell}\left(  \mathfrak{s},\mathfrak{D}_{\ell}\right)
\rightarrow\gamma_{\ell}$ as $\mathfrak{D}_{\ell}\rightarrow\infty$, which
establishes the familiar result for a non-truncated geometric distribution
used in the main paper. Under the truncated geometric distribution, we have%
\begin{align*}
&  \text{ \ \ }E\left[  \zeta_{i\ell,t+1}\left(  t_{i\ell}^{\ast}\right)
\left\vert x_{i\ell,t},y_{i\ell,t},y_{i\ell,t-1},\ldots,y_{i\ell,t_{i\ell
}^{\ast}}\right.  \right] \\
&  =\Pr\left[  \zeta_{i\ell,t+1}\left(  t_{i\ell}^{\ast}\right)  =1\left\vert
x_{i\ell,t},y_{i\ell,t}=0,y_{i\ell,t-1}=0,\ldots,y_{i\ell,t_{i\ell}^{\ast}%
}=0\right.  \right] \\
&  =h_{\ell}\left(  t-t_{i\ell}^{\ast},\mathfrak{D}_{\ell}\right)  ,
\end{align*}
and the recovery process will be given by
\begin{align}
E\left(  R_{\ell,t+1}|R_{\ell t},C_{\ell t}\right)   &  =R_{\ell,t}+\sum
_{i=1}^{n_{\ell}}h_{\ell}\left(  t-t_{i\ell}^{\ast},\mathfrak{D}_{\ell
}\right)  \left(  1-y_{i\ell,t}\right)  x_{i\ell,t}\nonumber\\
&  =R_{\ell,t}+\sum_{i=1}^{n_{\ell}}\frac{\gamma_{\ell}\left(  1-\gamma_{\ell
}\right)  ^{t-t_{i\ell}^{\ast}-1}}{\left(  1-\gamma_{\ell}\right)
^{t-t_{i\ell}^{\ast}-1}-\left(  1-\gamma_{\ell}\right)  ^{\mathfrak{D}_{\ell}%
}}\left(  1-y_{i\ell,t}\right)  x_{i\ell,t}, \label{GenRec}%
\end{align}
which does not simplify to the standard recovery process used in the SIR
models, unless $\mathfrak{D}_{\ell}\rightarrow\infty$.

\subsection{Derivation of vaccine efficacy in the multigroup
model\label{Sup: VE}}

The main paper has established Eq. (\ref{mu_sol}), $\mu^{1}/\mu^{0}%
\approx1/\left(  1-\epsilon_{v}\right)  $, assuming a single group model. Here
we show that this result also holds in the multigroup model. Recall that
\begin{align*}
E\left(  x_{i\ell,t+1}\left\vert x_{i\ell,t}=0,\mu_{i\ell},\mathbf{i}%
_{t}\right.  \right)   &  =1-\prod_{\ell^{^{\prime}}=1}^{L}\left(
1-p_{\ell\ell^{^{\prime}}}+p_{\ell\ell^{^{\prime}}}e^{-\frac{\tau_{\ell}}%
{\mu_{i\ell}}}\right)  ^{I_{\ell^{\prime}t}}\\
&  \approx1-e^{\frac{\tau_{\ell}}{\mu_{i\ell}}\left(  \sum_{\ell^{\prime}%
=1}^{L}i_{\ell^{\prime}t}k_{\ell\ell^{\prime}}\right)  }\approx\frac
{\tau_{\ell}}{\mu_{i\ell}}\left(  \sum_{\ell^{\prime}=1}^{L}i_{\ell^{\prime}%
t}k_{\ell\ell^{\prime}}\right)  .
\end{align*}
Using this in (\ref{vacc_efficacy}) of the main paper and letting $i_{\ell
t}^{0}$ be the proportion of active cases in group $\ell$ when the vaccine is
introduced, we obtain%
\[
\frac{\tau_{\ell}\sum_{\ell^{\prime}=1}^{L}i_{\ell^{\prime}t}^{0}k_{\ell
\ell^{\prime}}}{\mu^{1}}=\left(  1-\epsilon_{v}\right)  \frac{\tau_{\ell}%
\sum_{\ell^{\prime}=1}^{L}i_{\ell^{\prime}t}^{0}k_{\ell\ell^{\prime}}}{\mu
^{0}},
\]
which simplifies to $\mu^{1}/\mu^{0}=1/\left(  1-\epsilon_{v}\right)  $.

\section{Calibration and simulation of the model\label{Sup: calibration}}

\subsection{Generating random networks\label{Sup: Power Law}}

This section describes how we generated random draws from the
Erd\H{o}s-R\'{e}nyi and power law networks in the case of single and
multigroup random networks used in our simulations.

First, in an Erd\H{o}s-R\'{e}nyi\ (ER) random graph, each edge has a fixed
probability of being present or not independently of all other edges.
Specifically, we generate the ER random network with $n$ nodes and a single
group by considering all possible edges and including an edge between each
distinct pair of nodes with probability $p=k/\left(  n-1\right)  $.

Second, we generate the power law random network in the case of a single group
following the standard procedure in the literature:\footnote{See, for example,
\cite{Kiss2017book}, p. 20.} at each time $t$, we first draw a degree sequence
from the (truncated) power law distribution given by (\ref{dist_power_law}),
and then generate a network with that degree sequence based on a configuration
model.\footnote{A\ configuration model is a model of a random graph with a
given degree sequence. The name "configuration" originates from
\cite{Bollobas1980configuration} meaning arrangements of edges in the model.}
Specifically, we draw a degree sequence $k_{i}\left(  t\right)  $ randomly and
independently over $i$ for $i=1,2,\cdots,n$, (with replacement), such that
$k_{i}\left(  t\right)  $ realizes with probability $p_{k_{i}}$. Then we
generate a configuration model with the degree sequence $\left\{  k_{i}\left(
t\right)  \right\}  $ by the standard algorithm -- first assign each node with
a number of stubs (half edges) that is equal to its degree, then match two
stubs uniformly at random to form an edge and continue until all stubs are
matched. Since the number of edges, denoted by $m\left(  t\right)  $, in a
graph satisfies $2m\left(  t\right)  =\sum_{i}k_{i}\left(  t\right)  $, the
generated degrees must add to an even number to be able to construct a graph.
If the generated degrees add to an even number, we simply throw them away and
generate another sequence. Also notice that this algorithm may produce
self-loops and multi-edges. This is not a concern if $n$ is sufficiently large
since the density of such problematic links is of order $O(n^{-1}%
).$\footnote{A proof can be found in \cite{Newman2018book}, pp. 373--375.} In
simulations, we discard self-loops and collapse multi-edges. The resulting
graph is used as the power law contact network for time $t$, and the same
procedure is repeated in each $t$ and over replications.

It follows from the degree distribution given by (\ref{dist_power_law}) that
the normalizing constant has the expression $C=\left(  \sum_{k_{\min}%
}^{k_{\max}}x^{-\alpha}\right)  ^{-1}$, and then the average degree of the
power law graph is%
\begin{equation}
k=E\left(  x\right)  =\sum_{k_{\min}}^{k_{\max}}xp\left(  x\right)
=C\sum_{k_{\min}}^{k_{\max}}x^{1-\alpha}=\left(  \sum_{k_{\min}}^{k_{\max}%
}x^{-\alpha}\right)  ^{-1}\left(  \sum_{k_{\min}}^{k_{\max}}x^{1-\alpha
}\right)  .\label{E(k)}%
\end{equation}
In simulations, the value of the exponent, $\alpha,$ is solved from
(\ref{E(k)}) such that $k=10$.

Given the degree sequence $\boldsymbol{k}\left(  t\right)  =\left[
k_{1}\left(  t\right)  ,k_{2}\left(  t\right)  ,\ldots,k_{n}\left(  t\right)
\right]  ^{^{\prime}}$, the (conditional) edge probability between node $i$
and node $j$ in the configuration model is\footnote{See, e.g.,
\cite{Newman2018book}, p. 373.}
\[
E\left[  d_{ij}\left(  t\right)  |\text{ }\boldsymbol{k}\left(  t\right)
\right]  =\frac{k_{i}\left(  t\right)  k_{j}\left(  t\right)  }{2m\left(
t\right)  -1},
\]
which in the limit of large $m\left(  t\right)  $ can be rewritten as%
\[
E\left[  d_{ij}\left(  t\right)  |\text{ }\boldsymbol{k}\left(  t\right)
\right]  =\frac{k_{i}\left(  t\right)  k_{j}\left(  t\right)  }{2m\left(
t\right)  }=\frac{k_{i}\left(  t\right)  k_{j}\left(  t\right)  }{\sum
_{r=1}^{n}k_{r}\left(  t\right)  }=\frac{k_{i}\left(  t\right)  k_{j}\left(
t\right)  }{nk}.
\]
Since $k_{i}\left(  t\right)  $ and $k_{j}\left(  t\right)  $ are independent
draws from the power law distribution with mean $k$, the (unconditional) edge
probability is%
\[
p_{ij}=E\left[  d_{ij}\left(  t\right)  \right]  =E\left\{  E\left[
d_{ij}\left(  t\right)  |\text{ }\boldsymbol{k}\left(  t\right)  \right]
\right\}  =\frac{k^{2}}{nk}=\frac{k}{n},
\]
which is the same as the edge probability in the ER random network.

Finally, the network with multigroup can be generated following the stochastic
block model (SBM), which is a popular random graph model for blocks (groups or
communities) in networks.\footnote{A recent review of the stochastic block
models is provided by \cite{LW2019reviewSBM}.} Recall our assumption that the
probability of contacts is homogeneous within groups but different across
groups. Node (or individual) $i$ in group $\ell$ is denoted by $\left(
i,\ell\right)  $. At each time $t$, we draw a network in which the edge
between each distinct pair of nodes, $\left(  i,\ell\right)  $ and $\left(
j,\ell^{^{\prime}}\right)  $, exists with probability $p_{\ell\ell^{^{\prime}%
}}.$ That is, the edge probabilities depend on the groups to which nodes
belong. We set the within-group probability $p_{\ell\ell}=k_{\ell\ell}/\left(
n_{\ell}-1\right)  \approx k_{\ell\ell}/n_{\ell}$, and the between-group
probability $p_{\ell\ell^{^{\prime}}}=k_{\ell\ell^{\prime}}/n_{\ell^{\prime}}%
$. By construction, we have $p_{\ell\ell^{^{\prime}}}=p_{\ell^{^{\prime}}\ell
}$ under the reciprocity condition, $n_{\ell}k_{\ell\ell^{\prime}}%
=n_{\ell^{\prime}}k_{\ell^{\prime}\ell}$. Note that if $p_{\ell\ell^{^{\prime
}}}=p$ for all groups $\ell$ and $\ell^{^{\prime}}$, the SBM reduces to the ER
random graph. If $p_{\ell\ell^{^{\prime}}}$ are not all identical, the SBM
generates ER random graphs within each group and random bipartite graphs
between groups. Accordingly, the degree distribution of the generated network
is a mixture of Poisson degree distributions. To create heavy-tailed degree
distributions or other types of degree heterogeneity, one can generalize the
SBM analogous to the configuration model or consider the degree-corrected SBM,
but these generalizations are beyond the scope of the current
paper.\footnote{See, for example, \cite{Newman2018book}, Section 12.11.16, for
a discussion.}

\subsection{Simulated properties of the model \label{Sup: properties}}

\begin{figure}[t]%
\caption
{Simulated number of new cases using a single group model with ${\mathcal
{R}_0}=3$ under different population sizes}%
\vspace{-0.2cm}%
\label{fig: diff_n}%

\begin{footnotesize}%

\begin{center}%
\hspace*{-0.5cm}%
\begin{tabular}
[c]{ccccc}%
\begin{normalsize}%
$n=10,000$%
\end{normalsize}%
&  &
\begin{normalsize}%
$n=50,000$%
\end{normalsize}%
&  &
\begin{normalsize}%
$n=100,000$%
\end{normalsize}%
\vspace{0.15cm}%
\\%
{\includegraphics[
height=1.7071in,
width=2.1283in
]%
{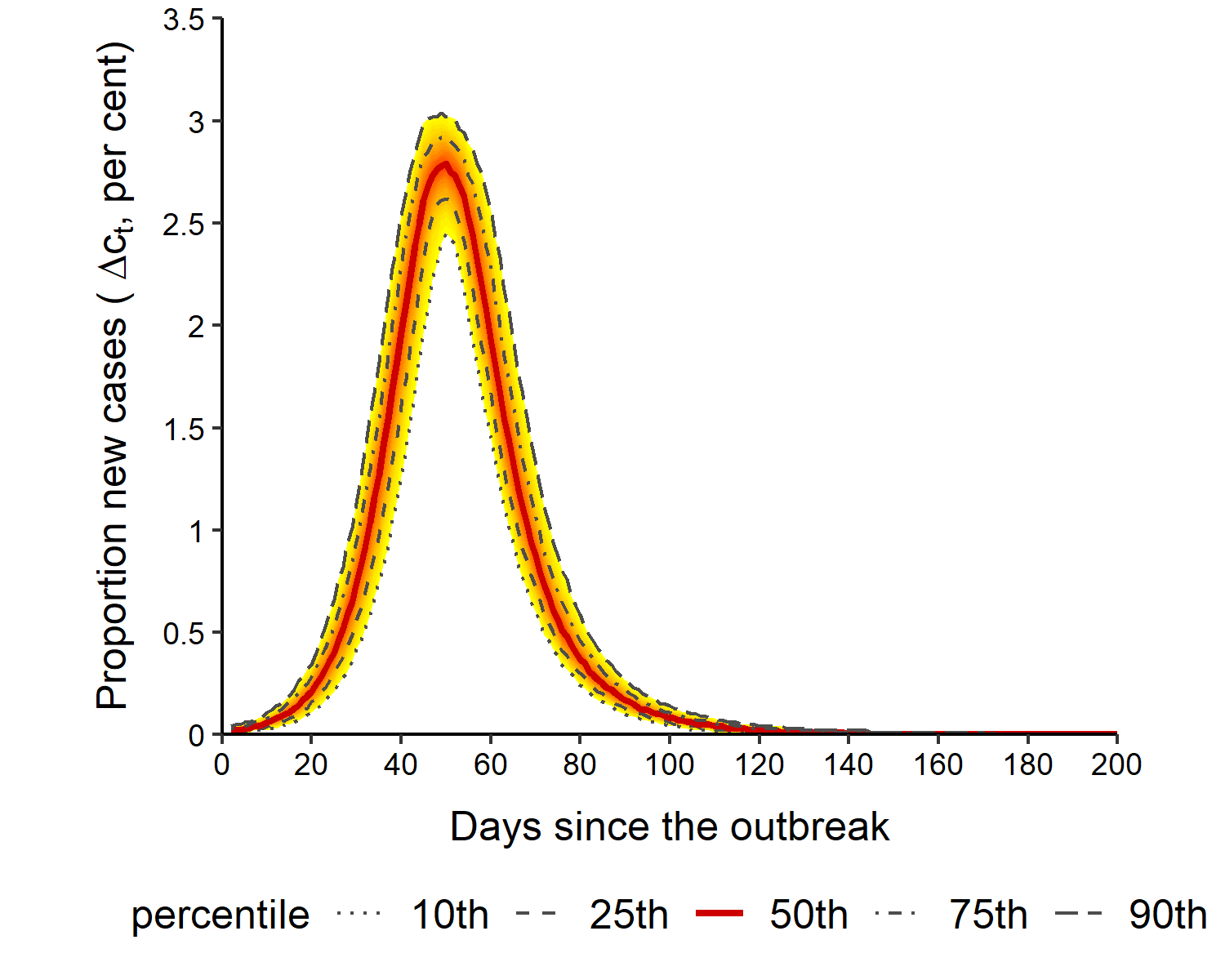}%
}
&  &
{\includegraphics[
height=1.7071in,
width=2.1283in
]%
{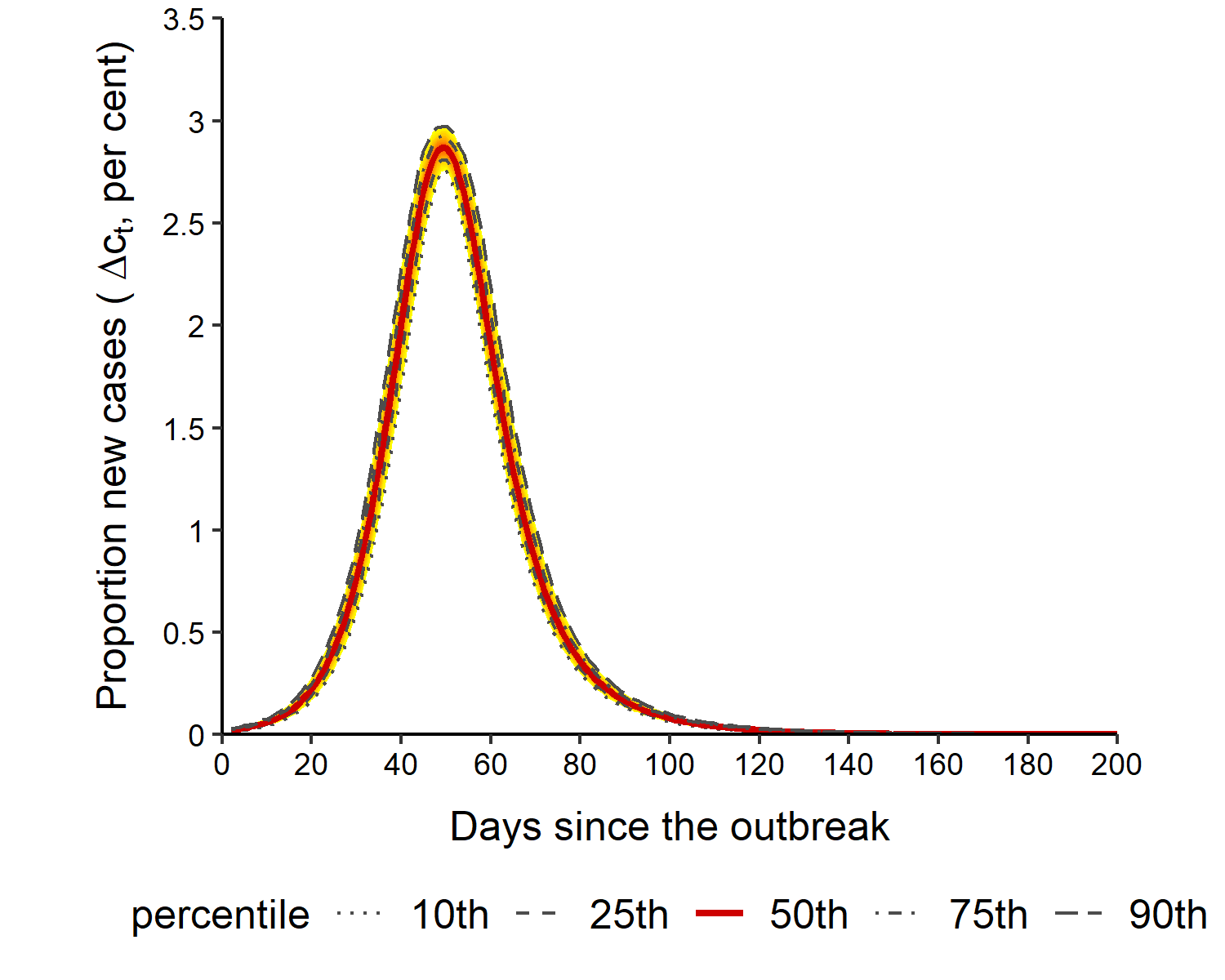}%
}
&  &
{\includegraphics[
height=1.7071in,
width=2.1283in
]%
{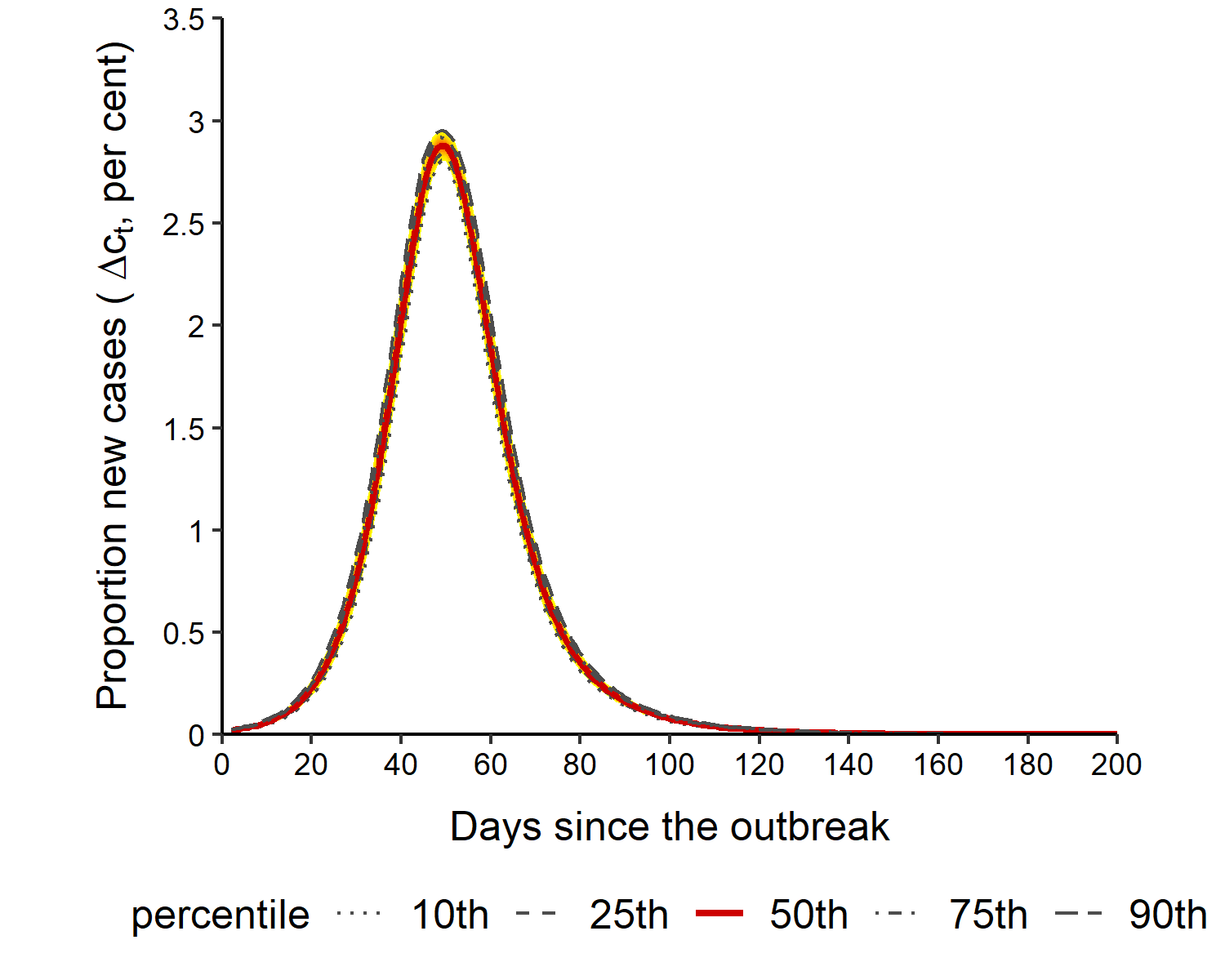}%
}
\end{tabular}

\end{center}

Notes: We set $1/1000$ of the population randomly infected on day $1$ and use
the Erd\H{o}s-R\'{e}nyi random network with mean contact number $k=10$. The
recovery rate is $\gamma=1/14$. The exposure intensity parameter is
$\tau=\gamma\mathcal{R}_{0}/k$. $c^{\ast}=B^{-1}\sum_{b=1}^{B}\max_{t}%
c_{t}^{(b)}=0.94$ in all three cases. The number of replications is $B=1,000$.%

\end{footnotesize}%
%

\end{figure}

First, we consider how the simulated properties of the proposed model vary as
we increase the population size, $n$. Specifically, we carry out simulations
with $n=10,000$, $50,000$, and $100,000$, assuming a fixed transmission rate
and a single group model, where the parameters take values\ $k=10$,
$\gamma=1/14$, $\mathcal{R}_{0}=3$, $\beta=\gamma\mathcal{R}_{0}=3/14$, and
$\tau=\beta/k$. For each replication, the simulation is initialized with
$1/1000$ of the population randomly infected on day $1$. The number of
replications is set to $B=1,000$ for all experiments. Figure \ref{fig: diff_n}
displays the proportions of new cases. As can be seen, the simulated cases are
hardly affected by the choice of $n$ in the range of ($10,000$, $100,000$).
The maximum proportion of infected, $c^{\ast}=B^{-1}\sum_{b=1}^{B}\max
_{t}c_{t}^{(b)}$, equals $0.94$ in all three cases. The time at which new
cases peak is also almost the same across $n$. Although uncertainty in the
simulation results decreases with larger $n$, the interquartile range with
$n=10,000$ is quite tight.%

\begin{figure}[tbh]%
\caption
{The average number of aggregate new cases using the single- and multi-group models with ${\mathcal
{R}_0}=3$}%
\vspace{-0.3cm}%
\label{fig: compare_multi_agg}

\begin{center}%
\begin{tabular}
[c]{c}%
{\includegraphics[
height=1.9951in,
width=2.6524in
]%
{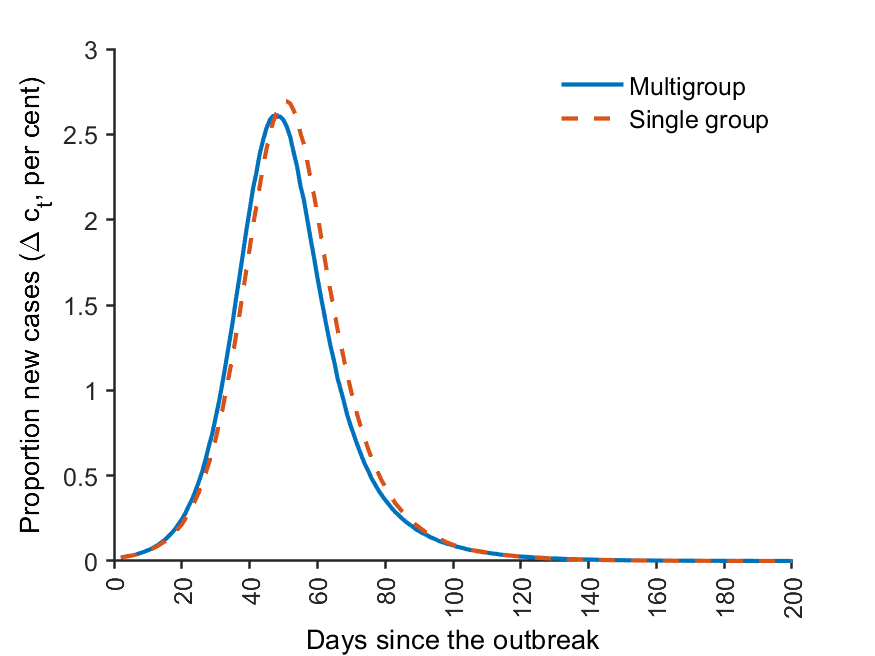}%
}
\end{tabular}

\end{center}

%

\vspace{-0.3cm}%
\footnotesize
{}Notes: The average proportion of new cases over $1,000$ replications is
displayed. The average number of new cases is very close to the median,
although not shown. Population size is $n=10,000$.$\ $In the case of a single
group, the Erd\H{o}s-R\'{e}nyi random network with mean contact number $k=10$
was used. $c^{\ast}=B^{-1}\sum_{b=1}^{B}\max_{t}c_{t}^{(b)}=0.94$, and the
duration of the epidemic is $T^{\ast}=212$ days. In the case of the multigroup
model, $c^{\ast}=0.90$, and $T^{\ast}=215$ days.%

\end{figure}

Figure \ref{fig: compare_multi_agg} compares the simulated aggregate new cases
averaged over $1,000$ replications obtained by the single group model and the
multigroup model with five age groups detailed in Section
\ref{Sec: properties} of the main paper. What stands out in the figure is the
similarity of the epidemic outcomes, including the peak of new cases, the
maximum proportion of infected, and the duration of the epidemic. This result
suggests that the aggregate outcomes do not seem to be affected by the number
of groups used in the simulations.%

\begin{figure}[t!]%
\caption{Examples of Erdos-Renyi and power law networks}%
\vspace{-0.2cm}%
\label{fig: network_examples}%
\begin{footnotesize}%

\begin{center}%
\hspace*{-0.4cm}%
\begin{tabular}
[c]{ccc}%
Erd\H{o}s-R\'{e}nyi random network &  &
\hspace*{-0.8cm}%
(Truncated) power law network\\%
{\includegraphics[
trim=0.000000in 0.600961in 0.000000in 0.478393in,
height=2.252in,
width=2.7164in
]%
{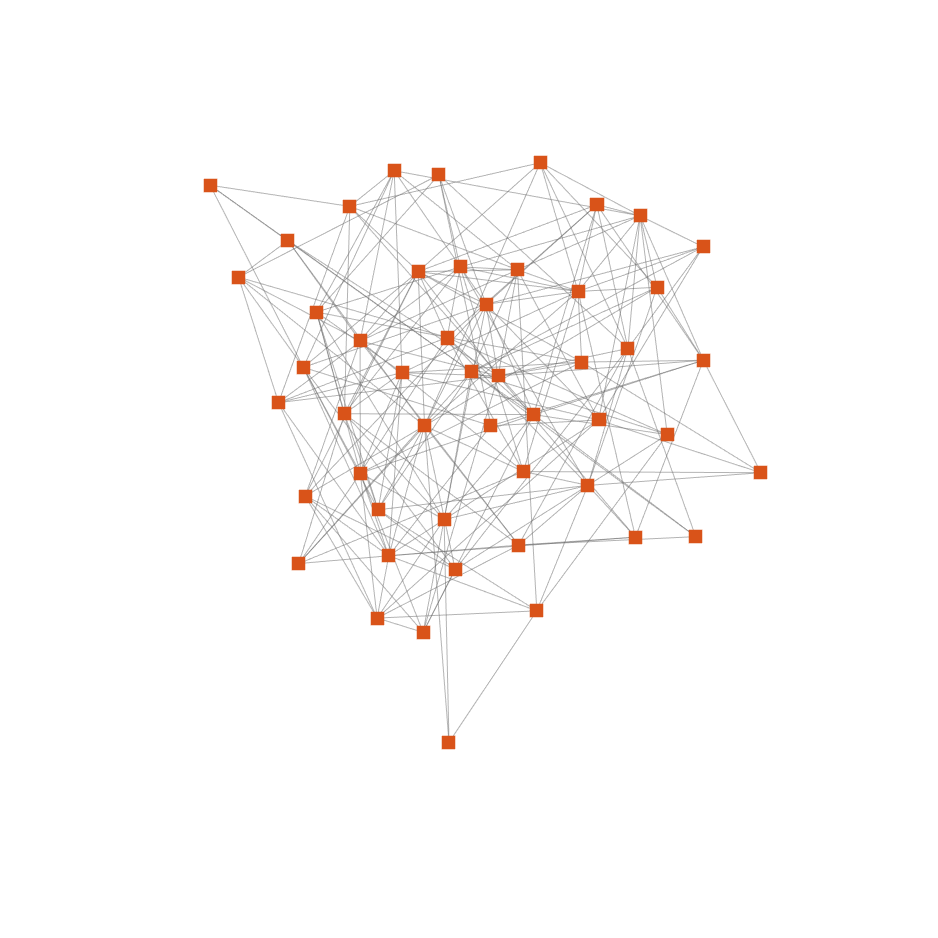}%
}
&  &
\hspace*{-0.8cm}%
\raisebox{-0.0415in}{\includegraphics[
trim=0.000000in 0.600961in 0.000000in 0.597835in,
height=2.2018in,
width=2.7164in
]%
{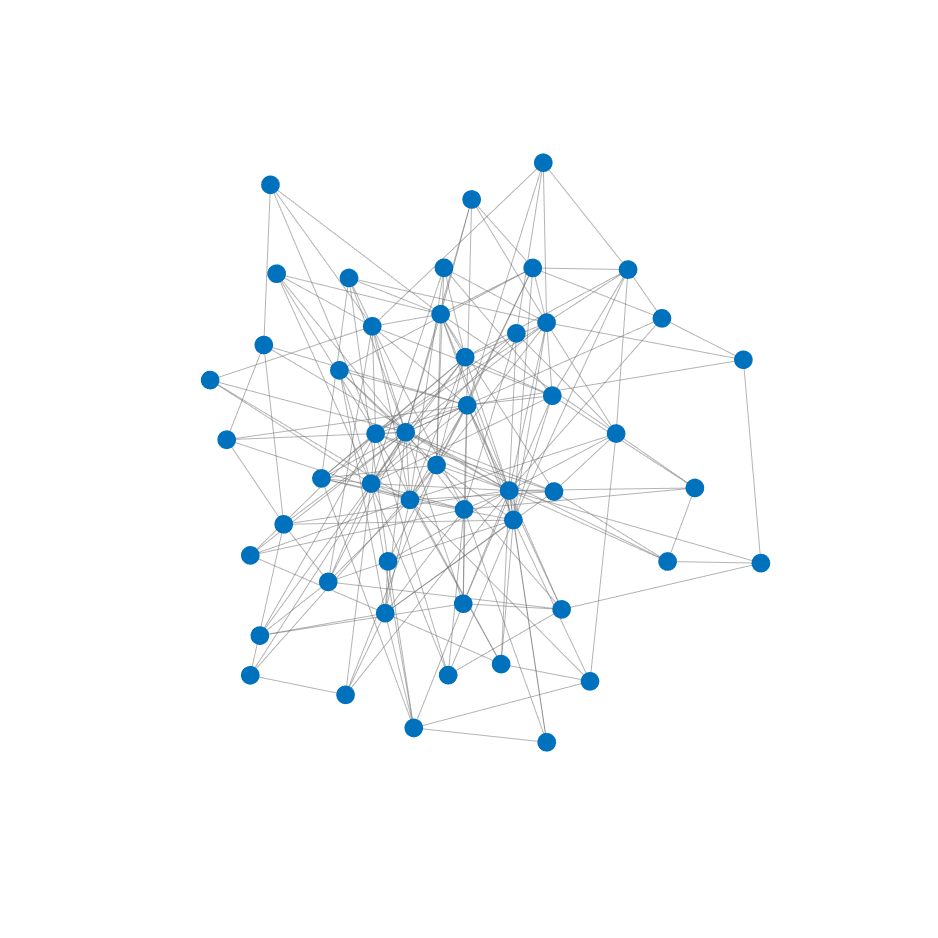}%
}
\end{tabular}

\end{center}

%

\vspace{-0.2cm}%
{}Notes: $n=50.$ Mean degree\ is $k=10$ in both networks. The degree
distribution in the power law network follows $p_{x}=Cx^{-2.43},$ for
$x=5,6,\ldots,49.$%

\end{footnotesize}%
%

\end{figure}

We next examine the effect of network topology on the simulation results. In
particular, we consider two widely used random networks -- the Erd\H
{o}s-R\'{e}nyi (ER) and the power law random networks. For simplicity, we
examine the single group model. Recall that in an ER random graph, each pair
of the nodes are connected at random with a uniform probability $p=k/\left(
n-1\right)  $. In the limit of large $n$ (with the mean degree $k$ fixed), the
ER random network has a Poisson degree distribution, which may depart from
real-world contact networks in which a small number of individuals (such as
school-aged children, medical professionals, delivery drivers, and sales
workers) may have a relatively high number of daily contacts. In other words,
the degree distribution of the contact networks may be heavy-tailed
(right-skewed). The power law random network is a popular choice to model this
phenomenon. In a (truncated) power law graph, the degree distribution follows
the power-law distribution:
\begin{equation}
p_{x}=Cx^{-\alpha},\text{ \ \ }x=k_{\min},k_{\min}+1,\ldots,k_{\max
},\label{dist_power_law}%
\end{equation}
where $p_{x}$ is the fraction of nodes in the graph with degree $x$, $k_{\min
}$ ($k_{\max}$) is the minimum (maximum) degree, $\alpha>1$ is a constant
known as the power law exponent, and $C$ is a normalization constant such that
$\sum_{k_{\min}}^{k_{\max}}p_{x}=1$. Figure \ref{fig: network_examples}
illustrates the two networks with $n=50$ nodes and the same average degree,
$k=10$.\ It is assumed that the minimum and maximum degrees of the power law
network are $k_{\min}=5$ and $k_{\max}=49$, respectively. The networks were
generated following the algorithms described in Section \ref{Sup: Power Law}
of this online supplement. It can be seen from the figure that most nodes in
the ER random network have comparable degrees with the mean degree of $10$
approximately. In contrast, the power law network has a heavy-tailed degree
distribution, and there are many small-degree nodes as well as a few highly
connected nodes in the graph.

\begin{figure}[ht!]%
\begin{footnotesize}%
\caption
{Simulated number of new cases using a single group model with ${\mathcal
{R}_0}=3$ under different network topologies}%
\vspace{-0.2cm}%
\label{fig: network_compare}

\begin{center}%
\hspace*{-0.5cm}%
\begin{tabular}
[c]{ccc}%
Random network & Power law network & Comparison of means%
\vspace{0.1cm}%
\\%
{\includegraphics[
height=1.7071in,
width=2.1283in
]%
{figs/theory_ER_N10000_dcT.png}%
}
&
{\includegraphics[
height=1.7071in,
width=2.1283in
]%
{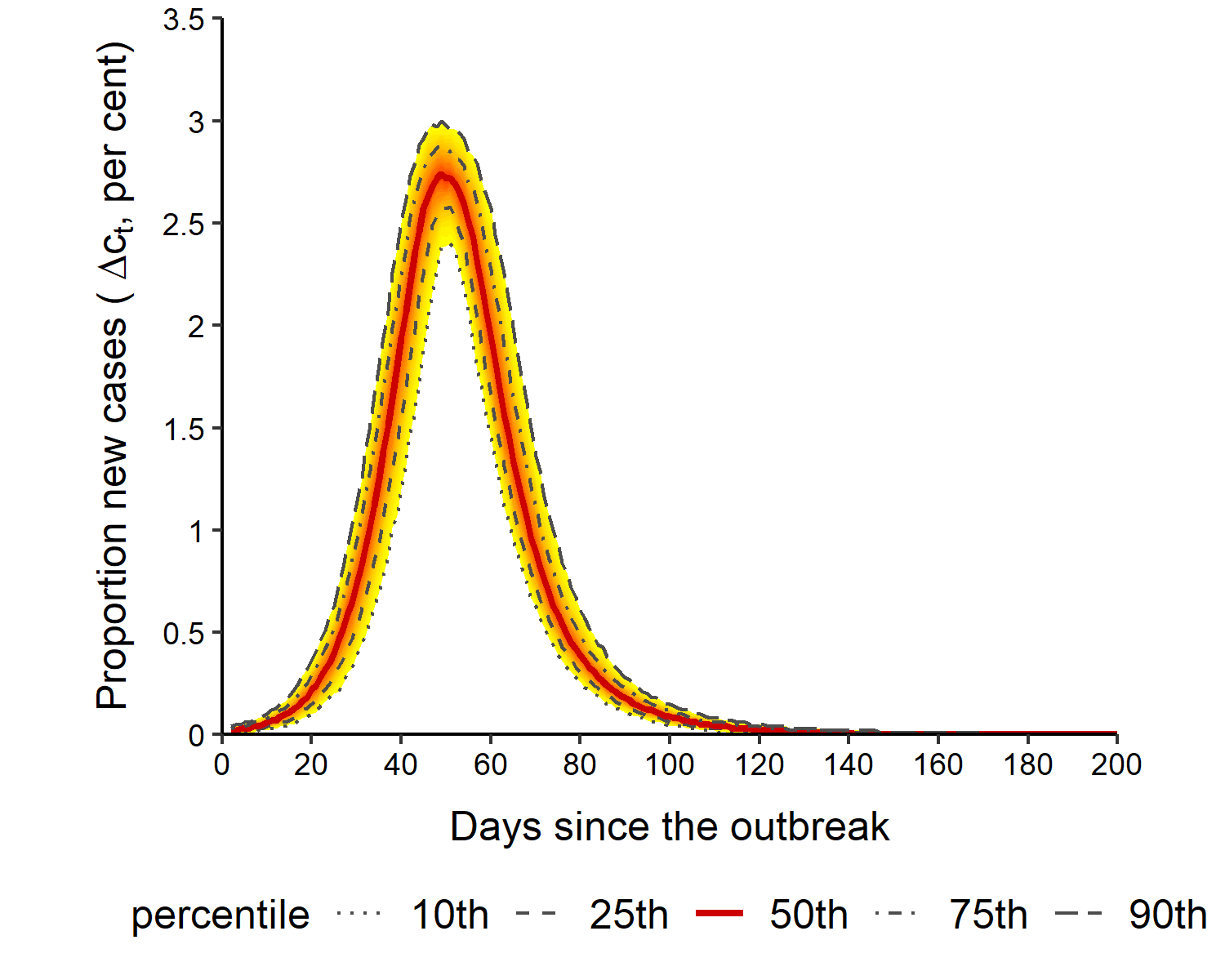}%
}
&
{\includegraphics[
trim=0.000000in -0.417660in 0.000000in 0.306138in,
height=1.6431in,
width=2.1283in
]%
{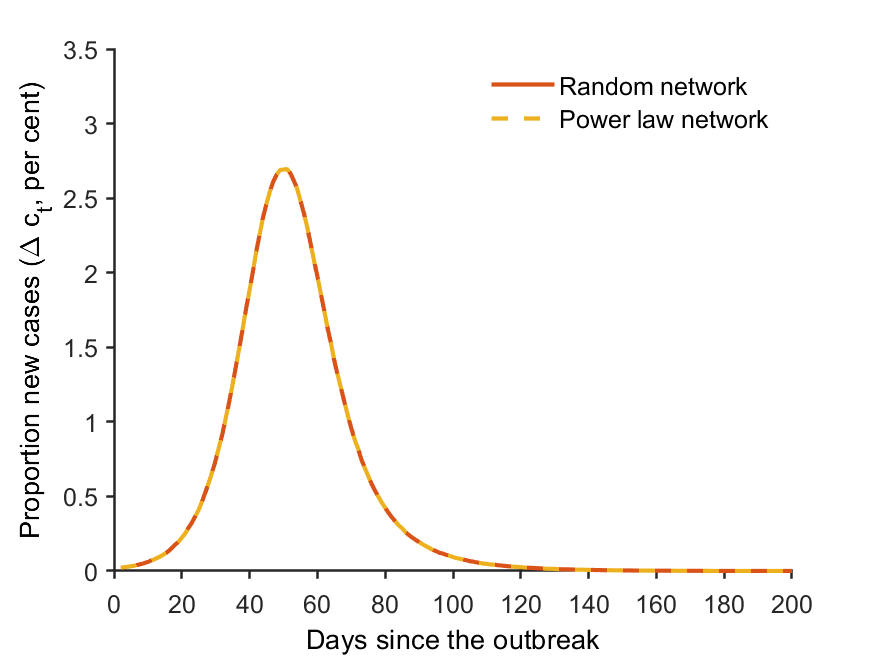}%
}
\vspace{0.1cm}%
\end{tabular}

\end{center}

%

\vspace{-0.3cm}
Notes: We set $1/1000$ of the population randomly infected on day $1$. Both
networks have mean degrees $k=10$. The recovery rate is $\gamma=1/14$. The
exposure intensity parameter is $\tau=\gamma\mathcal{R}_{0}/k$. Population
size is $n=10,000$. $c^{\ast}=B^{-1}\sum_{b=1}^{B}\max_{t}c_{t}^{(b)}=0.94$
using both networks. The number of replications is $B=1,000.$%

\end{footnotesize}%
%

\end{figure}

Figure \ref{fig: network_compare} compares the simulation results obtained
using the two random networks with the same average degree of $10$. We set
$k_{\min}=5$ and $k_{\max}=50$ for the power law networks. The values of
$\gamma$, $\mathcal{R}_{0}$, and $\tau$, and initialization of the simulation
process are as given above. We plot the proportion of new cases with
uncertainty bands and the mean values across $1,000$ replications for easy
comparison. It is clear from Figure \ref{fig: network_compare} that the mean
epidemic curves obtained by the two different random networks overlap.
Although not shown, the median simulation results are very close to the mean
values for both types of networks. We therefore focus on using the random
network in our simulation and calibration exercises.%

%

\section{Estimation of transmission and recovery rates
\label{Sup: estimation both}}

\subsection{Estimation of transmission rates \label{Sup: estimation}}

This section first provides further evidence on the performance of the rolling
estimators of the transmission rates assuming no measurement errors, and then
describes the method that estimates the transmission rate and the
multiplication factor jointly. Table \ref{tab: R0_3W}, which complements Table
\ref{tab: R0_2W} in the main paper, reports the finite sample properties of
the 3-weekly rolling estimates of $\mathcal{R}_{0}$ in the case where it is
fixed at $\mathcal{R}_{0}=3$. The simulated data were obtained under the same
set-up as that for Table \ref{tab: R0_2W}, and are based on a single group
model with the random network and the parameter values $k=10$, $\gamma=1/14$,
and $\beta=3/14$. Table \ref{tab: R0_3W} presents the bias and root mean
square error (RMSE) of the rolling estimates, $\mathcal{\hat{R}}_{0}\left(
W\right)  =\hat{\beta}_{t}\left(  W\right)  /\gamma$, where the window size
$W=3$ weeks, and $\hat{\beta}_{t}\left(  W\right)  $ is computed based on
(\ref{betathat}) of the main paper. The results refer to averages computed
over the four non-overlapping $3$-weekly sub-samples covering the
$4^{th}-15^{th}$ weeks since the outbreak.

As to be expected, the bias remains small and similar over different
sub-samples. The RMSE is smaller in the middle of the epidemic than at the
beginning and end stages, where $i_{t}$ is very close to zero. Overall, the
average RMSE of $\mathcal{\hat{R}}_{0}$ is reasonably small compared to the
true value of $3$. Compared with Table \ref{tab: R0_2W} of the main paper, the
properties of the $2$-weekly and $3$-weekly rolling estimates are very close,
with the $3$-weekly rolling estimates having slightly smaller RMSE than the
$2$-weekly estimates.%

\begin{table}[tbh]%
\caption
{Finite sample properties of the 3-weekly rolling estimates of $\mathcal
{R}_0$, in the case where it is fixed at ${\mathcal{R}_0}=3$}%
\label{tab: R0_3W}%
\renewcommand{\arraystretch}{1.1}%
\vspace{-0.35cm}%

\begin{center}%
\begin{tabular}
[c]{llcccc}\hline
&  & \multicolumn{4}{c}{3-weekly sub-samples}\\\cline{3-6}%
\multicolumn{2}{r}{Weeks since the outbreak} & $4^{th}-6^{th}$ &
$7^{th}-9^{th}$ & $10^{th}-12^{th}$ & $13^{th}-15^{th}$\\\hline
Population &  & \multicolumn{1}{r}{} & \multicolumn{1}{r}{} &
\multicolumn{1}{r}{} & \multicolumn{1}{r}{}\\
$n=10,000$ & Bias & \multicolumn{1}{r}{-0.0119} & \multicolumn{1}{r}{-0.0037}
& \multicolumn{1}{r}{-0.0024} & \multicolumn{1}{r}{0.0026}\\
& RMSE & \multicolumn{1}{r}{0.0966} & \multicolumn{1}{r}{0.0488} &
\multicolumn{1}{r}{0.0737} & \multicolumn{1}{r}{0.1687}\\
$n=50,000$ & Bias & \multicolumn{1}{r}{-0.0019} & \multicolumn{1}{r}{-0.0002}
& \multicolumn{1}{r}{-0.0009} & \multicolumn{1}{r}{-0.0005}\\
& RMSE & \multicolumn{1}{r}{0.0395} & \multicolumn{1}{r}{0.0218} &
\multicolumn{1}{r}{0.0332} & \multicolumn{1}{r}{0.0750}\\
$n=100,000$ & Bias & \multicolumn{1}{r}{-0.0003} & \multicolumn{1}{r}{0.0005}
& \multicolumn{1}{r}{0.0000} & \multicolumn{1}{r}{-0.0006}\\
& RMSE & \multicolumn{1}{r}{0.0275} & \multicolumn{1}{r}{0.0150} &
\multicolumn{1}{r}{0.0229} & \multicolumn{1}{r}{0.0544}\\\hline
\end{tabular}

\end{center}

%

\footnotesize\flushleft
{}Notes: The true value of $\mathcal{R}_{0}$ is set to $\beta/\gamma$, where
$\beta=3/14$ and $\gamma=1/14$ so that $\mathcal{R}_{0}=3$. We fix $\gamma$
and estimate $\beta$ using (\ref{cagg_L1_copy}) in the main paper. The number
of replications is $B=1,000$.%

\end{table}%

Next, to allow for time-varying under-reporting of cases, Section
\ref{Sec: estimation R with MF} of the main paper proposes a method that
jointly estimates the transmission rate and the multiplication factor (MF).
Here we give detailed steps for the joint estimation. Let $\tilde{c}^{0}$
denote a small threshold value and $\hat{m}_{\left(  j\right)  }$ denote the
$j^{th}$ estimate of MF, for $j=1,2,\ldots$. We propose the following algorithm.

\begin{itemize}
\item In the initial period of the epidemic when $\tilde{c}_{t}\leq\tilde
{c}^{0}$, $t=1,2,\ldots,t^{0}$, carry out the rolling estimation of $\beta
_{t}$ with a guess value of MF. Then simulate the stochastic network model
using the $\left\{  \hat{\beta}_{t}\right\}  _{t=1}^{t^{0}}$, and compute the
first estimate of MF as the ratio of the mean calibrated cases to realized
cases at the end of the initial period, namely, $\hat{m}_{\left(  1\right)
}=\bar{c}_{t^{0}}/\tilde{c}_{t^{0}}$, where $\bar{c}_{t^{0}}=B^{-1}\sum
_{b=1}^{B}c_{t^{0}}^{\left(  b\right)  }.$

\item When $\tilde{c}_{t}>\tilde{c}^{0}$, we jointly estimate $\beta_{t}$ and
$m_{t}$ by the two equations below:%
\begin{equation}
\hat{\beta}_{t}\left(  W_{\beta}\right)  =\text{Argmin}_{\beta}\sum
_{\tau=t-W_{\beta}+1}^{t}\left[  \frac{1-\hat{m}_{\tau}(W_{m})\tilde{c}_{\tau
}}{1-\hat{m}_{\tau-1}(W_{m})\tilde{c}_{\tau-1}}-e^{-\beta\text{ }\hat{m}%
_{\tau-1}(W_{m})\tilde{\imath}_{\tau-1}}\right]  ^{2}. \label{joint_est_beta}%
\end{equation}%
\begin{equation}
\hat{m}_{t}(W_{m})=\frac{W_{m}^{-1}\sum_{\tau=t-W_{m}+1}^{t}\left(
1-B^{-1}\sum_{b=1}^{B}e^{-\hat{\beta}_{\tau-1}\left(  W_{\beta}\right)  \text{
}i_{\tau-1}^{\left(  b\right)  }}\right)  }{W_{m}^{-1}\sum_{\tau=t-W_{m}%
+1}^{t}\left[  \tilde{c}_{\tau}-\left(  B^{-1}\sum_{b=1}^{B}e^{-\hat{\beta
}_{\tau-1}\left(  W_{\beta}\right)  \text{ }i_{\tau-1}^{\left(  b\right)  }%
}\right)  \tilde{c}_{\tau-1}\right]  }, \label{joint_est_MF}%
\end{equation}
where $W_{\beta}$ and $W_{m}$ are the rolling window sizes. Specifically,

\begin{itemize}
\item From $t^{0}+1$ to $t_{2}=t^{0}+W_{m}$, carry out rolling estimation of
$\beta_{t}$ by (\ref{joint_est_beta}) using $\hat{m}_{\left(  1\right)  }$.
Then continue the simulations using $\left\{  \hat{\beta}_{t}\right\}
_{t=t^{0}+1}^{t_{2}}$ from the stored status, and compute $\hat{m}_{\left(
2\right)  }$ by (\ref{joint_est_MF}) at $t_{2}$.

\item From $t_{2}+1$ to $t_{3}=t_{2}+W_{m}$, carry out rolling estimation of
$\beta_{t}$ by (\ref{joint_est_beta}) using $\hat{m}_{\left(  2\right)  }$.
Then continue the simulations using $\left\{  \hat{\beta}_{t}\right\}
_{t=t_{2}+1}^{t_{3}}$ and compute $\hat{m}_{\left(  3\right)  }$ by
(\ref{joint_est_MF}) at $t_{3}$.

\item Continue the above steps to obtain $\hat{m}_{\left(  4\right)  }$,
$\hat{m}_{\left(  5\right)  }$, $\ldots,$ until the end of the sample.
\end{itemize}
\end{itemize}

In practice, MF varies slowly, and it is reasonable to consider $W_{\beta
}=W_{m}=2$ or $3$ weeks. We apply the above procedure to Covid-19 data in a
number of European countries and the US. The results are presented in Section
\ref{Sec: empirical} of the main paper and Sections \ref{Sup: empirical_Re}
and \ref{Sup: empirical_MF} of this supplement.

\subsection{Estimation of the recovery rate \label{Sup: recovery}}

As noted in the main paper, with reliable data on the number of removed
(recovered or dead), the recovery rate, $\gamma$, can be estimated using the
moment condition given by (\ref{rlmoments}) of the main paper. In reality,
however, it is hard to measure $R_{t}$ accurately. We do not estimate $\gamma$
in the calibration exercise because the data on recovery are\ either
unavailable or problematic in the countries we considered. In the current
section, we demonstrate that the recovery rate can be estimated very precisely
using simulated data. To simplify the exposition, we consider a single group
($L=1$) and suppose that the time to recovery follows a geometric distribution
as in the standard SIR\ model. The same aggregate outcome follows in the
multigroup case if the probability of recovery is the same across all groups.
Under these conditions, the aggregate moment condition for recovery can be
written as%
\begin{equation}
\Delta R_{t+1}=\gamma I_{t}+u_{n,t+1}, \label{Rt_random}%
\end{equation}
where $\Delta R_{t+1}=R_{t+1}-R_{t}$ and $u_{n,t+1}$ is a martingale
difference process with respect to $I_{t}$ and $R_{t}$. ($I_{t}=C_{t}-R_{t}$).
Recall from (\ref{Rl}) and (\ref{Il}) of the main paper that $R_{t}=\sum
_{i=1}^{n}y_{it}$ and $I_{t}=\sum_{i=1}^{n}z_{it}$, we note that $u_{n,t+1}$
is an aggregated error, namely, $u_{n,t+1}=\sum_{i=1}^{n}u_{i,t+1}.$ Dividing
both sides of (\ref{Rt_random}) by $n$ yields
\begin{equation}
\Delta r_{t+1}=\gamma i_{t}+\bar{u}_{n,t+1}, \label{reg_gamma}%
\end{equation}
where $\Delta r_{t+1}=r_{t+1}-r_{t}$, $r_{t}=R_{t}/n$, $i_{t}=I_{t}/n$ and
$\bar{u}_{n,t+1}=n^{-1}\sum_{i=1}^{n}u_{i,t+1}$. For sufficiently large $n$
and assuming that the individual differences in recovery are cross-sectionally
weakly correlated, we have $\bar{u}_{t+1}=O_{p}(n^{-1/2})$. It follows that
$\gamma$ can be consistently estimated from (\ref{reg_gamma}) by ordinary
least squares (OLS) regression of $\Delta r_{t+1}$ on $i_{t}$. Note that $T$
is finite as $n\rightarrow\infty$. Due to the presence of $O_{p}\left(
n^{-1/2}\right)  $ in (\ref{reg_gamma}), it is expected that as $n$ increases,
the randomness will diminish and estimates of $\gamma$ become increasingly
precise. In the limit we would expect $\Delta r_{t+1}-\gamma i_{t}%
=O_{p}(n^{-1/2})$.

To examine the finite sample properties of the OLS\ estimator of $\gamma$, we
simulate our model assuming a homogeneous recovery rate and compute the
aggregate time series for $B=1,000$ replications under a given population
size, $n$. Denote the recovery and infection time series of the $b^{th}$
replication by $r_{t+1}^{(b)}$ and $i_{t}^{(b)}$, respectively, for
$b=1,2,\ldots,B.$ For each replication, we obtain $\hat{\gamma}^{(b)}$ by
regressing $\Delta r_{t+1}^{(b)}$ on $i_{t}^{(b)}$, without an intercept. The
true value of $\gamma$ in the experiment is set to $1/14$.%

\begin{table}[tb]%
\caption{Finite sample properties of the rolling estimates of $\gamma$}%
\label{tab: gamma}%
\renewcommand{\arraystretch}{1.1}%
\vspace{-0.35cm}%

\begin{center}%
\begin{tabular}
[c]{llcccc}\hline
&  & \multicolumn{4}{c}{3-weekly sub-samples}\\\cline{3-6}%
\multicolumn{2}{r}{Weeks since the outbreak} & $4^{th}-6^{th}$ &
$7^{th}-9^{th}$ & $10^{th}-12^{th}$ & $13^{th}-15^{th}$\\\hline
\multicolumn{2}{l}{$2$\textbf{-weekly rolling estimates}} &  &  &  & \\
Population &  &  &  &  & \\
$n=10,000$ & Bias$\left(  \times100\right)  $ & 0.0917 & 0.0111 & 0.0040 &
0.0072\\
& RMSE$\left(  \times100\right)  $ & 0.9077 & 0.2465 & 0.1315 & 0.1833\\
$n=50,000$ & Bias$\left(  \times100\right)  $ & 0.0247 & 0.0006 & 0.0000 &
0.0023\\
& RMSE$\left(  \times100\right)  $ & 0.3716 & 0.1034 & 0.0602 & 0.0840\\
$n=100,000$ & Bias$\left(  \times100\right)  $ & 0.0116 & 0.0012 & -0.0002 &
0.0008\\
& RMSE$\left(  \times100\right)  $ & 0.2536 & 0.0730 & 0.0430 & 0.0601\\\hline
\multicolumn{2}{l}{$3$\textbf{-weekly rolling estimates}} &
\multicolumn{1}{r}{} & \multicolumn{1}{r}{} & \multicolumn{1}{r}{} &
\multicolumn{1}{r}{}\\
Population &  & \multicolumn{1}{r}{} & \multicolumn{1}{r}{} &
\multicolumn{1}{r}{} & \multicolumn{1}{r}{}\\
$n=10,000$ & Bias$\left(  \times100\right)  $ & 0.0495 & 0.0082 & 0.0041 &
0.0067\\
& RMSE$\left(  \times100\right)  $ & 0.5614 & 0.1681 & 0.1104 & 0.1639\\
$n=50,000$ & Bias$\left(  \times100\right)  $ & 0.0092 & -0.0019 & 0.0005 &
0.0020\\
& RMSE$\left(  \times100\right)  $ & 0.2302 & 0.0713 & 0.0511 & 0.0760\\
$n=100,000$ & Bias$\left(  \times100\right)  $ & 0.0055 & 0.0008 & -0.0002 &
0.0009\\
& RMSE$\left(  \times100\right)  $ & 0.1589 & 0.0512 & 0.0366 & 0.0542\\\hline
\end{tabular}

\end{center}

%

\footnotesize\flushleft
Notes: The true value of $\gamma$ is $1/14$. The estimating equation is given
by (\ref{reg_gamma}). The number of replications is $1,000$.%

\end{table}%

Table \ref{tab: gamma} reports the bias and RMSE of the OLS estimator of
$\gamma$ averaged over the four non-overlapping $3$-weekly sub-samples during
the $4^{th}-15^{th}$ weeks after the outbreak. Even though the bias and RMSE
in the table have been multiplied by $100$, they are very small in magnitude.
It is evident that we can estimate $\gamma$ very precisely even with short
time series samples and population size $n=10,000$. The RMSE\ declines as $n$
increases, lending support to the theory. As in the case of estimating the
transmission rates, the RMSEs are relatively larger in the early and late
stages of the epidemic when $i_{t}$ is small. The 2-weekly and 3-weekly
estimates are similar, with the 3-weekly estimator having some improvement as
the outbreak amplifies into an epidemic.

\section{Estimates of the effective reproduction numbers
\label{Sup: empirical_Re}}

\subsection{Estimates for selected European
countries\label{Sup: empirical_Re_Euro}}

This section provides additional estimation results of the effective
reproduction numbers ($\mathcal{R}_{et}$) and transmission rates ($\beta_{t}$)
for the six European countries considered in the main paper. First, Figure
\ref{fig: Euro_dc_TR_fixMF} presents the realized daily new cases (7-day
average per $100,000$ people) and the 2-weekly rolling estimates of the
transmission rates assuming a fixed MF $=3$ and $5$. It can be seen that the
estimates of the transmission rate under different values of MF are virtually
the same when $c_{t}$ is small. This observation clearly shows that MF is not
identified in the early stage of the epidemic. It is also clear that the daily
number of infections rapidly rises when $\beta_{t}$ is high.%

\begin{figure}[!p]%
\caption
{Realized new cases and two-weekly rolling estimates of the transmission rates for selected European countries}%
\label{fig: Euro_dc_TR_fixMF}%
\vspace{-0.3cm}%

\begin{center}%
\hspace*{-0.2cm}%
\begin{tabular}
[c]{cc}%
\multicolumn{2}{c}{{\footnotesize Austria}}\\%
{\includegraphics[
height=1.7763in,
width=3.5293in
]%
{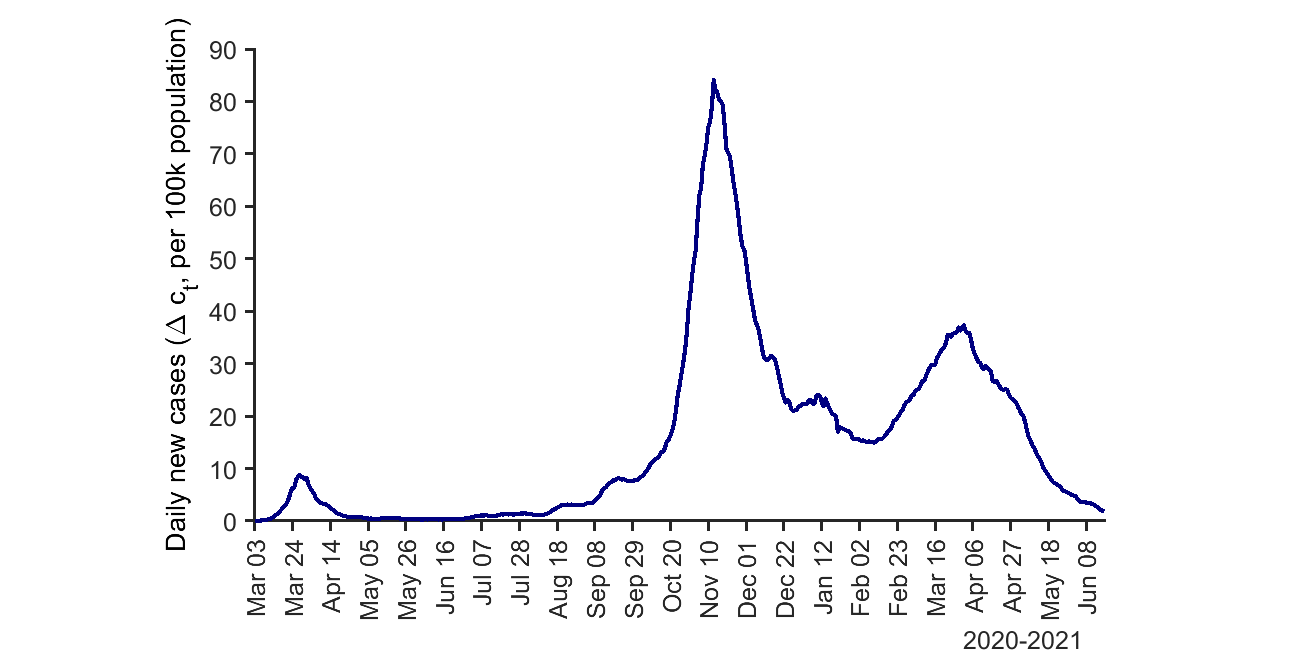}%
}
&
{\includegraphics[
height=1.7763in,
width=3.5293in
]%
{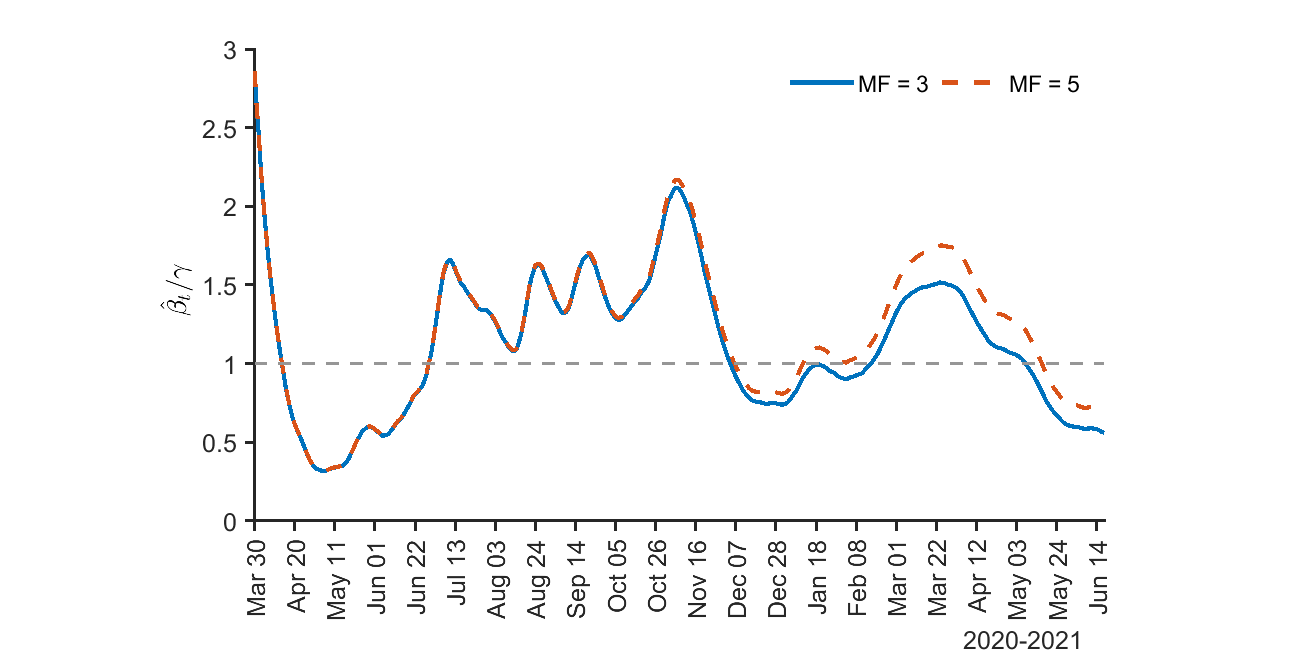}%
}
\\
& \\
\multicolumn{2}{c}{{\footnotesize France}}\\%
{\includegraphics[
height=1.7763in,
width=3.5293in
]%
{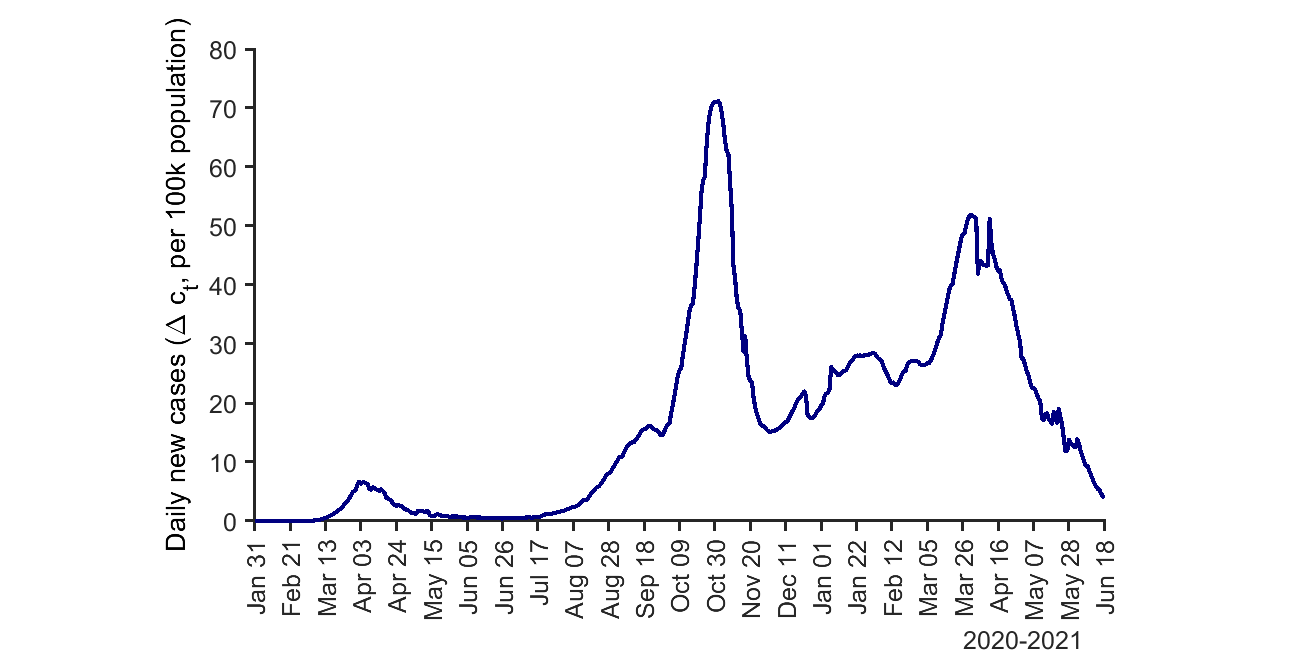}%
}
&
{\includegraphics[
height=1.7763in,
width=3.5293in
]%
{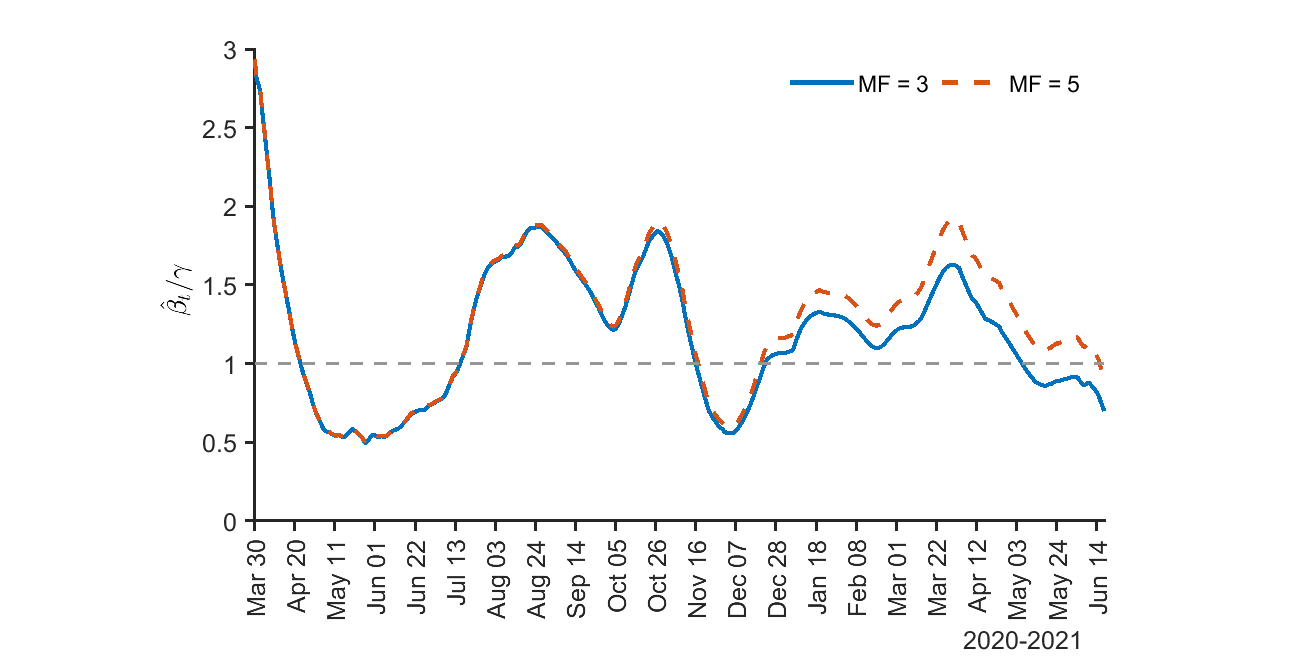}%
}
\\
& \\
\multicolumn{2}{c}{{\footnotesize Germany}}\\%
{\includegraphics[
height=1.7763in,
width=3.5293in
]%
{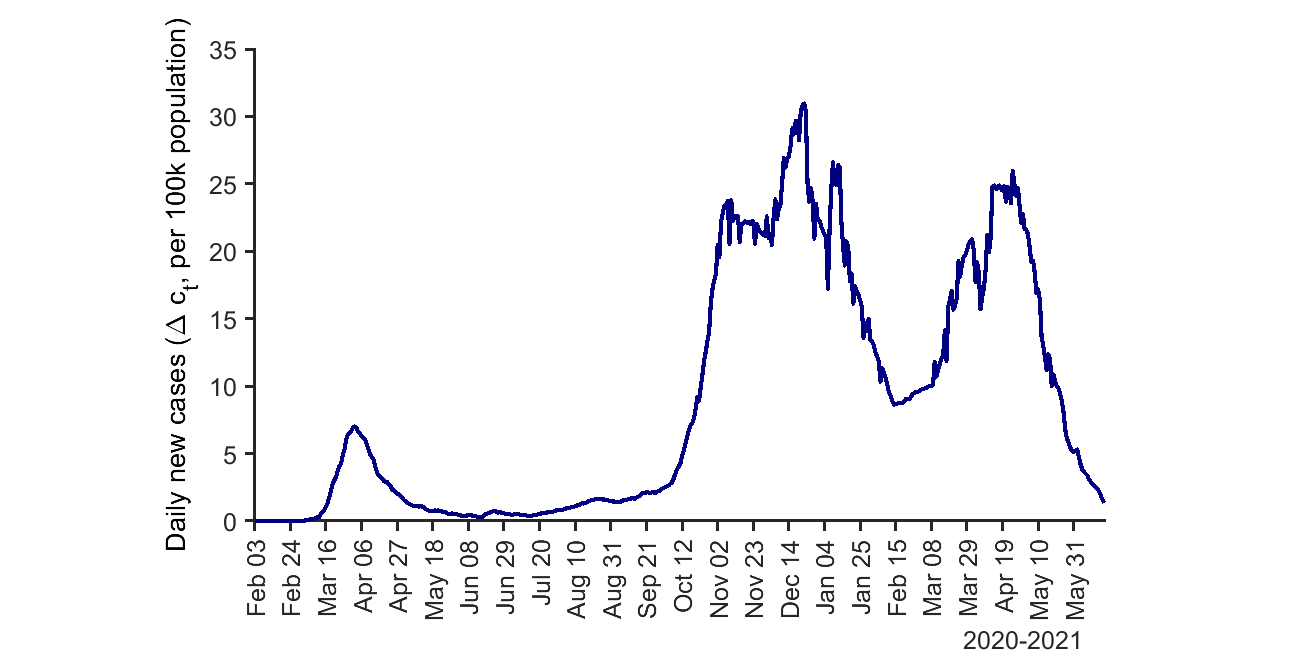}%
}
&
{\includegraphics[
height=1.7763in,
width=3.5293in
]%
{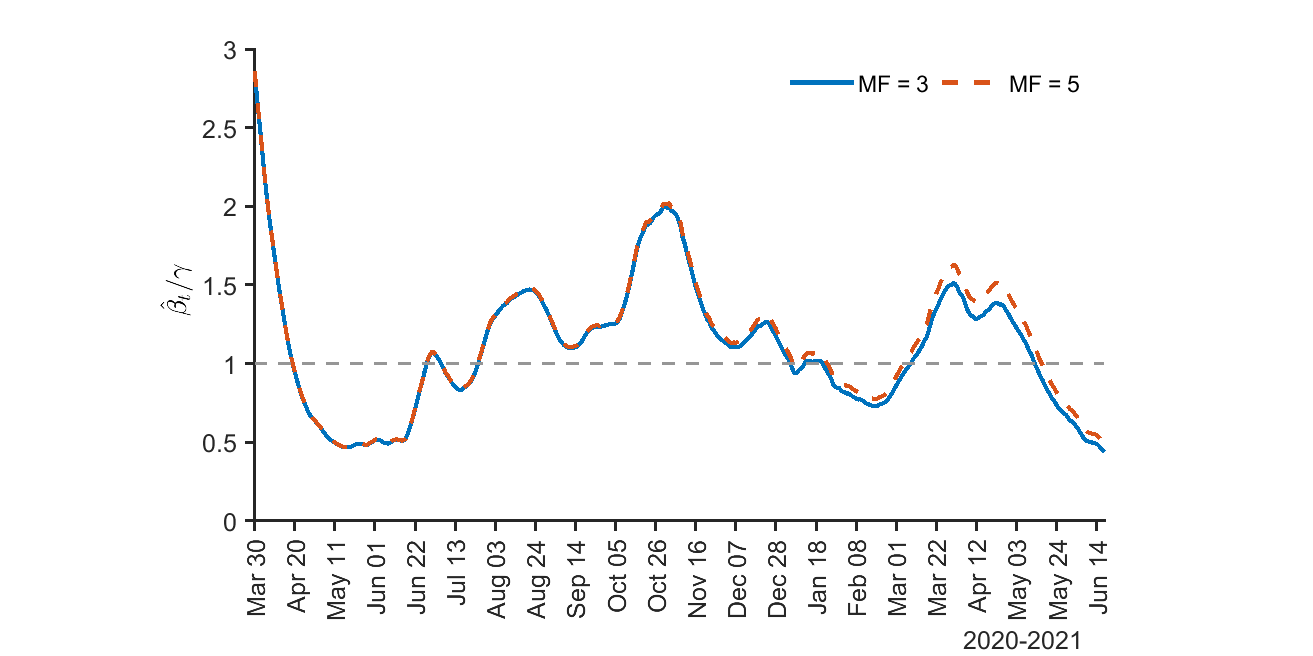}%
}
\end{tabular}

\end{center}

%

\vspace{-0.4cm}%
\footnotesize
{}Notes: The figure plots the 7-day moving average of the reported number of
new cases per 100k population and the 2-weekly rolling estimates of the
transmission rate, $\beta_{t}/\gamma$, where $\gamma=1/14$ and the
multiplication factor (MF) is fixed at $3$ and $5$.%

\end{figure}%
%

\addtocounter{figure}{-1}%
%

\begin{figure}[!p]%
\caption
{(Continued) Realized new cases and two-weekly rolling estimates of the transmission rate for selected European countries}%
\vspace{-0.3cm}%

\begin{center}%
\hspace*{-0.2cm}%
\begin{tabular}
[c]{cc}%
\multicolumn{2}{c}{{\footnotesize Italy}}\\%
{\includegraphics[
height=1.7763in,
width=3.5293in
]%
{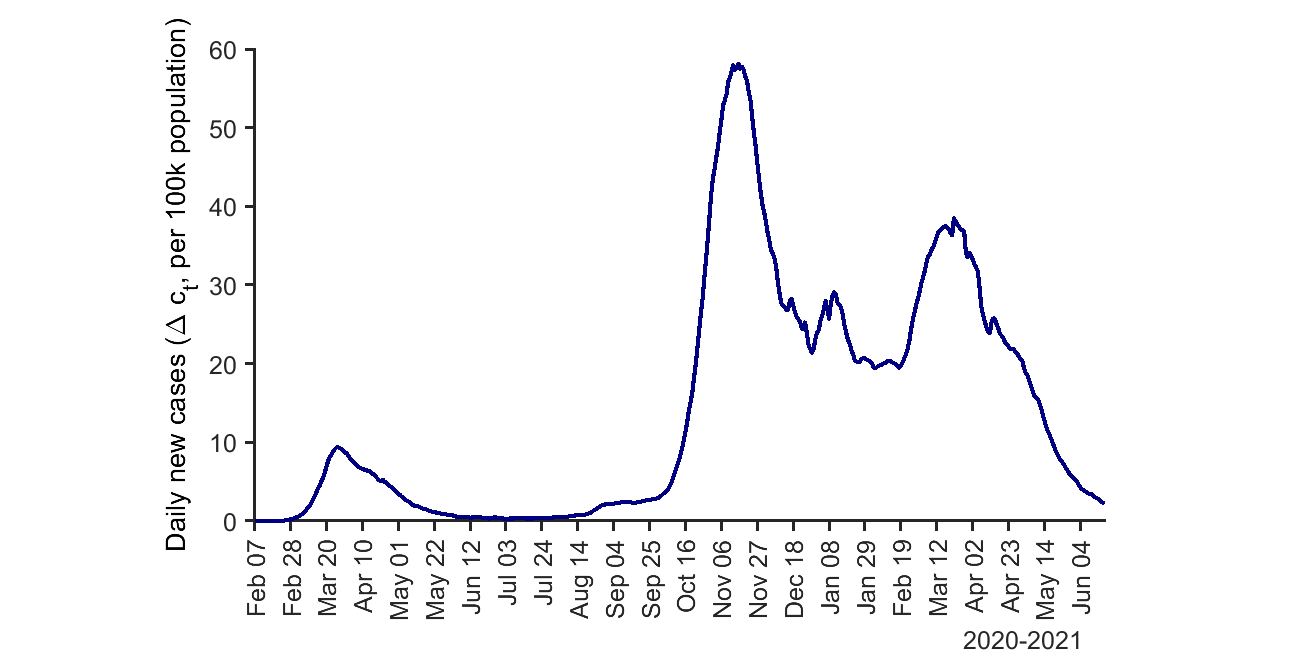}%
}
&
{\includegraphics[
height=1.7763in,
width=3.5293in
]%
{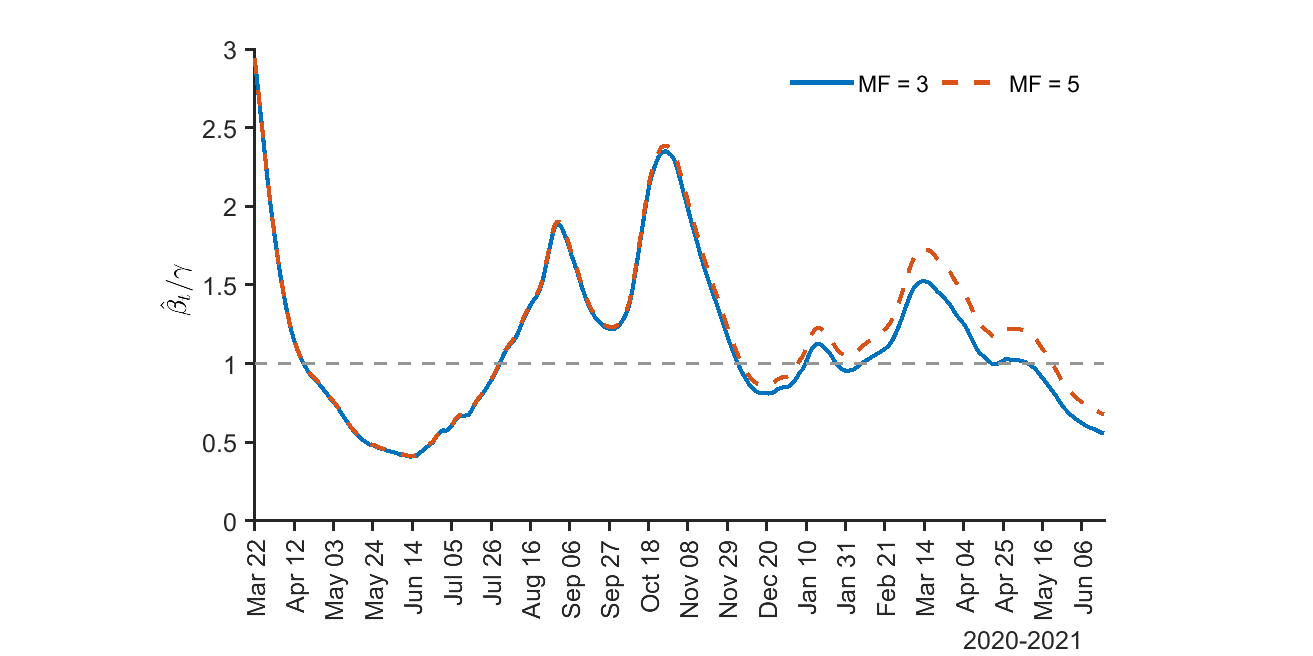}%
}
\\
& \\
\multicolumn{2}{c}{{\footnotesize Spain}}\\%
{\includegraphics[
height=1.7763in,
width=3.5293in
]%
{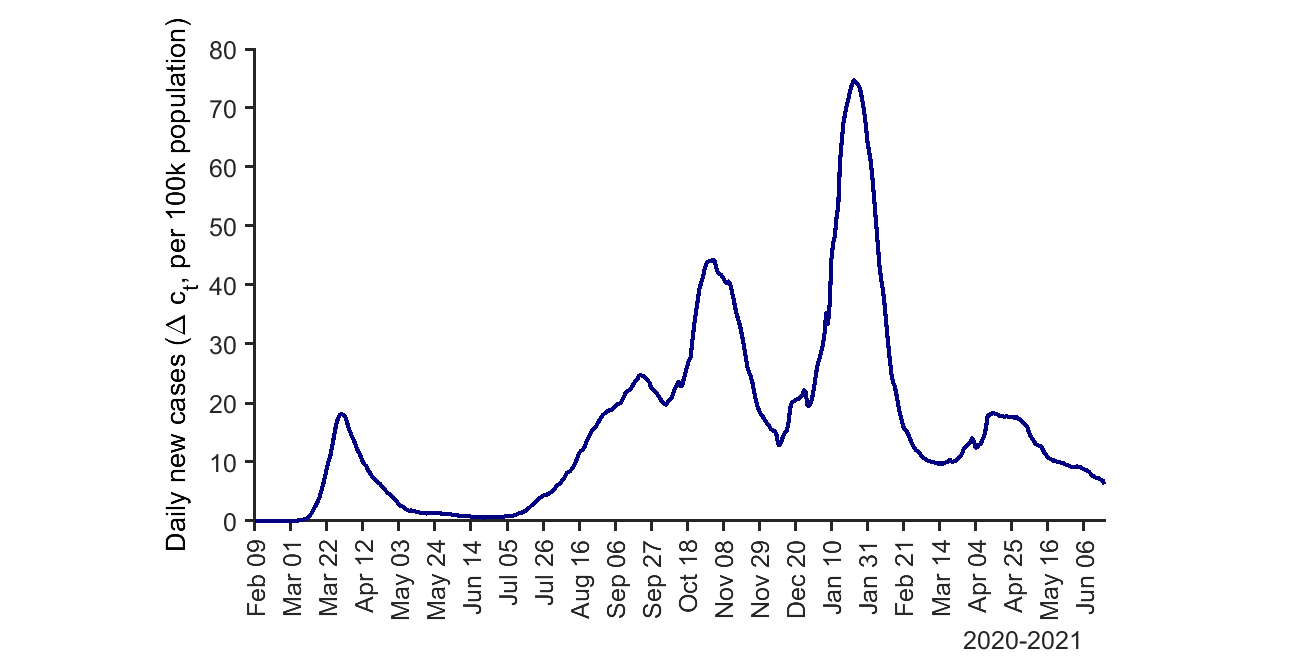}%
}
&
{\includegraphics[
height=1.7763in,
width=3.5293in
]%
{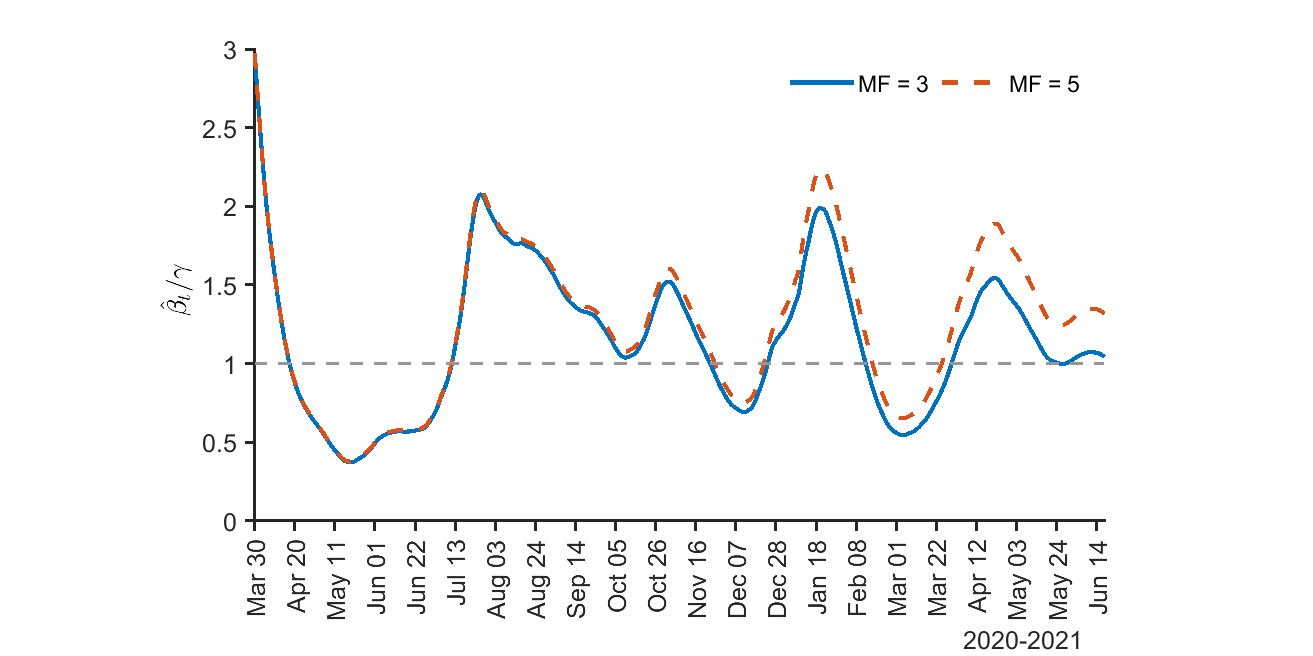}%
}
\\
& \\
\multicolumn{2}{c}{{\footnotesize UK}}\\%
{\includegraphics[
height=1.7763in,
width=3.5293in
]%
{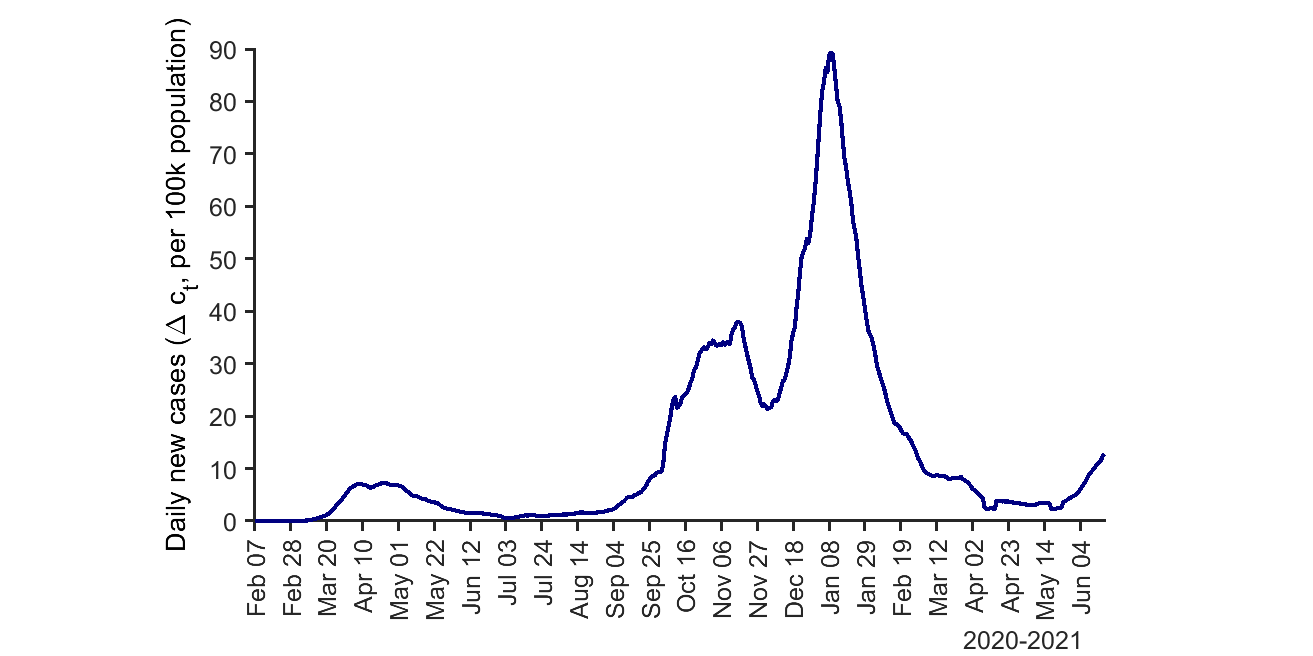}%
}
&
{\includegraphics[
height=1.7763in,
width=3.5293in
]%
{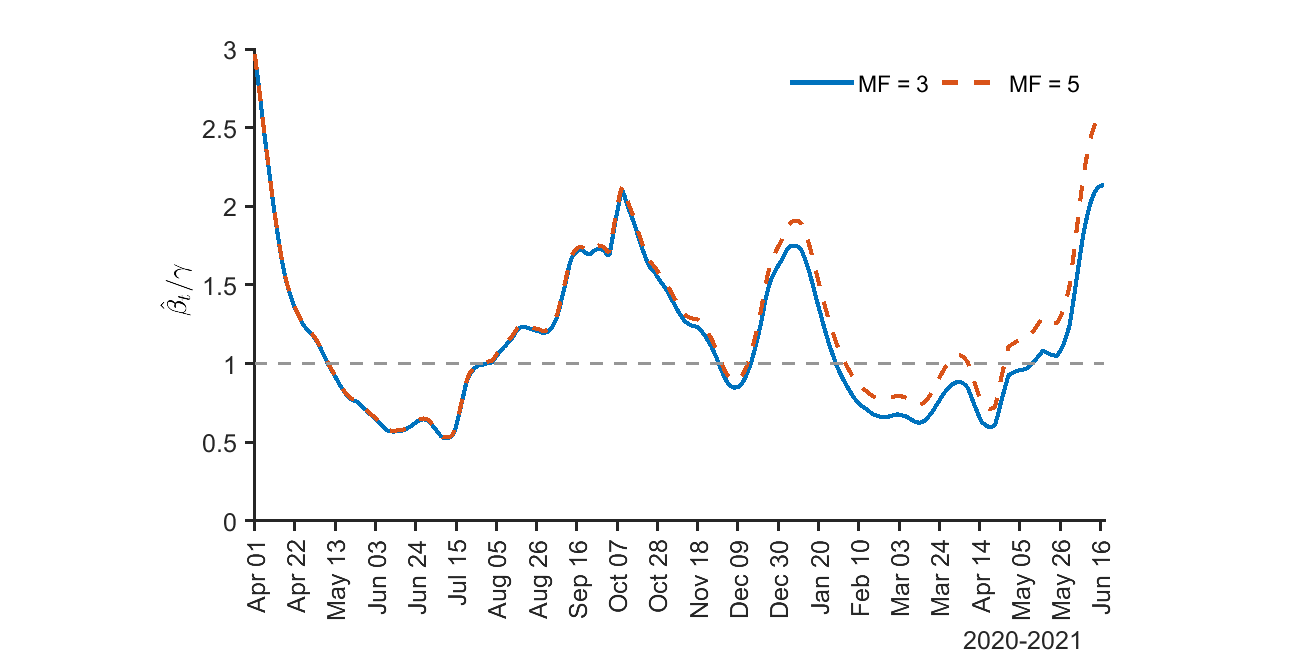}%
}
\end{tabular}

\end{center}

%

\end{figure}%
%

\begin{figure}[tp]%
\caption{Comparison of ${\hat{\mathcal{R}}_{et}}$ and ${\hat{\beta}_t}%
/{\gamma}$ for selected European countries}%
\vspace{-0.2cm}%
\label{fig: Euro_TR_Re}%

\begin{footnotesize}%

\begin{center}%
\begin{tabular}
[c]{ccc}%
Austria &  & France\\%
{\includegraphics[
height=1.9951in,
width=2.6524in
]%
{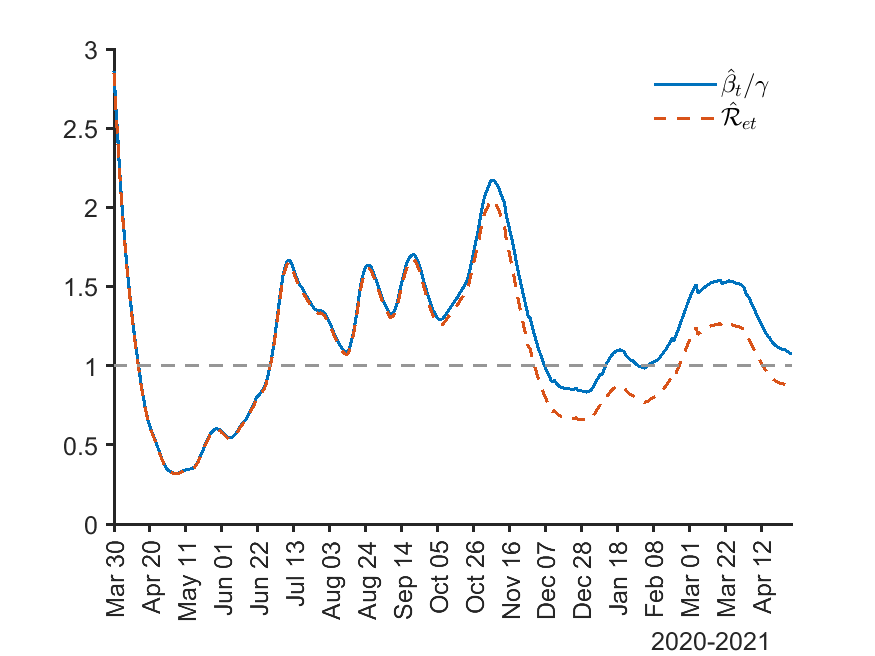}%
}
&  &
{\includegraphics[
height=1.9951in,
width=2.6524in
]%
{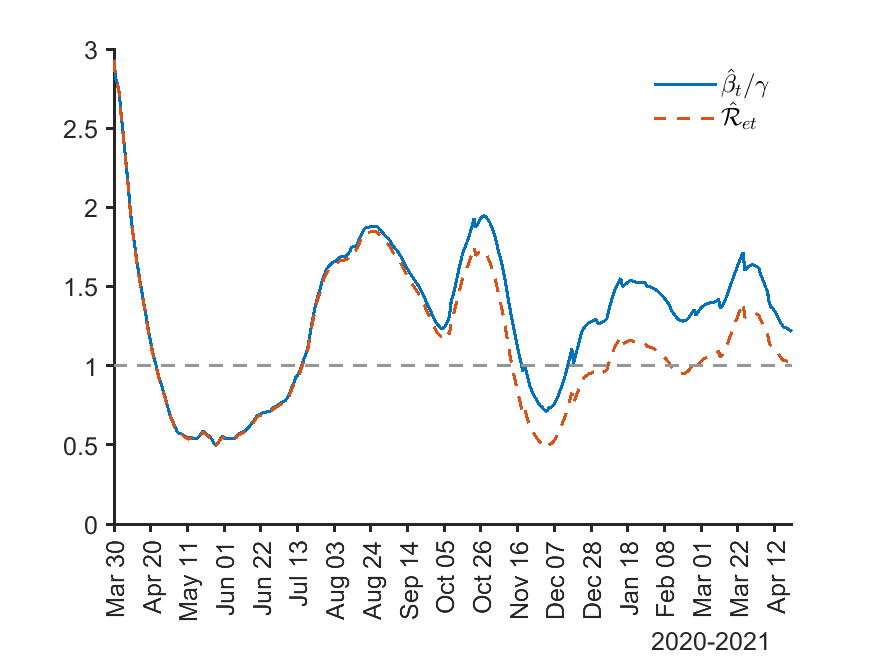}%
}
\\
&  & \\
Germany &  & Italy\\%
{\includegraphics[
height=1.9951in,
width=2.6524in
]%
{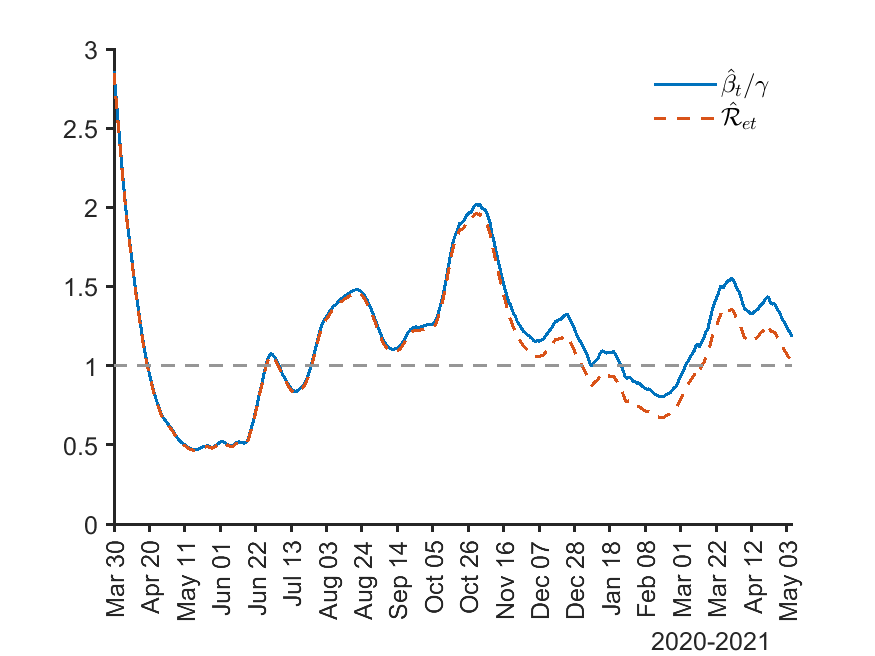}%
}
&  &
{\includegraphics[
height=1.9951in,
width=2.6524in
]%
{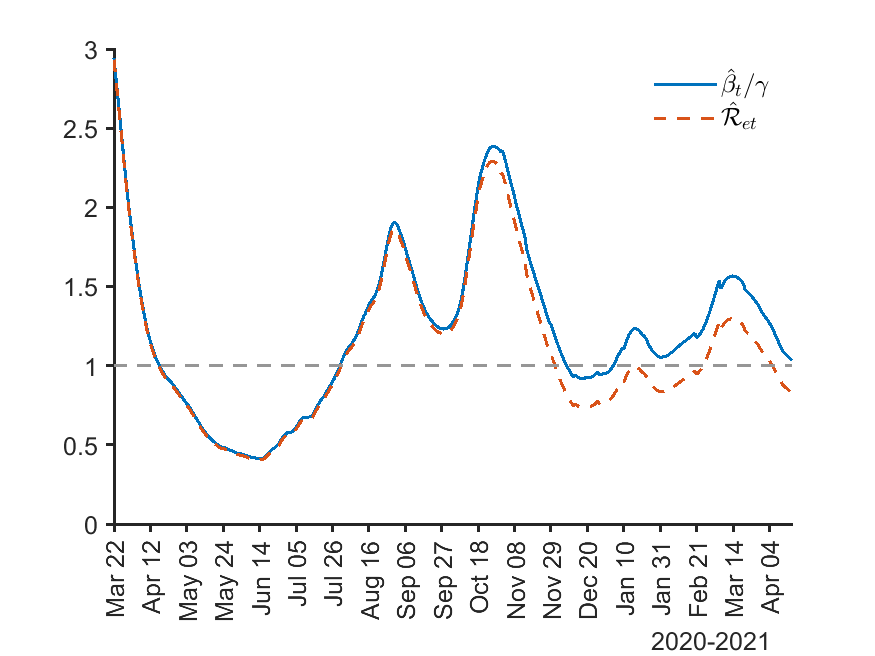}%
}
\\
&  & \\
Spain &  & UK\\%
{\includegraphics[
height=1.9951in,
width=2.6524in
]%
{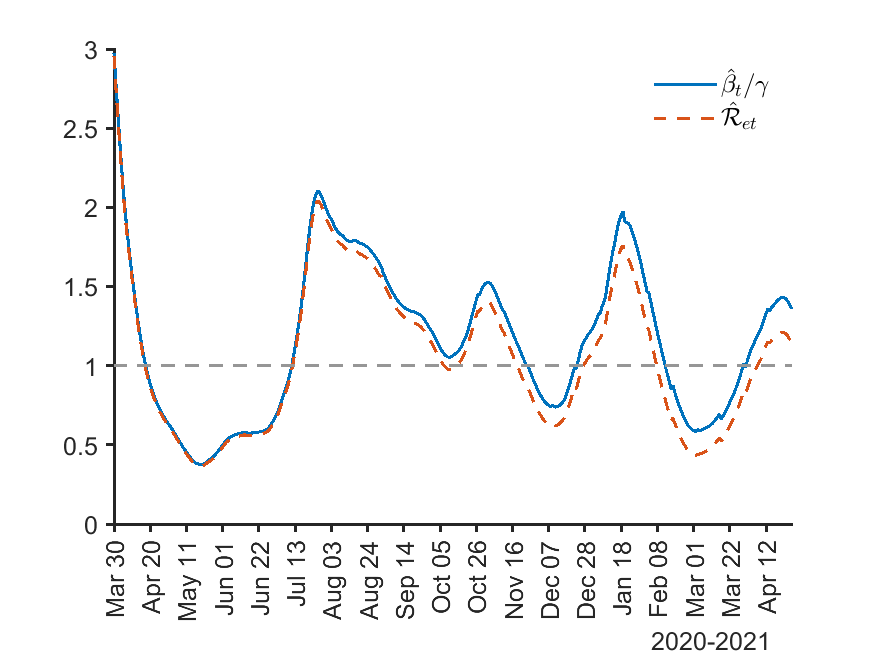}%
}
&  &
{\includegraphics[
height=1.9951in,
width=2.6524in
]%
{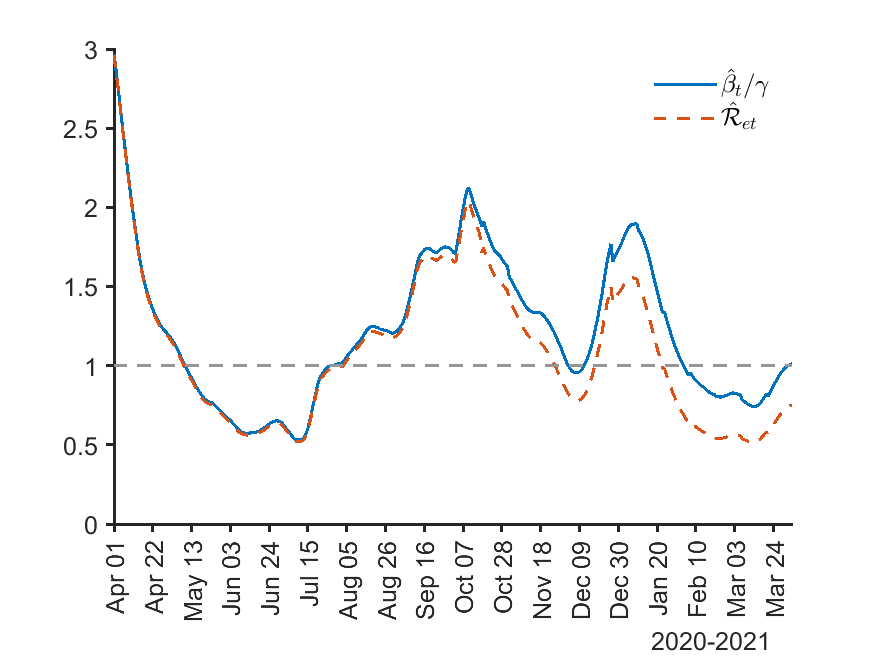}%
}
\end{tabular}

\end{center}

%

\vspace{-0.1cm}%
Notes: $\mathcal{\hat{R}}_{et}=\left(  1-\hat{m}_{t}\tilde{c}_{t}\right)
\hat{\beta}_{t}/\gamma$, where $\tilde{c}_{t}$ is the reported number of
infections per capita and $\gamma=1/14$. $W_{\beta}=W_{m}=2$ weeks. The joint
estimation starts when $\tilde{c}_{t}>0.01$. The initial guess estimate of the
multiplication factor is $5$. The simulation uses the single group model with
the random network and population size $n=50,000$. The number of replications
is $500$. The number of removed (recoveries + deaths) is estimated recursively
using $\tilde{R}_{t}=\left(  1-\gamma\right)  \tilde{R}_{t-1}+\gamma\tilde
{C}_{t-1}$ for all countries, with $\tilde{C}_{1}=\tilde{R}_{1}=0$.%

\end{footnotesize}%
%

\end{figure}%

To illustrate the relationship between the transmission rate and the effective
reproduction number, Figure \ref{fig: Euro_TR_Re} plots $\hat{\beta}%
_{t}/\gamma$ together with $\mathcal{\hat{R}}_{et}=\left(  1-\hat{m}_{t}%
\tilde{c}_{t}\right)  \hat{\beta}_{t}/\gamma$ for the six countries, where
$\hat{\beta}_{t}$ and $\hat{m}_{t}$ were estimated jointly using 2-weekly
rolling windows. The values of $\mathcal{\hat{R}}_{et}$ and $\hat{m}_{t}$ are
displayed in Figures \ref{fig: Euro_Re} and \ref{fig: Euro_MF} of the main
paper. It is worth noting that $\mathcal{\hat{R}}_{et}$ is almost the same as
$\hat{\beta}_{t}/\gamma$ in the early stage of the epidemic, since the
proportion of infected is very small even after taking account of
under-reporting. As infected cases grow, small differences between $\hat
{\beta}_{t}/\gamma$ and $\mathcal{\hat{R}}_{et}$ start to become visible.%

\begin{figure}[tp]%
\caption{Rolling estimates of the effective reproduction numbers ($\mathcal
{R}_{et}$) using the 2- and 3-weekly
rolling windows for selected European countries}%
\vspace{-0.2cm}%
\label{fig: Euro_Re_cmp_2W_3W}%

\begin{footnotesize}%

\begin{center}%
\begin{tabular}
[c]{ccc}%
Austria &  & France\\%
{\includegraphics[
height=1.9951in,
width=2.6524in
]%
{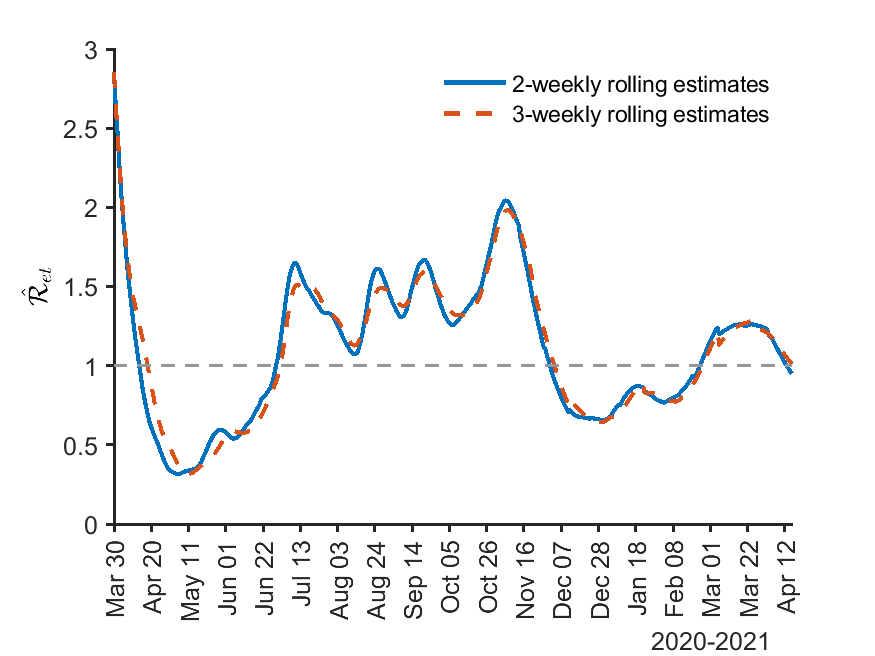}%
}
&  &
{\includegraphics[
height=1.9951in,
width=2.6524in
]%
{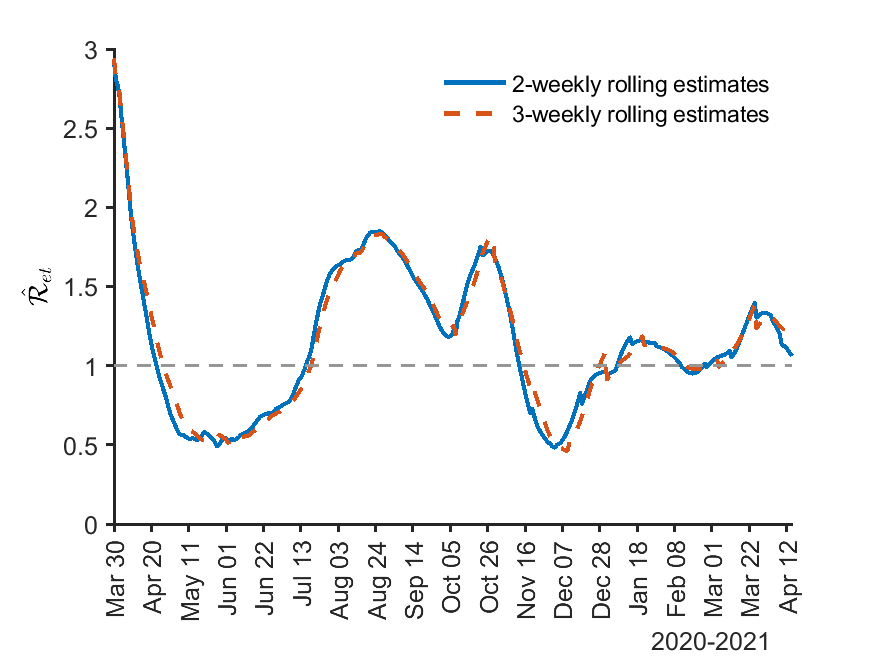}%
}
\\
&  & \\
Germany &  & Italy\\%
{\includegraphics[
height=1.9951in,
width=2.6524in
]%
{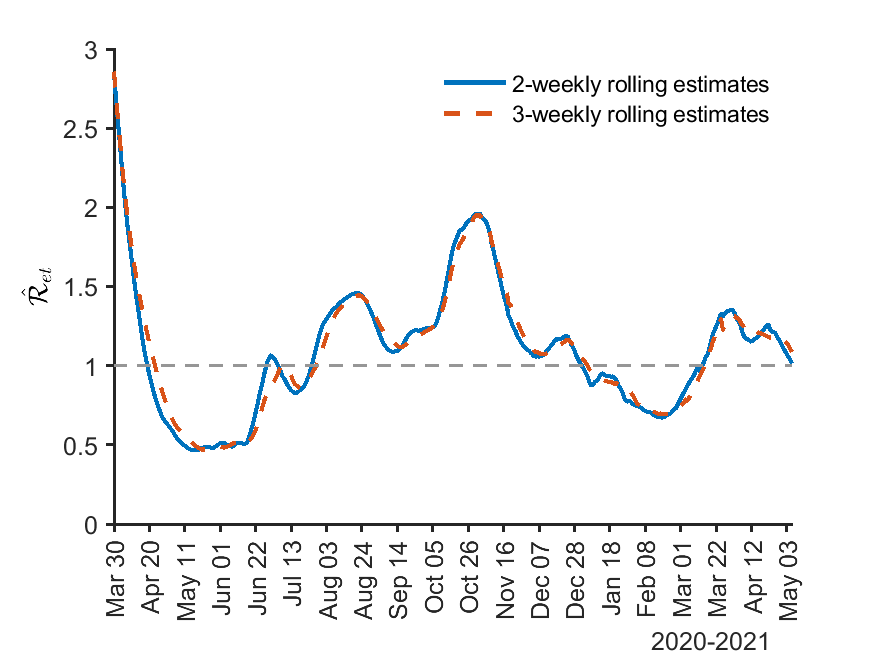}%
}
&  &
{\includegraphics[
height=1.9951in,
width=2.6524in
]%
{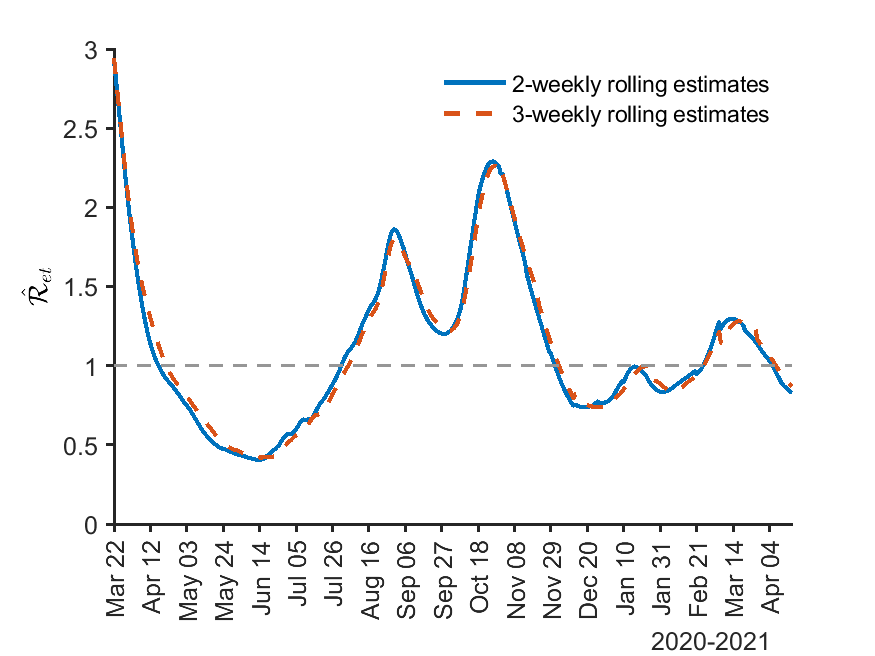}%
}
\\
&  & \\
Spain &  & UK\\%
{\includegraphics[
height=1.9951in,
width=2.6524in
]%
{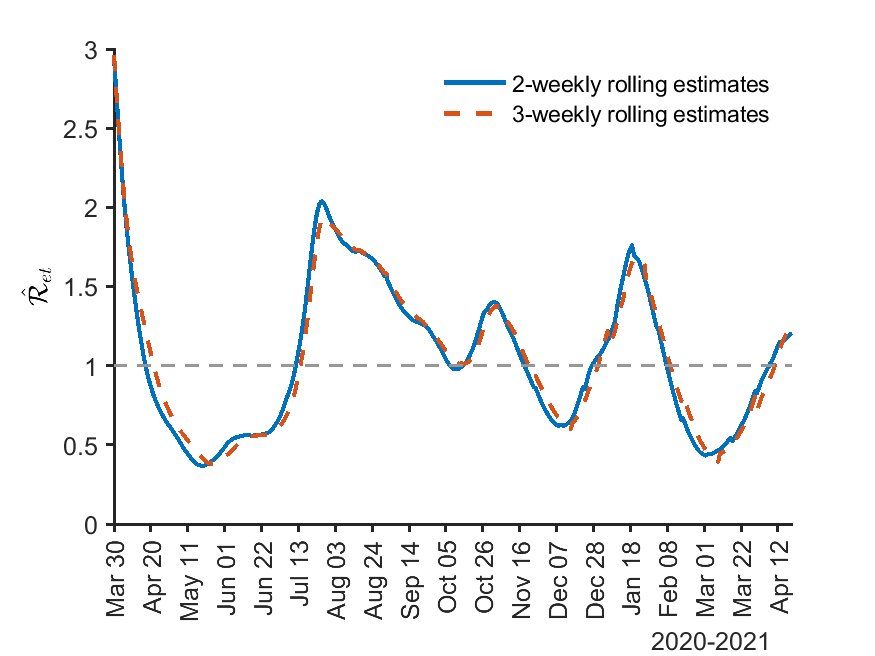}%
}
&  &
{\includegraphics[
height=1.9951in,
width=2.6524in
]%
{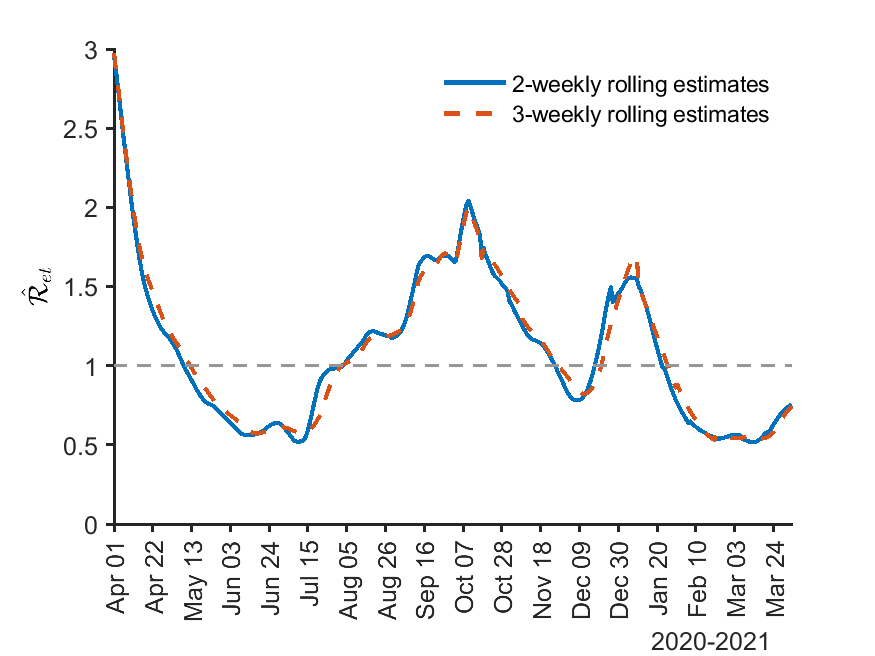}%
}
\end{tabular}

\end{center}

%

\vspace{-0.1cm}%
Notes: $\mathcal{\hat{R}}_{et}=\left(  1-\hat{m}_{t}\tilde{c}_{t}\right)
\hat{\beta}_{t}/\gamma$, where $\tilde{c}_{t}$ is the reported number of
infections per capita and $\gamma=1/14$. $W_{\beta}=W_{m}=2$ and $3$ weeks.
The joint estimation starts when $\tilde{c}_{t}>0.01$. The initial guess
estimate of the multiplication factor is $5$. The simulation uses the single
group model with the random network and population size $n=50,000$. The number
of replications is $500$. The number of removed (recoveries + deaths) is
estimated recursively using $\tilde{R}_{t}=\left(  1-\gamma\right)  \tilde
{R}_{t-1}+\gamma\tilde{C}_{t-1}$ for all countries, with $\tilde{C}_{1}%
=\tilde{R}_{1}=0$.%

\end{footnotesize}%
%

\end{figure}%

Figure \ref{fig: Euro_Re_cmp_2W_3W} compares the 2- and 3-weekly rolling
estimates of $\mathcal{R}_{et}$ obtained by the joint estimation procedure. It
is apparent that using 2- and 3-weekly rolling windows produces very similar
results. Therefore, our conclusions are unaltered if we adopt the 3-weekly
estimates of the transmission rates in calibrating the model to the empirical evidence.

\subsection{Estimates for the US\label{Sup: empirical_Re_US}}

\begin{figure}[tbh]%
\caption
{Realized new cases and two-weekly rolling estimates of the effective reproduction numbers ($\mathcal
{R}_{et}$)  for the US}%
\vspace{-0.2cm}%
\label{fig: US_Re}%

\begin{footnotesize}%

\begin{center}%
\begin{tabular}
[c]{ccc}%
Daily new cases &  & Estimates of $\mathcal{R}_{et}$\\%
{\includegraphics[
height=1.7763in,
width=3.5293in
]%
{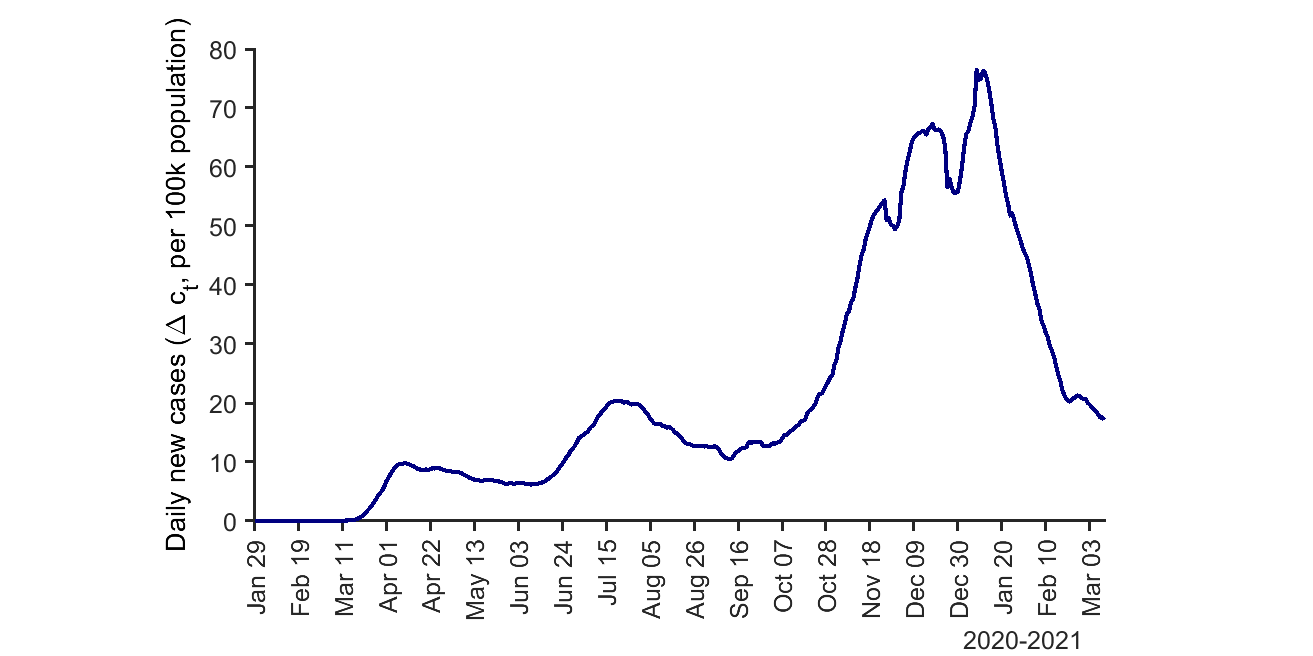}%
}
&  &
{\includegraphics[
height=1.9951in,
width=2.6524in
]%
{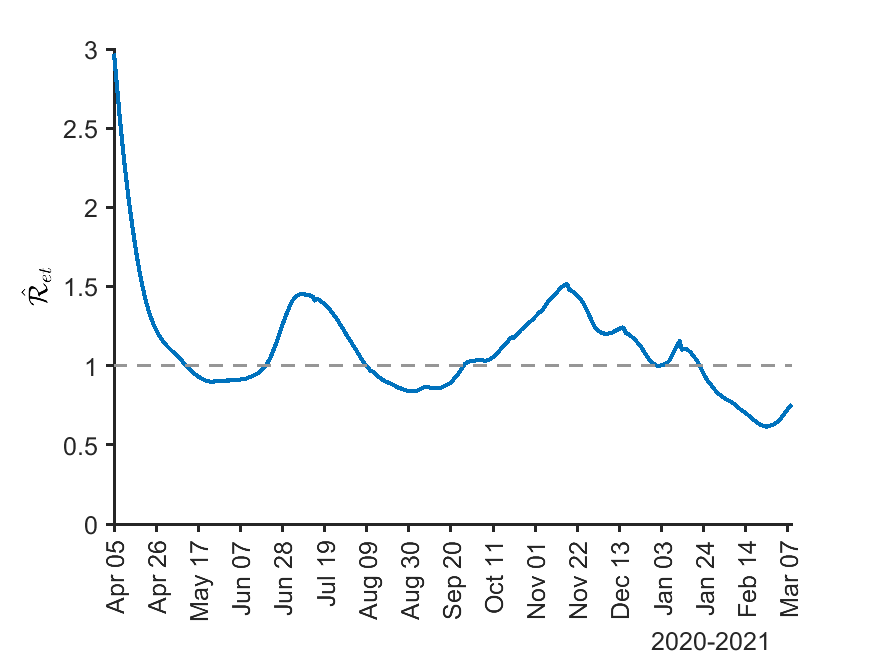}%
}
\end{tabular}

\end{center}

%

\vspace{0cm}%
\footnotesize
Notes: The reported daily new cases (7-day moving average) are displayed on
the left. $\mathcal{\hat{R}}_{et}=\left(  1-\hat{m}_{t}\tilde{c}_{t}\right)
\hat{\beta}_{t}/\gamma$, where $\tilde{c}_{t}$ is the reported number of
infections per capita and $\gamma=1/14.$ $W_{\beta}=W_{m}=2$ weeks. The joint
estimation starts when $\tilde{c}_{t}>0.01$. The initial guess estimate of the
multiplication factor is $5$. The simulation uses the single group model with
population size $n=50,000$. The number of replications is $500$. The number of
removed (recoveries + deaths) is estimated recursively using $\tilde{R}%
_{t}=\left(  1-\gamma\right)  \tilde{R}_{t-1}+\gamma\tilde{C}_{t-1}$, with
$\tilde{C}_{1}=\tilde{R}_{1}=0$.%

\end{footnotesize}%
%

\end{figure}

It is also interesting to examine how the reproduction numbers have evolved in
the US. This section presents estimates of $\mathcal{R}_{et}$ for the US at
the country and state levels. Figure \ref{fig: US_Re} presents the reported
daily new cases (per $100,000$ people) and the 2-weekly rolling estimates of
$\mathcal{R}_{et}$ obtained by the joint estimation method for the US over the
period of March 2020 to March 10, 2021 (when the share of the population fully
vaccinated reached 10 percent). The results show that $\mathcal{\hat{R}}_{et}$
briefly dipped below $1$ in May and then again in August 2020. In contrast
with the estimates in the European countries, $\mathcal{R}_{et}$ in the US
never decreased to a level as low as $0.5$, resulting in a higher number of
cases per capita.%

Figure \ref{fig: Re_states} presents the 2-weekly rolling estimates of
$\mathcal{R}_{et}$ for the $48$ contiguous states and the District of Columbia
over the period March 2020 to August 7, 2021. For simplicity, we used a fixed
MF $=3$ in the estimation.\footnote{Some states had large-scale retrospective
reporting or correction that resulted in drastic changes in the estimates. See
the figure notes and also the readme file at the CSSE repository for more
details.} It can be seen that the estimates share similar comovements but also
display interesting patterns of heterogeneity across states. Overall, we
observe that $\mathcal{R}_{et}$ rises above one around July and November 2020
and then increases quite rapidly again in July 2021 in most states. The peaks
of the estimates were largely around $1.5$--$1.7$ in the first two hikes. In
contrast, the latest surge in $\mathcal{R}_{et}$ was very rapid, reaching
nearly two in most states in early August. This newest wave occurred right
after many states had brought down $\mathcal{R}_{et}$ to $0.5$, the lowest
level for many places in the US since the pandemic began.%

\begin{figure}[!p]%
\caption
{Two-weekly rolling estimates of the effective reproduction numbers ($\mathcal
{R}_{et}$) for the contiguous US, by state}%
\vspace{-0.3cm}%
\label{fig: Re_states}

\begin{center}%
\hspace*{-0.2cm}%
\begin{tabular}
[c]{cc}%
{\footnotesize Alabama} & {\footnotesize Arizona}\\%
{\includegraphics[
height=1.7772in,
width=3.5293in
]%
{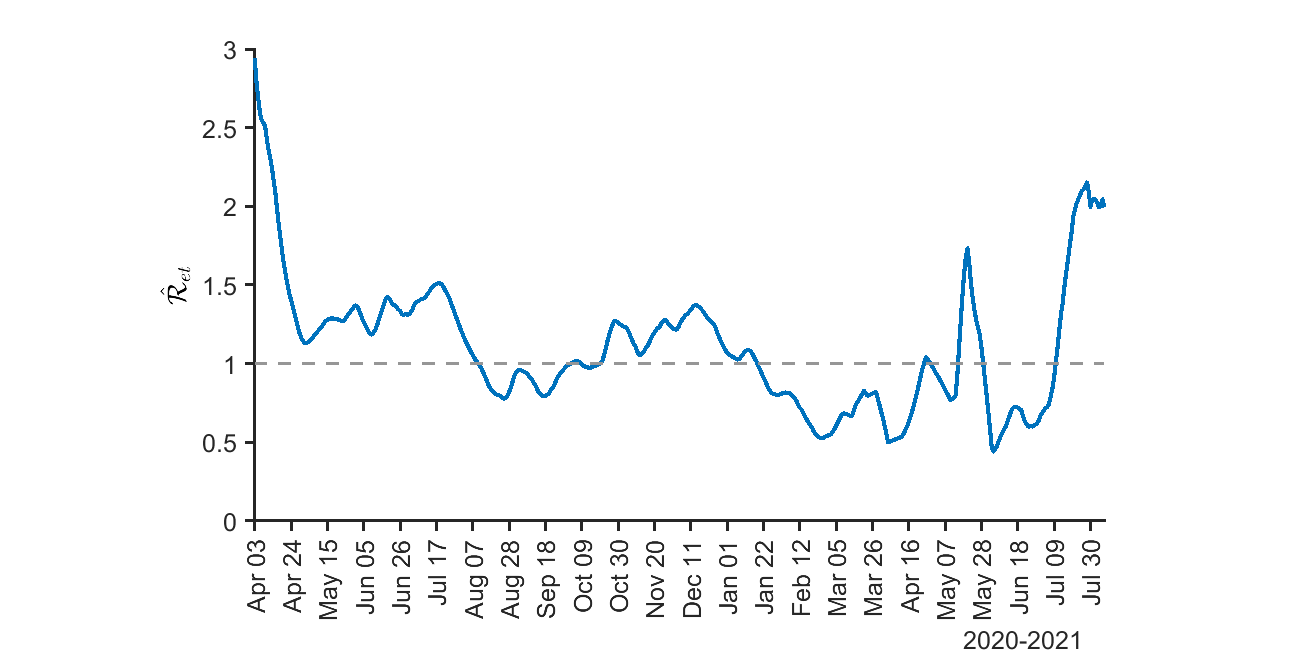}%
}
&
{\includegraphics[
height=1.7772in,
width=3.5293in
]%
{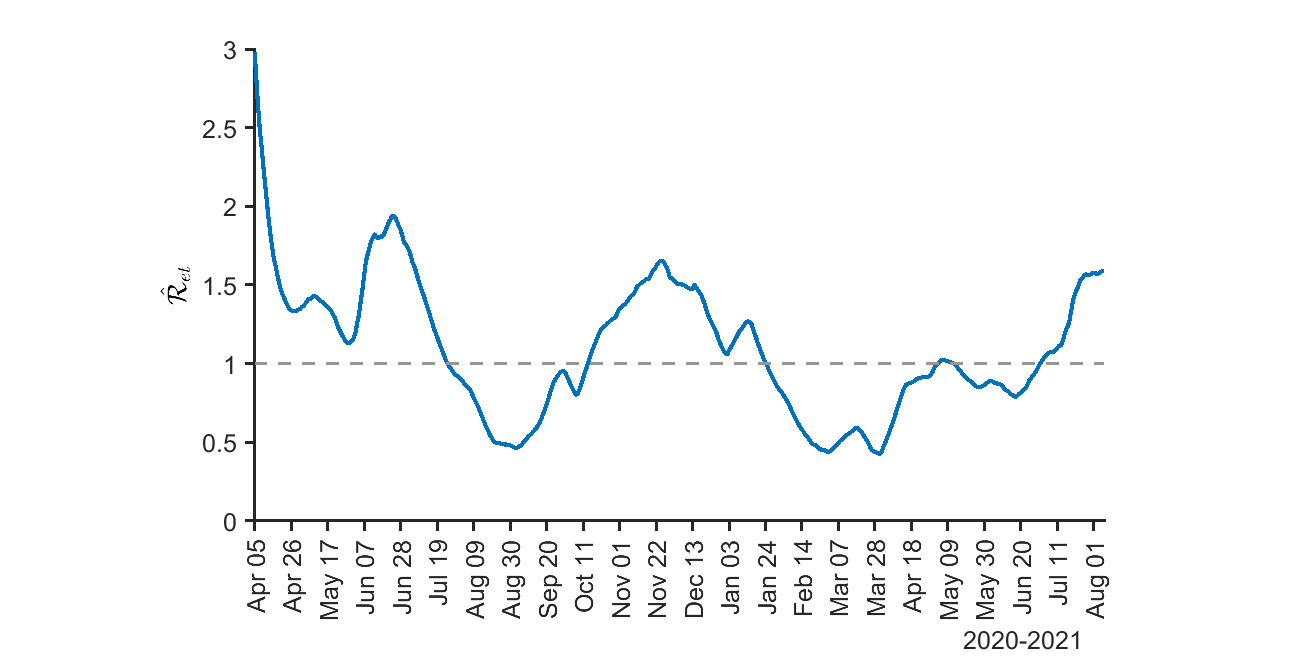}%
}
\\
& \\
{\footnotesize Arkansas} & {\footnotesize California}\\%
{\includegraphics[
height=1.7763in,
width=3.5293in
]%
{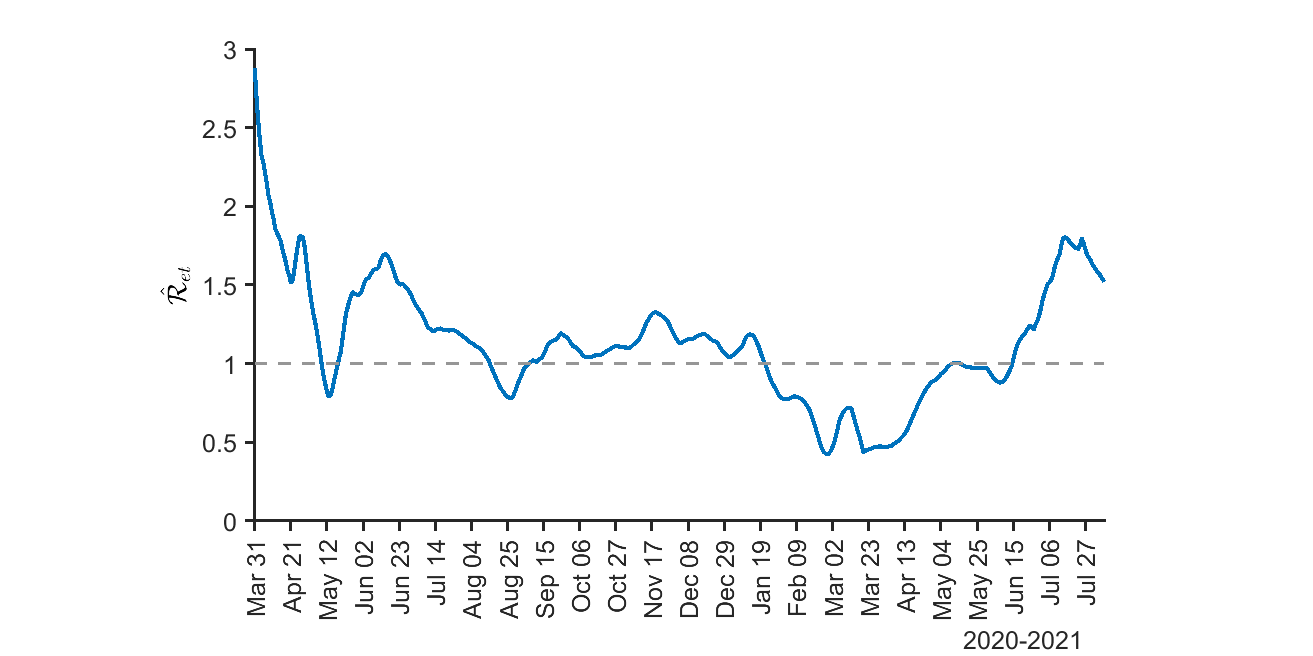}%
}
&
{\includegraphics[
height=1.7763in,
width=3.5293in
]%
{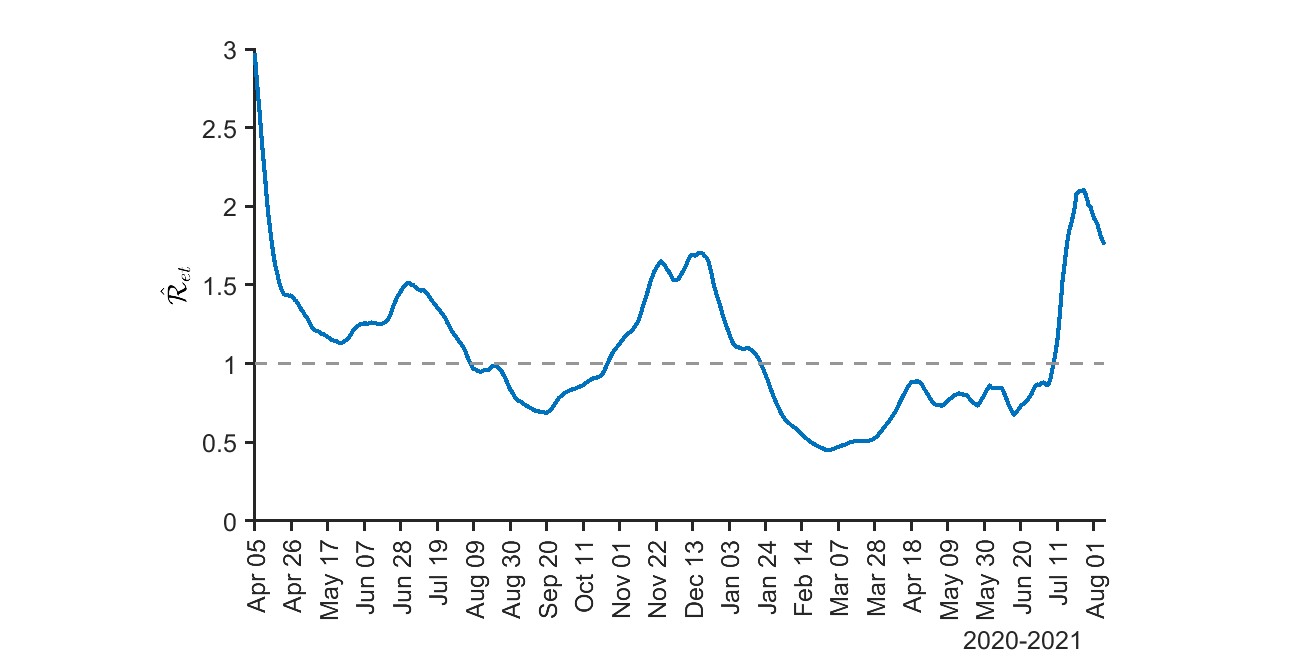}%
}
\\
& \\
{\footnotesize Colorado} & {\footnotesize Connecticut}\\%
{\includegraphics[
height=1.7763in,
width=3.5293in
]%
{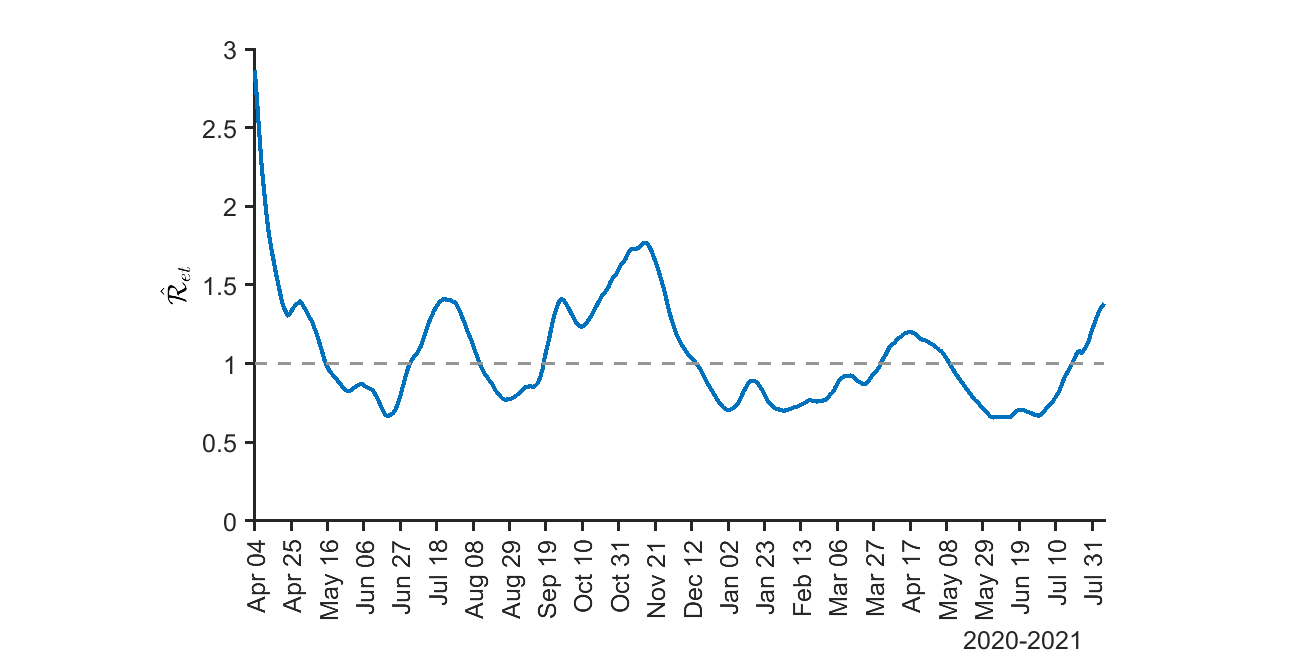}%
}
&
{\includegraphics[
height=1.7763in,
width=3.5293in
]%
{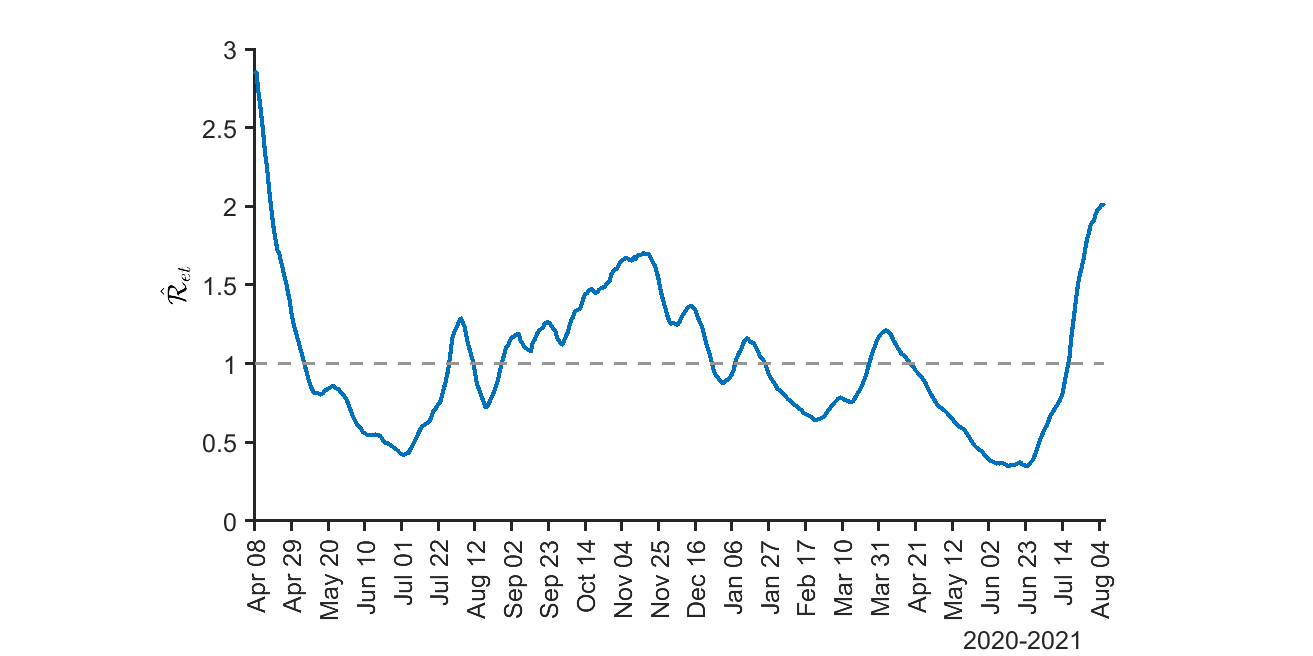}%
}
\\
& \\
{\footnotesize Delaware} & {\footnotesize District of Columbia}\\%
{\includegraphics[
height=1.7763in,
width=3.5293in
]%
{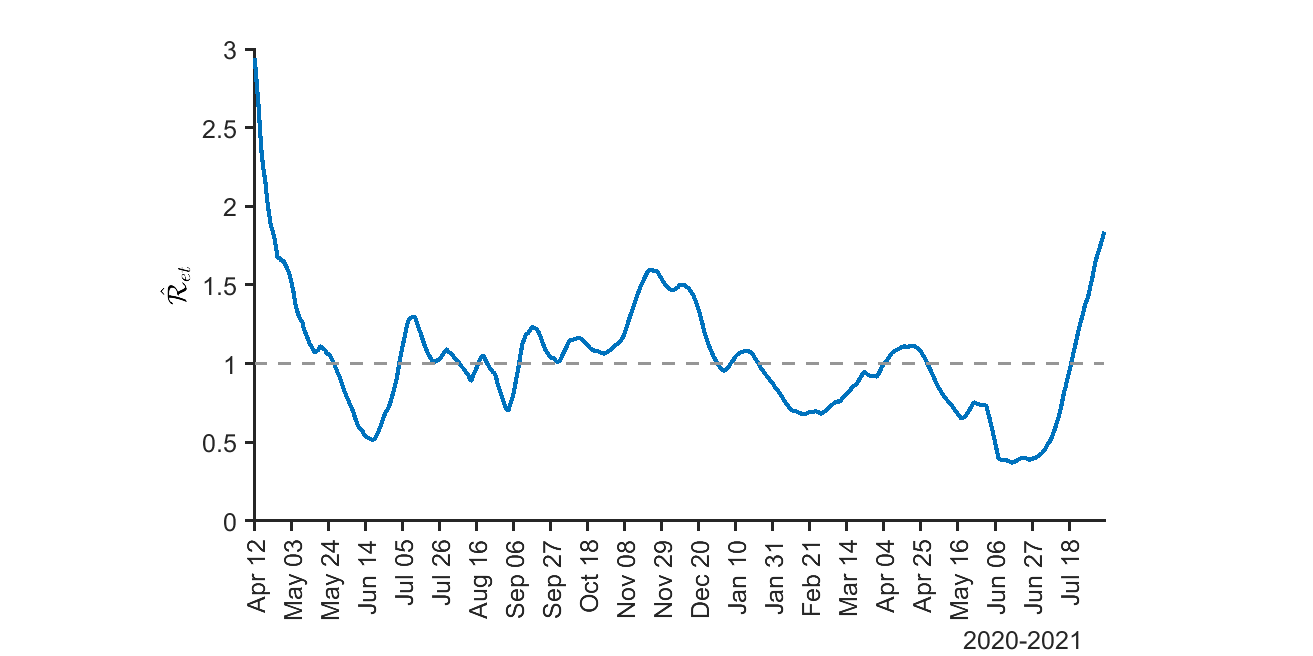}%
}
&
{\includegraphics[
height=1.7763in,
width=3.5293in
]%
{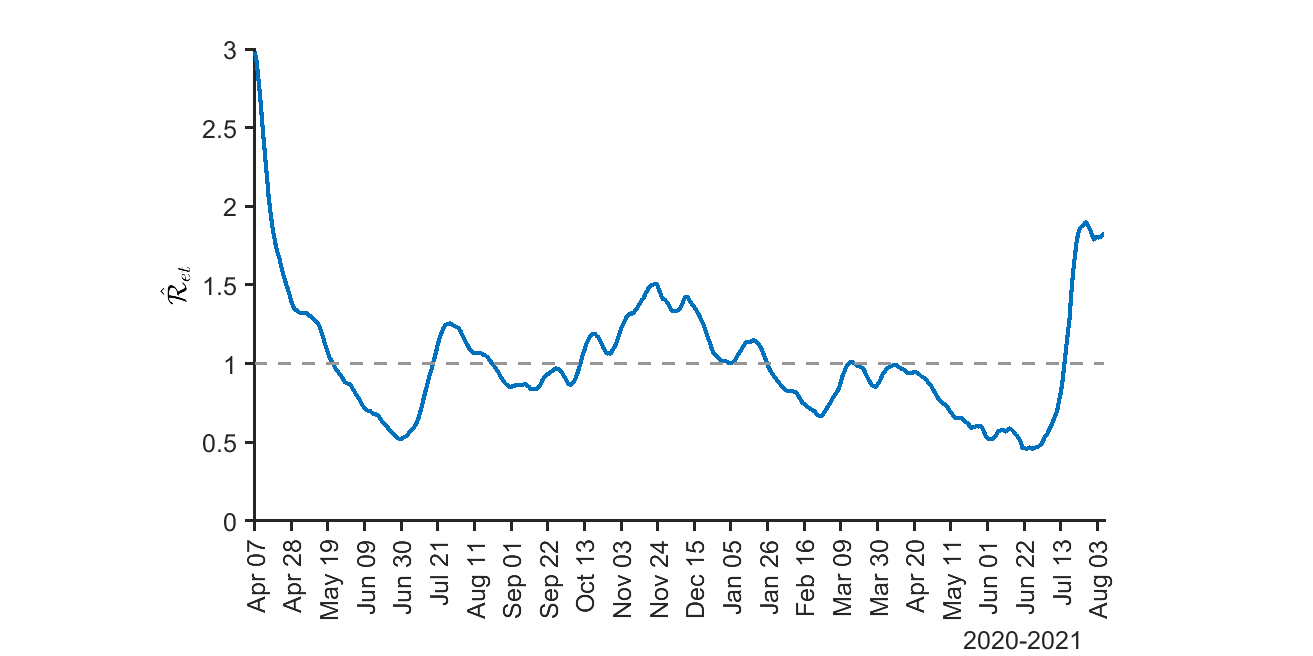}%
}
\end{tabular}

\end{center}

%

\vspace{-0.4cm}%
\footnotesize
{}Notes: $\mathcal{\hat{R}}_{et}=\left(  1-\text{MF}\tilde{c}_{t}\right)
\hat{\beta}_{t}/\gamma$, where MF $=3$, $\tilde{c}_{t}$ is the reported number
of infections per capita, and $\gamma=1/14$. The number of removed (recoveries
+ deaths) is estimated recursively using $\tilde{R}_{t}=\left(  1-\gamma
\right)  \tilde{R}_{t-1}+\gamma\tilde{C}_{t-1}$ for all states, with
$\tilde{C}_{1}=\tilde{R}_{1}=0$. Alabama included $306$, $4,877$, and $1,235$
backlogged cases on May 13--15, 2021.%

\end{figure}%
%

\addtocounter{figure}{-1}%
%

\begin{figure}[!p]%
\caption
{(Continued) Two-weekly rolling estimates of the effective reproduction numbers ($\mathcal
{R}_{et}$) for the contiguous US, by state}%
\vspace{-0.3cm}%

\begin{center}%
\hspace*{-0.2cm}%
\begin{tabular}
[c]{cc}%
{\footnotesize Florida} & {\footnotesize Georgia}\\%
{\includegraphics[
height=1.7763in,
width=3.5293in
]%
{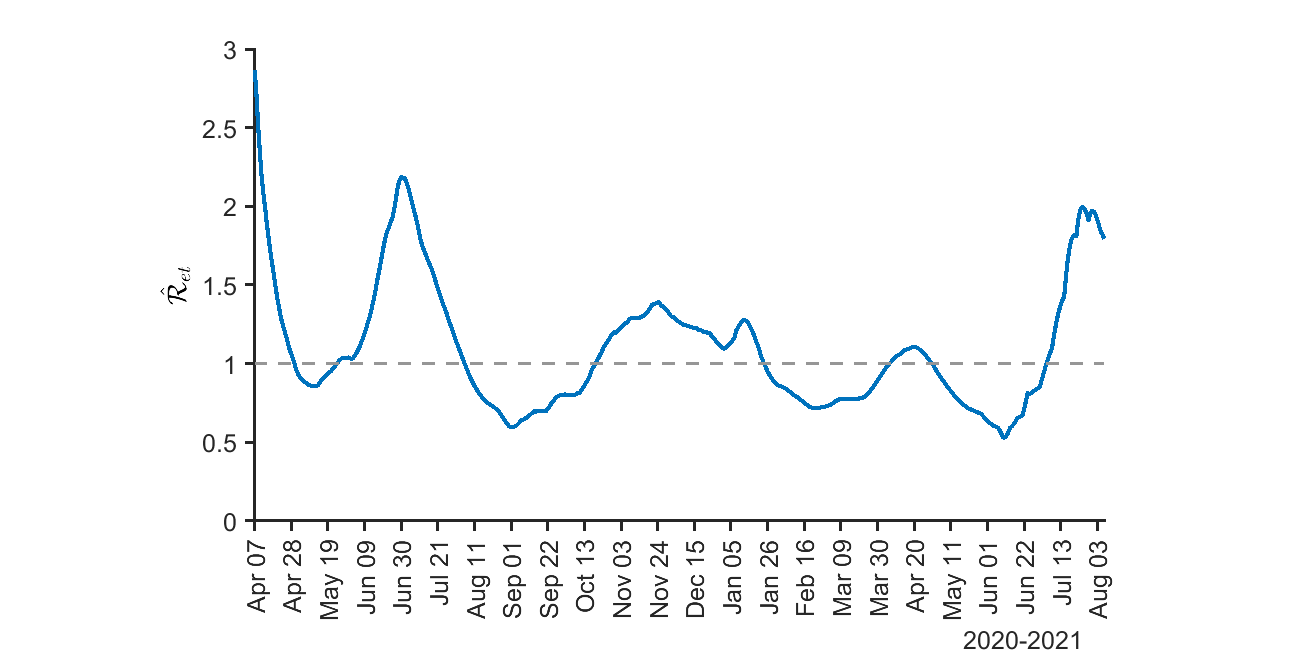}%
}
&
{\includegraphics[
height=1.7763in,
width=3.5293in
]%
{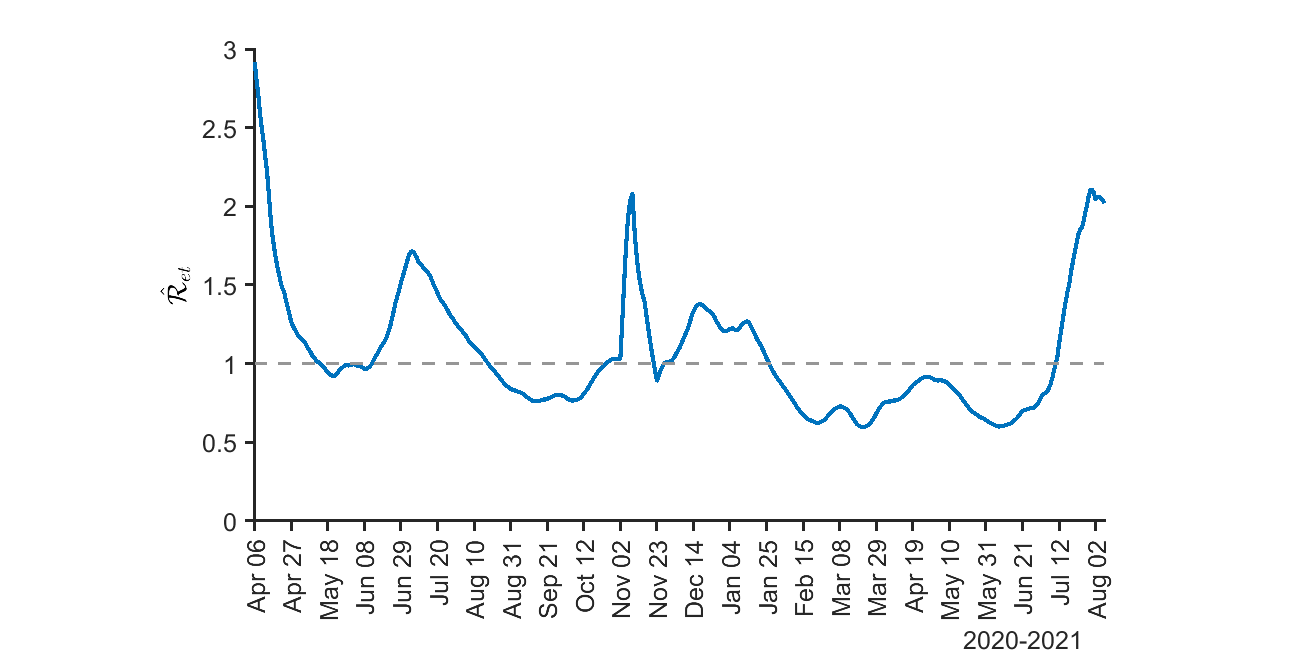}%
}
\\
& \\
{\footnotesize Idaho} & {\footnotesize Illinois}\\%
{\includegraphics[
height=1.7763in,
width=3.5293in
]%
{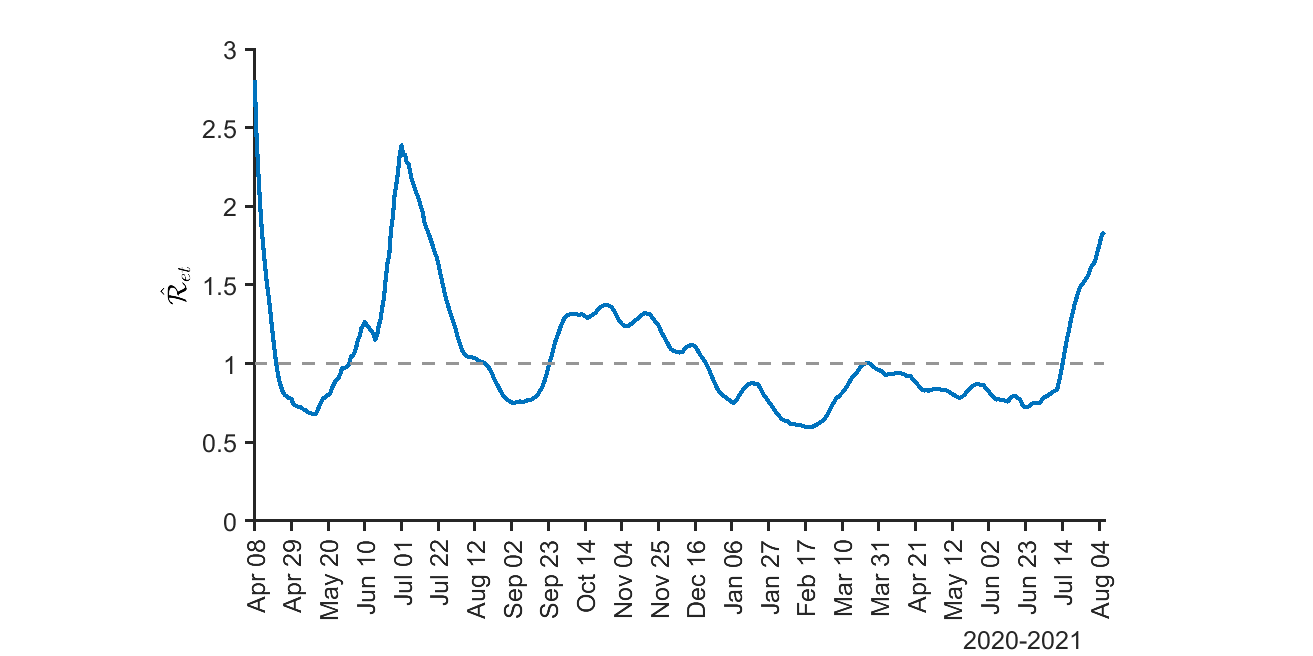}%
}
&
{\includegraphics[
height=1.7763in,
width=3.5293in
]%
{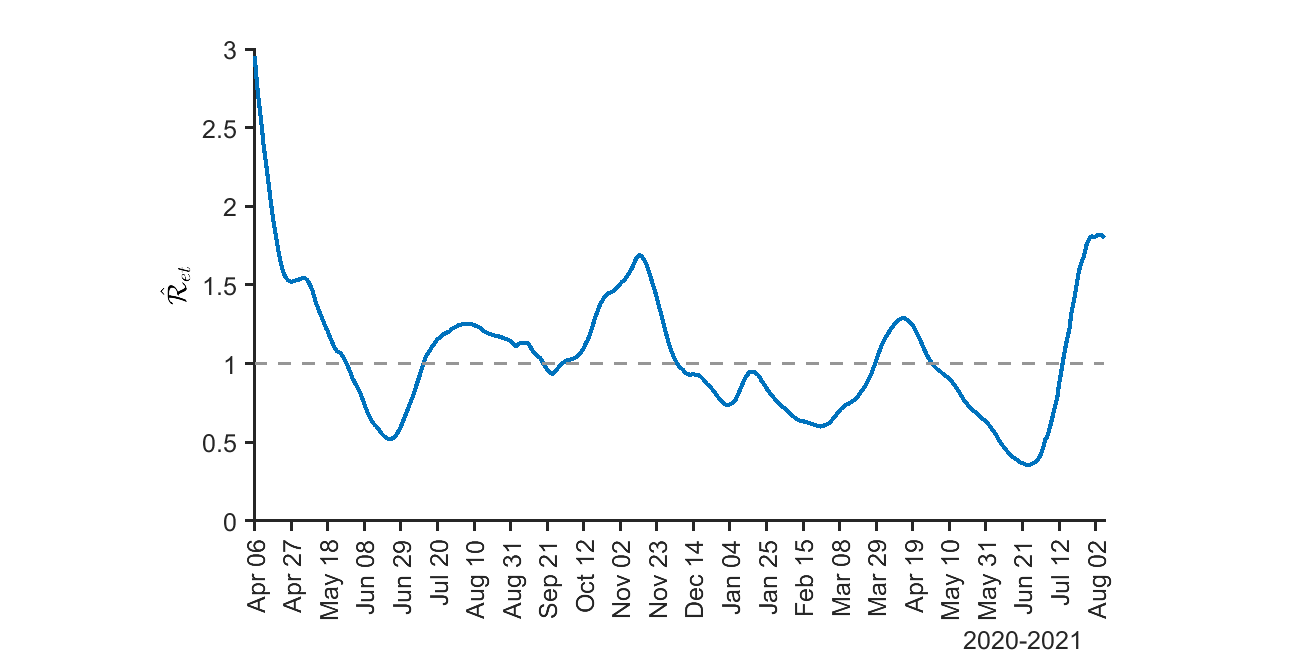}%
}
\\
& \\
{\footnotesize Indiana} & {\footnotesize Iowa}\\%
{\includegraphics[
height=1.7763in,
width=3.5293in
]%
{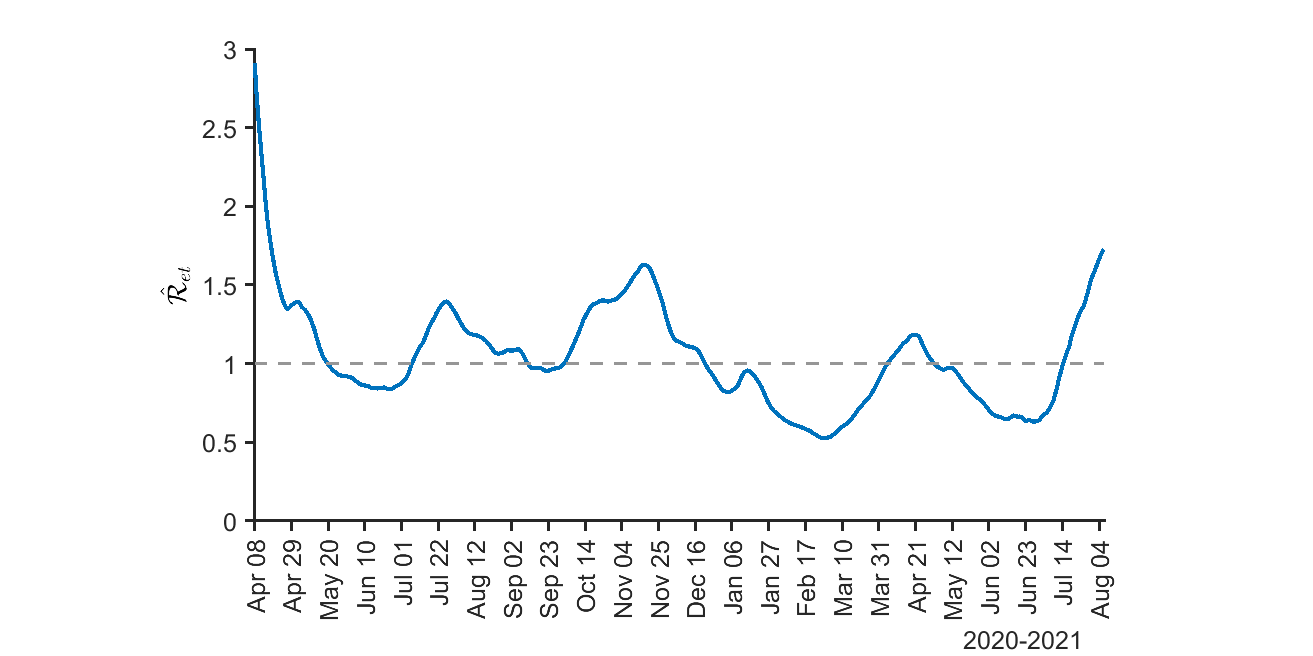}%
}
&
{\includegraphics[
height=1.7763in,
width=3.5293in
]%
{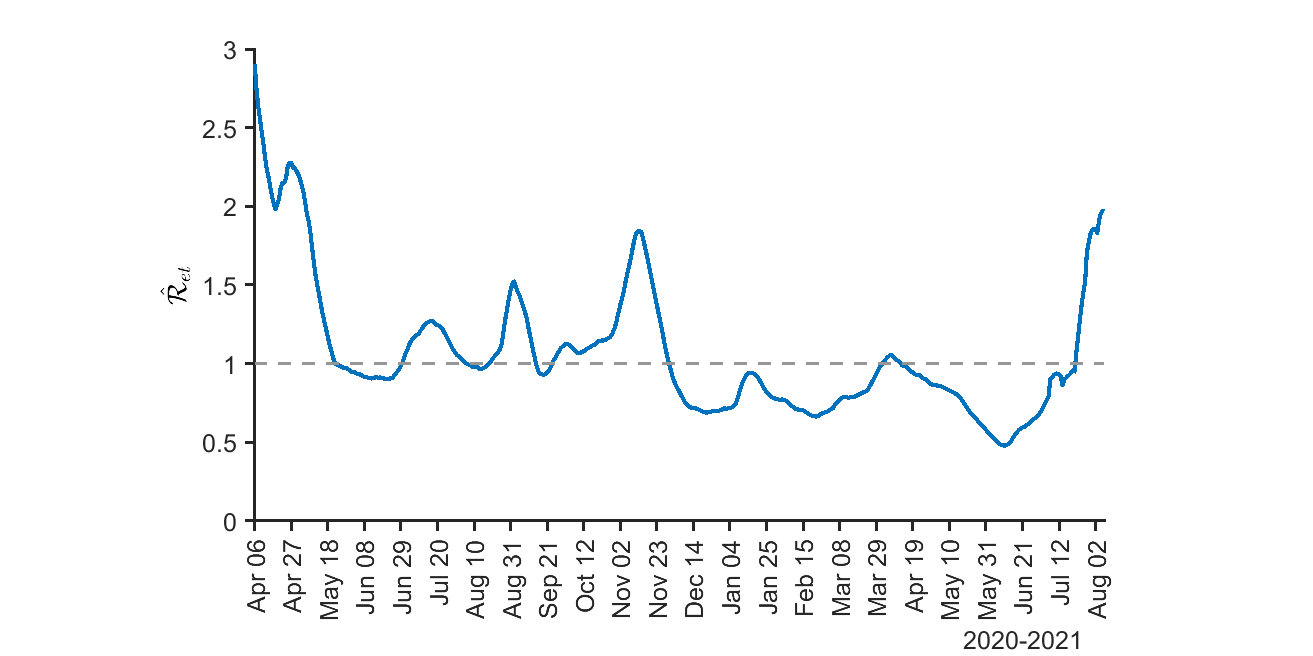}%
}
\\
& \\
{\footnotesize Kansas} & {\footnotesize Kentucky}\\%
{\includegraphics[
height=1.7763in,
width=3.5293in
]%
{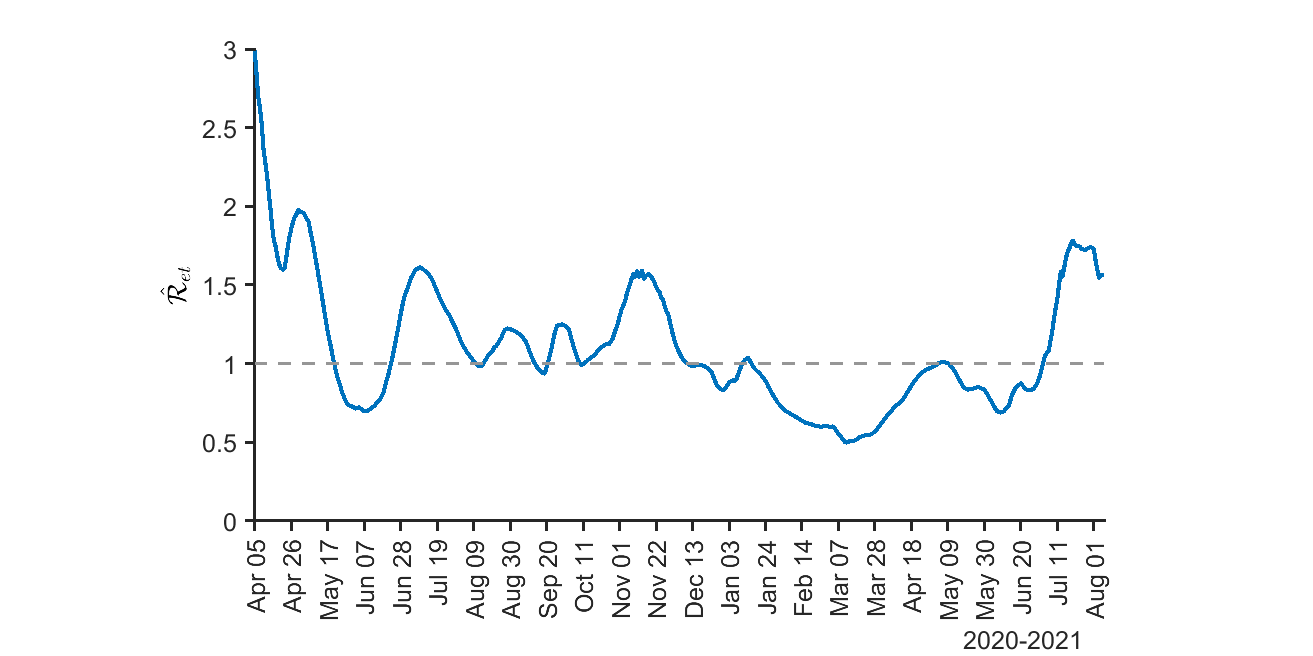}%
}
&
{\includegraphics[
height=1.7763in,
width=3.5293in
]%
{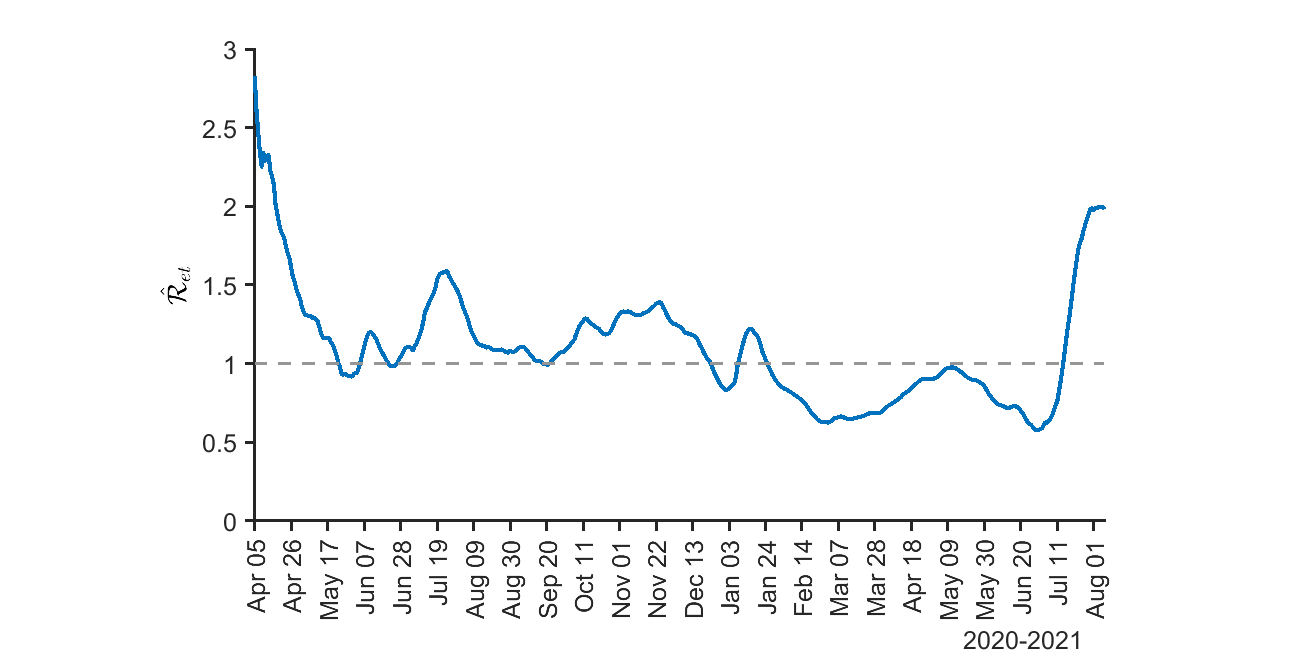}%
}
\end{tabular}

\end{center}

%

\vspace{-0.4cm}%
\footnotesize
{}Notes: Florida stopped publishing daily case numbers on June 7, 2021.
Georgia added $29,937$ antigen positive cases on November 3, 2020.%

\end{figure}%
%

\addtocounter{figure}{-1}%
%

\begin{figure}[!p]%
\caption
{(Continued) Two-weekly rolling estimates of the effective reproduction numbers ($\mathcal
{R}_{et}$) for the contiguous US, by state}%
\vspace{-0.3cm}%

\begin{center}%
\hspace*{-0.2cm}%
\begin{tabular}
[c]{cc}%
{\footnotesize Louisiana} & {\footnotesize Maine}\\%
{\includegraphics[
height=1.7763in,
width=3.5293in
]%
{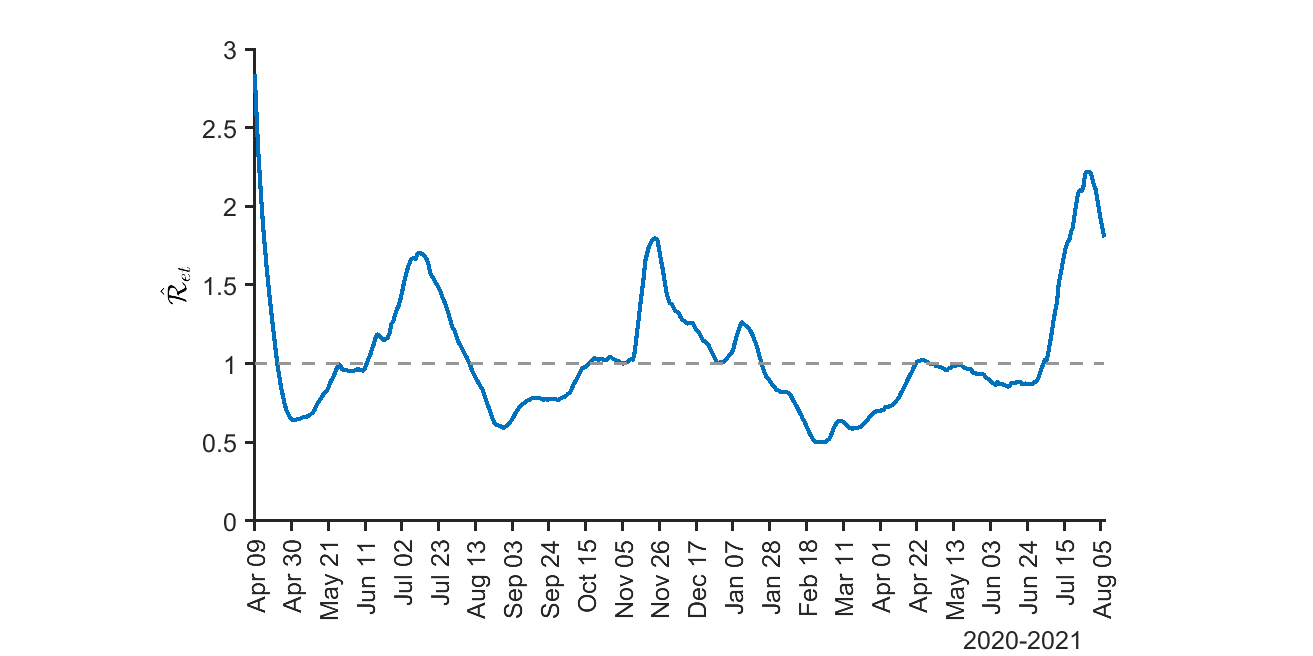}%
}
&
{\includegraphics[
height=1.7763in,
width=3.5293in
]%
{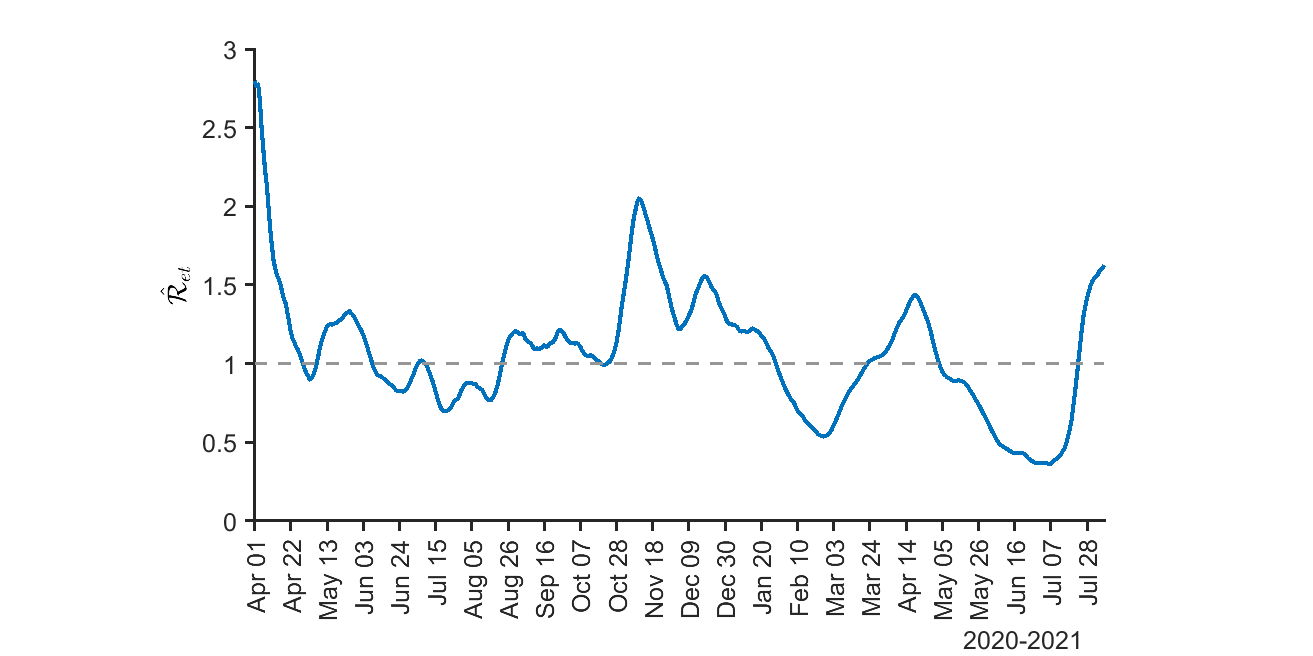}%
}
\\
& \\
{\footnotesize Maryland} & {\footnotesize Massachusetts}\\%
{\includegraphics[
height=1.7763in,
width=3.5293in
]%
{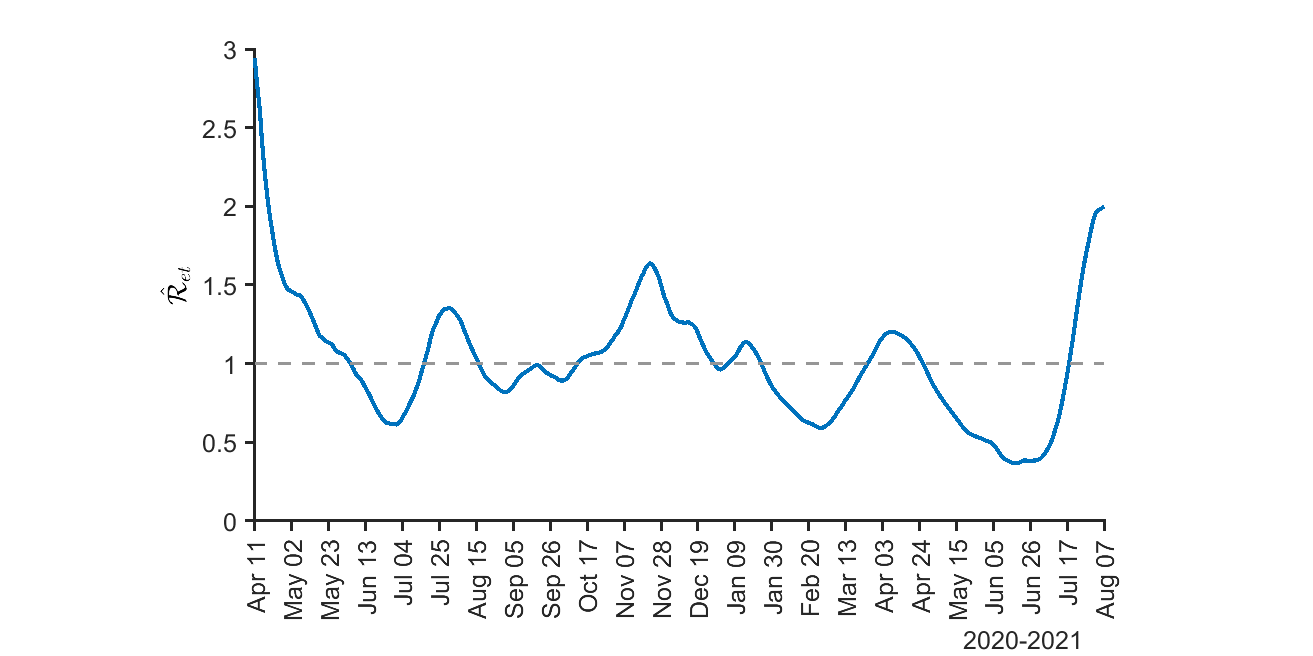}%
}
&
{\includegraphics[
height=1.7763in,
width=3.5293in
]%
{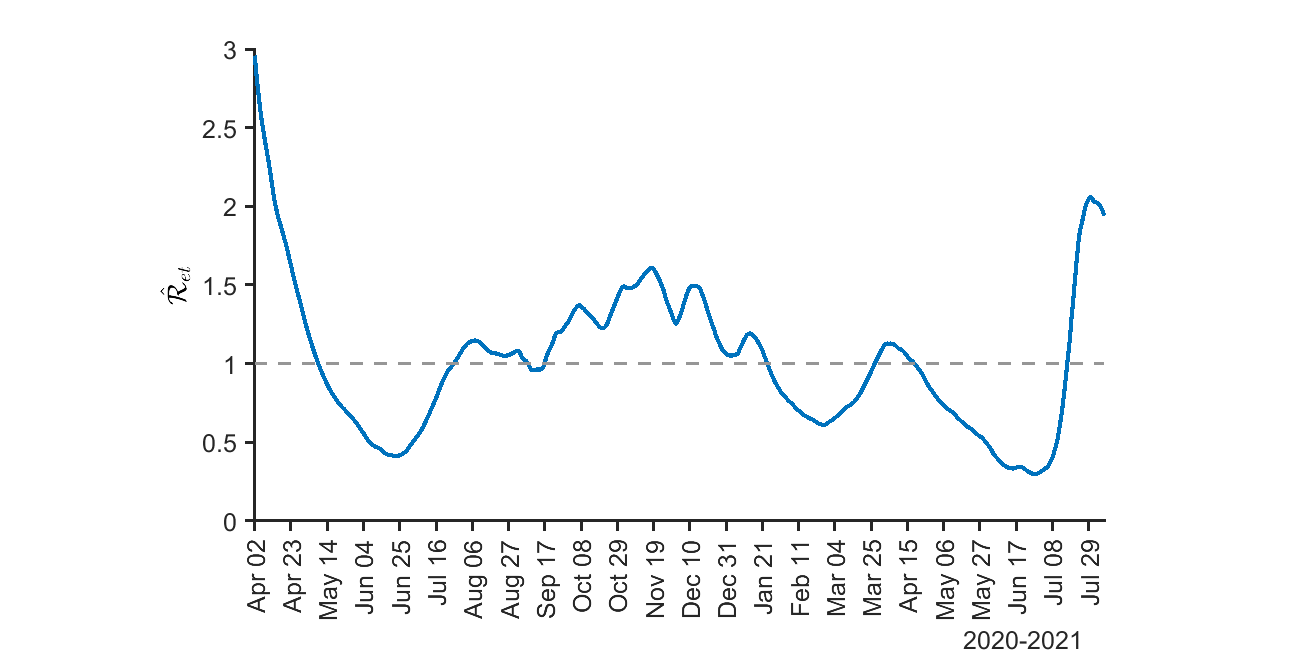}%
}
\\
& \\
{\footnotesize Michigan} & {\footnotesize Minnesota}\\%
{\includegraphics[
height=1.7763in,
width=3.5293in
]%
{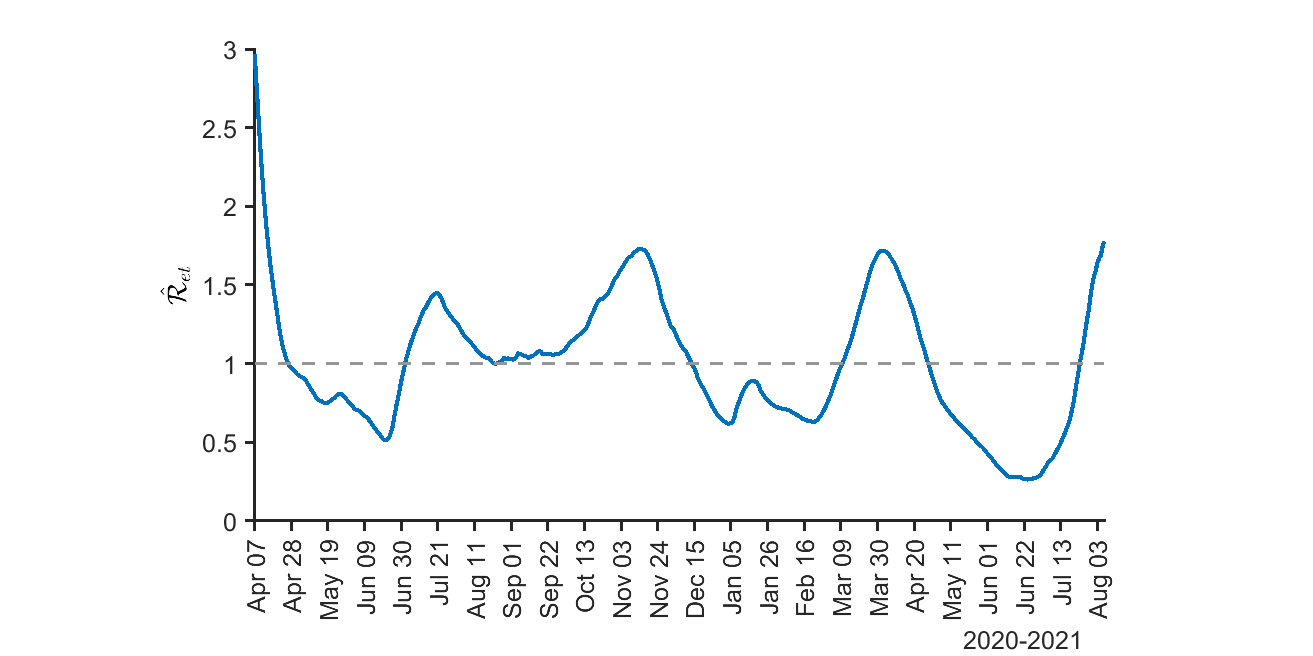}%
}
&
{\includegraphics[
height=1.7763in,
width=3.5293in
]%
{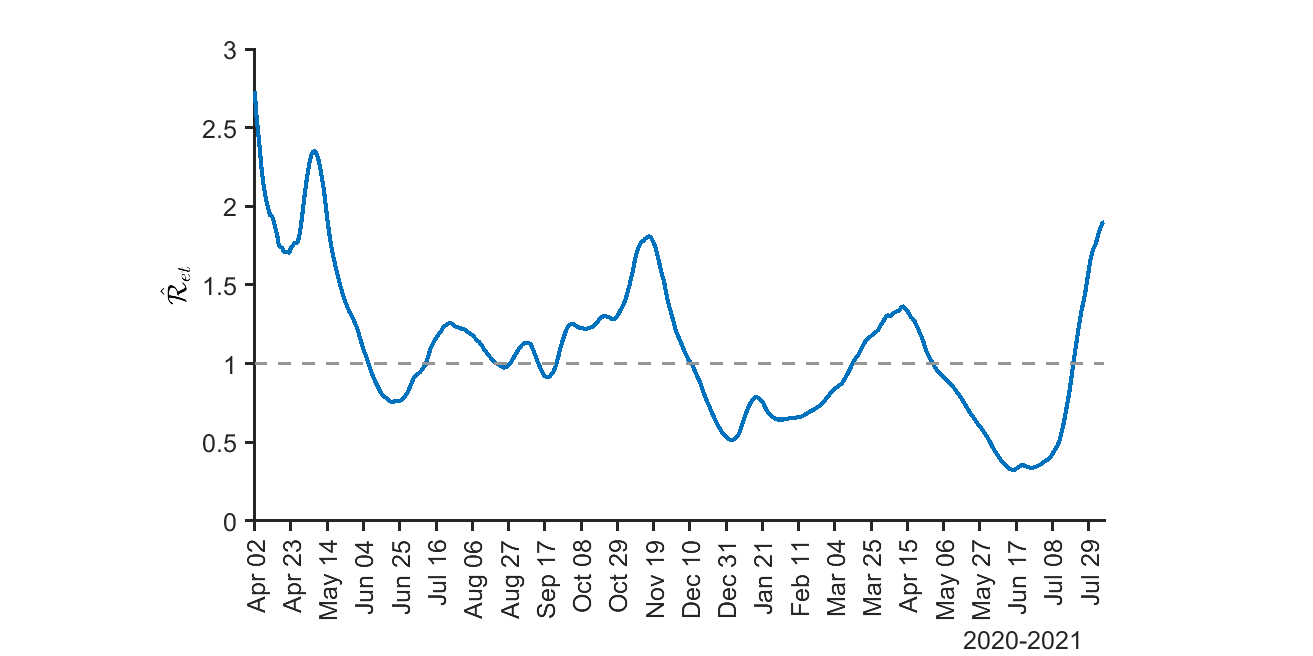}%
}
\\
& \\
{\footnotesize Mississippi} & {\footnotesize Missouri}\\%
{\includegraphics[
height=1.7763in,
width=3.5293in
]%
{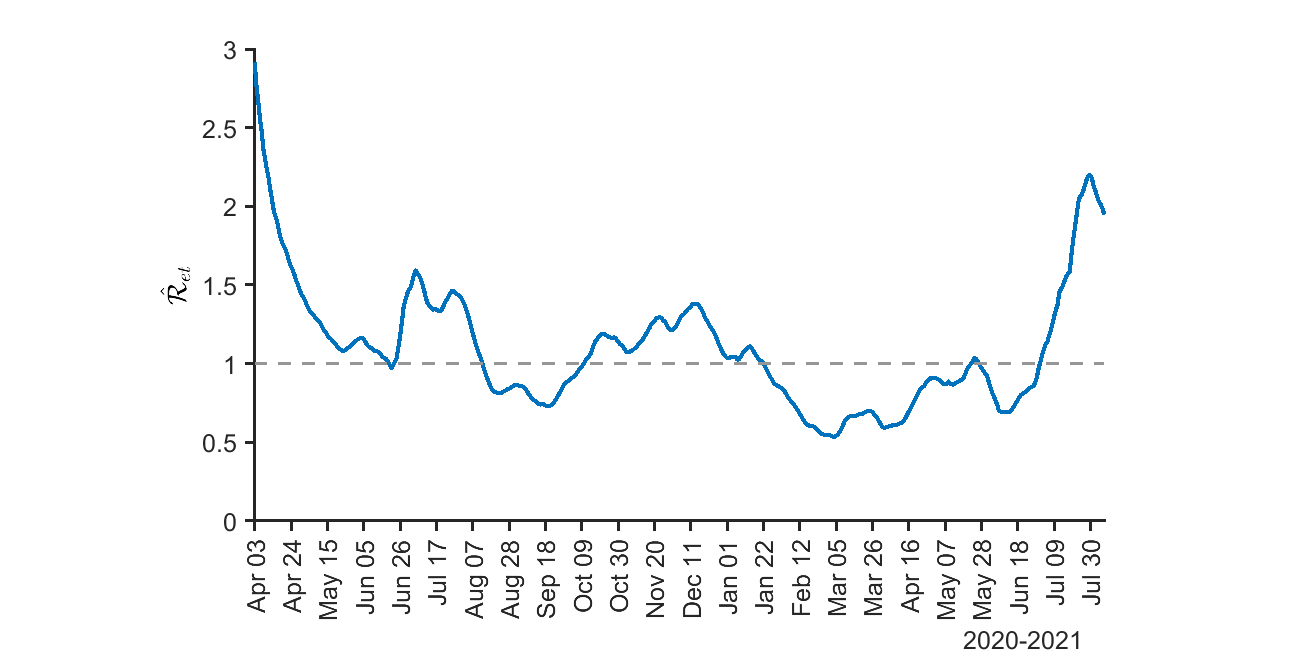}%
}
&
{\includegraphics[
height=1.7763in,
width=3.5293in
]%
{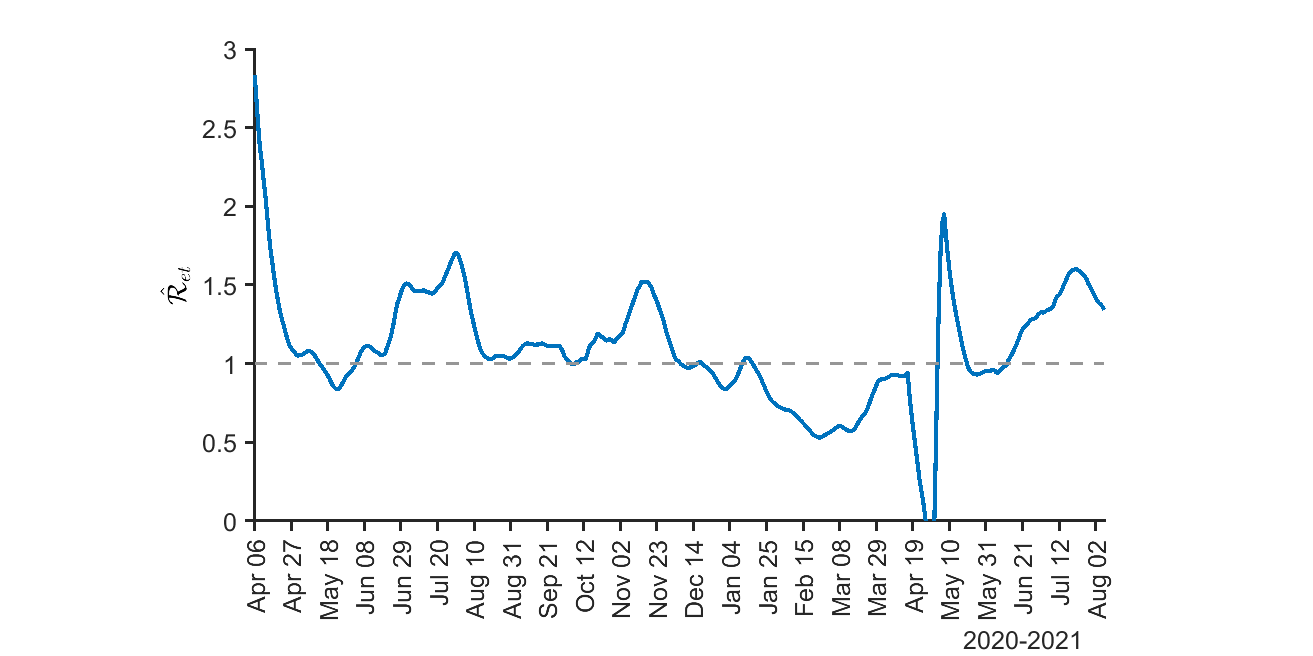}%
}
\end{tabular}

\end{center}

%

\vspace{-0.4cm}%
\footnotesize
{}Notes: Missouri removed $11,454$ double counted cases on April 17, 2021.%

\end{figure}%
%

\addtocounter{figure}{-1}%
%

\begin{figure}[!p]%
\caption
{(Continued) Two-weekly rolling estimates of the effective reproduction numbers ($\mathcal
{R}_{et}$) for the contiguous US, by state}%
\vspace{-0.3cm}%

\begin{center}%
\hspace*{-0.2cm}%
\begin{tabular}
[c]{cc}%
{\footnotesize Montana} & {\footnotesize Nebraska}\\%
{\includegraphics[
height=1.7763in,
width=3.5293in
]%
{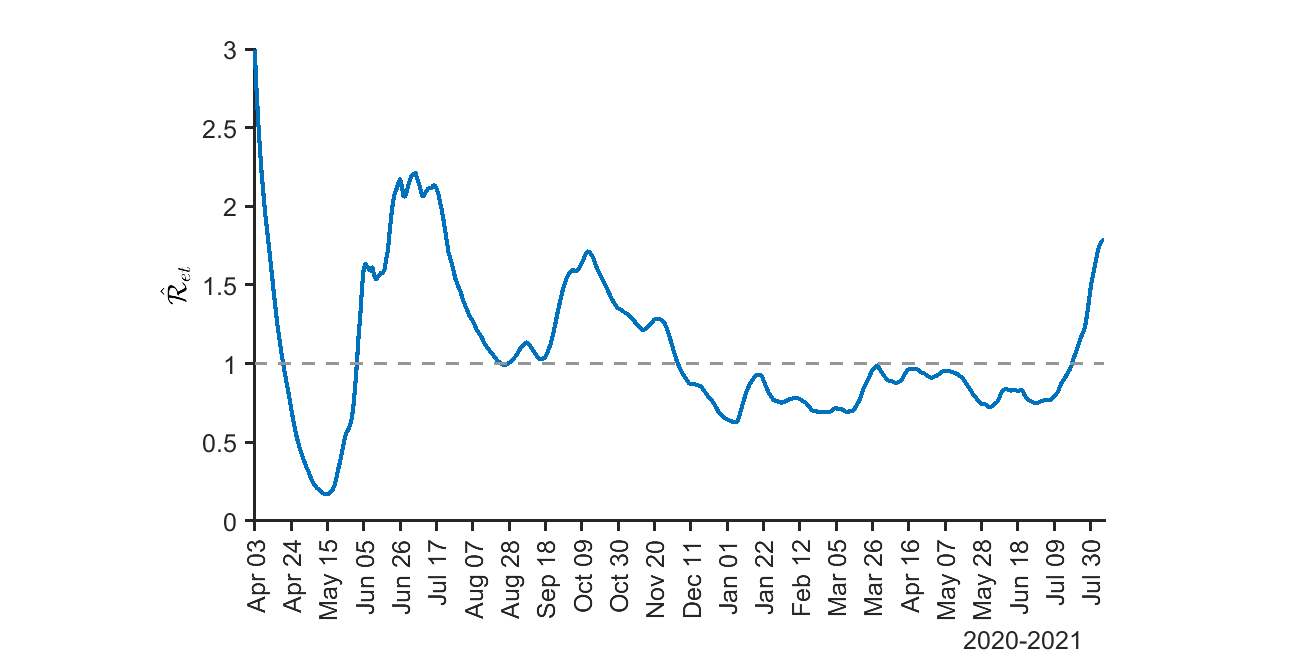}%
}
&
{\includegraphics[
height=1.7763in,
width=3.5293in
]%
{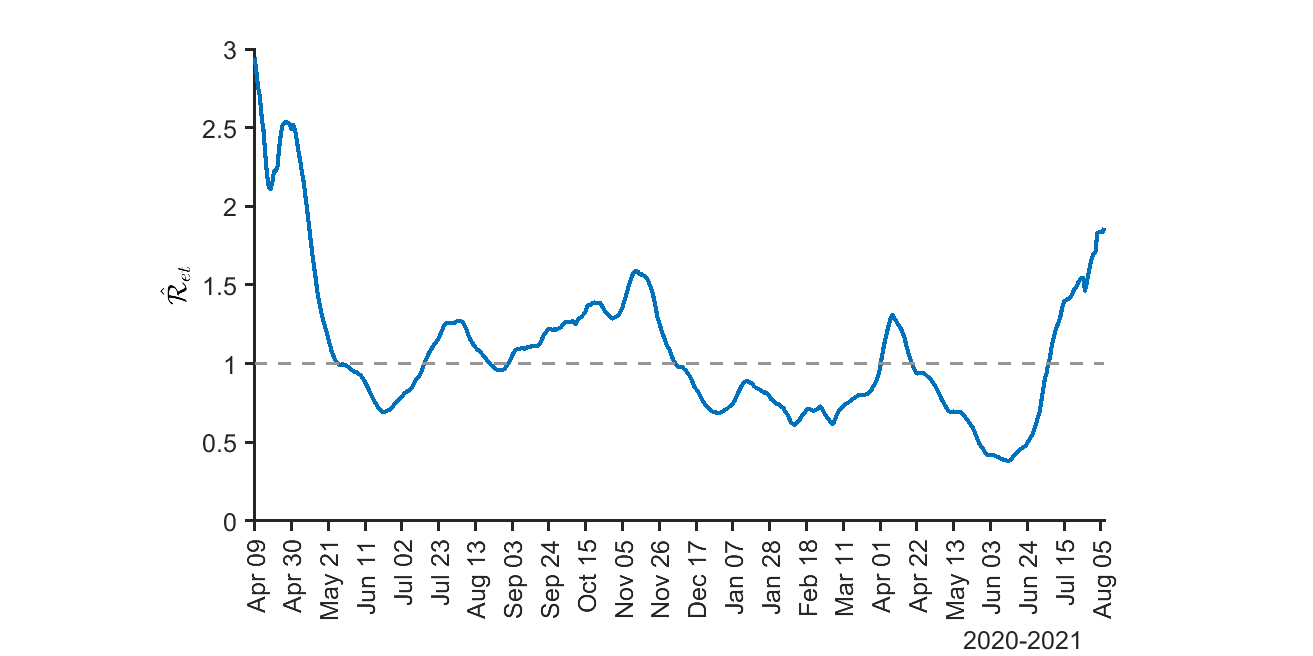}%
}
\\
& \\
{\footnotesize Nevada} & {\footnotesize New Hampshire}\\%
{\includegraphics[
height=1.7763in,
width=3.5293in
]%
{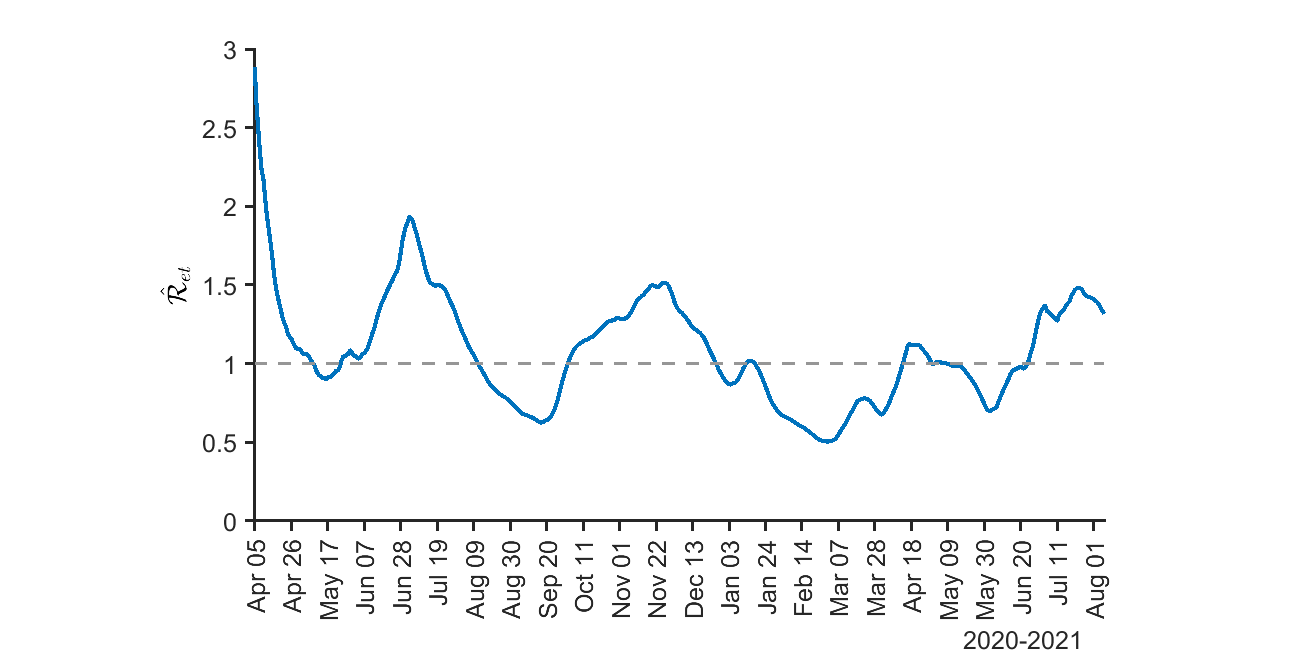}%
}
&
{\includegraphics[
height=1.7763in,
width=3.5293in
]%
{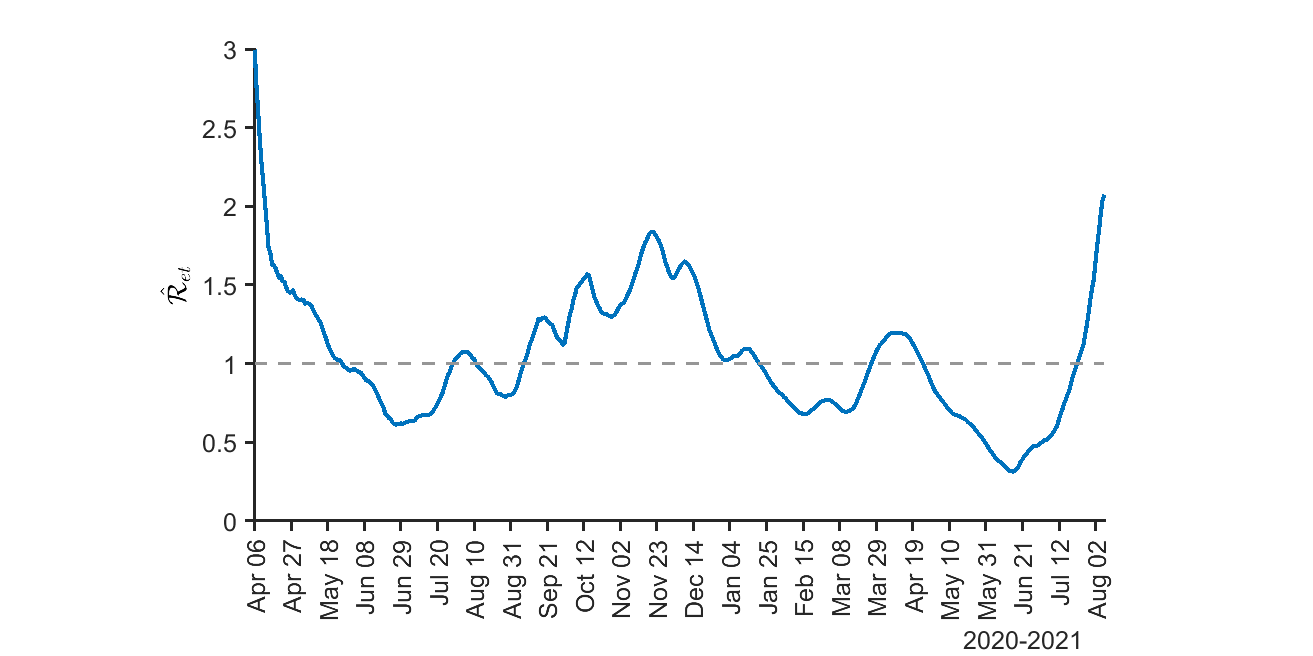}%
}
\\
& \\
{\footnotesize New Jersey} & {\footnotesize New Mexico}\\%
{\includegraphics[
height=1.7763in,
width=3.5293in
]%
{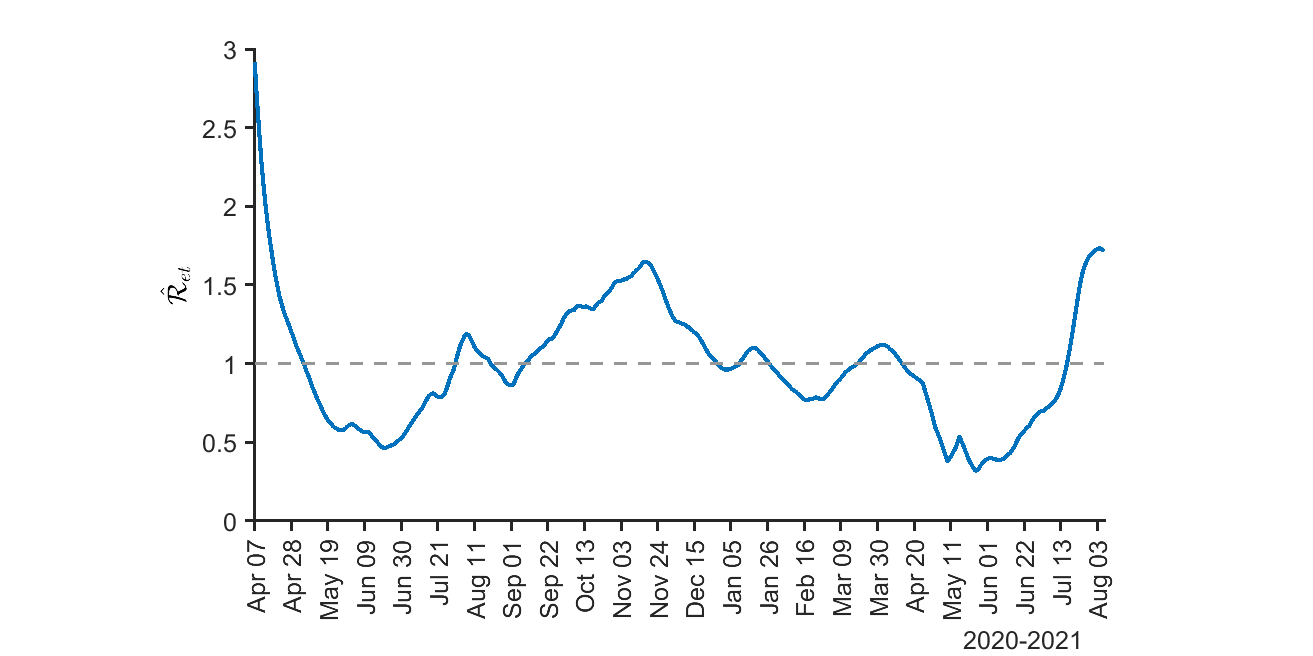}%
}
&
{\includegraphics[
height=1.7763in,
width=3.5293in
]%
{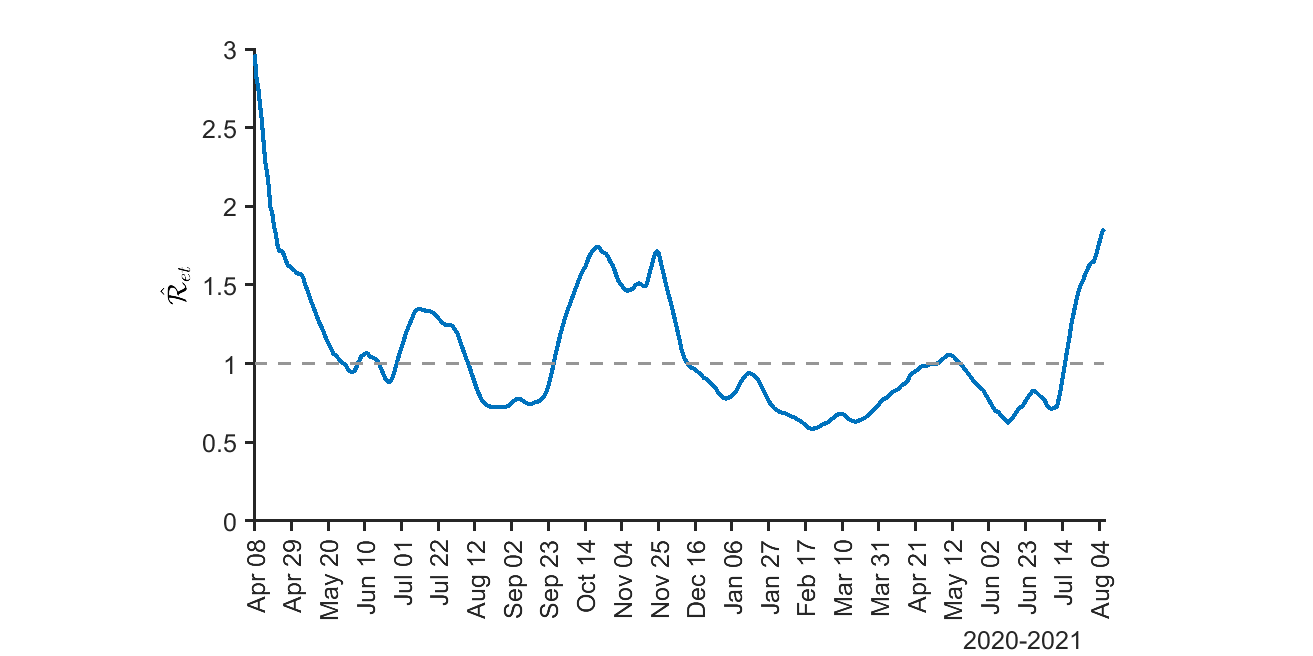}%
}
\\
& \\
{\footnotesize New York} & {\footnotesize North Carolina}\\%
{\includegraphics[
height=1.7763in,
width=3.5293in
]%
{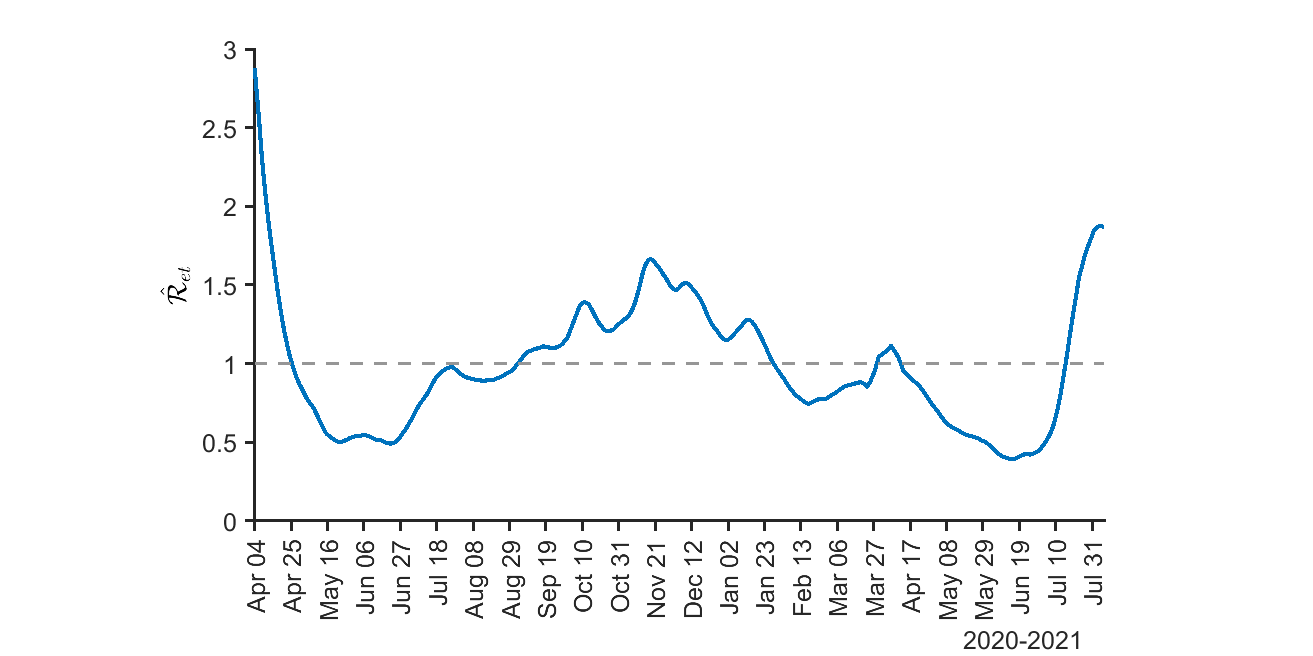}%
}
&
{\includegraphics[
height=1.7763in,
width=3.5293in
]%
{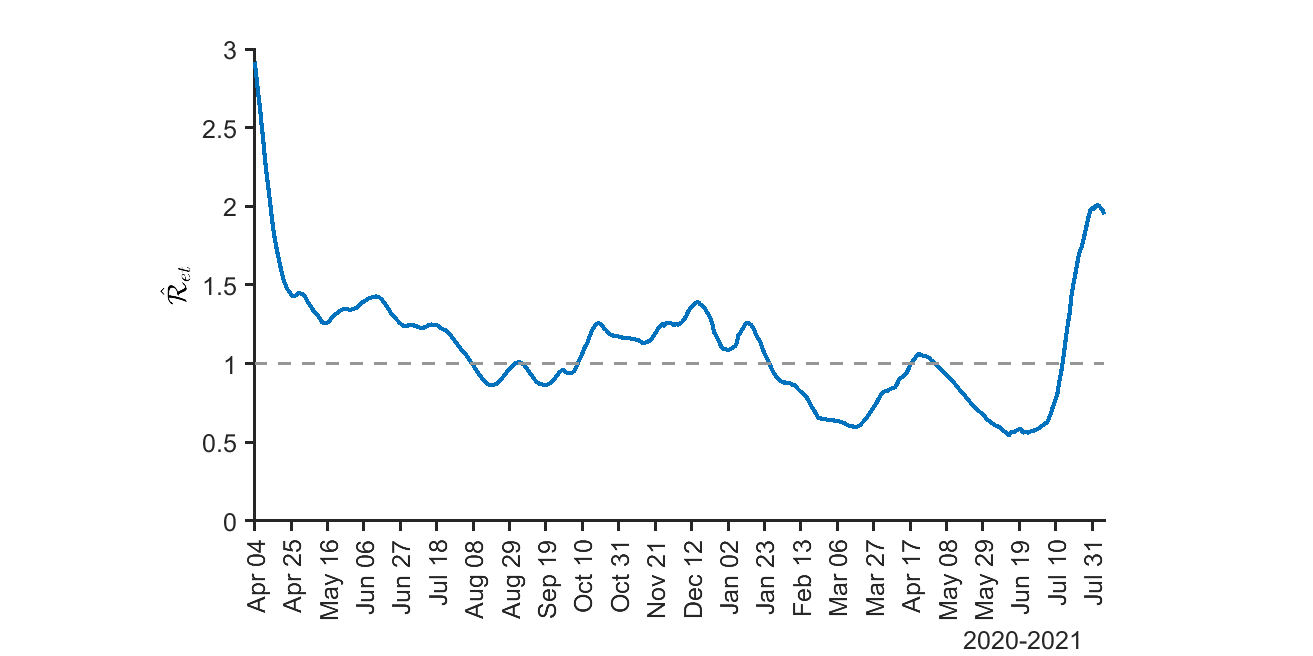}%
}
\end{tabular}

\end{center}

%

\end{figure}%
%

\addtocounter{figure}{-1}%
%

\begin{figure}[!p]%
\caption
{(Continued) Two-weekly rolling estimates of the effective reproduction numbers ($\mathcal
{R}_{et}$) for the contiguous US, by state}%
\vspace{-0.3cm}%

\begin{center}%
\hspace*{-0.2cm}%
\begin{tabular}
[c]{cc}%
{\footnotesize North Dakota} & {\footnotesize Ohio}\\%
{\includegraphics[
height=1.7763in,
width=3.5293in
]%
{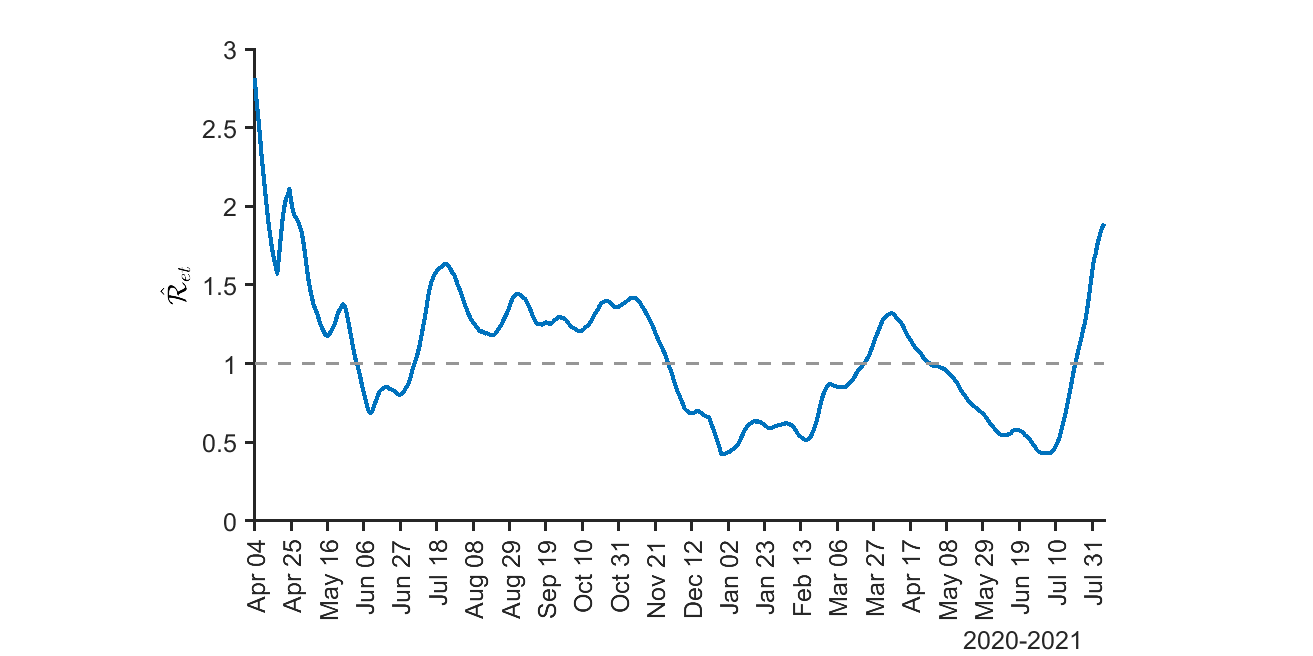}%
}
&
{\includegraphics[
height=1.7763in,
width=3.5293in
]%
{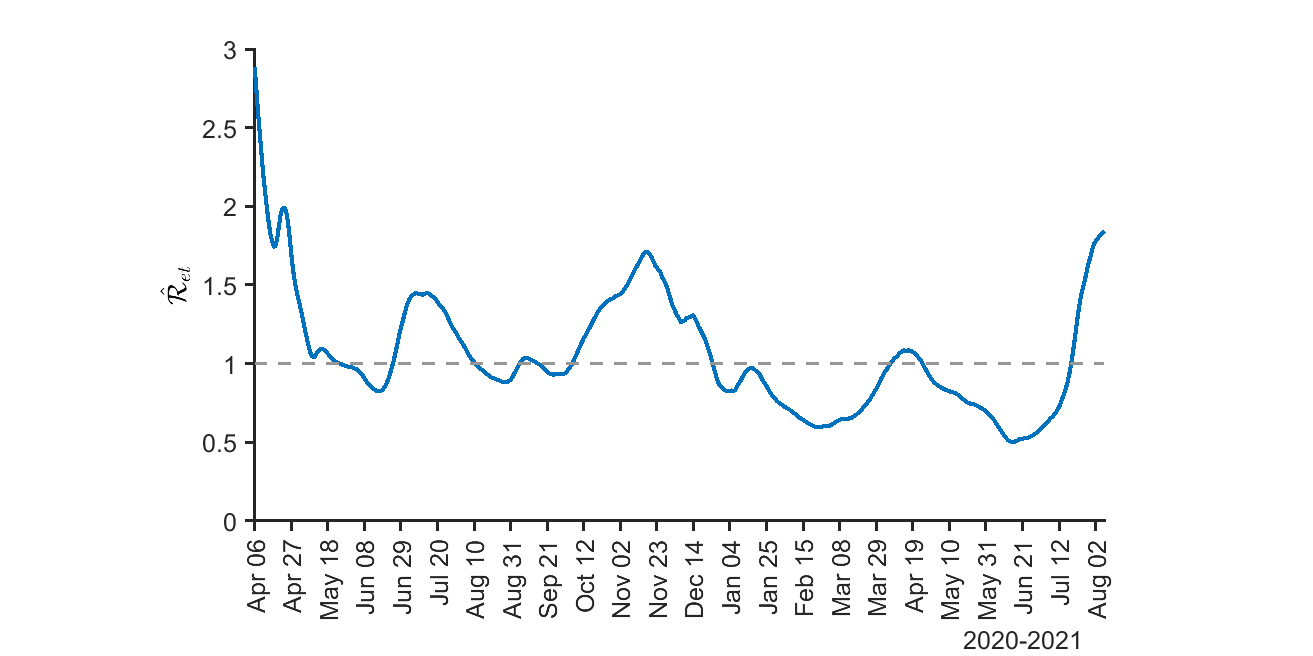}%
}
\\
& \\
{\footnotesize Oklahoma} & {\footnotesize Oregon}\\%
{\includegraphics[
height=1.7763in,
width=3.5293in
]%
{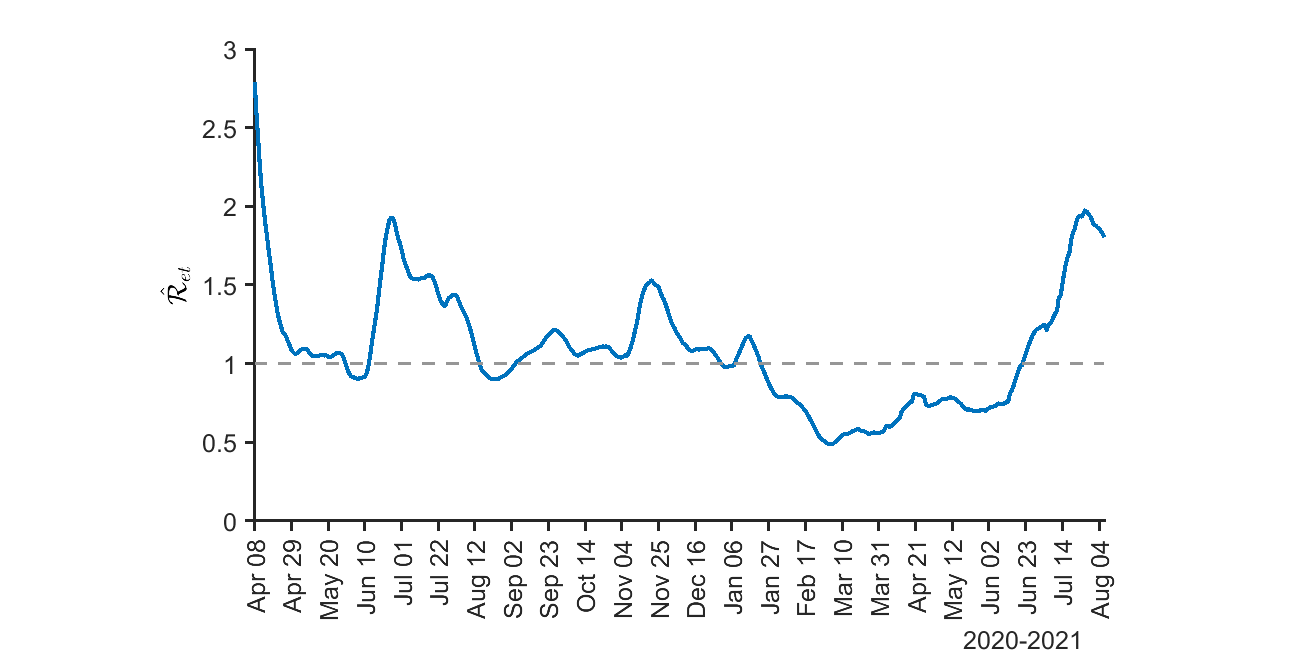}%
}
&
{\includegraphics[
height=1.7763in,
width=3.5293in
]%
{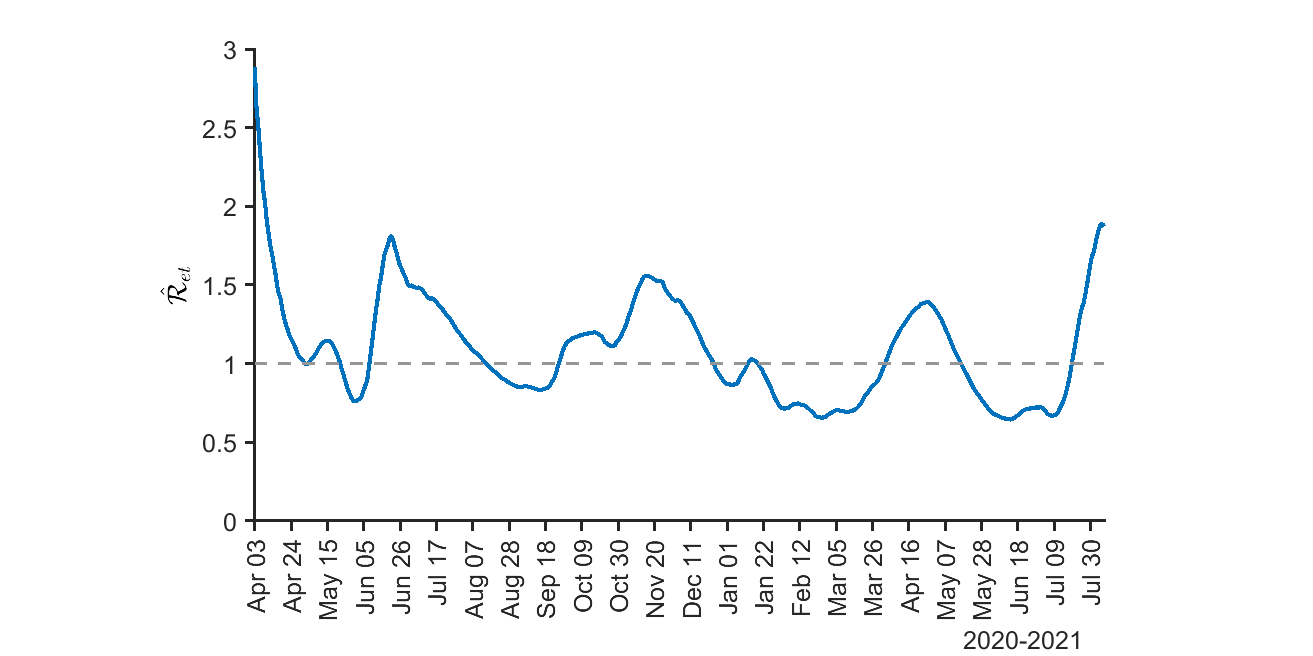}%
}
\\
& \\
{\footnotesize Pennsylvania} & {\footnotesize Rhode Island}\\%
{\includegraphics[
height=1.7763in,
width=3.5293in
]%
{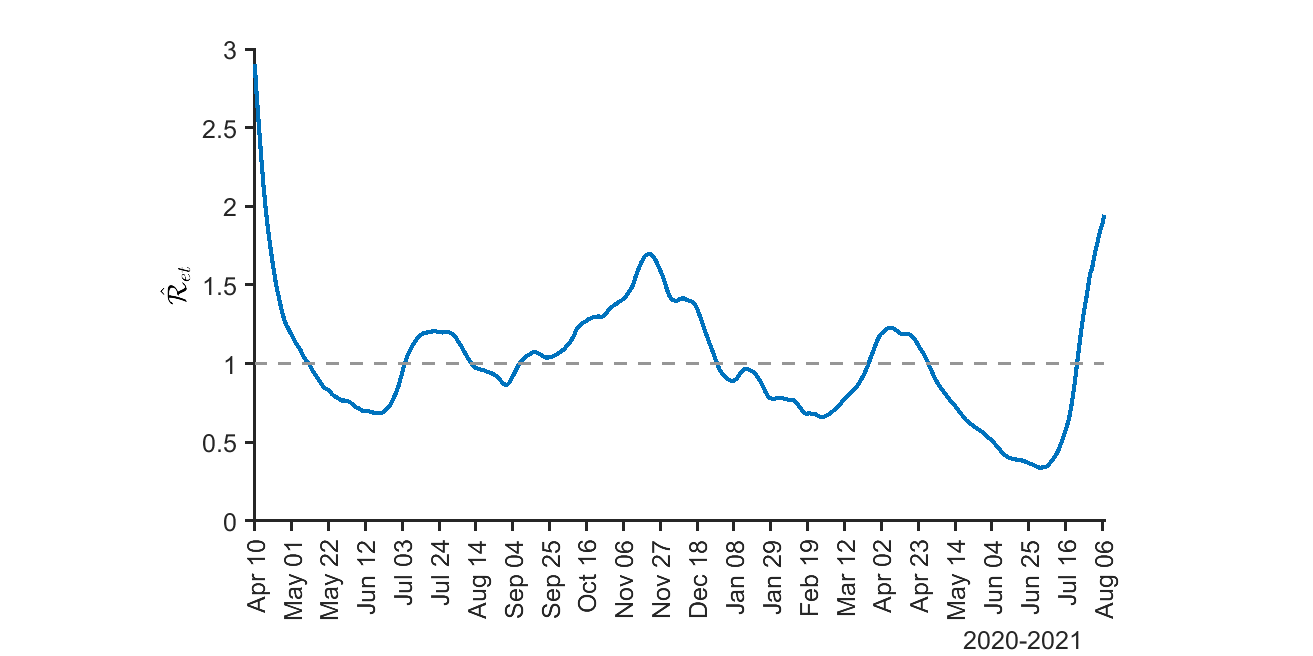}%
}
&
{\includegraphics[
height=1.7763in,
width=3.5293in
]%
{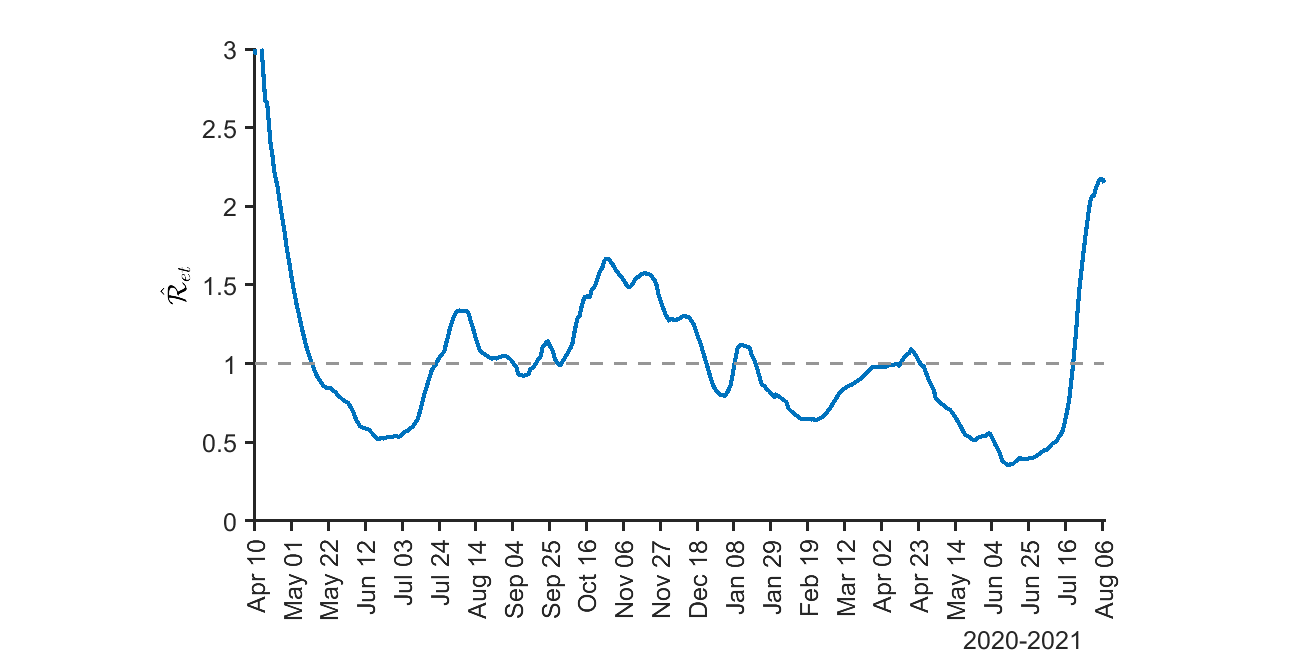}%
}
\\
& \\
{\footnotesize South Carolina} & {\footnotesize South Dakota}\\%
{\includegraphics[
height=1.7763in,
width=3.5293in
]%
{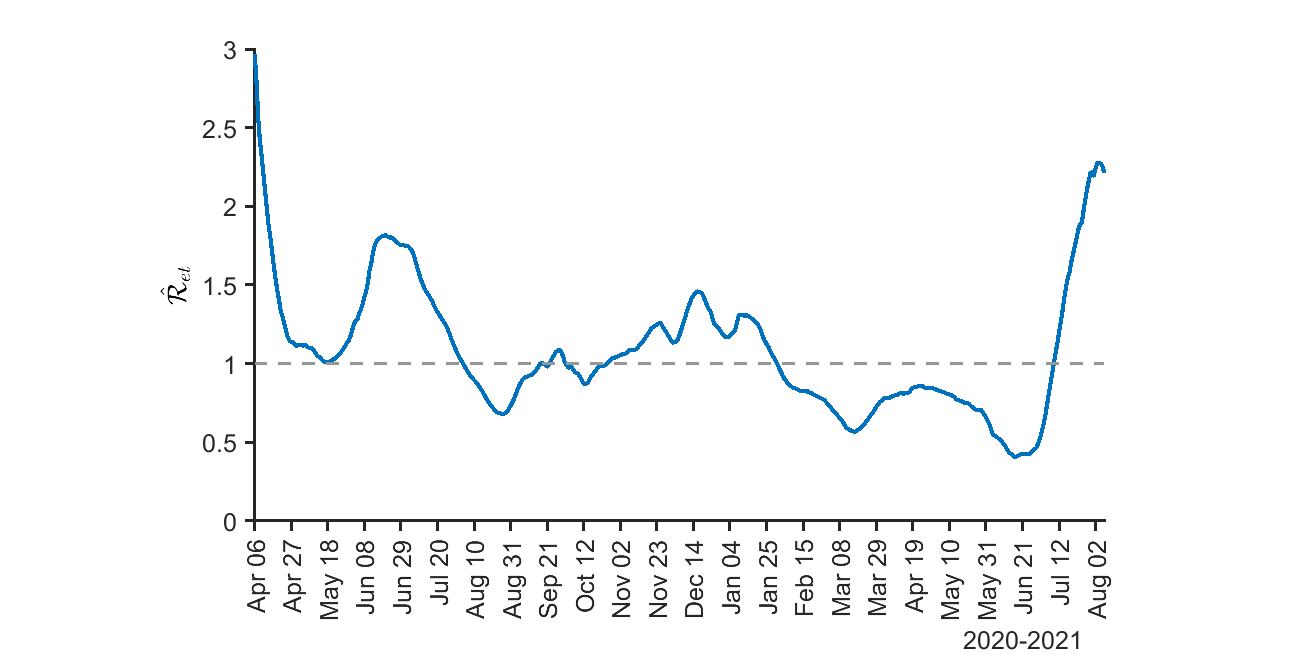}%
}
&
{\includegraphics[
height=1.7763in,
width=3.5293in
]%
{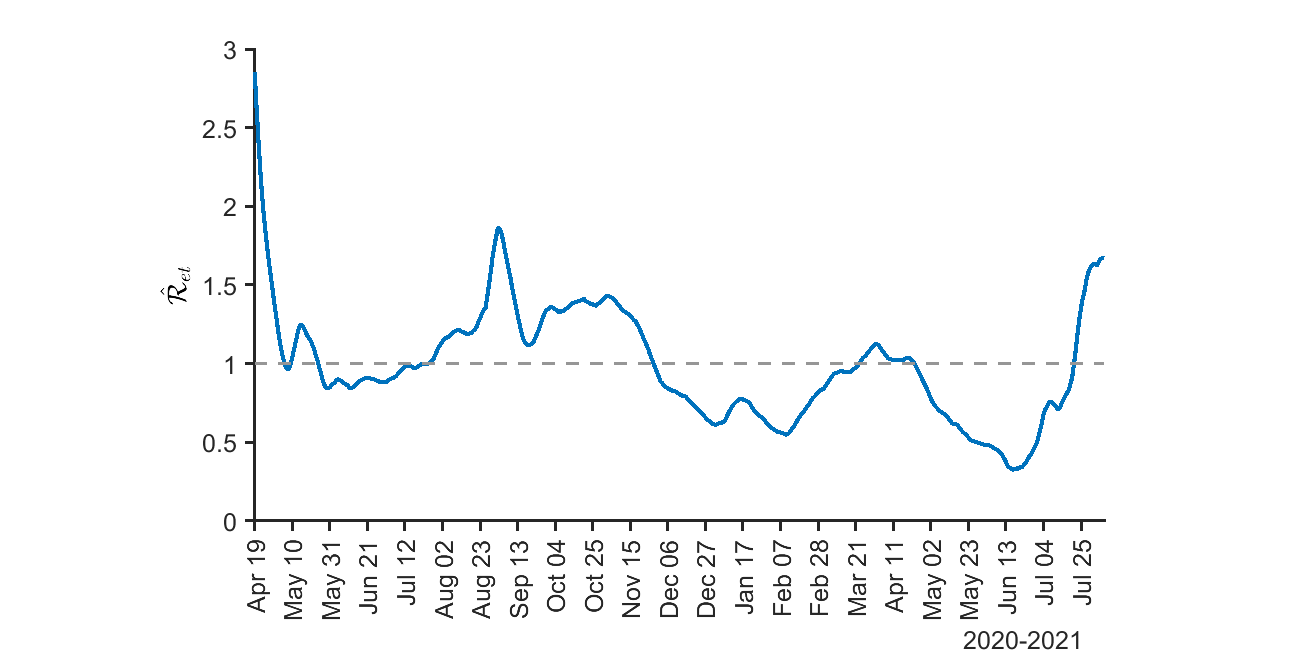}%
}
\end{tabular}

\end{center}

%

\vspace{-0.4cm}%
\footnotesize
{}Notes: Ohio stopped publishing daily case numbers on June 2, 2021.%

\end{figure}%
%

\addtocounter{figure}{-1}%
%

\begin{figure}[!p]%
\caption
{(Continued) Two-weekly rolling estimates of the effective reproduction numbers ($\mathcal
{R}_{et}$) for the contiguous US, by state}%
\vspace{-0.3cm}%

\begin{center}%
\hspace*{-0.2cm}%
\begin{tabular}
[c]{cc}%
{\footnotesize Tennessee} & {\footnotesize Texas}\\%
{\includegraphics[
height=1.7763in,
width=3.5293in
]%
{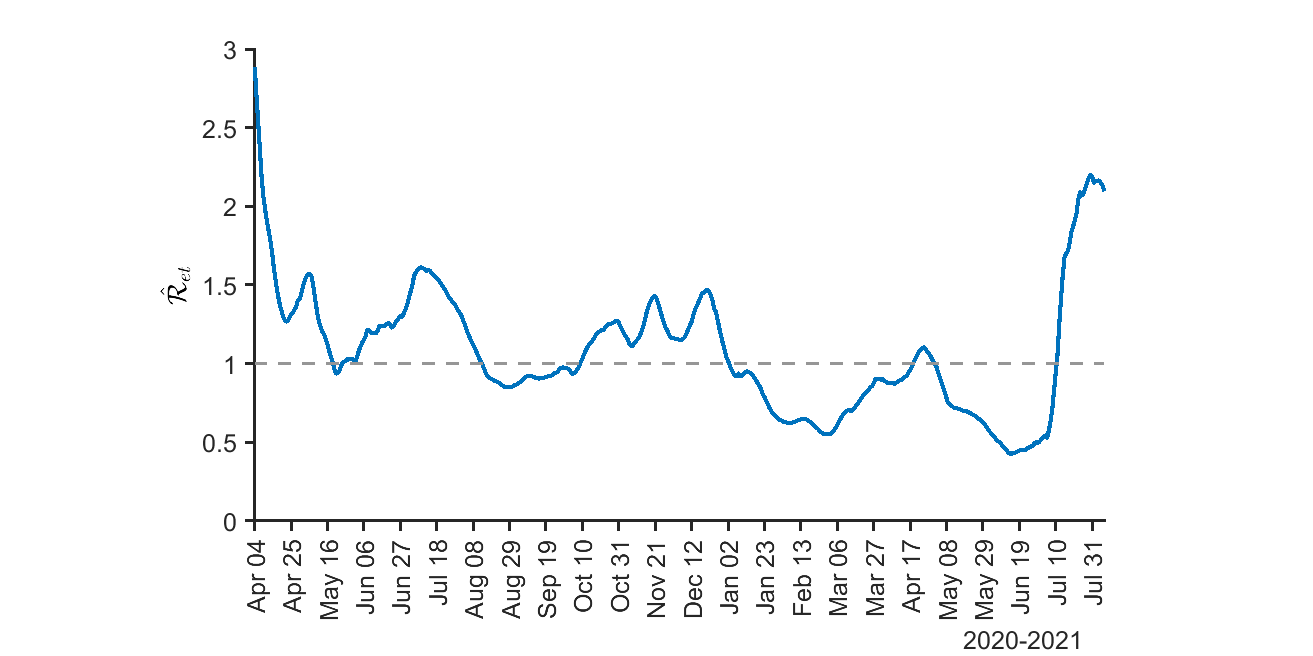}%
}
&
{\includegraphics[
height=1.7763in,
width=3.5293in
]%
{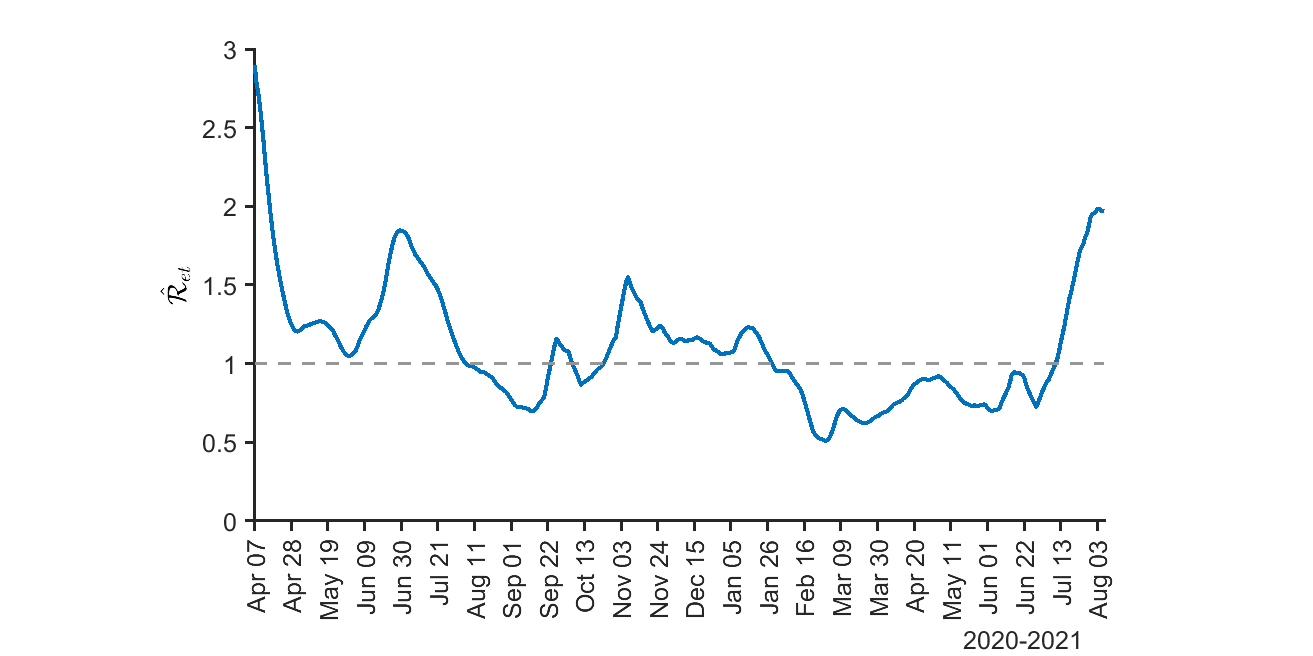}%
}
\\
& \\
{\footnotesize Utah} & {\footnotesize Vermont}\\%
{\includegraphics[
height=1.7763in,
width=3.5293in
]%
{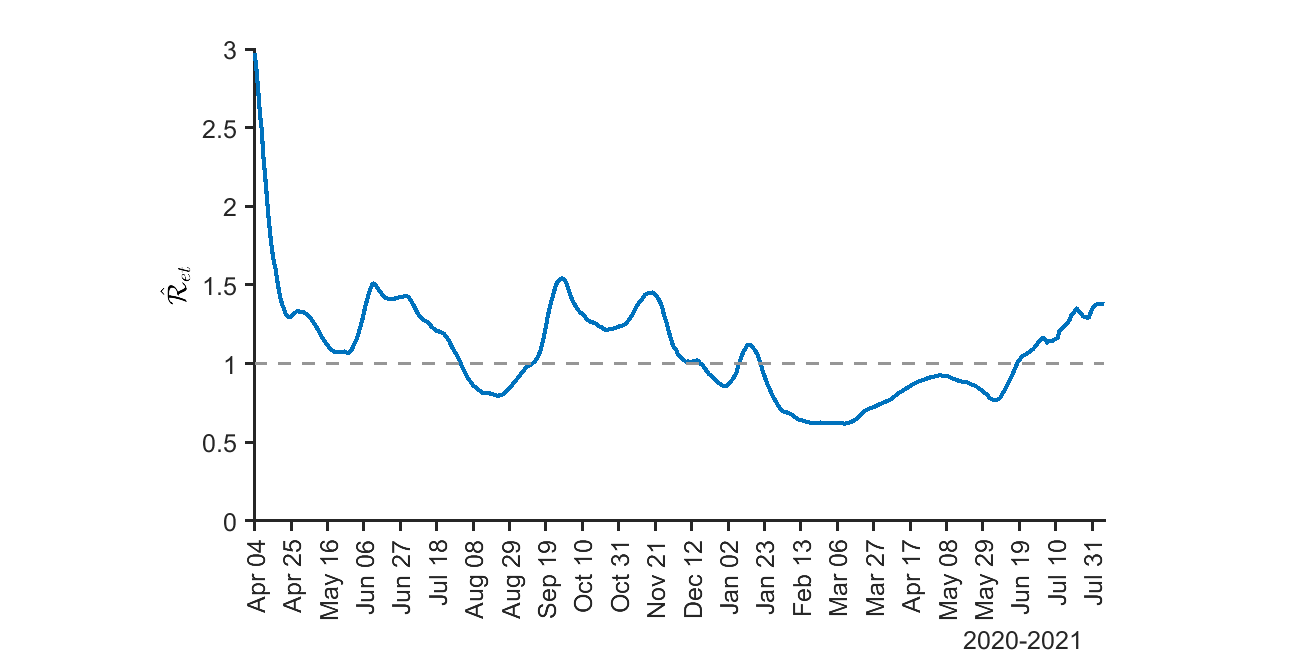}%
}
&
{\includegraphics[
height=1.7763in,
width=3.5293in
]%
{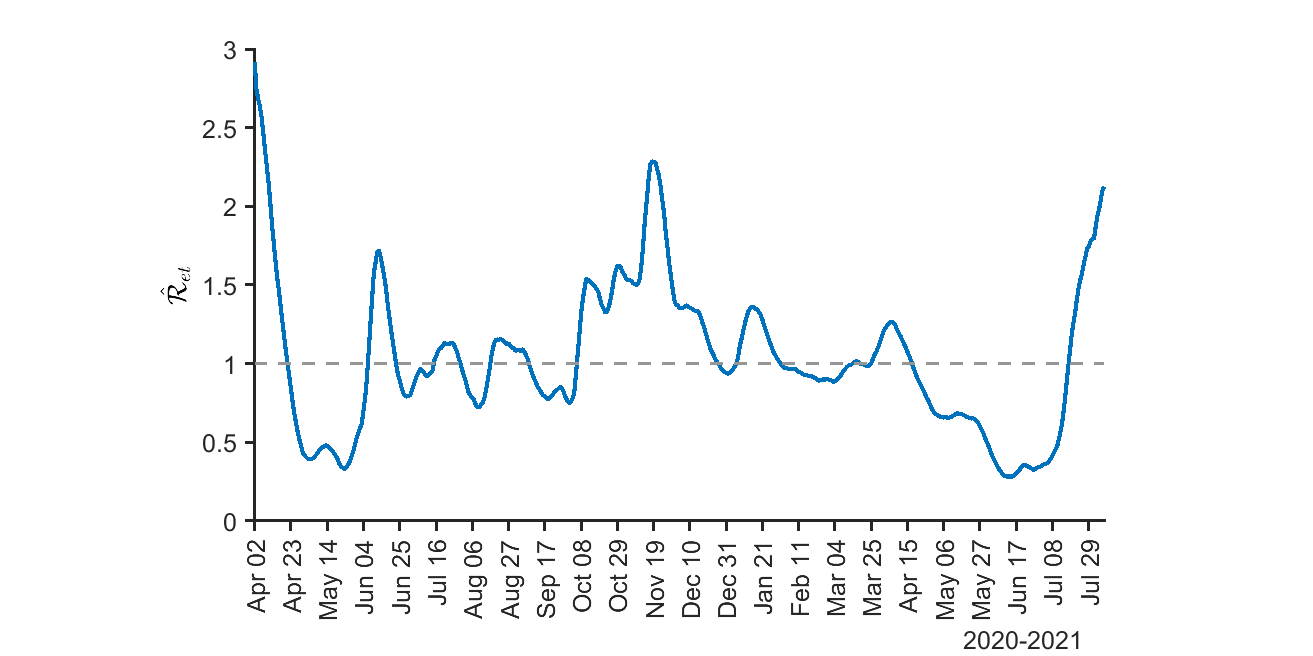}%
}
\\
& \\
{\footnotesize Virginia} & {\footnotesize Washington}\\%
{\includegraphics[
height=1.7763in,
width=3.5293in
]%
{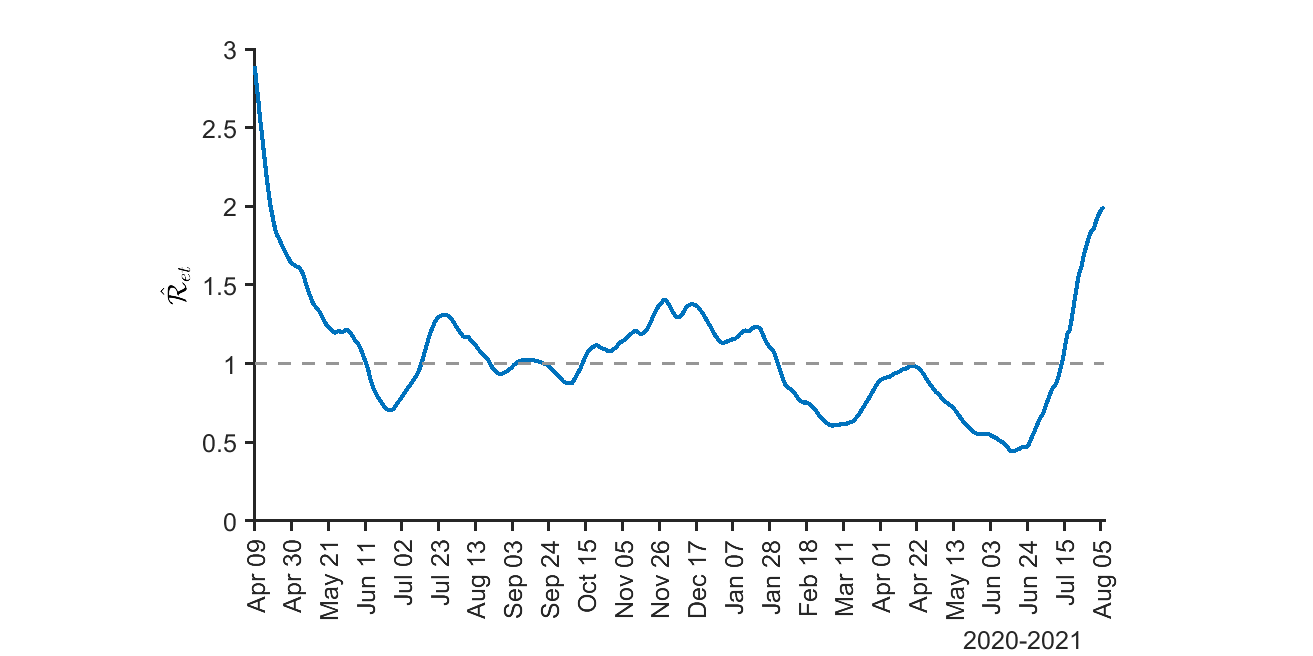}%
}
&
{\includegraphics[
height=1.7763in,
width=3.5293in
]%
{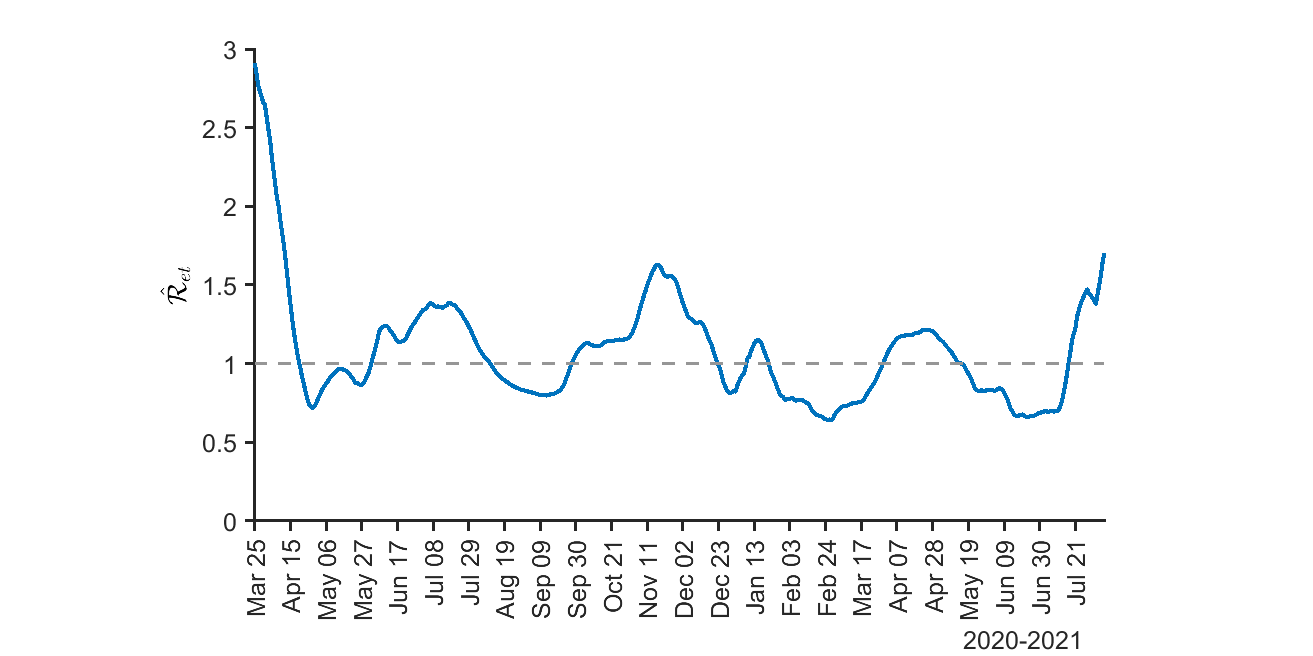}%
}
\\
& \\
{\footnotesize West Virginia} & {\footnotesize Wisconsin}\\%
{\includegraphics[
height=1.7763in,
width=3.5293in
]%
{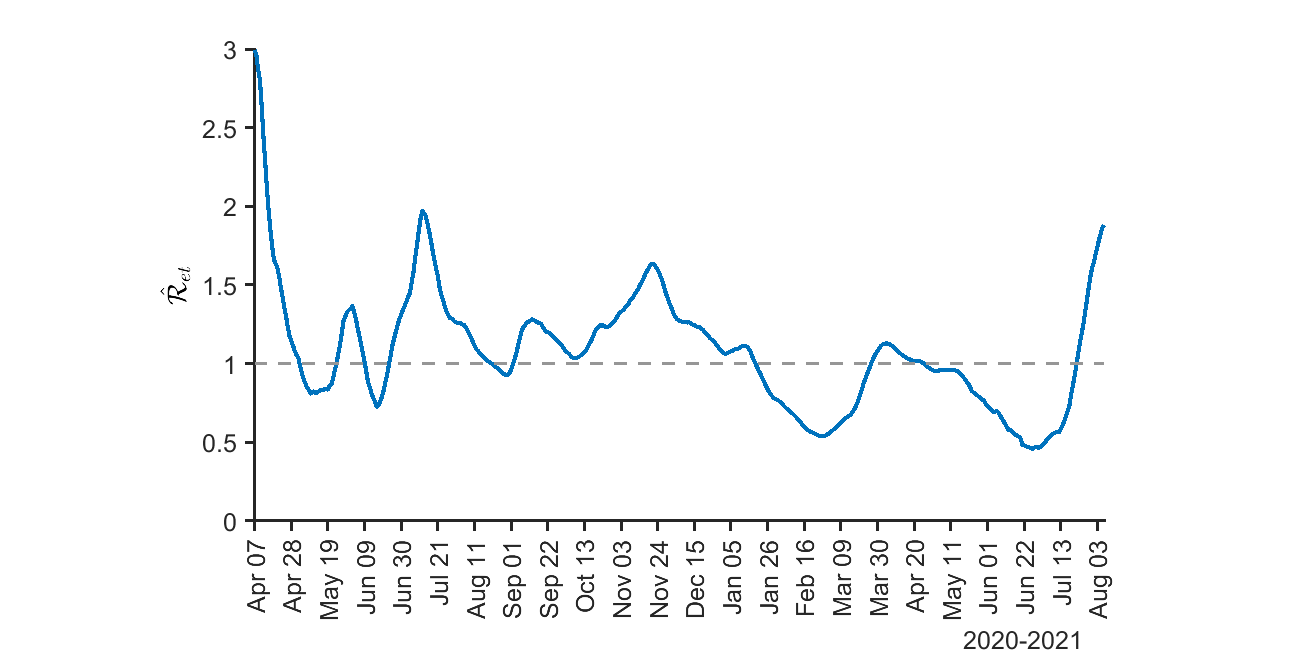}%
}
&
{\includegraphics[
height=1.7763in,
width=3.5293in
]%
{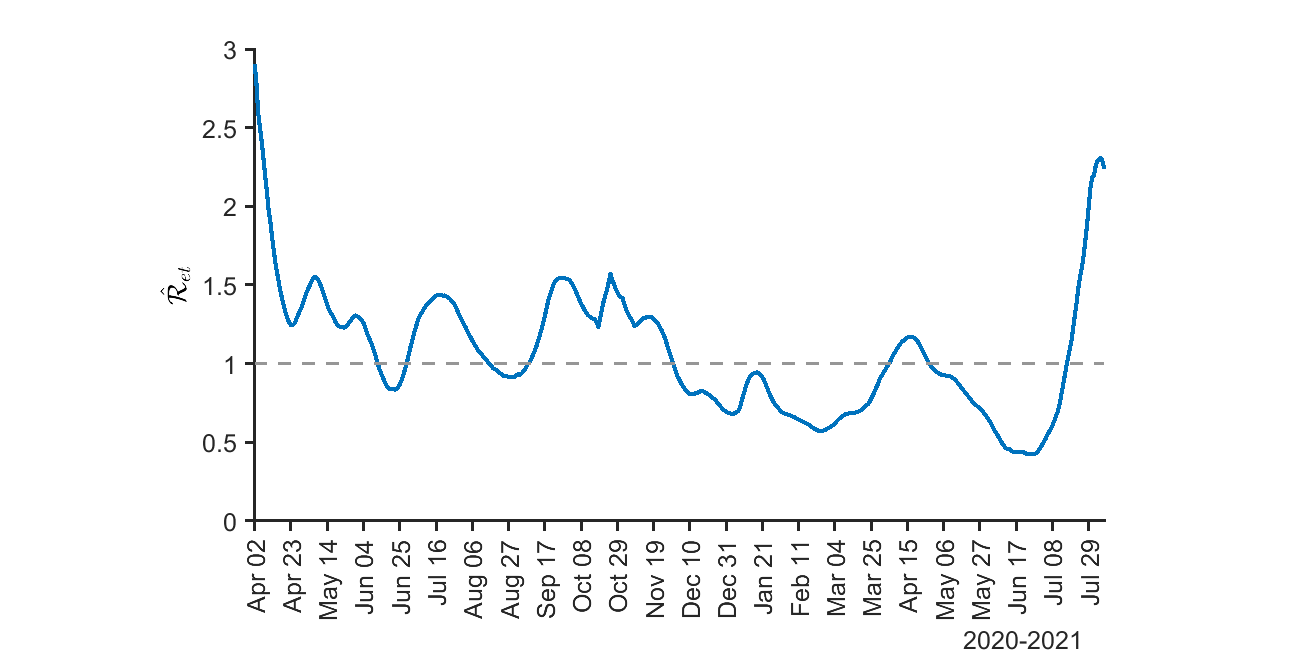}%
}
\end{tabular}

\end{center}

%

\end{figure}%
%

\addtocounter{figure}{-1}%
%

\begin{figure}[!th]%
\caption
{(Continued) Two-weekly rolling estimates of the effective reproduction numbers ($\mathcal
{R}_{et}$) for the contiguous US, by state}%
\vspace{-0.3cm}%

\begin{center}%
\hspace*{-0.2cm}%
\begin{tabular}
[c]{c}%
{\footnotesize Wyoming}\\%
{\includegraphics[
height=1.7763in,
width=3.5293in
]%
{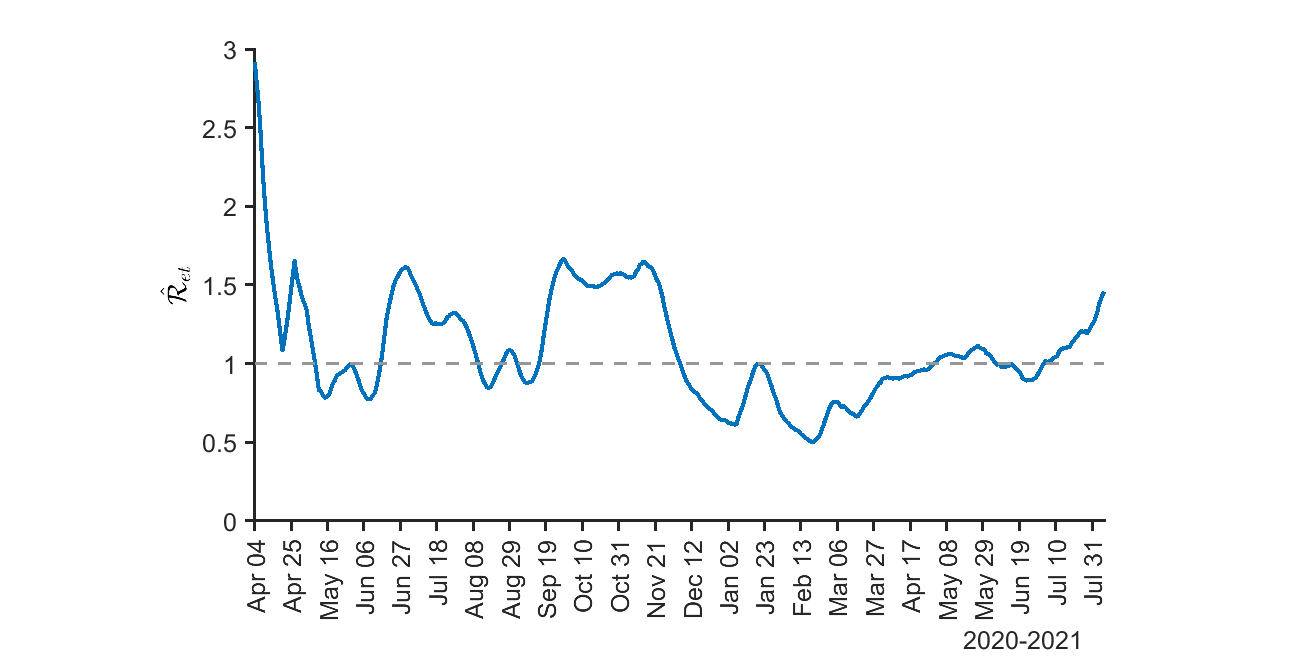}%
}
\end{tabular}

\end{center}

%

\end{figure}%
%

\FloatBarrier

\section{Estimates of the multiplication factor\label{Sup: empirical_MF}}

This section presents additional estimation results of the multiplication
factor for the selected European countries and the US. It also compares the
number of cases per capita with and without adjusting for under-reporting.

\begin{figure}[tp]%
\caption
{Rolling estimates of the multiplication factor using the 2- and 3-weekly
rolling windows for selected European countries}%
\vspace{-0.2cm}%
\label{fig: Euro_MF_cmp_2W_3W}%

\begin{footnotesize}%

\begin{center}%
\begin{tabular}
[c]{ccc}%
Austria &  & France\\%
{\includegraphics[
height=1.9951in,
width=2.6524in
]%
{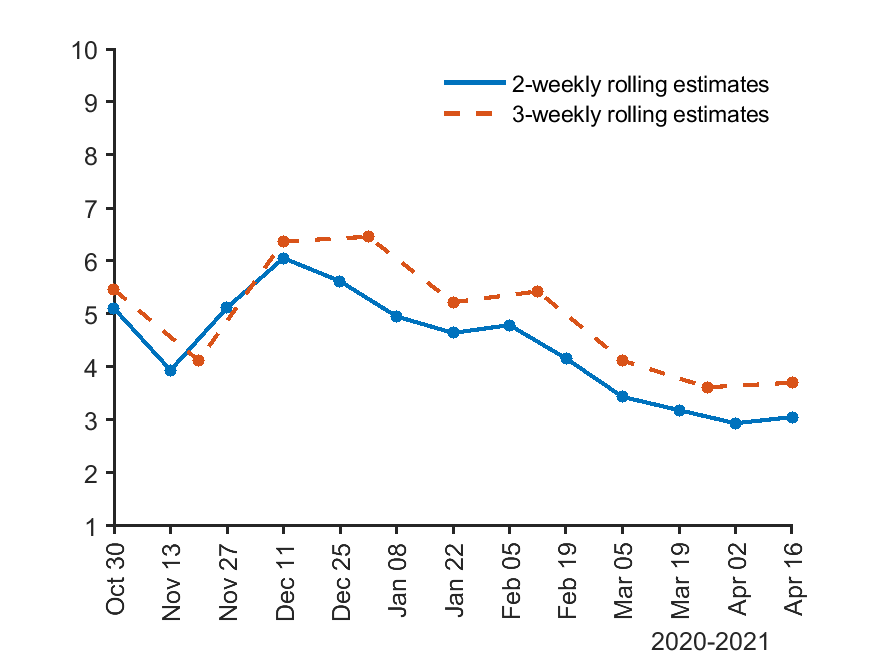}%
}
&  &
{\includegraphics[
height=1.9951in,
width=2.6524in
]%
{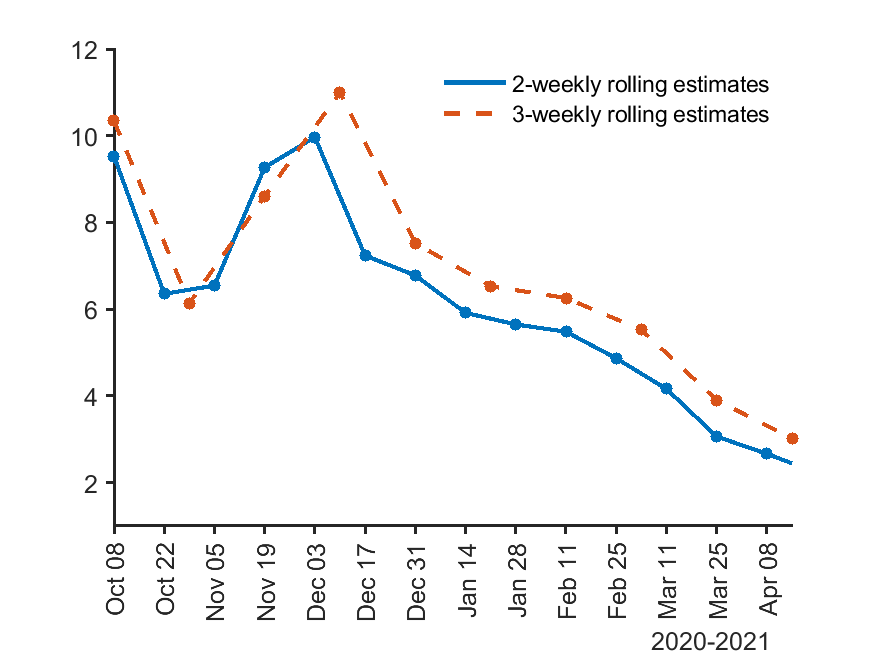}%
}
\\
&  & \\
Germany &  & Italy\\%
{\includegraphics[
height=1.9951in,
width=2.6524in
]%
{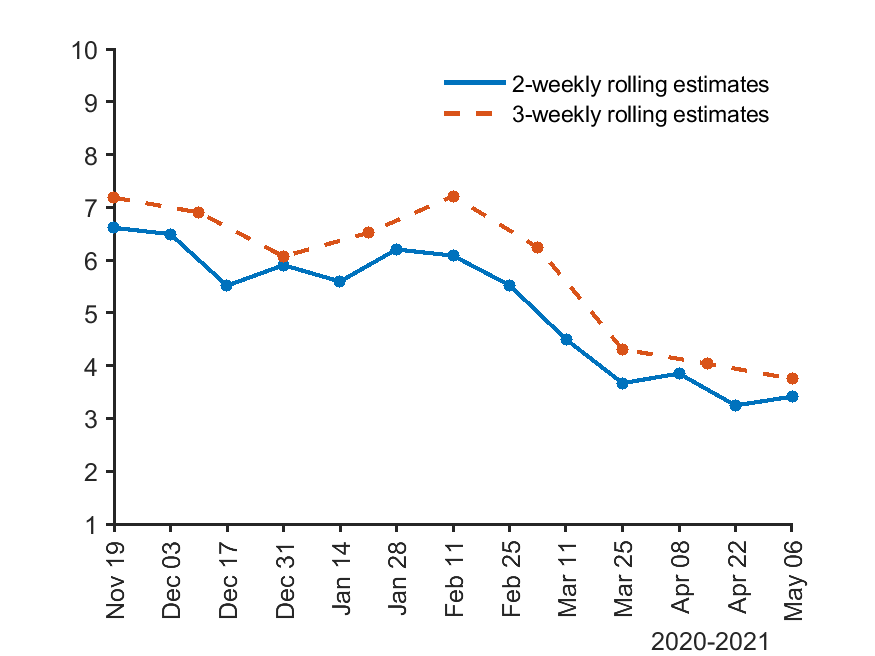}%
}
&  &
{\includegraphics[
height=1.9951in,
width=2.6524in
]%
{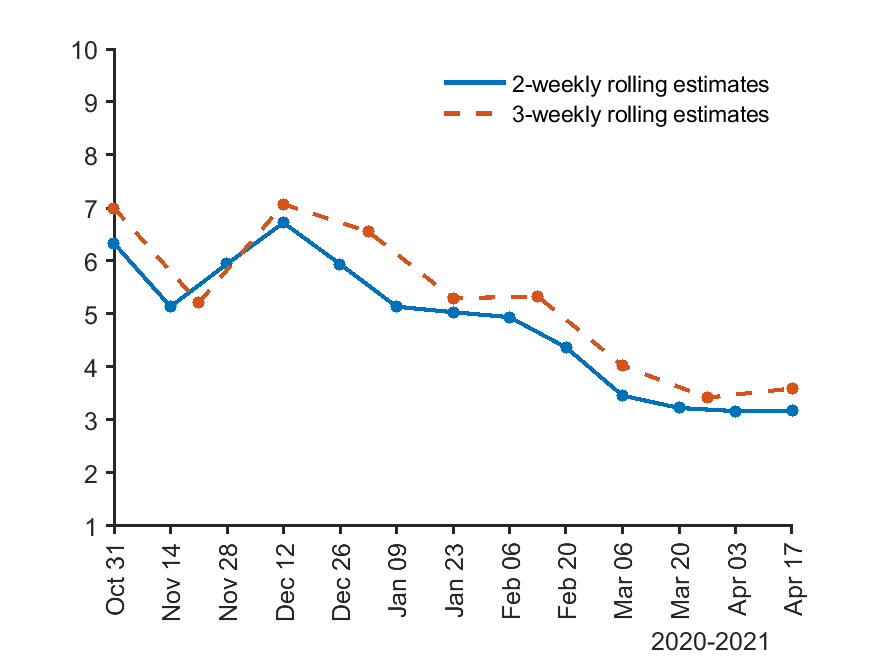}%
}
\\
&  & \\
Spain &  & UK\\%
{\includegraphics[
height=1.9951in,
width=2.6524in
]%
{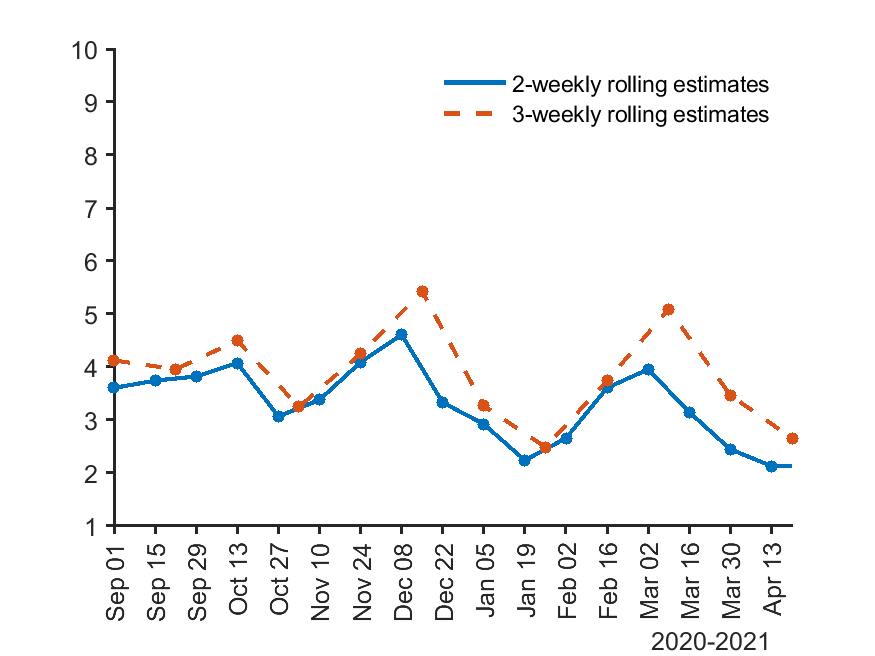}%
}
&  &
{\includegraphics[
height=1.9951in,
width=2.6524in
]%
{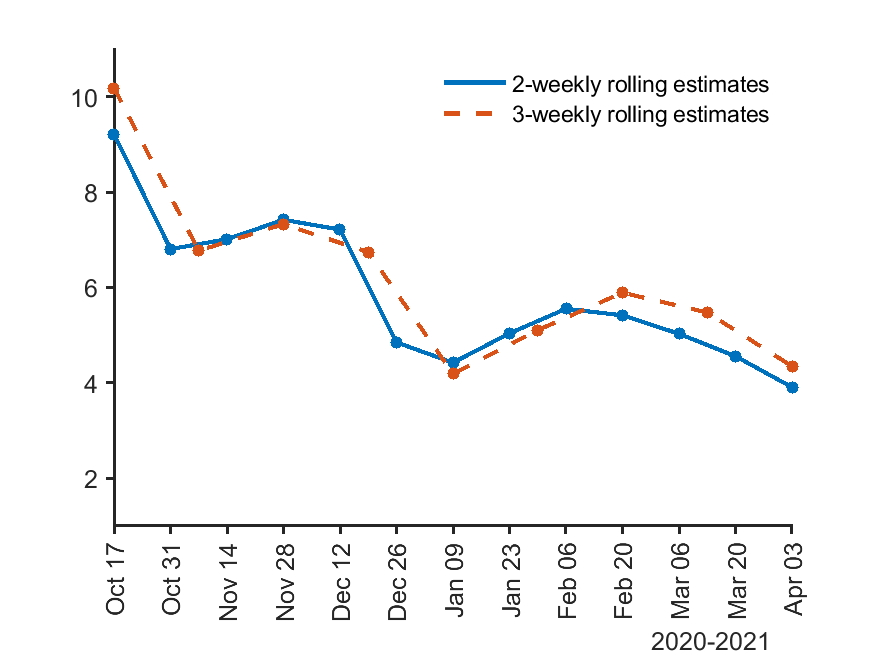}%
}
\end{tabular}

\end{center}

%

\vspace{0cm}%
\footnotesize
Notes: See the notes to Figure \ref{fig: Euro_Re_cmp_2W_3W}.%

\end{footnotesize}%
%

\end{figure}

Figure \ref{fig: Euro_MF_cmp_2W_3W} compares the 2- and 3-weekly estimates of
MF for the six European countries considered in the main paper. As the figure
shows, the 3-weekly estimates of MF are slightly higher than the 2-weekly
estimates. Still, overall they are very close and lead to negligible
differences in the estimates of transmission rates, as we have seen in Figure
\ref{fig: Euro_Re_cmp_2W_3W}.%

\begin{figure}[tb]%
\caption
{Total number of infected cases for selected European countries, without and with adjusting for under-reporting}%
\vspace{-0.2cm}%
\label{fig: Euro_ct}%

\begin{footnotesize}%

\begin{center}%
\begin{tabular}
[c]{ccc}%
Reported number of cases ($c_{t}$) &  & $c_{t}$ after adjusting for
under-reporting\\%
{\includegraphics[
height=1.9951in,
width=2.6524in
]%
{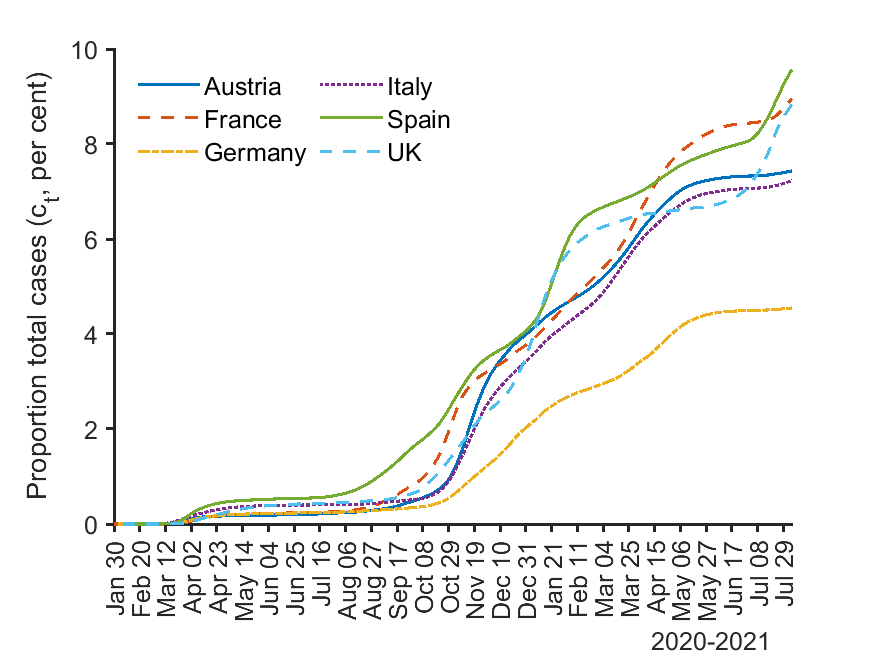}%
}
&  &
{\includegraphics[
height=1.9951in,
width=2.6524in
]%
{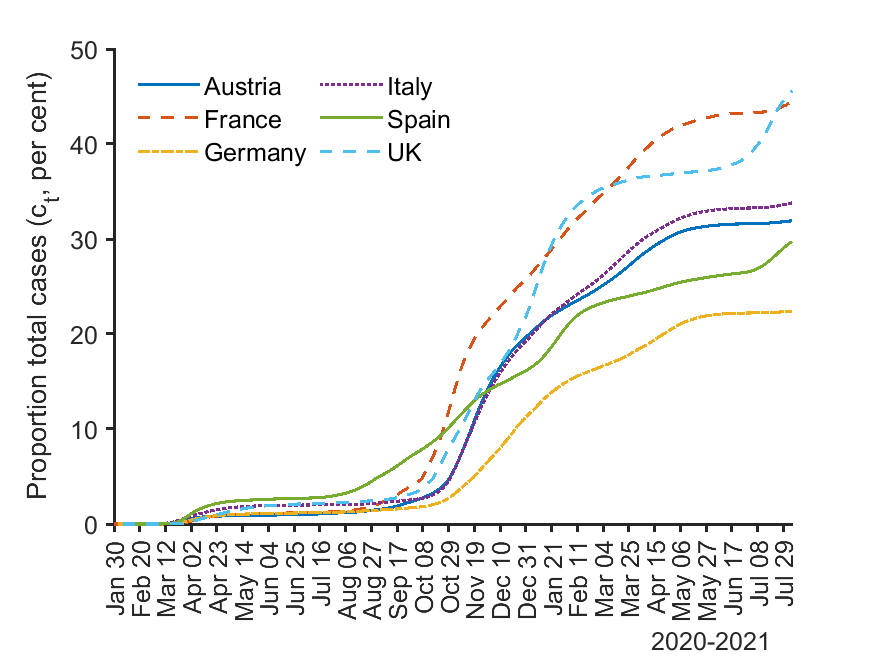}%
}
\end{tabular}

\end{center}

%

\vspace{-0.1cm}%
Notes: All series are 7-day moving averages. The figure on the right shows the
adjusted number of total cases computed by accumulating $\hat{m}_{t}%
\times\Delta\tilde{c}_{t}$ from the outbreak to August 5, 2021, where the
multiplication factor, $m_{t}$, was updated every two weeks with values in
between obtained by linear interpolation.%

\end{footnotesize}%
%

\end{figure}

It is interesting to compare the reported number of total cases with the case
numbers after adjusting for under-reporting using the 2-weekly estimates of
the MF, which are displayed in Figure \ref{fig: Euro_MF} of the main paper.
Figure \ref{fig: Euro_ct} plots the 7-day moving average of infected cases
(per $100,000$ people) for the six countries using the more recent data as of
August 5, 2021. The left panel displays the raw data. The right panel shows
the MF-adjusted total cases computed by accumulating $\hat{m}_{t}\times
\Delta\tilde{c}_{t}$ from the start of the epidemic, where the MF estimates,
$\hat{m}_{t}$, was updated every two weeks with values in between obtained by
linear interpolation, and $\Delta\tilde{c}_{t}$ is the reported number of
daily new cases (per 100k population). The MF is fixed at the last estimate
for the period after the joint estimation ends.\footnote{We also considered
stopping the joint estimation when the share of the population fully
vaccinated reaches 15 percent. The results are quite similar and available
upon request.} The figure clearly shows that it is important to adjust the
case counts over time by the time-varying MF. We find that, as of August 5,
2021, the number of total cases may be underestimated by three times in Spain,
four times in Austria, and five times in the other four countries. After
adjusting for under-reporting, we see that Spain ranks fifth in the case rate
instead of first. France and the UK have the highest number of cases per
capita after adjustment among these countries, approaching $40$ percent in
comparison to $8$ percent without adjustment. In contrast, Germany did the
best job controlling the total cases even after taking under-reporting into account.%

\begin{figure}[tbh]%
\caption
{Estimates of the multiplication factor and comparison of realized and calibrated new cases for the US}%
\vspace{-0.2cm}%
\label{fig: US_MF}%

\begin{footnotesize}%

\begin{center}%
\begin{tabular}
[c]{ccc}%
Estimates of MF &  & Realized and calibrated new cases\\
&  & \\%
{\includegraphics[
height=1.9951in,
width=2.6524in
]%
{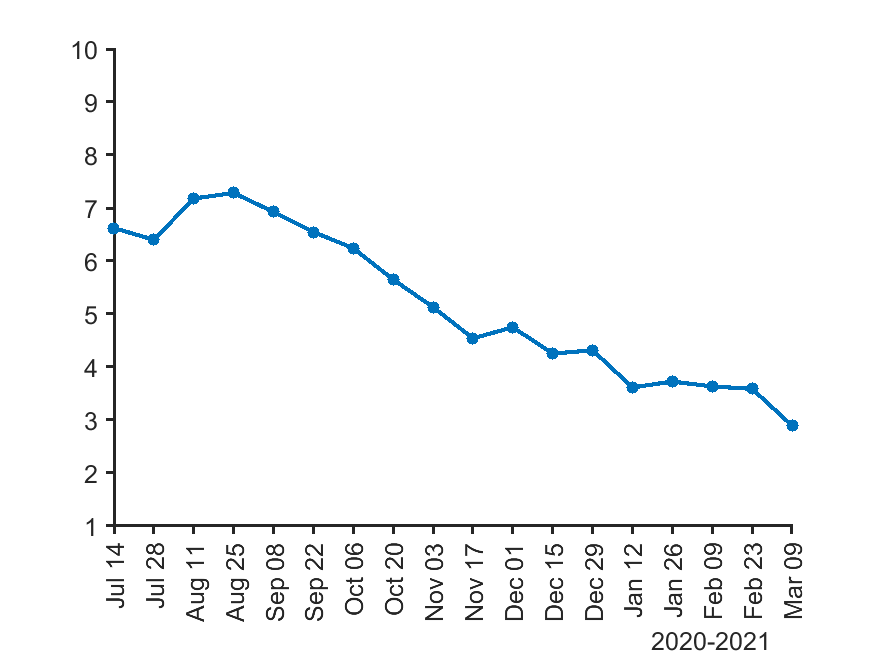}%
}
&  &
{\includegraphics[
height=1.8273in,
width=2.2779in
]%
{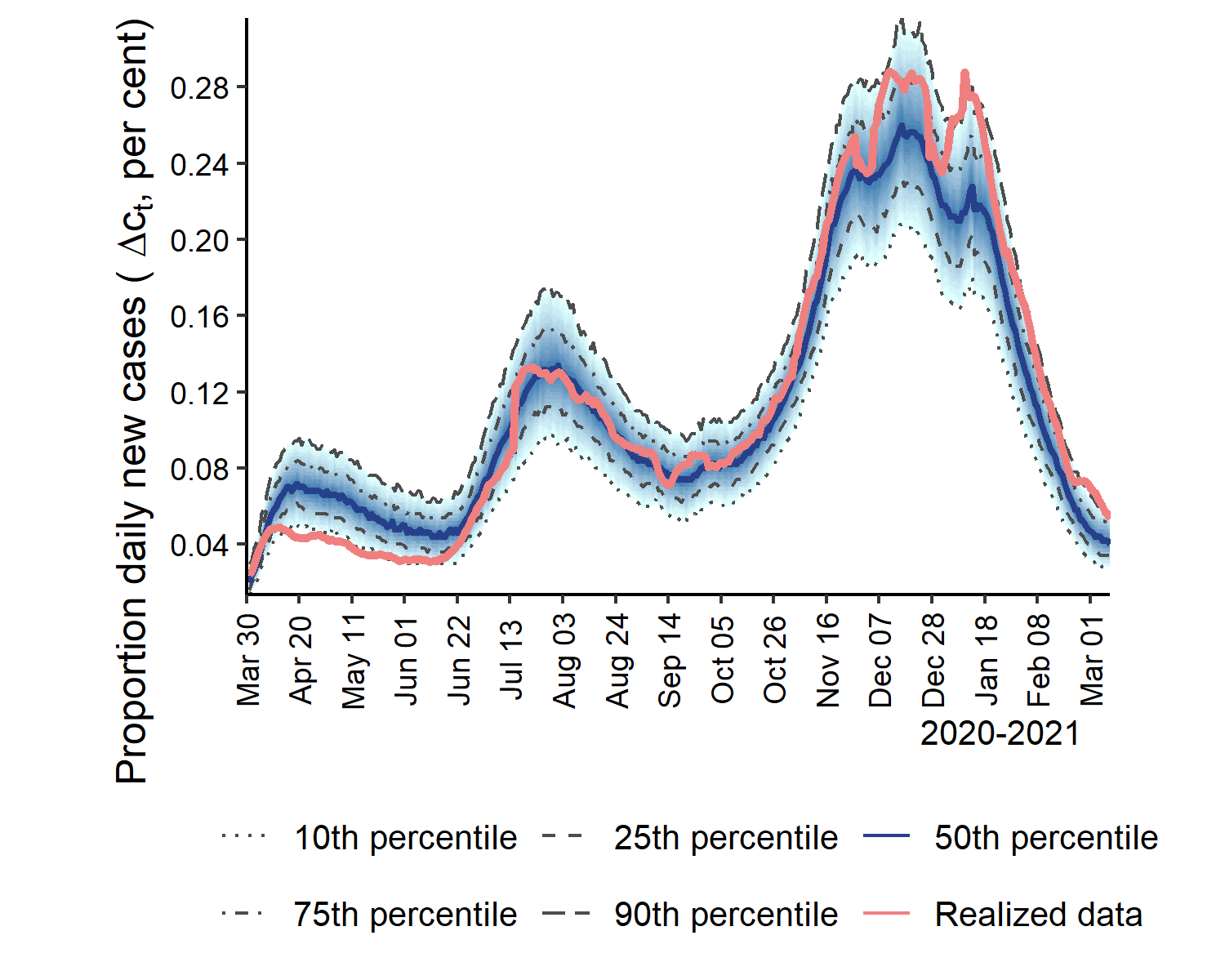}%
}
\end{tabular}

\end{center}

%

\vspace{0cm}%
\footnotesize
Notes: $\mathcal{\hat{R}}_{et}=\left(  1-\hat{m}_{t}\tilde{c}_{t}\right)
\hat{\beta}_{t}/\gamma$, where $\tilde{c}_{t}$ is the reported number of
infections per capita and $\gamma=1/14.$ $W_{\beta}=W_{m}=2$ weeks. The joint
estimation starts when $\tilde{c}_{t}>0.01$. The initial guess estimate of the
multiplication factor is $5$. The simulation uses the single group model with
population size $n=50,000$. The number of replications is $500$. The number of
removed (recoveries + deaths) is estimated recursively using $\tilde{R}%
_{t}=\left(  1-\gamma\right)  \tilde{R}_{t-1}+\gamma\tilde{C}_{t-1}$, with
$\tilde{C}_{1}=\tilde{R}_{1}=0$. Realized series (7-day moving average)
multiplied by the estimated multiplication factor is displayed in red.%

\end{footnotesize}%
%

\end{figure}

We next turn to the estimates of MF for the US. The left panel of Figure
\ref{fig: US_MF} presents the 2-weekly estimates of MF obtained by the joint
estimation method over the period of March 2020 to March 2021 (when the share
of the population fully vaccinated reached $10$ percent). The results show
that the estimated MF gradually declined from $7$ to $3$ from July 2020 to
March 2021. This finding is in line with an estimate of $7$ times
under-reporting by mid-July in the US based on antibody tests
\citep{Kalish2021undiagnosed}%
. The right panel of Figure \ref{fig: US_MF} shows the calibrated new cases
compared with the realized cases that have been multiplied by the estimated
MF. We can immediately observe that the calibrated cases match the several
waves of Covid in the US reasonably well.%

\section{Additional counterfactual exercises\label{Sup: counterfactual}}

\subsection{Social distancing and vaccination\label{Sup: vacc}}

\begin{figure}[tbh]%
\caption{Time profile of the transmission rate under social distancing}%
\label{fig: TR_dist}

\begin{center}%
\begin{tabular}
[c]{c}%
{\includegraphics[
height=1.9951in,
width=2.6524in
]%
{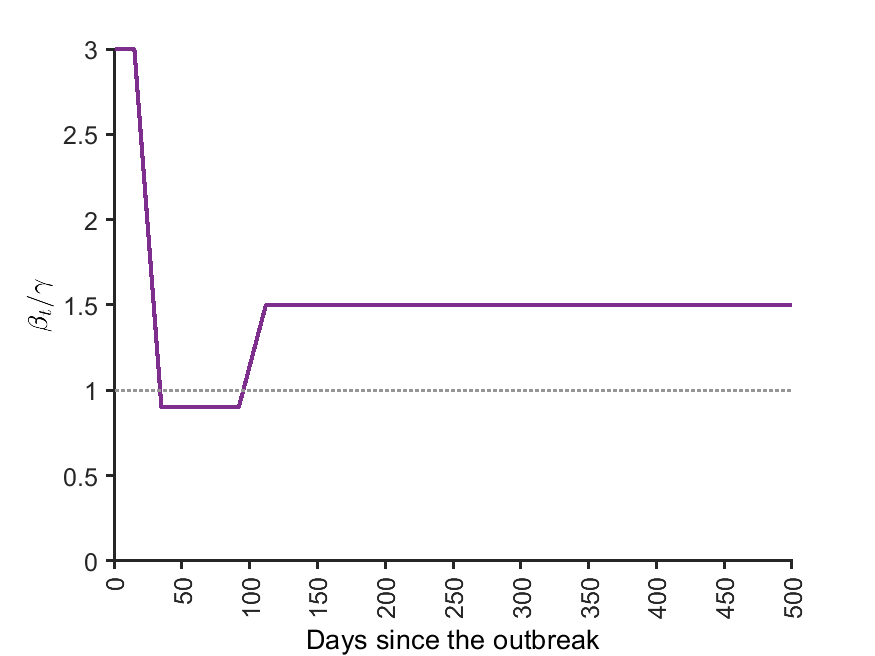}%
}
\end{tabular}

\end{center}

%

\end{figure}

This section presents the results of additional counterfactual experiments of
social distancing and vaccination. We consider the same social distancing
policy as described in the main paper: the (scaled) transmission rate,
$\beta_{t}/\gamma$, equals $3$ in the first two weeks, falls to $0.9$ linearly
over the next three weeks, and remains at $0.9$ for eight weeks. When social
distancing is relaxed, the transmission rate increases to $1.5$ linearly over
the next three weeks and remains at $1.5$ afterward. Figure \ref{fig: TR_dist}
displays the time profile of the transmission rate under this social
distancing policy.%

Figure \ref{fig: vacc_suppl} complements Figure \ref{fig: dist_vacc} of the
main paper and presents additional simulation outcomes under different
vaccination coverages, start times, speeds of delivery, and vaccine efficacies
under the random vaccination scheme. Specifically, Figure
\ref{fig: vacc_suppl}(a) displays the simulated new cases under $50$ and $75$
percent vaccination coverages, which are assumed to take $8$ and $12$ weeks,
respectively. The epidemic lasts much longer if the vaccine uptake is lower,
although the number of daily new cases and total cases end up very similar. In
both cases, the second epidemic wave is successfully prevented. Two important
reasons are that the vaccination started early enough (during the last month
of social distancing), and the vaccine is highly effective. If the vaccination
begins at the end of social distancing (as shown in Figure
\ref{fig: vacc_suppl}(b)), there will be a resurgence of cases, resulting in a
longer duration of the epidemic and a greater number of total cases. Figure
\ref{fig: vacc_suppl}(c) shows that if $75$ percent of the population gets
vaccinated over $8$ rather than $12$ weeks, the epidemic could end within
$200$ days, and there will not be any uptick in new cases when the social
distancing is relaxed. Finally, Figure \ref{fig: vacc_suppl}(d) shows that if
the vaccine has $66$ percent efficacy instead of $95$ percent, one would
expect to see a small second wave of cases, and the epidemic would last for
$65$ days longer.\footnote{The Pfizer-BioNTech, Moderna, and Johnson \&
Johnson vaccines reported efficacy rates of 95\%, 94.1\%, and 66.3\%,
respectively, in preventing symptomatic Covid-19 infection (Oliver et al.,
\citeyear{Pfizer2020efficacy}%
,
\citeyear{Moderna2021efficacy}%
, and
\citeyear{Janssen2021efficacy}%
).}%

\begin{figure}[tp]%
\caption
{Simulated average number of new cases for different random vaccination experiments, with the same social distancing policy}%
\label{fig: vacc_suppl}%
\begin{footnotesize}%

\begin{center}%
\begin{tabular}
[c]{c}%
(a) Comparing different vaccination coverages%
\vspace{0.2cm}%
\\%
{\includegraphics[
height=1.9951in,
width=2.6524in
]%
{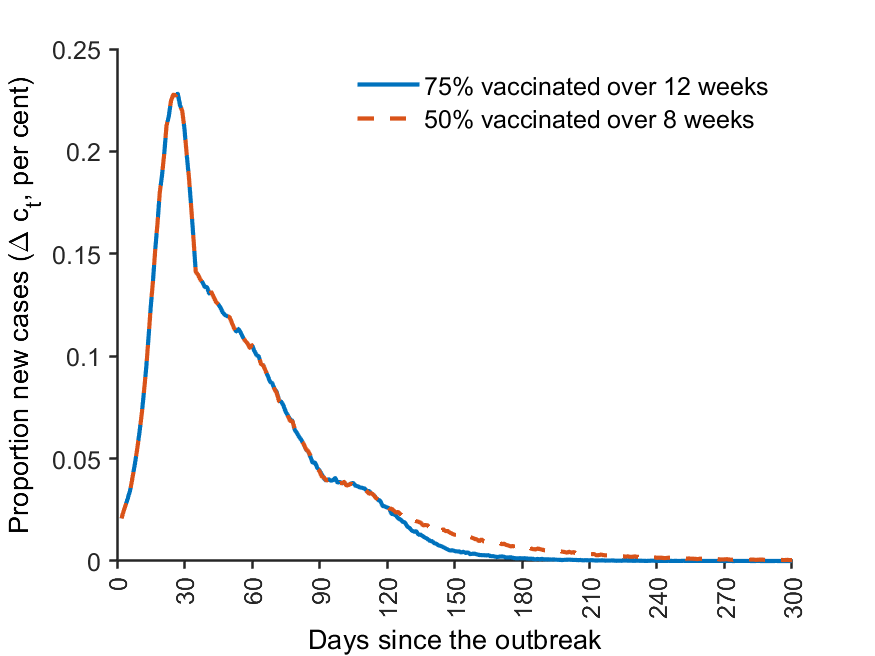}%
}
\end{tabular}

\end{center}

%

\vspace{-0.2cm}%
Notes: The average number of new cases over $B=1,000$ replications is
displayed. Population size is $n=10,000$. The time profile of $\beta
_{t}/\gamma$ under social distancing is displayed in Figure \ref{fig: TR_dist}%
. The vaccination starts during the last month of social distancing (i.e., the
$10^{th}$ week after the outbreak). The vaccine efficacy is $\epsilon
_{v}=0.95$. If $75$ percent of the population is randomly vaccinated over $12$
weeks, $c^{\ast}=B^{-1}\sum_{b=1}^{B}\max_{t}c_{t}^{(b)}=0.12$, and the
duration of the epidemic is $T^{\ast}=215$ days. If $50$ percent of the
population is randomly vaccinated over $8$ weeks, $c^{\ast}=0.12$, and
$T^{\ast}=270$ days.%

\vspace{0.3cm}%

\begin{center}%
\begin{tabular}
[c]{c}%
(b) Comparing different vaccination start times%
\vspace{0.2cm}%
\\%
{\includegraphics[
height=1.9951in,
width=2.6524in
]%
{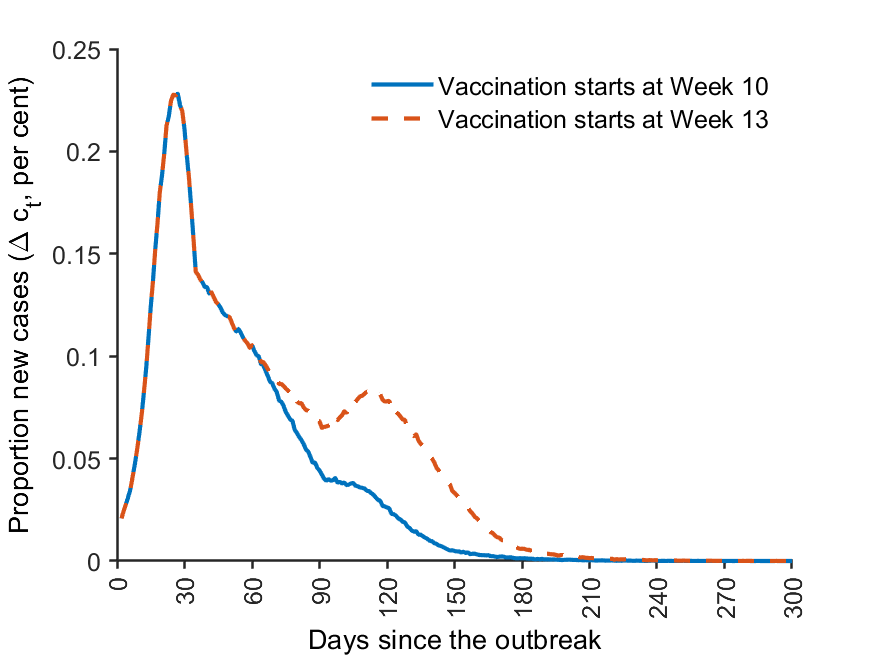}%
}
\end{tabular}

\end{center}

%

\vspace{-0.2cm}%
Notes: The average number of new cases over $B=1,000$ replications is
displayed. Population size is $n=10,000$. The time profile of $\beta
_{t}/\gamma$ under social distancing is displayed in Figure \ref{fig: TR_dist}%
. The vaccine efficacy is $\epsilon_{v}=0.95$. $75$ percent of the population
is randomly vaccinated over $12$ weeks. If the vaccination starts during the
last month of social distancing (i.e., the $10^{th}$ week after the outbreak),
$c^{\ast}=B^{-1}\sum_{b=1}^{B}\max_{t}c_{t}^{(b)}=0.12$, and the duration of
the epidemic is $T^{\ast}=215$ days. If the vaccination starts at the end of
social distancing (i.e., the $13^{th}$ week after the outbreak), $c^{\ast
}=0.15$, and $T^{\ast}=248$ days.%

\end{footnotesize}%
%

\end{figure}%
%

\addtocounter{figure}{-1}%
%

\begin{figure}[tp]%
\caption
{(Continued) Simulated average number of new cases for different random vaccination experiments, with the same social distancing policy}%
\begin{footnotesize}%

\begin{center}

\begin{tabular}
[c]{c}%
(c) Comparing different vaccination speeds%
\vspace{0.2cm}%
\\%
{\includegraphics[
height=1.9951in,
width=2.6524in
]%
{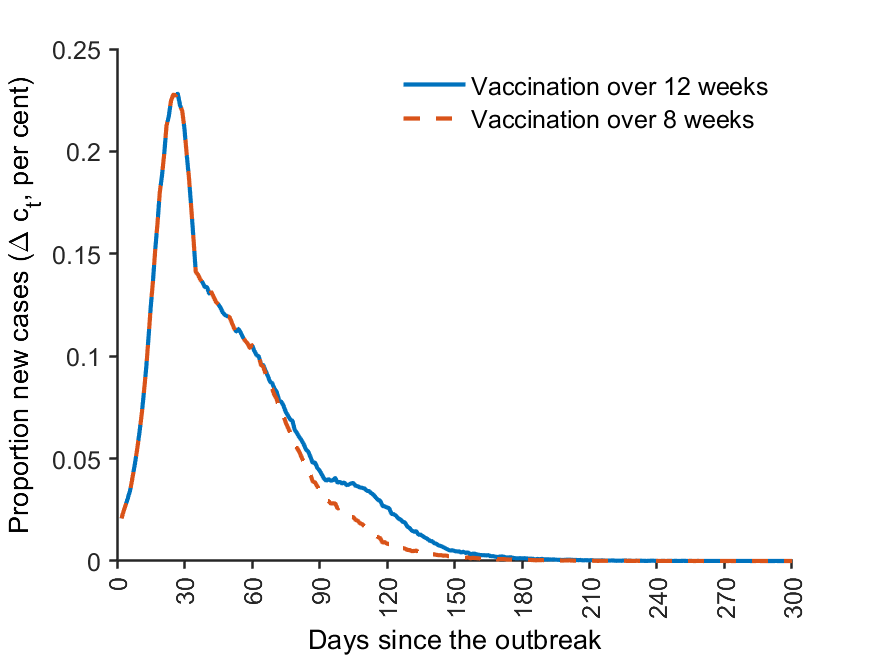}%
}
\end{tabular}

\end{center}

%

\vspace{-0.2cm}%
Notes: The average number of new cases over $B=1,000$ replications is
displayed. Population size is $n=10,000$. The time profile of $\beta
_{t}/\gamma$ under social distancing is displayed in Figure \ref{fig: TR_dist}%
. The vaccination starts during the last month of social distancing (i.e., the
$10^{th}$ week after the outbreak). The vaccine efficacy is $\epsilon
_{v}=0.95$. $75$ percent of the population is randomly vaccinated. If the
vaccination is administered over $12$ weeks, $c^{\ast}=B^{-1}\sum_{b=1}%
^{B}\max_{t}c_{t}^{(b)}=0.12$, and the duration of the epidemic is $T^{\ast
}=215$ days. If the vaccination is administered over $8$ weeks, $c^{\ast
}=0.11$, and $T^{\ast}=197$ days.%

\vspace{0.3cm}%

\begin{center}

\begin{tabular}
[c]{c}%
(d) Comparing different vaccine efficacies%
\vspace{0.2cm}%
\\%
{\includegraphics[
height=1.9951in,
width=2.6524in
]%
{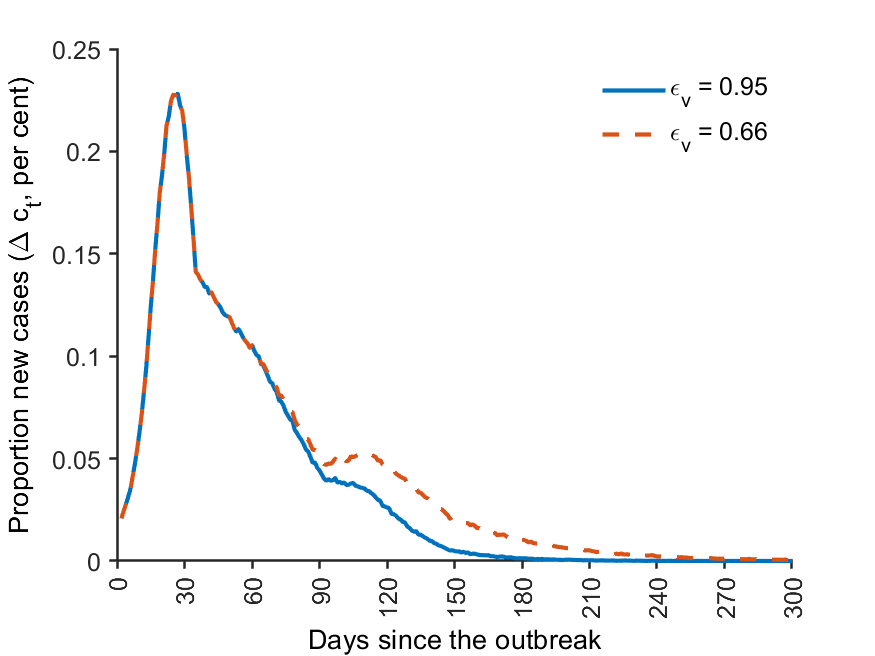}%
}
\end{tabular}

\end{center}

%

\vspace{-0.2cm}%
Notes: The average number of new cases over $B=1,000$ replications is
displayed. Population size is $n=10,000$. The time profile of $\beta
_{t}/\gamma$ under social distancing is displayed in Figure \ref{fig: TR_dist}%
. The vaccination starts during the last month of social distancing (i.e., the
$10^{th}$ week after the outbreak). $75$ percent of the population is randomly
vaccinated over $12$ weeks. If the vaccine efficacy is $\epsilon_{v}=0.95$,
$c^{\ast}=B^{-1}\sum_{b=1}^{B}\max_{t}c_{t}^{(b)}=0.12$, and the duration of
the epidemic is $T^{\ast}=215$ days. If $\epsilon_{v}=0.66$, then $c^{\ast
}=0.13$, and $T^{\ast}=280$ days.%

\end{footnotesize}%
%

\end{figure}%
%

\begin{figure}[hp]%
\caption
{Simulated average number of group-specific and aggregate new cases, assuming social distancing combined with random vaccination or vaccination in decreasing age order}%
\label{fig: vacc_priority_pct50}%
\vspace{-0.2cm}%
\footnotesize

\begin{center}%
\begin{tabular}
[c]{ccc}%
\textbf{Group 1: [0, 15)} &  & \textbf{Group 2: [15, 30)}\\%
{\includegraphics[
height=1.7772in,
width=2.3618in
]%
{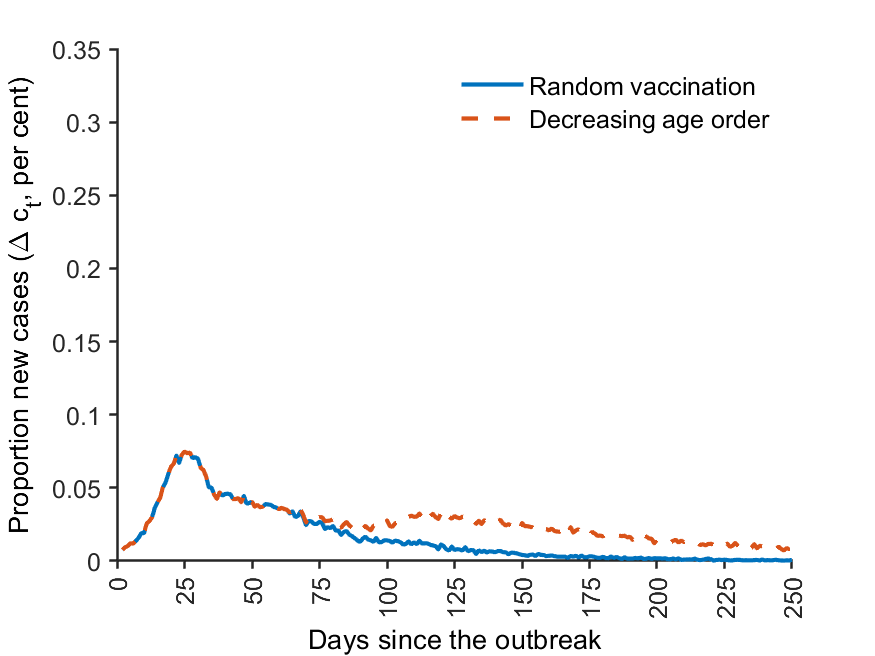}%
}
&  &
{\includegraphics[
height=1.7772in,
width=2.3618in
]%
{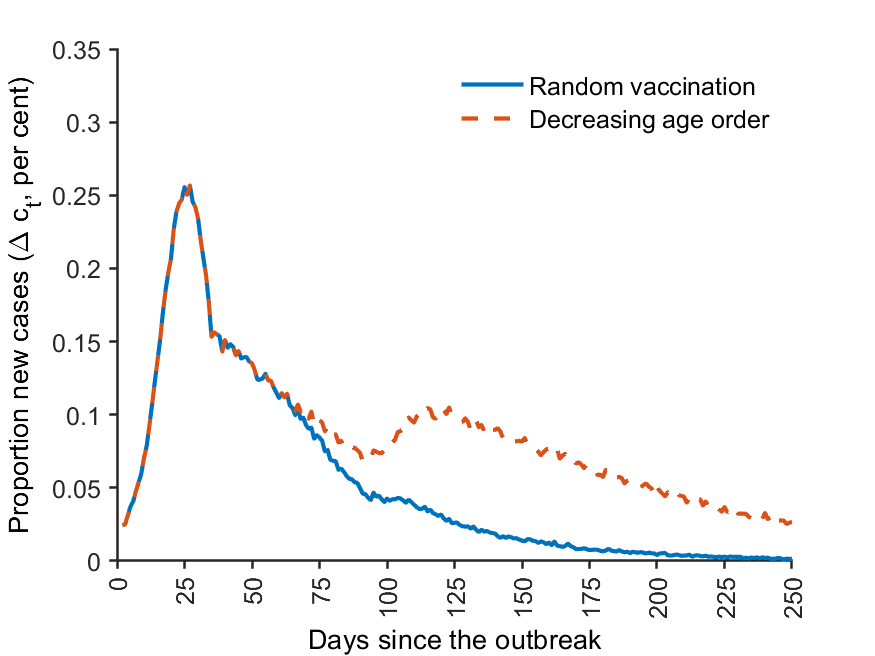}%
}
\\
Random: $c_{1}^{\ast}=0.04$ &  & Random: $c_{2}^{\ast}=0.14$\\
By age: \ $\ c_{1}^{\ast}=0.07$ &  & By age: \ $\ c_{2}^{\ast}=0.24$\\
&  & \\
\textbf{Group 3: [30, 50)} &  & \textbf{Group 4: [50, 65)}\\%
{\includegraphics[
height=1.7772in,
width=2.3618in
]%
{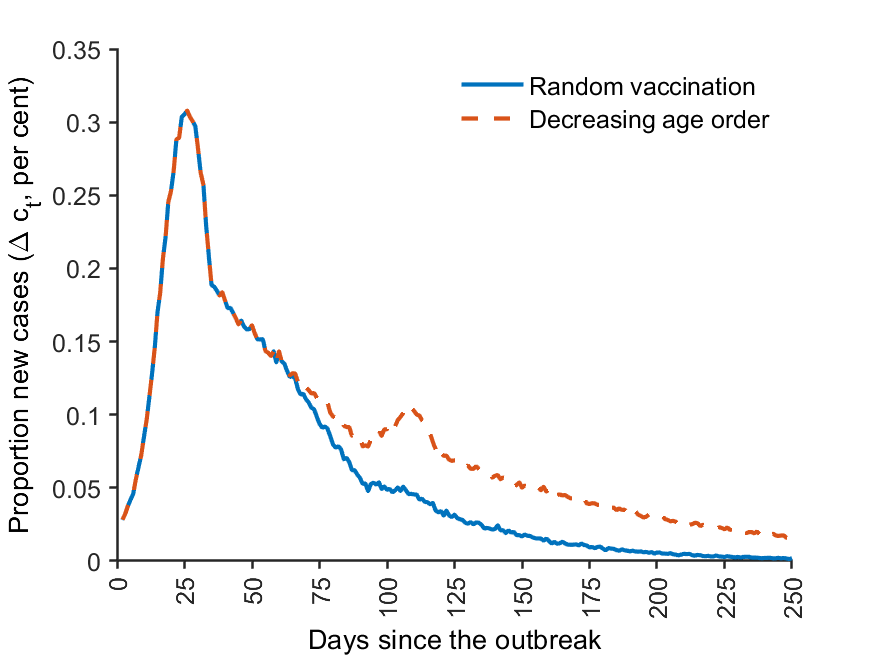}%
}
&  &
{\includegraphics[
height=1.7772in,
width=2.3618in
]%
{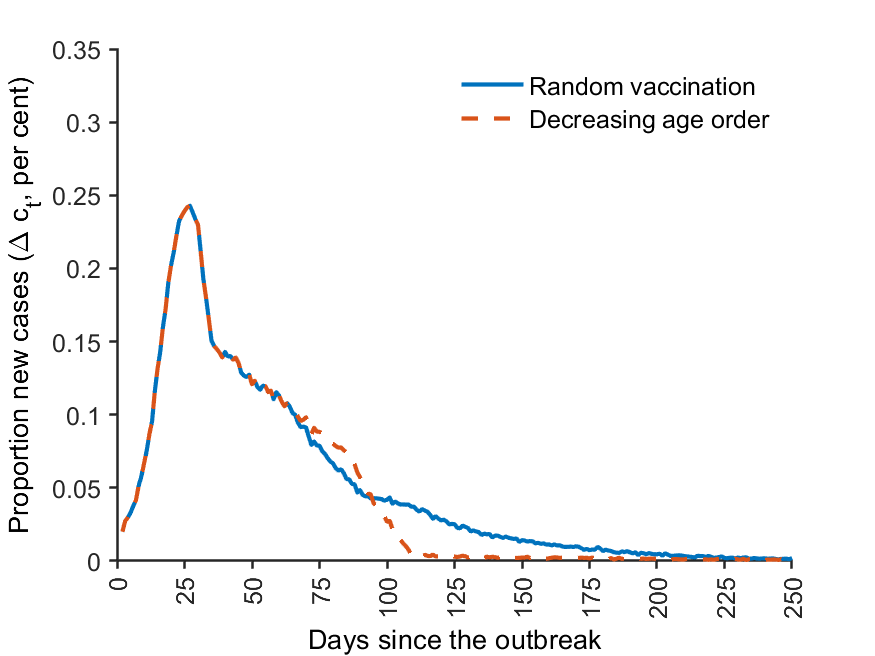}%
}
\\
Random: $c_{3}^{\ast}=0.16$ &  & Random: $c_{4}^{\ast}=0.13$\\
By age: \ $\ c_{3}^{\ast}=0.23$ &  & By age: \ $\ c_{4}^{\ast}=0.12$\\
&  & \\
&  & \\
\textbf{Group 5: 65+} &  & \textbf{Aggregate}\\%
{\includegraphics[
height=1.7772in,
width=2.3618in
]%
{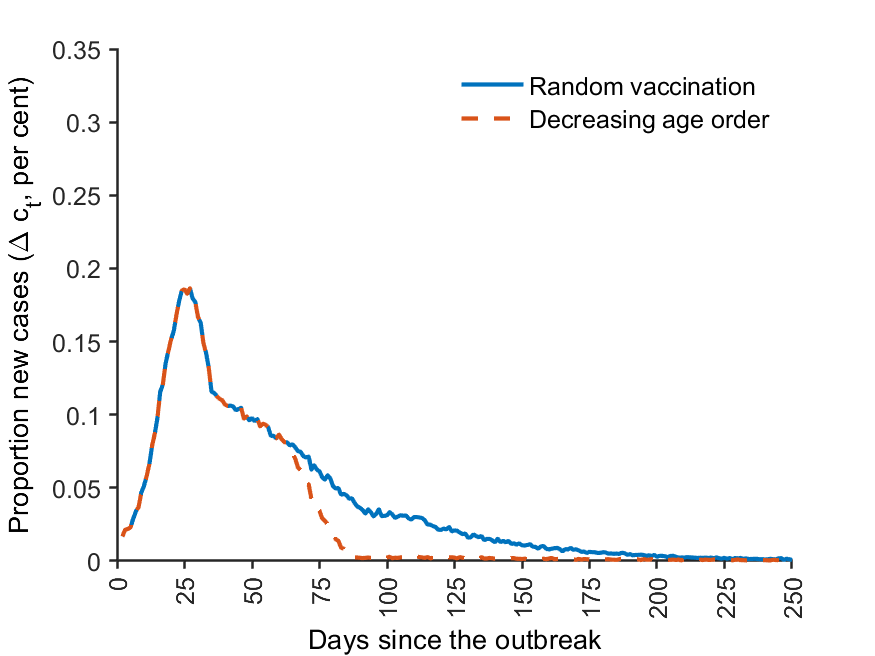}%
}
&  &
{\includegraphics[
height=1.7772in,
width=2.3618in
]%
{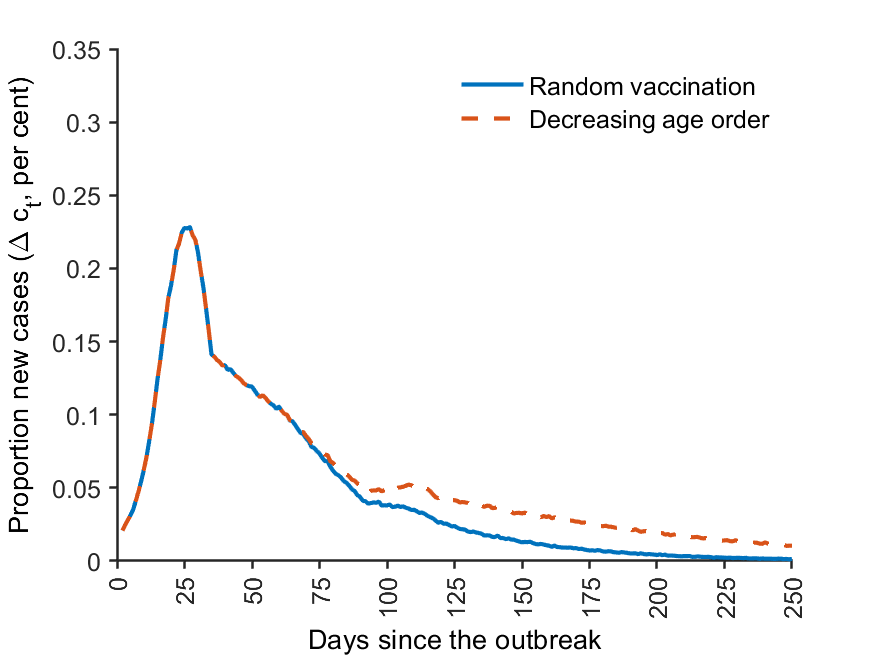}%
}
\\
Random: $c_{5}^{\ast}=0.10$ &  & Random: $c^{\ast}=0.12$\\
By age: \ $\ c_{5}^{\ast}=0.08$ &  & By age: \ $\ c^{\ast}=0.15$\\
&  &
\end{tabular}

\end{center}

%

\vspace{-0.2cm}%
Notes: The average number of new cases over $B=1,000$ replications is
displayed. Population size is $n=10,000$. The time profile of $\beta
_{t}/\gamma$ under social distancing is displayed in Figure \ref{fig: TR_dist}%
. The vaccination starts during the last month of social distancing (i.e., the
$10^{th}$ week after the outbreak). $50$ percent of the population is
vaccinated over eight weeks. The vaccine efficacy is $\epsilon_{v}=0.95.$ The
duration of the epidemic is $T^{\ast}=270$ days under random vaccination, and
$T^{\ast}=380$ days under vaccination by decreasing age order$.$ $c_{\ell
}^{\ast}=B^{-1}\sum_{b=1}^{B}\max_{t}c_{\ell t}^{(b)}$, for $\ell
=1,2,\ldots,5$, and $c^{\ast}=B^{-1}\sum_{b=1}^{B}\max_{t}c_{t}^{(b)}.$%

\end{figure}%

Figure \ref{fig: vacc_priority_pct50} compares the simulated group-specific
and aggregate outcomes under random vaccination and vaccination in decreasing
age order. It is assumed that $50$ percent of the population is vaccinated
over $8$ weeks, as opposed to $75$ percent vaccinated over $12$ weeks as
considered in the main paper. Comparing Figure \ref{fig: vacc_priority_pct50}
with Figure \ref{fig: vacc_priority} of the main paper reveals that if the
vaccine coverage is lower, prioritizing the elderly would lead to a much
higher level of infections for the younger age groups. The proportions of
infected in Groups 2 and 3 could reach $24$ and $23$ percent, respectively.
The age-based vaccination would also substantially increase the duration of
the epidemic from $233$ to $380$ days if the vaccine coverage decreases from
$75$ to $50$ percent. By contrast, the random vaccination strategy would
increase the duration from $215$ to $270$ days.

\subsection{Counterfactual outcomes of early interventions in UK and Germany}

\begin{figure}[tbh]%
\caption
{Counterfactual number of infected and active cases for Germany and UK under different lockdown scenarios}%
\vspace{-0.2cm}%
\label{fig: Germany_UK_2w}

\begin{center}%
\hspace*{-0.6cm}%
\begin{tabular}
[c]{ccc}%
\multicolumn{3}{c}{%
\begin{footnotesize}%
\textbf{What if the German lockdown was delayed two weeks?}%
\end{footnotesize}%
\vspace{0.2cm}%
}\\%
\hspace*{-0.2cm}%
{\footnotesize Infected cases} & {\footnotesize Active cases} &
\hspace*{-0.8cm}%
\begin{footnotesize}%
$\mathcal{\hat{R}}_{et}$%
\end{footnotesize}%
\\%
{\includegraphics[
height=1.8273in,
width=2.8392in
]%
{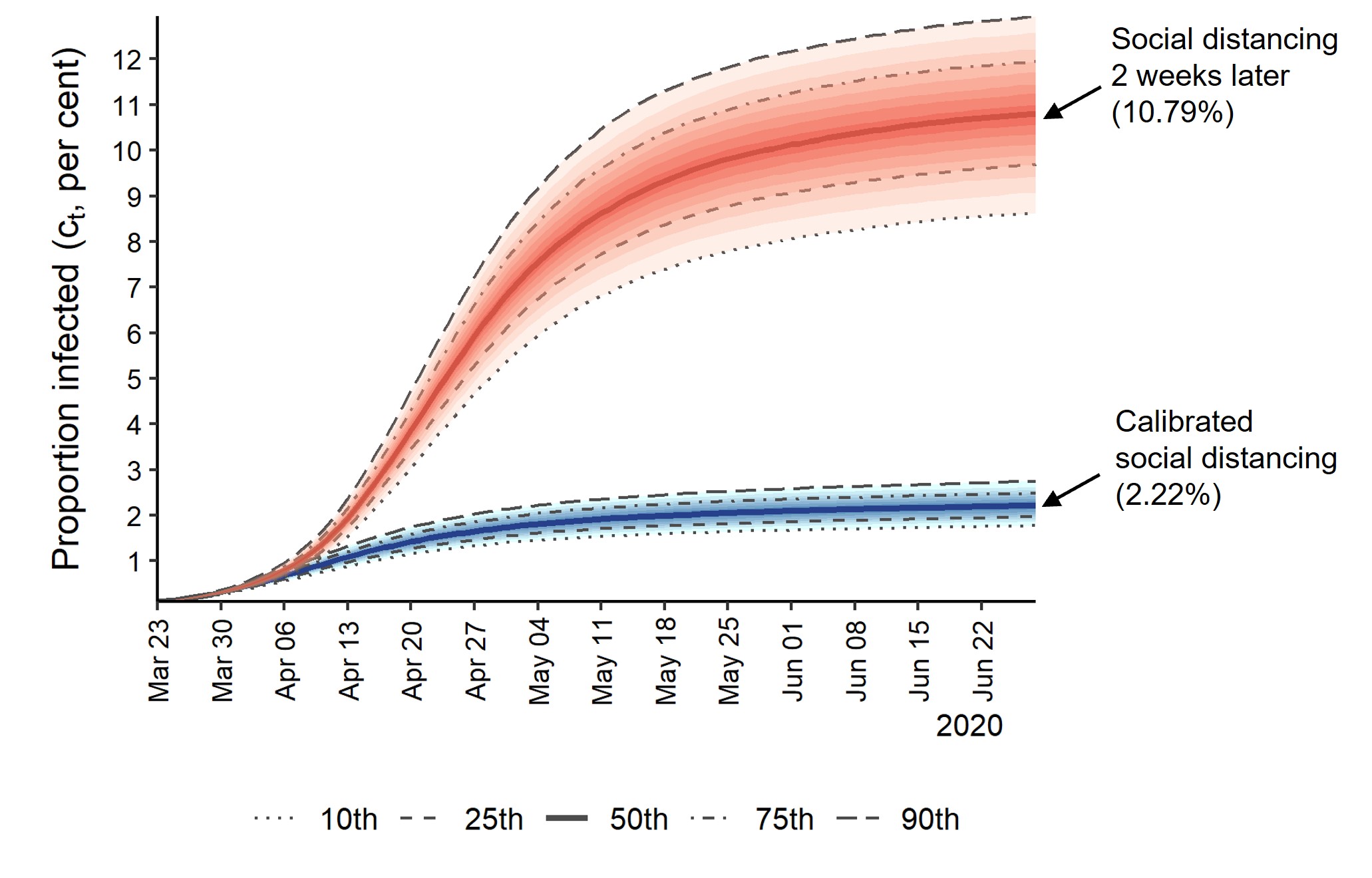}%
}
&
{\includegraphics[
height=1.8542in,
width=2.284in
]%
{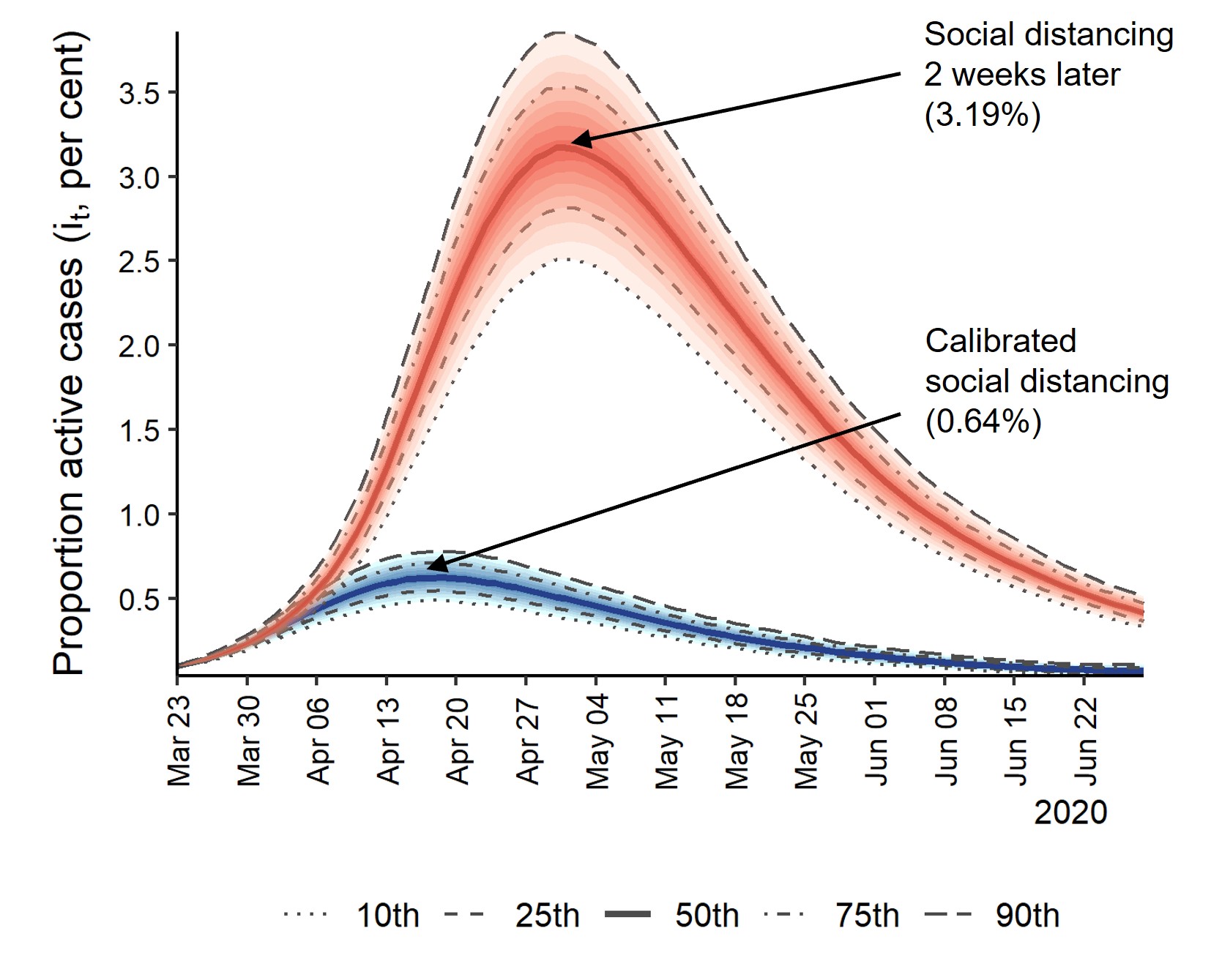}%
}
&
\hspace*{-0.5cm}%
{\includegraphics[
trim=0.000000in -0.660383in 0.000000in 0.000000in,
height=1.9398in,
width=2.2451in
]%
{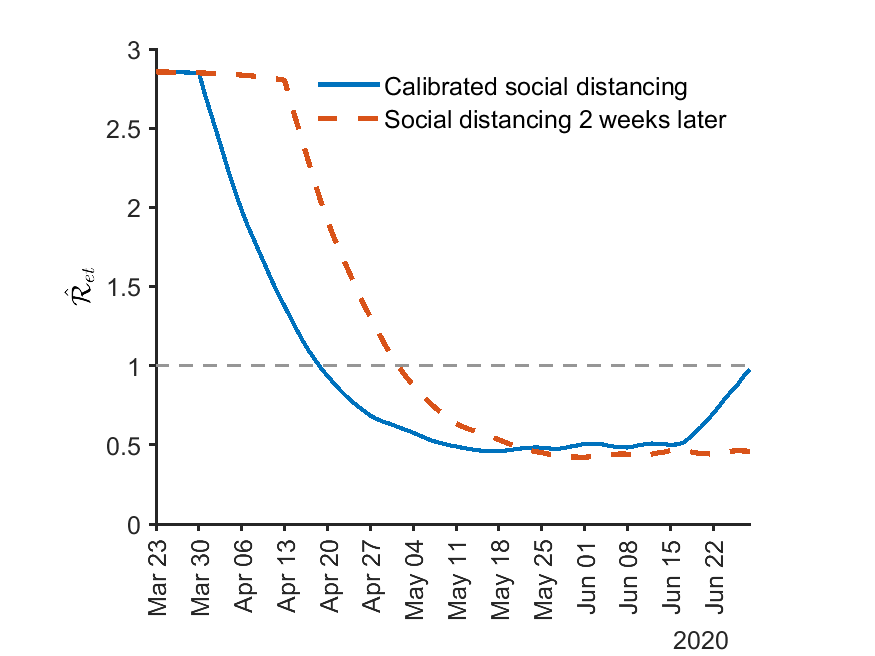}%
}
\\
&  & \\
\multicolumn{3}{c}{%
\begin{footnotesize}%
\textbf{What if the UK lockdown was brought forward two weeks?}%
\end{footnotesize}%
\vspace{0.2cm}%
}\\%
\hspace*{-0.2cm}%
{\footnotesize Infected cases} & {\footnotesize Active cases} &
\hspace*{-0.8cm}%
\begin{footnotesize}%
$\mathcal{\hat{R}}_{et}$%
\end{footnotesize}%
\\%
{\includegraphics[
height=1.8273in,
width=2.8288in
]%
{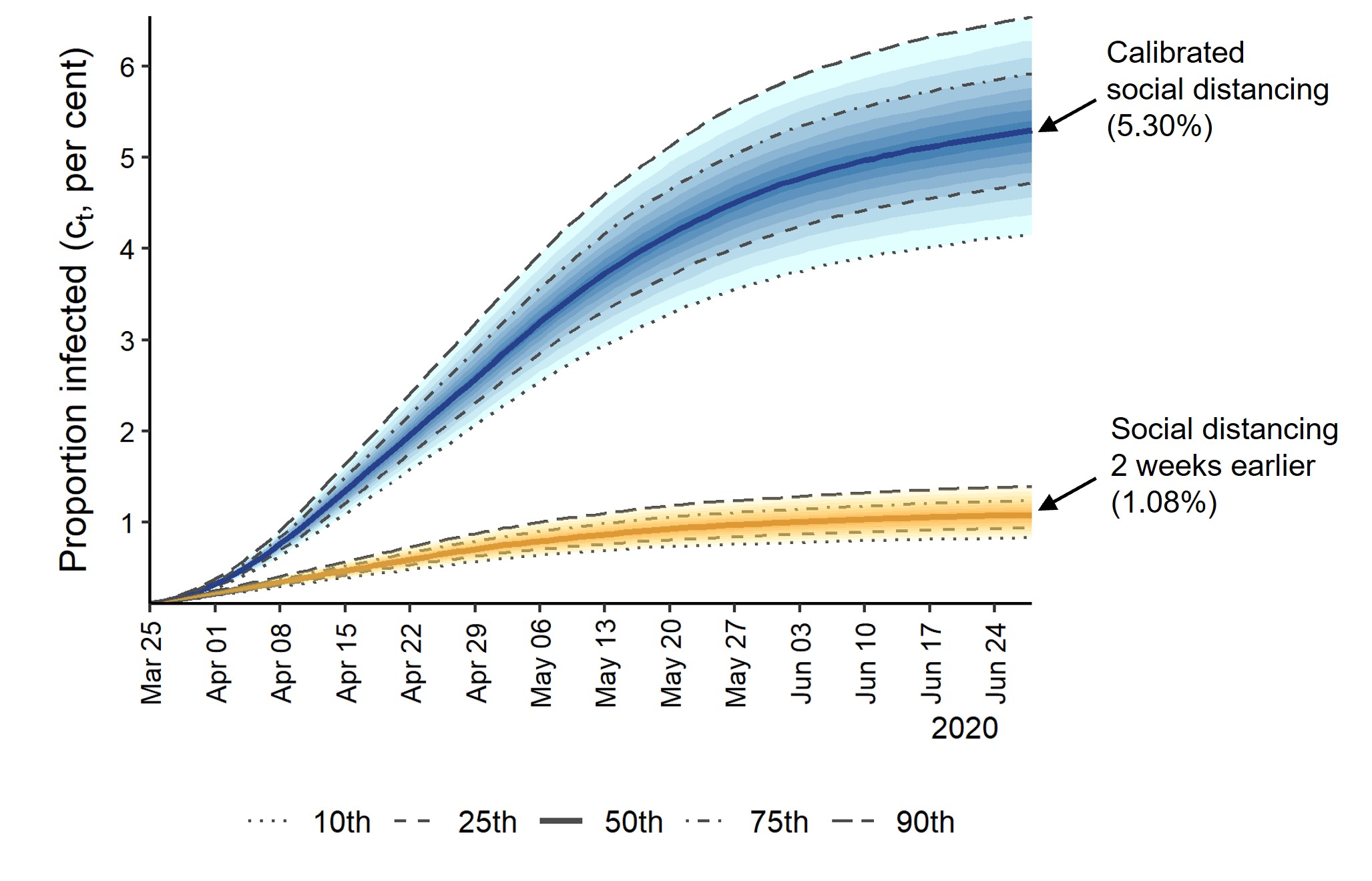}%
}
&
{\includegraphics[
height=1.8403in,
width=2.322in
]%
{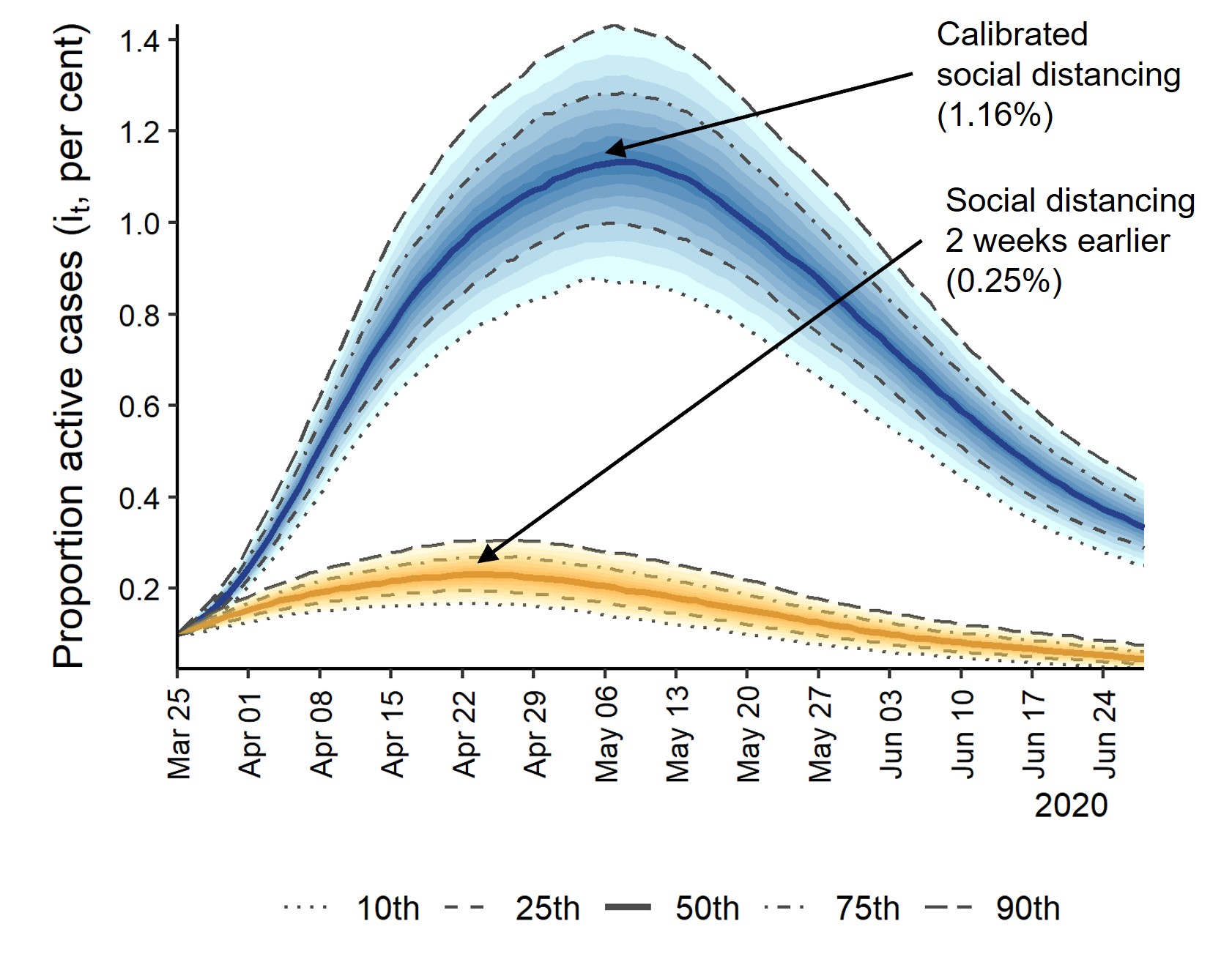}%
}
&
\hspace*{-0.5cm}%
{\includegraphics[
trim=0.000000in -0.660383in 0.000000in 0.000000in,
height=1.9398in,
width=2.2451in
]%
{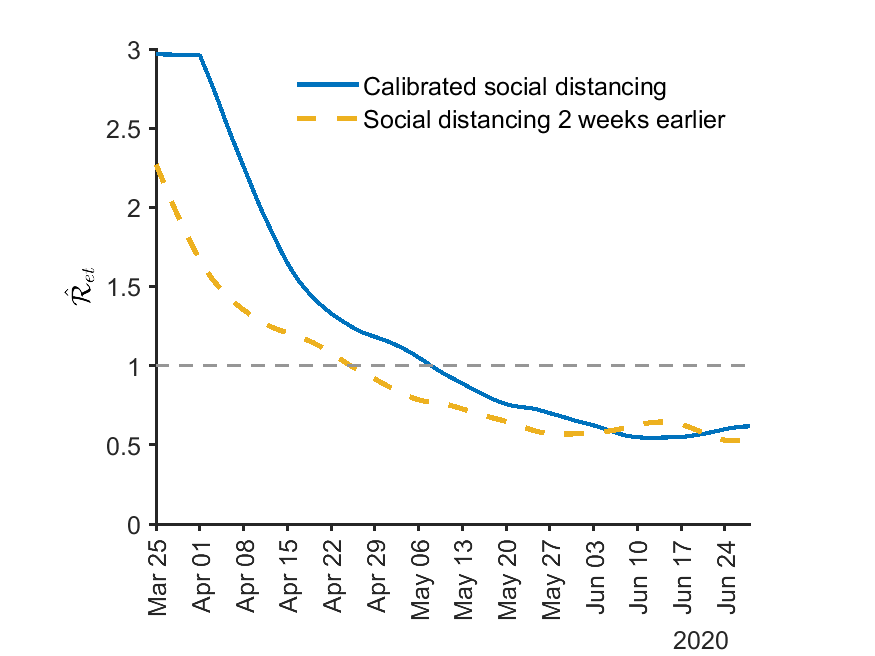}%
}
\end{tabular}

\end{center}

%

\vspace{-0.3cm}%
\footnotesize
{}Notes: The simulation uses the single group model with the
Erd\H{o}s-R\'{e}nyi random network and begins with $1/1000$ of the population
randomly infected on day $1$. The population size used in the simulation is
$n=50,000$. The recover rate is $\gamma=1/14$. The number of removed
(recoveries + deaths) is estimated recursively using $\tilde{R}_{t}=\left(
1-\gamma\right)  \tilde{R}_{t-1}+\gamma\tilde{C}_{t-1}$ for both countries,
with $\tilde{C}_{1}=\tilde{R}_{1}=0$, where $\tilde{C}_{t}$ is the reported
number of infections. $\hat{\beta}_{t}$ is the $2$-weekly rolling estimate
computed by (\ref{betathat}) assuming MF $=5$. The mean of $\mathcal{\hat{R}%
}_{et}^{(b)}=\left(  1-c_{t}^{(b)}\right)  \hat{\beta}_{t}/\gamma$, for
$b=1,2,\ldots,1000$ replications, is displayed in the last column.%

\end{figure}

To complement Figure \ref{fig: Germany_UK_1w} presented in the main paper
assuming that the German (UK) lockdown had been delayed (brought forward) one
week, we further examine the potential outcomes if the lockdown had been
delayed or advanced two weeks. As shown in Figure \ref{fig: Germany_UK_2w}, if
the German lockdown had been delayed one week, there would have been a
whopping five-fold increase in both infected and active cases. By contrast, if
the UK lockdown had been implemented two weeks earlier, both infected and
active cases could have been one-fifth of the realized level. These results
further highlight the importance of taking mitigation actions early in an
epidemic outbreak.%

\section{Data Sources \label{Sup: data}}

This section provides sources of all the data used in our study. For the
multigroup model, the latest population estimates by age for Germany are
sourced from the database of the Federal Statistical Office of Germany at
\url{https://www-genesis.destatis.de/genesis/online}. The large-scale social
contact surveys by \cite{Mossong2008} provides detailed information on the
contact patterns in Germany, and the age-specific contact matrix can be
conveniently constructed using the Social Contact Rates (SOCRATES) Data Tool
by \cite{Willem2020socrates} available at
\url{https://lwillem.shinyapps.io/socrates_rshiny/}. The data on Germany's
Covid-19 cases by age group are retrieved from the website of the Robert Koch
Institute at \url{http://www.rki.de/covid-19-altersverteilung}.

In matching the model with empirical evidence, the primary data source for the
Covid-19 cases is the repository by the Center for Systems Science and
Engineering (CSSE) at Johns Hopkins University available at
\url{https://github.com/CSSEGISandData/COVID-19}. The Covid-19 cases for each
state in the US were aggregated from the county-level data, also available at
the CSSE's repository. Since the CSSE data for France and Spain contain
negative new cases at the time of our access, for these two countries we used
the data compiled by the World Health Organization available at
\url{https://covid19.who.int/WHO-COVID-19-global-data.csv}. The population
data (for year 2019) are obtained from the World Bank database at
\url{https://data.worldbank.org/indicator/SP.POP.TOTL}. The lockdown dates
across countries can be found at \url{https://en.wikipedia.org/wiki/COVID-19_pandemic_lockdowns}.

The efficacy rates of the Pfizer-BioNTech, Moderna, and Johnson \& Johnson
vaccines are reported in the CDC's Morbidity and Mortality Weekly Reports by
\cite{Pfizer2020efficacy}, \cite{Moderna2021efficacy}, and
\cite{Janssen2021efficacy}, respectively. The shares of people fully
vaccinated are sourced from the Our World in Data Covid vaccination dataset at \url{https://github.com/owid/covid-19-data/tree/master/public/data/vaccinations/}.%

\FloatBarrier
%

\begin{singlespace}%
%

\footnotesize
\bibliographystyle{chicago}
\bibliography{epidemic}
%

\end{singlespace}%
\pagebreak
\end{document}